
\voffset=-1.5truecm\hsize=16.5truecm    \vsize=24.truecm
\baselineskip=14pt plus0.1pt minus0.1pt \parindent=12pt

\lineskip=4pt\lineskiplimit=0.1pt      \parskip=0.1pt plus1pt 
\def\st{\scriptstyle} 
 
 
\input blackdvi   

 
\let\a=\alpha \let\b=\beta  \let\d=\delta \let\e=\varepsilon 
\let\f=\varphi \let\g=\gamma \let\h=\eta    \let\k=\kappa \let\l=\lambda 
\let\m=\mu \let\n=\nu       
\let\r=\rho \let\s=\sigma   
\let\y=\upsilon \let\x=\xi \let\z=\zeta 
\let\D=\Delta  \let\G=\Gamma \let\L=\Lambda

 
\global\newcount\numsec\global\newcount\numfor 
\gdef\profonditastruttura{\dp\strutbox} 
\def\senondefinito#1{\expandafter\ifx\csname#1\endcsname\relax} 
\def\SIA #1,#2,#3 {\senondefinito{#1#2} 
\expandafter\xdef\csname #1#2\endcsname{#3} \else 
\write16{???? il simbolo #2 e' gia' stato definito !!!!} \fi} 
\def\etichetta(#1){(\veroparagrafo.\veraformula) 
\SIA e,#1,(\veroparagrafo.\veraformula) 
 \global\advance\numfor by 1 
 \write16{ EQ \equ(#1) ha simbolo #1 }} 
\def\etichettaa(#1){(A\veroparagrafo.\veraformula) 
 \SIA e,#1,(A\veroparagrafo.\veraformula) 
 \global\advance\numfor by 1\write16{ EQ \equ(#1) ha simbolo #1 }} 
\def\BOZZA{\def\alato(##1){ 
 {\vtop to \profonditastruttura{\baselineskip 
 \profonditastruttura\vss 
 \rlap{\kern-\hsize\kern-1.2truecm{$\scriptstyle##1$}}}}}} 
\def\alato(#1){} 
\def\veroparagrafo{\number\numsec}\def\veraformula{\number\numfor} 
\def\Eq(#1){\eqno{\etichetta(#1)\alato(#1)}} 
\def\eq(#1){\etichetta(#1)\alato(#1)} 
\def\Eqa(#1){\eqno{\etichettaa(#1)\alato(#1)}} 
\def\eqa(#1){\etichettaa(#1)\alato(#1)} 
\def\equ(#1){\senondefinito{e#1}$\clubsuit$#1\else\csname e#1\endcsname\fi}

 
\def\bb{\hbox{\vrule height0.4pt width0.4pt depth0.pt}}\newdimen\u 
\def\pp #1 #2 {\rlap{\kern#1\u\raise#2\u\bb}} 
 
\def\ins #1 #2 #3 {\rlap{\kern#1\u\raise#2\u\hbox{$#3$}}}

\def\pallina{{\kern-0.4mm\raise-0.02cm\hbox{$\scriptscriptstyle\bullet$}}} 
\def\palla{{\kern-0.6mm\raise-0.04cm\hbox{$\scriptstyle\bullet$}}} 
\def\pallona{{\kern-0.7mm\raise-0.06cm\hbox{$\displaystyle\bullet$}}} 
\def\bull{\vrule height .9ex width .8ex depth -.1ex } 
 
\def\data{\number\day/\ifcase\month\or gennaio \or febbraio \or marzo \or 
aprile \or maggio \or giugno \or luglio \or agosto \or settembre 
\or ottobre \or novembre \or dicembre \fi/\number\year} 
 
\setbox200\hbox{$\scriptscriptstyle \data $} 
 
\newcount\pgn \pgn=1 
\def\foglio{\number\numsec:\number\pgn 
\global\advance\pgn by 1} 
\def\foglioa{a\number\numsec:\number\pgn 
\global\advance\pgn by 1} 
 
\footline={\rlap{\hbox{\copy200}\ $\st[\number\pageno]$}\hss\tenrm \foglio\hss} 
 
 
\def\sqr#1#2{{\vcenter{\vbox{\hrule height.#2pt 
\hbox{\vrule width.#2pt height#1pt \kern#1pt 
\vrule width.#2pt}\hrule height.#2pt}}}}

\let\ciao=\bye  
 \def\\{\noindent}

  \let\io=\infty  
 
\let\dpr=\partial

\def\SS{{\cal S}}  
 \def\AA{{\cal A}}

\def\tende#1{\vtop{\ialign{##\crcr\rightarrowfill\crcr 
              \noalign{\kern-1pt\nointerlineskip} 
              \hskip3.pt${\scriptstyle #1}$\hskip3.pt\crcr}}} 
\def\otto{{\kern-1.truept\leftarrow\kern-5.truept\to\kern-1.truept}} 
 
\font\smfnt=cmr8 scaled\magstep0 
\font\myfonta=msbm10 scaled \magstep0 
\def\math#1{\hbox{\myfonta #1}}


\vglue.5truecm 
{\centerline{\bf THE GLOBAL RENORMALIZATION GROUP TRAJECTORY IN A CRITICAL} }
{\centerline{\bf SUPERSYMMETRIC FIELD THEORY ON THE LATTICE 
$\math{Z}^{3}$}}
\vglue1.truecm 
{\centerline{{\bf P. K. Mitter}$^{1}$, {\bf B. Scoppola}$^{2}$}} 
\vglue.5truecm 
{\centerline{\smfnt $^1$Laboratoire de Physique Theorique et Astroparticules, 
CNRS-IN2P3-Universit\'e Montpellier 2}} 
{\centerline{\smfnt Place E. Bataillon, Case 070,  
34095 Montpellier Cedex 05 France}} 
{\centerline{\smfnt e-mail: pkmitter@LPTA.univ-montp2.fr}} 
\vglue.5truecm 
{\centerline{\smfnt $^2$Dipartimento di Matematica,  
Universit\'a ``Tor Vergata'' di Roma}} 
{\centerline{\smfnt Via della Ricerca Scientifica, 00133 Roma, Italy}} 
{\centerline{\smfnt e-mail: scoppola@mat.uniroma2.it}} 
\vglue1cm

\\{\bf Abstract}: We consider an Euclidean supersymmetric field theory
in $\math{Z}^{3}$ given by a supersymmetric $\Phi^{4}$ perturbation of an
underlying massless Gaussian measure on scalar bosonic and Grassmann fields 
with covariance the Green's function of
a (stable) L\'evy random walk in  $\math{Z}^{3}$. The Green's function 
depends on the L\'evy-Khintchine parameter $\a={3+\e\over 2}$ with 
$0<\a<2$. For $\a ={3\over 2}$ the $\Phi^{4}$ interaction is marginal. 
We prove for $\a-{3\over 2}={\e\over 2}>0$ sufficiently small and 
initial parameters held in an appropriate domain the existence of a global
renormalization group trajectory uniformly bounded on all renormalization group
scales and therefore on lattices which become arbitrarily fine. At the
same time we establish the
existence of the critical (stable) manifold. The 
interactions are uniformly bounded away from zero  on all scales and 
therefore we 
are constructing a non-Gaussian supersymmetric field theory on all scales.    
The interest of this theory
comes from the easily established fact that the Green's function
of a (weakly) self-avoiding L\'evy walk in $\math{Z}^{3}$
is a second moment  (two point correlation 
function) of the supersymmetric measure governing this model. The rigorous 
control of the critical renormalization group trajectory is a preparation 
for the study of the critical exponents of the 
(weakly) self-avoiding L\'evy walk in $\math{Z}^{3}$.

\vglue.5truecm
\\{\bf 0.  Introduction} 
\numsec=0\numfor=0 
\vglue.5truecm

It was observed long ago, [PS, McK ], that the Green's function of weakly 
self avoiding simple random 
walks (SAW) on a lattice $\math{Z}^d$  can be expressed as a 
correlation function in a
supersymmetric field theory. This can be shown rigorously by the same 
derivation as
in [BEI, BI1, BI2] for SAWs on hierarchical lattices. 
Consider instead of simple random walks the more
general case of continuous time (stable) L\'evy walks whose scaling 
limits are stable L\'evy
distributions, [KG, F]. Such walks can be realized as jump processes with
probability distributions permitting long range jumps, [F]. 
Their characteristic functions are given by the L\'evy-Khintchine formula
with characteristic exponent $\a$, $0<\a\le 2$, [F], $\a=2$ corresponding to 
simple random walks. The Green's function of continuous time weakly 
self avoiding L\'evy walks (SALW) can also be realized as a two point 
correlation function in a supersymmetric field theory by the same derivation 
as in [ BEI, BI1, BI2].
This paper is concerned with proving the existence of 
a critical uniformly bounded renormalization group (RG) trajectory for the 
interactions in the 
underlying supersymmetric field theory corresponding to the class of 
SALWs where $\a={3+\e\over 2}$ with $0<\e<1$ and $\e$ held small. The case
$\a={3\over 2}$ corresponds to mean field theory. Uniformity
is with respect to the lattice scale which changes with each step of the
renormalization group map. We find that the interactions are non-vanishing
at all renormalization group scales, which is the lattice version of a 
non-Gaussian fixed point. This gives the foundation for the study of the 
Green's function of SALWs in the scaling limit which is postponed to the 
sequel. Ultimately one would like to be able to extract from this
the end-to-end distance behaviour for SALWs.  

\vglue0.2cm

The supersymmetric field theory in question is a lattice supersymmetric
generalization of the model considered in [BMS]. We describe it informally
here and leave the details for the next section. 
Let $\D$ be the standard
Laplacian in $\math{Z}^3$. Then for $x,\> y\in\math{Z}^3$, and $0<\a<2$,
$C(x-y)=(-\D)^{-\a/2}(x-y)$ is the Green's function  of a stable L\'evy walk.
Let $\f_1,\>\f_2$ be independent identically distributed Gaussian random 
fields in $\math{Z}^3$ with covariance ${1\over 2}C$. 
Let $\f=\f_1 +i\f_2$ and $\bar\f$ its complex conjugate. Introduce a pair of 
Grassmann fields $\psi, \bar\psi$ of degree $1$ and $-1$ respectively.
Let $\Phi=(\f,\psi)$ and 
$\bar\Phi=(\bar\f,\bar\psi)$. The inner product is given by
$(\Phi,\Phi)= \Phi\bar\Phi  =\f\bar\f +\psi\bar\psi$. Let $\L\subset\math{Z}^3$
be a finite subset. Define 

$$V_{0}(\L,\Phi)= g_{0}\int_{\L}dx (\Phi\bar\Phi)^{2}(x)+ \tilde\m_{0}
\int_{\L}dx \Phi\bar\Phi(x)  \Eq(0.61)  $$ 

\\where the coupling constant $g_{0}>0$ and $dx$ is the counting measure in 
${\math{Z}^3}$. Then our model in finite volume $\L$ is defined by the 
supermeasure 

$$d\m_{\L}(\Phi)= d\m_{C_{\L}}(\Phi) e^{-V_{0}(\L,\Phi)} \Eq(0.62)$$

\\where $C_{\L}$ is the restriction of $C$ to the points of $\L$ and
$d\m_{C_{\L}}(\Phi)$ is the Gaussian supermeasure

$$d\m_{C_{\L}}(\Phi)=\prod_{x\in \L} d\f_1(x)d\f_2(x)d\psi(x)d\bar\psi (x)
\> e^{-(\Phi,C_{\L}^{-1}\bar\Phi)_{L^{2}(\L)}} \Eq(0.63)$$

\\Integration over the Grassmann fields is Berezin integration and 
$d\m_{\L}(\Phi)$ is interpreted as a linear functional on the Grassman algebra
(generated by the $\psi,\bar\psi$ with coefficients which are functionals of 
the $\f,\bar\f$).
An important
fact is that the potential $V_{0}(\L,\Phi)$ is supersymmetric ( supersymmetry
in this context and some of its consequences are given in the Section~1.1).
As a consequence we have that the supermeasure $d\m_{\L}(\Phi)$ is normalized :

$$\int d\m_{\L}(\Phi)\> 1 =1 \Eq(0.64)$$

\textRed
\\The parameters of the supermeasure  $\m_{\L}$ defined in \equ(0.62)
correspond to those of SALWs. Thus
$g_{0}$ measures the strength of self-repulsion 
and $\tilde\m_{0}$ the killing rate of a weakly self-avoiding L\'evy walk.
The reader will get a full dictionary in [BEI, BI1, BI2] where the 
end-to-end distance behaviour was studied for SAWs in a four dimensional
hierarchical lattice with the help of supersymmetry. \textBlack

\vglue0.2cm

We give an informal description of the results of this paper.
We will choose $\a={3+\e \over 2}$ with $0<\e <1$ , in particular we hold
$\e>0$ very small. We will take $\L$ to be
a very large cube. By successive RG transformations we will get a sequence
of measures ( the RG trajectory of measures) living in smaller and smaller 
cubes in 
finer and finer lattices till we arrive at a fixed small cube in a very 
fine lattice. This will take $\log \L$ steps. At every step the measure is
a new gaussian measure times a new supersymmetric density. The Gaussian measure
is characterized by a covariance and the sequence of covariances converge to
a smooth continuum covariance.
The supersymmetric
density incorporates the interactions. The principal information is in the
local interactions incorporated in local potentials of the above type albeit
with new parameters (coupling constants) and on a finer lattice. The other 
interactions are contracting ( irrelevant)in an appropriate sense and are 
expressed in the form of polymer activities.
The coupling constants and polymer activities
give coordinates of the RG trajectory. 
\textRed These coordinates provide  Banach spaces
of interactions which permit a rigorous study of the Wilson RG [WK] 
avoiding real space renormalization group pathologies, [GP1], [GP2],  
related to the Griffiths singularity problem in disordered systems, [G]. 
See [BKL] for a review of these pathologies.  \textBlack
The goal of this paper is to study the
RG trajectory of these coordinates in the infinite 
volume limit which makes sense for these coordinates. The true infinite volume
limit and the scaling limit will be taken at the level of correlation 
functions.
 
In section~1 we define the model, introduce supersymmetry and develop some of 
its consequences. The RG analysis of this paper is based on the finite range
multiscale expansion of covariances of [BGM]. We summarize the basic results
of [BGM] pertinent to this paper in Theorem~1.1.
This is an alternative to the Kadanoff-
Wilson block spin RG developed extensively by 
Gawedzki and Kupiainen [GK 1,2], and 
Balaban [Bal 1,2,3]. A crucial simplification arises due
to the finite range of the fluctuation covariances: Cluster expansions are no 
longer needed in the control of the fluctuation integration which is an 
essential step of  RG transformations. As a result all estimates are local in 
character. In this section we also define lattice
polymers and polymer activities. 

\\In section~2 we introduce norms which will
measure the size of polymer activities. These norms  
are suggested by those in
the continuum analysis of [BMS]
\footnote{$^{(1)}$}{A. Abdesselam in [A] 
corrected an error which occurs in [BDH-eps, BMS] where it 
is  wrongly asserted that certain normed spaces are complete. This problem
was resolved in [A] by some changes in definitions which fortunately are such 
that, as noted in [A], the estimates, theorems and proofs of [BDH-eps, BMS] 
remain true without change. 
On the lattice however the function space subtleties encountered in [A] 
disappear.} but now take account of the presence of Grassmann fields. 
\textRed  The choice of these norms was inspired by discussions with David
Brydges. They are closely related to norms which will appear
in the forthcoming study of self-avoiding simple random walks in four 
dimensions by Brydges and Slade [BS].  
 
\textBlack

\\In section~3 we define the RG map as we will use it and
in section~4 apply it to our model. In particular we 
develop second order perturbation theory. The task is to control 
the contributions from the remainder and this is taken up in the next section. 

\\Section~5 gives the basic estimates that we will need for the control
of the RG trajectory. \textRed
These estimates are extensions of those in section~5 of
[BMS]. The latter paper studied a critical bosonic theory in the continuum. 
Our present estimates take 
account of the presence of Grassmann variables as well as the lattice which
has led to a considerable number of new details. \textBlack 
The upshot is Theorem~5.1. 

\textRed
\\Section~6 is devoted to the proof of existence of the stable manifold: there
exists an initial critical mass $\tilde\m_0$ which is a Lipshitz continuous 
function of the coupling constant $g_0$  such that RG trajectory is 
bounded uniformly on all scales. The proof is established by a combination 
of three theorems, namely Theorems~6.2, 6.4, and 6.6. \textBlack

\\Finally we observe that the coupling constant $g_n$ is uniformly bounded 
away from $0$ at all scales $n\ge 0$ . As a result the
global RG trajectory gives rise to a non-Gaussian field theory. We remark that
in a continuum version of this model with a cutoff modelled on that of [BMS]
one can prove more: the continuum RG trajectory ends at a non-trivial fixed 
point. But the notion of a fixed point is devoid of meaning for lattice field
theories because the RG map even in infinite volume does not give an 
autonomous action on a fixed Banach space. 

\vglue.5truecm
\\{\bf 1.1  Definitions, model, supersymmetry} 
\numsec=1\numfor=1 
\vskip.5truecm

Let $e_1,e_2,e_3$ be the standard basis of unit vectors specifying the 
orientation of $\math{Z}^{3}$. We let $\dpr_{\m}$ denote the forward lattice 
derivative in direction $e_{\m}$ and $\dpr_{\m}^{*}$ its $L^{2}(\math{Z}^{3})$
adjoint. The latter is the backward derivative.
Then the lattice Laplacian $\D$ in $\math{Z}^{3}$ is defined by  

$$-\D = \sum_{\m=1}^{3}\dpr_{\m}^{*}\dpr_{\m} \Eq(0.1) $$

\\Let $\hat\D (p)$ be the Fourier transform of the
integral kernel of $\D$ in $\math{Z}^{3}$, namely

$$\hat\D (p) = 2\sum_{\m =1}^{3} (\cos(p_{\m})-1)\Eq(0.2)$$.

\\Let $\a={3+\e\over 2} $ be a real number with $0 < \a <2$. Then the 
Green's function 
of a (stable) L\'evy walk in  $\math{Z}^{3}$ is given by 

$$C(x-y)= (-\D)^{-\a/2}(x-y)= \int_{[-\pi,\pi]^3} {d^{3}p\over (2\pi)^3}\> 
e^{ip(x-y)}\> (-\hat\D (p))^{-\a/2} \Eq(0.3)$$

\\$C$ is positive-definite and therefore qualifies as the covariance of a 
Gaussian random field in $\math{Z}^{3}$. We introduce a pair of independent
identically distributed Gaussian random fields $\f_{1},\> \f_{2}$ with mean $0$
and covariance

$$E(\f_{j}(x)\f_{j}(y))={1\over 2} C(x-y)\Eq(0.4) $$
for  $j=1,2.$

\\Let $\f(x)=\f_1(x)+i\f_2(x) $ be a complex scalar field and $\bar\f(x)$ its
compex conjugate. On the space of functionals of $\f,\bar\f$ we have the
Gaussian probability measure

$$d\m_{C}(\phi)= d\m_{{1\over 2}C}(\phi_{1})d\m_{{1\over 2}C}(\phi_{2})
\Eq(0.4a) $$

\\Then each of $\f, \bar\f$ has zero covariance and

$$E(\bar\f(x)\f(y))= C(x-y)\Eq(0.5)  $$

\vglue0.3cm
\\{\it Grassmann algebra and integration}

\\Let $\L \subset \math{Z}^{3}$ be a bounded subset. ${\cal F}_{\L}$ represents
the algebra of $\math{C}$ valued functionals of the fields 
$\{\phi,\bar\phi :\> \L\rightarrow \math{C}\}$. 
Let $\psi(x), \bar\psi(x)$
for all $x\in \L$ be the ( anti-commuting) Grassman elements of degree 
$1, -1$ respectively. Following standard usage we will refer to them as 
(scalar) {\it fermions}. We denote by
${\Omega}_{\L}$ the Grassman algebra generated by the 
$\psi(x), \bar\psi(y)$
by multiplication and linear sums for all $x,y\in \L$ with coefficients in 
${\cal F}_{\L}$.
The Grassmann algebra is naturally graded 
${\Omega}_{\L}= \oplus_{p} {\Omega}_{\L}^{p}$ where the integer $p$ is the 
degree and each ${\Omega}_{\L}^{p}$ is a ${\cal F}_{\L}$ module. 
${\Omega}_{\L}^{0}$ is an algebra. Because of the 
anticommuting property of the generators, and because $\L$ is a finite lattice
an element of ${\Omega}_{\L}^{p}$ is a finite sum of degree $p$ elements
with coefficients in ${\cal F}_{\L}$. \textRed
For example, an element $F_{\L}$ of $\Omega_{\L}^{0}$
can be uniquely represented as

$$F_{\L}(\f,\psi)= \sum_{p\ge 0}\int_{\L^{2p}} \prod_{j=1}^{p}dx_{j}dy_{j}\>
F_{\L,2p}(\f;x_1,..,x_p,y_1,..,y_p )\prod_{j=1}^{p}\psi(x_{j})\bar\psi(y_j)
\Eq(0.81)$$ 

\\where $dx$ is the counting measure in $\math{Z}^{3}$. The coefficients, 
$F_{\L,2p}(\f;x_1,..,x_p,y_1,..,y_p )$, are antisymmetric in $(x_1,...,x_p)$
and in $(y_1,...,y_p)$.
When $\L$ is a finite subset the above multiple sum is finite. 
In the following we will often refer to the coefficients
$F_{\L,2k}$ above as {\it bosonic coefficients}. Here and in the
following we suppress
indicating the dependence of the bosonic coefficients on $\bar\f$. 

\textBlack
\\These considerations 
are of course valid for a lattice $(\d\math{Z})^{3}$ for any lattice spacing
$\d$ with the corresponding notations $\L_{\d},{\cal F}_{\L_{\d}}, 
{\Omega}_{\L_{\d}}$, $dx$ being  ${\d}^{3}$ times the counting measure in
$(\d\math{Z})^{3}$. 

\vglue0.3cm
\\Now we define  fermionic  expectation (integration) using Berezin 
integration which we review briefly and set up our conventions.
Berezin integration 
is a linear map $\Omega_{\L}\rightarrow \Omega_{\L}$ which satisfies

$$\int d\psi(x) F_{\L}(\psi, \bar\psi,\phi,\bar\phi)=
\pi^{-1/2}{\dpr\over\dpr\psi (x)}F_{\L} 
\Eq(0.8)$$ 

\\where $ F_{\L} \in {\Omega}_{\L} $ and the fermionic derivative  
${\dpr\over\dpr\psi (x)}$ is an antiderivation: \textRed
If ${f\in \Omega^{p}_{\L}}$ and
${g\in \Omega^{q}_{\L}}$ then  

$${\dpr\over\dpr\psi(x) }(fg)={\dpr\over\dpr\psi}f g + 
(-1)^{p}f{\dpr\over\dpr\psi(x)}g $$

\textBlack
\\Integration with respect to $ d\bar\psi(x)$ is given by the same formula 
with $\dpr\over\dpr\bar\psi (x)$ on the right hand side. Multiple integration
is repeated integration using the above rule, keeping in mind that
fermionic derivatives anticommute.

\\Define $C_{\L}(x-y)= C(x-y)\> :\> x,y \in \L$. We consider this as a 
$|\L|\times |\L|$ dimensional positive definite symmetric matrix with 
$x,y$ labelling the entries. Then we define the 
fermionic expectation $E_{f,\L}$ 
as a linear map ${\Omega}_{\L}\rightarrow {\cal F}_{\L}$ as follows:
Let $F_{\L}\in{\Omega}_{\L}$. We adopt the convention 
$F_{\L}(\psi,\phi)\equiv F_{\L}(\psi,\bar\psi,\phi,\bar\phi)$. Then

$$E_{f,\L}(F_{\L})= \int d\m_{C_\L}(\psi) F_{\L}(\psi,\phi)  \Eq(0.9)$$       

\\where

$$\int d\m_{C_\L}(\psi) F_{\L}(\psi,\phi)) = (det\>\pi C_{\L})^{|\L|} 
\int \prod_{x\in \L} (d\psi(x)d\bar\psi (x))\> 
e^{-(\psi,C_{\L}^{-1}\bar\psi)_{L^{2}(\L)}}F_{\L}(\psi,\phi) \Eq(0.10)$$

\\We call $d\m_{C_\L}(\psi)$ a fermionic Gaussian measure and use the 
terminology measure and expectation interchangeably. It is not difficult to
show that we have a fermionic counterpart of the bosonic gaussian formula,
namely

$$\int d\m_{C_\L}(\psi) F_{\L}(\psi,\phi)= e^{\int_{\L\times\L}dx dy
C_{\L}(x-y) {\dpr\over \dpr\psi(x)}{\dpr\over \dpr\bar\psi(y)}}
F_{\L}(\psi,\phi)\vert_{\psi=\bar\psi=0} \Eq(0.10a) $$ 

\\where $dx$ is the counting measure.
The fermionic expectation above annihilates the component of 
$F_{\L}\notin \Omega_{\L}^{0}$.

\vglue0.3cm
\\Note that the expectation of a product of two $\psi$ or of two $\bar\psi$ 
vanishes whereas if $x,y \in \L$

$$E_{f,\L}(\bar\psi(x)\psi(y))= C(x-y) \Eq(0.11)  $$

\\More generally, if $x_{j}, y_{j} \in \L,\> j=1,2,..n$

$$E_{f,\L}(\> \prod_{j=1}^{n}\bar\psi(x_{j})\psi(y_{j}))= 
{\rm det}\> ( C(x_{j}-y_{k}))_{j,k=1}^{n} \Eq(0.11a)$$

\vglue0.3cm
\\We define the field  $\Phi(x)$ (called superfield in anticipation)  
as the pair     

$$\Phi(x)=(\f(x),\psi(x)) \Eq(0.12) $$  

\\with the scalar product 

$$(\Phi(x),\Phi(y))= \Phi(x)\bar\Phi(y)=\f(x) \bar\f(y) +\psi(x)\bar\psi(y) 
\Eq(0.13)$$ 

\\More generally if $A(x,y)$ is a matrix for $x,y\in\L$ we define

$$(\Phi,A\Phi)_{L^{2}(\L)}= \int_{\L\times\L}dxdy\> \Phi(x)A(x,y)\bar\Phi(y)
=\int_{\L\times\L}dxdy\>(\f(x)A(x,y)\bar\f(y) +\psi(x)A(x,y)\bar\psi(y)) 
\Eq(0.14)$$

\vglue0.3cm

\\Let $F_{\L}(\Phi)$ belong to $\Omega_{\L}$. $F_{\L}$ also depends on 
$\bar\Phi$ but here and in the following this is not explicitly indicated.
Since $F_{\L}(\Phi)\in \Omega_{\L}$ it has the representation
\equ(0.81). We define the  expectation $E_{\L}$ as a linear map 
$\Omega_{\L}\rightarrow  \bf C$ obtained by combining the bosonic and 
fermionic expectations: If $F_{\L}\in\Omega_{\L}$ with $\m_{C}$ integrable
bosonic coefficients then 

$$E_{\L}(F_{\L}(\Phi))=\int d\m_{C_{\L}}(\Phi)F_{\L}(\Phi)=
\int d\m_{C_{\L}}(\phi)d\m_{C_{\L}}(\psi)F_{\L}(\Phi) \Eq(0.14.1)$$

\\Thus

$$E_{\L}(F_{\L}(\Phi))= \int\prod_{x\in \L} d\f(x)d\bar\f(x)
\prod_{x\in \L} d\psi(x)\prod_{x\in \L} d\bar\psi (x)
\> e^{-(\Phi,C_{\L}^{-1}\bar\Phi)_{L^{2}(\L)}}F_{\L}(\Phi) 
\Eq(0.15)$$
 
\\Notice that the determinant in the fermionic integration formula \equ(0.10)
has cancelled out with the inverse of the same determinant which appears 
in the bosonic integration measure. 

\\The expectation defined above is normalized. In other words if 
$1_{\L}(\Phi)$ is the indicator function of $\Omega_{\L}$ then

$$E_{\L}(1_{\L}(\Phi))=1  \Eq(0.17) $$

\\We have the natural order relation ${\Omega}_{\L}\subset 
{\Omega}_{\L'}$ if $\L\subset\L'$. Moreover if $\L\subset\L'$ and 
$F_{\L}\in {\Omega}_{\L}$ then $E_{\L'}(F_{\L})= E_{\L}(F_{\L})$ as is 
not difficult to show. We
define ${\Omega}$ as the inductive limit of the ${\Omega}_{\L}$ as 
$\L\subset \math{Z}^{3} $ varies over increasing subsets tending to 
$\math{Z}^{3}$ respecting 
the order relation above. The $\{E_{\L},{\Omega}_{\L}\}$ constitute a 
projective family. We denote by $E$ the projective limit: Let 
$F\in {\Omega}$ with $\m_{C}$ integrable
bosonic coefficients.  We have

$$E(F)=\int d\m_{C}(\Phi)\>F(\Phi)=
\lim_{\L'\uparrow \math{Z}^{3}}E_{\L'}(F) \Eq(0.101) $$

\\and this limit exists since $F\in {\Omega}_{\L}$ for some finite set
$\L$ and therefore $E(F)=E_{\L}(F)$ which exists. 

\\{\it Remark} : The above construction is motivated by analogous 
considerations in [BEI].

\vglue0.3cm

\\{\it Lattice integration } : In the following and throughout this paper 
we will represent lattice sums as integrals where for the $(\d\math{Z})^{3}$ 
lattice the integration measure is the counting measure in $(\d\math{Z})^{3}$
times a factor $\d^{3}$. Thus if $f$ is a function on $(\d\math{Z})^{3}$ we 
define

$$\int_{(\d\math{Z})^{3}}dx\> f(x)=\d^{3}\sum_{x\in (\d\math{Z})^{3}} f(x) 
\Eq(0.18a) $$

\vglue0.3cm

\\We now define a Laplacian acting on functionals in $\Omega$

$${\bf\Delta}_{C}=\int_{\math{Z}^{3}\times 
\math{Z}^{3}} dx dy\> C(x-y){\partial\over \partial\Phi(x)}\cdot 
{\partial\over \partial\bar\Phi(y)}=$$ 
$$=\int_{\math{Z}^{3}\times 
\math{Z}^{3}}  dx dy\> C(x-y)[{\partial\over \partial\phi(x)}{\partial\over \partial\bar\phi(y)}+{\partial\over \partial\psi(x)}{\partial\over \partial\bar\psi(y)}] 
\Eq(0.18)$$ 
\textRed
\\These integrals on $\math{Z}^3$ 
automatically restricts to $\L \times \L$ when applied to 
functionals of $\Phi$ which live in a bounded subset $\L$ of $\math{Z}^3$. 
It follows from \equ(0.10a) and its bosonic counterpart that
if $F_{\L}(\Phi)\in \Omega_{\L}^{0}$ with $\m_{C}$
integrable bosonic coefficients then

$$E(F_{\L}(\Phi))=
e^{{\bf\Delta}_{C}}F_{\L}(\Phi)\vert_{\f=\bar\f=\psi=\bar\psi=0} 
\Eq(1.16a)$$

\\Note that the action of $e^{{\bf\Delta}_{C}}$ is well defined. In fact since
$F_{\L}(\Phi)$ is in $\Omega_{\L}^{0}$ and $\L$ is a finite lattice,
it can be expressed as a finite sum of Grassmann elements with 
coefficients in ${\cal F}_{\L}$. $e^{{\bf\Delta}_{C}}$ factorises into  
bosonic and Grassmann exponentials. The expansion of the Grassman exponential
acting on $F_{\L}(\Phi)$ evaluated at $\psi=\bar\psi=0$ thus terminates and 
we are left with the expectation of the bosonic coefficients
which is well defined since they are $\m_{C}$ integrable by assumption.
\textBlack

\\We have in particular

$$E(\Phi(x)\bar\Phi(y))=0   \Eq(1.16b) $$

\\and more generally

$$E(\prod_{j=1}^{n}\Phi(x_{j})\bar\Phi(y_{j}))=0    \Eq(1.16c) $$

\\This can be proved by computation or more simply using supersymmetry
(introduced later). The integrand is supersymmetric and Lemma~1.1 below gives
the result.

\\Wick polynomials  $P(\Phi)$ are defined by the formula 
$$:P(\Phi):_{C}=e^{-{\bf\Delta}_{C}}P(\Phi) \Eq(0.18b)$$ 

\\This implies in particular that 
$$:\Phi(x)\bar\Phi(y):_{C}=\Phi(x)\bar\Phi(y) \Eq(0.19)$$ 

\\and 

$$:(\Phi\bar\Phi)^{2}:_{C}(x)= (\Phi\bar\Phi)^{2}(x)- 
2C(0)(\Phi\cdot\bar\Phi)(x) \Eq(0.20)   $$

\\For future reference we note that for $\a=1,2$

$$:(\Phi\bar\Phi)\Phi_{\a}:_{C}(x)= 
(\Phi\bar\Phi)\Phi_{\a})(x)-C(0)\Phi_{\a}(x) \Eq(0.19a) $$

\\where $\Phi_{1}=\f,\> \Phi_{2}=\psi$.

\vglue0.1cm
\\{\it Remark} : The considerations from \equ(0.8) to \equ(0.20) remain
valid on a lattice $(\d\math{Z})^{3}$ if we replace in the above 
$\L$ by a bounded subset $\L_{\d}\subset (\d\math{Z})^{3}$ and the positive
definite matrix $C$ by an arbitrary positive definite matrix $C_{\d}(x,y)$ with
$x,y\in (\d\math{Z})^{3}$. The functional Laplacian $\D_{C}$ in \equ(0.18)
is replaced by $\D_{C_{\d}}$ with the integration over 
$(\d\math{Z})^{3}\times (\d\math{Z})^{3}$.

\vglue0.3cm

\\{\it The model} : 

\vglue0.3cm
\textRed

\\Let $L$ be a triadic integer, 
$L=3^p$ with integer $p\ge 2$. Let 
$\L_{N}= (-{L^{N}\over 2},{L^{N}\over 2})^{3}\subset \math{R}^{3}$, with $N$ 
large be a large open cube in $ \math{R}^{3}$. \textBlack 
Distances in  $ \math{R}^{3}$ and 
lattices $(\d\math{Z})^{3}$ will be measured in the norm

$$|x-y|= \max_{1\le j\le 3}|x_{j}-y_{j}| \Eq(0.21a)$$
 
\\ Define $\L_{N,0}= \L_{N}\cap \math{Z}^{3}$. This is a (large) cube in 
$\math{Z}^{3}$ of edge length $L^ {N}$.
The second index $0$ in $\L_{N,0}$ emphasizes that this is a cube in 
$\math{Z}^{3}$. The local potential  \equ(0.61) will be written in a 
$C$-Wick ordered form by using \equ(0.20) and \equ(0.19). This gives 

$$V_{0}(\L_{N,0},\Phi)= \int_{\L_{N,0}}dx\> g_{0} :(\Phi\bar\Phi)^{2}:_{C}(x)+
\m_{0}\int_{\L_{N,0}}dx\> :\Phi\bar\Phi:_{C}(x)\Eq(0.21) $$

\\where $\m_{0}= \tilde\m_{0} + 2C(0)g_{0}$.

\\Define

$${\cal Z}_{0}(\L_{N,0},\Phi)= e^{-V(\L_{N,0},\Phi)} \Eq(0.22) $$ 

\\We define the measure 

$$d\m_{N,0} (\Phi)= d\m_{C}(\Phi) {\cal Z}_{0}(\L_{N,0},\Phi) \Eq(0.221)  $$ 

\\Note that the measure is normalized

$$ \int d\m_{N,0} (\Phi) =1 \Eq(0.222)  $$ 

\\This follows from Lemma~1.1 below which exploits supersymmetry introduced
later. However heuristically this is evident if we formally expand the 
exponential, integrate term by term and use \equ(1.16c). This measure defines 
our model. 

\vglue.3truecm 
\\{\it Supersymmetry} 

\vskip.3truecm 
\\The density of the measure  $d\m_{N,0} (\Phi)$ as well as its RG evolution 
 have the important property of being {\it supersymmetric}. This will restrict
considerably the form of the evolved density.

\\A {\it supersymmetry} transformation 
${\cal Q} : \Omega_{\L}\rightarrow \Omega_{\L}$ is a derivation on
the bosonic fields and an antiderivation on the grassman fields which
acts on the fields as follows :

$$\eqalign{{\cal Q}\f&=\psi\cr 
{\cal Q}\bar\f&=-\bar\psi\cr 
{\cal Q}\psi&=\f\cr 
{\cal Q}\bar\psi&=\bar\f\cr}\Eq(3.3)$$ 

\\Let $F_{\L}(\Phi)=F_{\L}(\f,\bar\f,\psi,\bar\psi)$ belong to $\Omega_{\L}$ 
with bosonic coefficients differentiable in the bosonic fields 
$\f(x),\> x\in \L$.
Then the action of $\cal Q$ on $F_{\L}$ is given by a super vector field 
denoted by the same symbol ${\cal Q}$ 

$${\cal Q}F_{\L}= \int_{\L} dx\> \Bigl(\psi(x){\dpr\over \dpr\f(x)} -
\bar\psi (x){\dpr\over 
\dpr\bar\f(x)}+\f(x){\dpr\over \dpr\psi(x)} +\bar\f (x){\dpr\over
\dpr\bar\psi(x)}\Bigr)F_{\L} \Eq(3.3.2)  $$

\\We say that a functional $F_{\L}$ is supersymmetric if ${\cal Q}F=0$.  

\\{\it Remark}: A super vector field is not a vector field because fermionic
derivatives are antiderivations.

\vglue0.3cm
\\An (infinitesimal) {\it gauge transformation} 
${\cal G}: \Omega_{\L}\rightarrow \Omega_{\L} $ is a derivation
whose action is given by  
$$\eqalign{{\cal G}\f&=i\f\cr 
{\cal G}\bar\f&=-i\bar\f\cr 
{\cal G}\psi&=i\psi\cr 
{\cal G}\bar\psi&=-i\bar\psi\cr}\Eq(3.3.1)$$ 

\\This induces on an $\Omega_{\L}$ function $F_{\L}$ the action of a 
vector field denoted by the same symbol $\cal G$

$${\cal G}F_{\L}=
i\int_{\L} dx\> \Bigl(\f(x){\dpr\over \dpr\f(x)} -\bar\f(x){\dpr\over 
\bar\dpr\f(x)}+\psi(x){\dpr\over \dpr\psi(x)} +\bar\psi(x){\dpr\over
\dpr\bar\psi(x)}\Bigr)F_{\L} \Eq(3.3.12)  $$
  
\\We say that a functional $F_{\L}$ is gauge invariant if ${\cal G}F_{\L}=0$. 

\vglue0.3cm
\\From \equ(3.3) we see that ${\cal Q}^2$ engenders an infinitesimal 
gauge transformation \equ(3.3.1). Thus acting on 
gauge invariant functionals 

$${\cal Q}^2 =0\Eq(3.5)$$ 

\\An important property of the super vector field $\cal Q$ which we will 
exploit later is that it commutes with the super Laplacian ${\bf\Delta}_{C}$ 
defined in \equ(0.18):

$$ [\>{\cal Q},\>{\bf\Delta}_{C}\>]=0$$ 

\\as is easy to verify.

\\It is easy to verify that any polynomial in $\Phi\bar\Phi$ and their 
(lattice) derivatives is supersymmetric.
As a consequence we have ${\cal Q}V(\L,\Phi)=0$ where $V$ is given in
\equ(0.21) and thus the starting interaction potential is supersymmetric. 

\vglue0.3cm

\\Let $\G(x,y)$ be any positive definite symmetric matrix. Let 
${\bf\Delta}_{\G}$ be a super Laplacian given by \equ(0.18) with $C$ replaced
by $\G$. Let $F_{\L}(\Phi)$ be an $\Omega_{\L}$  functional with $\m_{C}$
integrable bosonic coefficients.
Let $\xi = (\z, \h)$ be another superfield. Define the convolution

$$\m_{\G}*F_{\L}(\Phi)=
\int d\m_{\G}(\xi)F_{\L}(\Phi+ \xi )=e^{{\bf\Delta}_{\G}}F_{\L}(\Phi)
\Eq(0.24)$$

\\Since $\cal Q$ commutes with ${\bf\Delta}_{\G}$, $\cal Q$  {\it also 
commutes with convolution with the measure} $\m_{\G}$ :

$$\m_{\G}*{\cal Q}F_{\L}(\Phi)= {\cal Q}\m_{\G}*F_{\L}(\Phi) \Eq(0.2411)$$

\\Therefore if $F_{\L}$ is supersymmetric so is $\m_{\G}*F_{\L}$.
This observation prefigures the supersymmetry invariance of the 
renormalization group map which we will introduce later.

\vskip0.2cm

\\It follows by evaluating \equ(0.2411) at $\Phi=0$ that

$$\int d\m_{\G}(\Phi)\> {\cal Q}F_{\L}(\Phi) =0 \Eq(0.241) $$

\\since the left hand side is given by 
$(\m_{\G}*{\cal Q}F_{\L}(\Phi))\Bigr|_{\Phi=0}$ and this vanishes by virtue of
\equ(0.2411) since the coefficients
of the super vector field ${\cal Q}$ vanish when the fields vanish. \bull

\vglue0.3cm

\\{\it Lemma~1.1} : Let $F_{\L}(\Phi)$ be a supersymmtric 
$\Omega_{\Lambda}$ functional with differentiable bosonic coefficients which 
are $\m_{\G}$ integrable. Then

$$\int d\m_{\G}(\Phi)\> F_{\L}(\Phi)=F_{\L}(0) \Eq(0.242)$$

\\{\it Proof} : $\l$ be a real parameter. Define

$$f(\l)=\int d\m_{\G}(\Phi)\> F_{\L}(\l\Phi)\Eq(0.243)$$

\\We will prove

$${d\over d\l}f(\l)=0 \Eq(0.244)$$

\\This implies that $ f(\l)$ is a constant and hence evaluating at $\l=0$ 
gives \equ(0.242).

\\Taking the $\l$ derivative in \equ(0.243) we get

$${d\over d\l}f(\l)=\int d\m_{\G}(\Phi)\> ({\cal D}F_{\L})(\l\Phi)\Eq(0.245)$$

\\where

$${\cal D}=\int_{\L}dx \Bigl(\phi(x){\dpr\over\dpr\phi(x)} +
\bar\phi(x){\dpr\over\dpr\bar\phi(x)}+\psi(x){\dpr\over\dpr\psi(x)}+
\bar\psi(x){\dpr\over\dpr\bar\psi(x)}\Bigr) \Eq(0.246)$$

\\Note that the four coefficients of ${\cal D}$ can also be written as 
$({\cal Q}\psi(x),{\cal Q}\bar\psi(x), {\cal Q}\phi(x), 
-{\cal Q}\bar\phi(x))$ which we have taken in the same order as above.
This suggests that we consider the operator

$${\cal L}= \int_{\L}dx \Bigl(\psi(x){\dpr\over\dpr\phi(x)} +
\bar\psi(x){\dpr\over\dpr\bar\phi(x)}+\phi(x){\dpr\over\dpr\psi(x)}-
\bar\phi(x){\dpr\over\dpr\bar\psi(x)}\Bigr) \Eq(0.247)$$

\\and act with ${\cal Q}$ on it. We also consider the action of ${\cal L}$
on ${\cal Q}$. A straight forward computation gives the nice formula

$${\cal D}= {1\over 2} ({\cal Q}{\cal L}+ {\cal L} {\cal Q}) \Eq(0.248) $$

\\We substitute for ${\cal D}$ in \equ(0.245) the right hand side of
\equ(0.248). The contribution of the first term vanishes by \equ(0.241).
The contribution of the second term vanishes because $F_{\L}$ is supersymmetric
by hypothesis. This proves \equ(0.244) and we are done. \bull   

\\{\it Remark} : The special case of Lemma~1.1 for a hierarchical lattice 
is Lemma~2.1 of [BI]. {\it This Lemma has the important 
consequence that no field independant relevant parts ( defined later) will 
arise in the renormalization group analysis to follow.}

\vglue0.5cm

\\{\bf 1.2 Lattice renormalization group transformations} 

\vglue0.3cm 

\\We say
that a function $f(x,y)$ has finite range $L$ if $f(x,y)=0\> :\> |x-y|\ge L$.
Lattice renormalization group transformations will be based on the finite
range multiscale expansion of the covariance $C$ established in [BGM]. 

\textRed
\\Let $L$ be a large triadic integer, $L=3^{p}$, $p\ge 2$. \textBlack 
Define $\d_{n}= L^{-n}$. We have
a sequence of compatible lattices $(\d_{n}\math{Z})^{3}\subset \math{R}^{3}$,
$(\d_{n}\math{Z})^{3}\subset (\d_{n+1}\math{Z})^{3}$, with $n=0,1,2,...$. 
$B_{\d_n}= [-{\pi\over \d_n},{\pi\over \d_n}]^3 $ denotes the first Brillouin
zone of the dual of the $\d_n$ lattice. We have the following theorem 
which gives
the multiscale expansion of the covariance $C$ on $\math{Z}^{3}$ as a sum of
finite range {\it fluctuation} covariances living on increasingly finer 
lattices, together with their properties which we will need later : 

\vglue0.3cm
\\{\it Theorem 1.1 (finite range multiscale expansion) : 
For $0<\a<2$ , $d_{s} ={(3-\a)\over 2}$ and $ n= 0, 1, 2,...$ there 
exist positive definite 
functions $\G_{n}(x)$ defined for $x \in (\d_{n}\math{Z})^{3}$ and a smooth
positive definite function $\G_{c,*}$ in $\math{R}^{3}$ such that 
for all $k\ge 0$, constants $c_{k,L}$ , $c_{L,m}$
independent of $n$ and \textRed $q= {1\over 2}$\textBlack 

$$\leqalignno{ C(x-y) =\sum_{n\ge 0} L^{-2nd_{s}} 
\G_{n}\Bigl({{x-y}\over L^{n}}\Bigr) & {\rm \> and \> the\> series\> 
converges \> in} \> L^{\infty}(\math{Z}^{3}) & (1) \cr 
\G_{n}(x)& =0 \quad {\rm for}\quad |x|\ge {L\over 2} & (2) \cr  
\Bigl|\hat\G_{n}(p)\Bigr| & \le c_{k,L} (1+p^{2})^{-2k} \quad {\rm for}\quad
p\in B_{\d_{n}},\quad\forall k\ge 0 & (3) \cr 
\hat\G_{c,*}(p) & = {\rm lim}_{n\rightarrow\infty}\hat\G_{n}(p)\quad 
{\rm exists\> pointwise\> in }\> p & (4a) \cr \textRed 
\Bigl|\hat\G_{n}(p)- \hat\G_{c,*}(p)\Bigr| \le c_{k,L}& (1+p^{2})^{-2k}
(1+{1\over p^2}) L^{-qn},
\> \forall n \ge 3,\>\forall k\ge 0,\> p\in B_{\d_{n}}\setminus {0}
\textBlack & (4b) \cr 
\Vert \dpr_{\d_{n}}^{m}\G_{n}\Vert_{L^{\infty}((\d_{n}\math{Z})^{3})} 
&\le c_{L,m}, \forall m\ge 0 & (5a) \cr
\dpr_{c}^{m} \G_{c,*} &= {\rm lim}_{n\rightarrow\infty}\dpr_{\d_{n}}^{m}\G_{n}
\> {\rm exists\> in}\>
L^{\infty}((\d_{l}\math{Z})^{3}) & (5b)\cr 
\hskip3cm\Vert \dpr_{\d_{n}}^{m}\G_{n}- \dpr_{c}^{m} \G_{c,*}
\Vert_{L^{\infty}((\d_{l}\math{Z})^{3})}& \le c_{L,m}L^{-qn},\>
\forall n\ge l\ge 3, \> \forall m\ge 0  & (5c) \cr}$$  

\\where $\dpr_{c}$ is a continuum partial derivative, $\dpr_{\d_{n}}$ is a 
forward lattice partial derivative in $(\d_{n}\math{Z})^{3}$ and 
the dependence on the  direction vectors have been suppressed. For 
$\dpr_{\d_{n}}^{m}$ and $\dpr_{c}^{m}$ a multi-index convention is implicit. 
}
\textRed
\vglue0.2cm
\\{\it Remark}: The theorem is for the most part a combination of results 
obtained in various theorems in [BGM]. Before we outline the proof 
note that in
[BGM], $L$ was a large dyadic integer whereas we have chosen here $L$ to be
triadic. The results of [BGM] remain unaffected provided we define the 
continuum cube $U_c(R)\subset \math{R}^{3}$ in section~1 of [BGM] to be
$(-{R\over 3},{R\over 3})^{3}$. This guarantees in particular
that if $R=R_{m}=L^{-(m-1)}$, $0\le m\le n $,
and $U_{\d_n}(R_{m})=U_{c}(R_{m})\cap (\d_{n}\math{Z})^{3}$ then the important
property $\dpr U_{\d_n}(R_{m})\subset \dpr U_{c}(R_{m})$ remains true. This
last property is invoked in section~6, page 439 of [BGM], in preparation for 
the convergence proof therein.

\textBlack
\vglue0.2cm
\\{\it Proof} : The multiscale expansion in part (1) and the
finite range property of part (2) were given in Section 4, [BGM].
\textRed  The factor 
$6$ in the range $6L$ of $\G_{n}$ is an artifact. By scaling down $R_m$
in the cube $U_{\d_n}(R_{m})$ by a factor of $3^{-4}$ and the range of the
function $g$ in section~1 to $3^{-6}L$ we get $\G_{n}$ to have range $L/2$.
\textBlack Convergence of (1) in 
$L^{\infty}(\math{Z}^{3})$ follows on using $d_{s} >0$, Corollary 5.6 and 
the Sobolev embedding inequality for lattice $L^{2}_k =H_k$ spaces
with $k$ in the Corollary 
sufficiently large. Part (3) follows from (5.10) of 
Theorem 5.5 by integration on $a$ with the measure $da\> a^{-\a/2}$ ( see 
(4.3) of section 4). Corollary 5.6 and lattice Sobolev embedding gives (5a).
Corollary 6.2 gives parts (4a) and (5b). The 
convergence rate estimates of parts (4b) and (5c) which were not given in
[BGM] also follow from the results therein. The proof is given elsewhere,
[BM].  \bull 

\vglue0.2cm
\textRed

\\{\it Remark}: (4b) is not necessarily the best possible estimate. The
left hand side has no singularity at $p=0$ whereas the right hand side does.
However it suffices for our purposes because (5c) above follows from (4b) and
it is (5c) which will be put to use later. 
In fact (4b) implies that for fixed $l\ge 3$ and all $n\ge l$, $k\ge0$,  

$$\Vert\G_{n}- \G_{c,*}\Vert_{L^{1}_{k}((\d_l\math{Z})^{3})}
 \le c_{k,L} L^{-qn}$$
 
\\where $L^{1}_{k}((\d_l\math{Z})^{3})$ is a lattice Sobolev space. The finite
range of $\G_{n},\>\G_{c,*}$ and lattice Sobolev embedding for $k\ge 3+ m$ 
implies (5c). The singularity at $p=0$ in the right hand side of (4b)
is integrable in $B_{\d_{n}}$. It thus turns out to be harmless.\textBlack 

\vglue0.3cm

\\Define for all $n\ge 0$ the positive definite functions $C_{n},\> C_{c,*}$
on $(\d_{n}\math{Z})^{3}$ and $\math{R}^3$ respectively \textRed
by the recursion relations

$$C_{n}(x)= \G_{n}(x) + L^{-2d_{s}} C_{n+1}\Bigl({{x}\over L}\Bigr)
\Eq(0.28) $$

$$C_{c,*}(x)= \G_{c,*}(x) + L^{-2d_{s}} C_{c,*}\Bigl({{x}\over L}\Bigr)
\Eq(0.28a) $$

\\Solving these relations by iteration gives \textBlack

$$C_{n}(x)=\sum_{j=0}^{\infty} L^{-2jd_{s}} \>
\G_{n+j}\Bigl({{x}\over L^{j}}\Bigr)     \Eq(0.25) $$

$$C_{c,*}(x)=\sum_{j=0}^{\infty} L^{-2jd_{s}} \>
\G_{c,*}\Bigl({{x}\over L^{j}}\Bigr)     \Eq(0.26) $$

\textRed

\\Note that $C_{0}=C$ as follows from (1) of Theorem~1.1. 

\textBlack

\vglue0.3cm

\\{\it Corollary 1.2 : The series \equ(0.25) for $C_n$ together with that 
for its multiple lattice derivatives in $(\d_{n}\math{Z})^{3}$
converge in $L^{\infty}((\d_{n}\math{Z})^{3})$. For every integer $m\ge 0$ we
have a constant $c_{L,m}$ such that

$$\Vert \dpr_{\d_n}^{m} C_{n}\Vert_{L^{\infty}((\d_n\math{Z})^{3})}\le c_{L,m}
\Eq(0.271) $$

\\The series \equ(0.25) defining $C_{c,*}$
and its multiple continuum derivatives of arbitrary order
converge in $L^{\infty}(\math{R}^3)$
so that $C_{c,*}$ is a smooth continuum function. For all $m\ge 0$ and
$\dpr_{c}$ the continuum partial derivative

$$\sup_{x\in \math{R}^3}\vert \dpr_{c}^{m}C_{c,*}(x)\vert\le c_{L,m}
\Eq(0.271a) $$

\\Moreover 
for  $n\ge l\ge 3$ with $l$ fixed  and $\forall m \ge 0$, there exists a 
constant $c_{L,m}$ such that 

$$\Vert \dpr_{\d_l}^{m}C_{n}- \dpr_{c}^{m} C_{c,*}
\Vert_{L^{\infty}((\d_l\math{Z})^{3})} \le c_{L,m}L^{-qn} \Eq(0.27) $$ }

\textRed

\\{\it Proof} : The first part together with the bound \equ(0.271) follow from
(5a) of Theorem 1.1. In fact from \equ(0.25) we have 

$$\dpr_{\d_n}^{m} C_{n}= \sum_{j=0}^{\infty} L^{-2jd_{s}} \> L^{-mj}
(\dpr_{\d_{n+j}}^{m} \G_{n+j})\Bigl({{x}\over L^{j}}\Bigr) $$

\\where we have used repeatedly ($m$-times) the identity
$\dpr_{\d_n}\G_{n+j}\Bigl({{x}\over L^{j}}\Bigr)= L^{-j}
(\dpr_{\d_{n+j}}\G_{n+j})\Bigl({{x}\over L^{j}}\Bigr) $
as is easy to show. Therefore 

$$\eqalign{\Vert \dpr_{\d_n}^{m} C_{n}\Vert_{L^{\infty}((\d_n\math{Z})^{3})}
&\le \sum_{j=0}^{\infty} L^{-2jd_{s}} L^{-mj} \> \sup_{x\in (\d_n\math{Z})^{3}}
\Bigl|(\dpr_{\d_{n+j}}^{m} \G_{n+j})\Bigl({{x}\over L^{j}}\Bigr)\Bigr|\cr
&\le \sum_{j=0}^{\infty} L^{-2jd_{s}}L^{-mj} \>
\sup_{y\in (\d_{n+j}\math{Z})^{3}}
\Bigl|(\dpr_{\d_{n+j}}^{m} \G_{n+j})(y)\Bigr|\cr} $$

\\Now use the bound in (5a) together with $d_s \ge {1\over 2}$ to get
\equ(0.271).  
To prove the next statement observe that the first part of
Theorem 6.1 of [BGM] together with Sobolev embedding implies that
$\Vert \dpr_{c}^{m}\G_{c,*}\Vert_{L^{\infty}(\math{R}^{3})} \le c_{L,m}$.
Using this \equ(0.271a) follows from \equ(0.26). 
Finally to prove the estimate \equ(0.27) observe that 

$$\eqalign{\Vert \dpr_{\d_l}^{m}C_{n}- \dpr_{c}^{m} C_{c,*}
\Vert_{L^{\infty}((\d_l\math{Z})^{3})} &\le
\sum_{j=0}^{\infty} L^{-2jd_{s}} L^{-mj} \> 
\sup_{x\in (\d_{l}\math{Z})^{3}}
\Bigl|(\dpr_{\d_{n+j}}^{m} \G_{n+j})\Bigl({{x}\over L^{j}}\Bigr)-
(\dpr_{c}^{m} C_{c,*})({x\over L^{j}})\Bigr|\cr
&\le \sum_{j=0}^{\infty} L^{-2jd_{s}}L^{-mj} \>
\sup_{y\in (\d_{l+j}\math{Z})^{3}}
\Bigl|(\dpr_{\d_{n+j}}^{m} \G_{n+j})(y)-\dpr_{c}^{m} C_{c,*}(y)\Bigr|\cr
&\le L^{-nq}c_{L,m} \sum_{j=0}^{\infty} L^{-2jd_{s}}L^{-mj}L^{-jq}\cr } $$

\\where in the last line we have used part (5c) of Theorem 1.1. \equ(0.27)
now follows with a new constant $c_{L,m}$. This also establishes that 
$\dpr_{\d_l}^{m}C_{n}\rightarrow \dpr_{c}^{m} C_{c,*}$ in
$L^{\infty}((\d_l\math{Z})^{3})$. \bull \textBlack
 
\vglue0.3cm

\\We consider the finite sequence of compatible lattices 
$\{(\d_{n}\math{Z})^{3}\} $ for $0\le n\le N$. The considerations in 
Section 1.1
for fields in $\math{Z}^{3}$ remain valid for 
every lattice $(\d_{n}\math{Z})^{3}$ provided for the expectations we replace 
the covariance $C$ by $C_{n}$. Let the fields $\f,\psi,\bar\psi$ be
defined in $(\d_{N}\math{Z})^{3}$. These fields restrict to the coarser lattices
$(\d_{n}\math{Z})^{3}$ for every $n$ with $0\le n\le N$.

\vglue0.2cm

\\We introduce a parameter $\e$ with $0<\e\le 1$ and define

$$\a={3+\e\over 2} \Eq(1.41a) $$

\\Let $x\in(\d_{n}\math{Z})^{3}$. For every $n\le N-1$ we define the scale 
transformation $S_{L}$ by 

$$S_{L}\Phi(x)= \Phi_{L^{-1}}(x)= L^{-d_{s}}\Phi({x\over L}) \Eq(1.41) $$

\\where

$$d_{s}={(3-\a)\over 2}= {3-\e\over 4}   \Eq(1.42) $$

\\is the {\it dimension} of the field $\Phi$. The fields  
$\f,\bar\f,\psi,\bar\psi$ are thus assigned the 
same dimension $d_{s}$ and the same transformation law \equ(1.41). Note that
the scale transformed fields now live in  $(\d_{n}\math{Z})^{3}$. 

\vglue0.1cm
\textRed
\\Let $\L \subset \math{R}^{3}$ and $\L_{\d_n}=\L\cap (\d_{n}\math{Z})^{3}$.
We define the scale transformation on functionals of fields by 

$$(S_{L}F)(L^{-1}\L_{\d_{n+1}},\Phi)=F(\L_{\d_n}, S_{L}\Phi) \Eq(1.422)$$

\textBlack
\\The $C_n$ and $\G_n$ are positive definite and therefore qualify as 
covariances of Gaussian measures. For $x,y\in (\d_{n}\math{Z})^{3}$ we define 
the scale transformation of the covariance $C_{n+1}$ by

$$S_{L}C_{n+1}(x-y)=L^{-2d_{s}} C_{n+1}({{x-y}\over L})\Eq(1.43)$$. 

\\which permits us to write \equ(0.28) as

$$C_{n}(x-y)= \G_{n}(x-y) + S_{L}C_{n+1}(x-y) \Eq(1.44) $$

\\ Let $\L_{\d_n} \subset (\d_{n}\math{Z})^{3}$ be a bounded subset. 
Then \equ(1.44) 
implies upon using \equ(1.16a) (with $C$ replaced by $C_{n}$)  that

$$\int d\m_{C_{n}}(\Phi) F(\L_{\d_n} ,\Phi)  = \int d\m_{S_{L}C_{n+1}}(\Phi) 
\int d\m_{\G_{n}}(\xi) F(\L_{\d_n} ,\xi + \Phi)   \Eq(1.45)$$

\vglue0.3cm
\textRed
\\Let $L=3^{p}$ with integer $p\ge 2$ and let
$\L_{m}= (-{L^{m}\over 2},{L^{m}\over 2})^{3}\subset \math{R}^{3}$  
be an open cube in $\math{R}^{3}$ centered at the origin. \textBlack 
We denote by 
$$\L_{m,n}= \L_{m}\cap (\d_{n}\math{Z})^{3} \Eq(1.45a)$$
the induced
cube of side length $L^{m}$ in  $(\d_{n}\math{Z})^{3}$ centered at the origin.
Let $F_{0}(\L_{N,0}\>,\Phi)$ be a 
functional of $ \Phi$ and ($\bar\Phi$) belonging to $\Omega^{0}(\L_{N,0})$. 
By virtue of \equ(1.45)  we have for $n=0$ 

$$\int d\m_{C_{0}}(\Phi)F_{0}( \L_{N,0}\>, \Phi)=
\int d\m_{C_{1}}(\Phi)F_{1}(\L_{N-1,1}\>,\Phi)                  \Eq(1.46) $$

\\where 

$$F_{1}(\L_{N-1,1}\>,\Phi):= (S_L\m_{\G_{0}}*F_{0})(\L_{N-1,1}\>,\Phi)=
\int d\m_{\G_{0}}(\xi) 
F_{0}(\L_{N,0}\>,\xi +S_{L}\Phi) \Eq(1.47) $$

\\The final scale transformation takes us to a finer lattice as well as
scaling down the size of the cube.

\\The iteration of \equ(1.47) using \equ(1.46) gives after $n$ steps

$$\int d\m_{C_{0}}F_{0}(\L_{N,0}\>,\Phi)=
\int d\m_{C_{n}}F_{n}(\L_{N-n,n}\>,\Phi)\Eq(1.47a) $$

\\where

$$F_{n}(\L_{N-n,n}\>,\Phi)
:= \m_{\G_{n-1}}*F_{n-1}(\L_{N-n+1,n-1},\> S_{L}\Phi)   \Eq(1.48) $$

\\\equ(1.48) defines for $N>0$ fixed and $1\le n\le N-1$ a 
sequence of maps 

$$T_{N-n,n}\> :\>  \Omega^{0}(\L_{N-n+1,n-1})\rightarrow 
\Omega^{0}(\L_{N-n,n}) \Eq(1.48a)$$ 

\\any member of which we call a 
{\it renormalization group (RG) transformation}. The map is clearly not 
autonomous. The first index refers to the cube whose size has gotten reduced
because of the rescaling. The second index refers to the lattice spacing
which has gotten finer because of the rescaling. In the following we will 
apply the RG  transformation iteratively to the (interaction) density 
${\cal Z}_{0}(\L_{N,0}\>,\Phi)$ of the measure $d\m_{N,0}(\Phi)$ defined in
\equ(0.221) generating thereby the sequence ${\cal Z}_{n}(\L_{N-n,n}\>,\Phi)$
for $0\le n\le N-1$. After $N-1$ steps we arrive at 
${\cal Z}_{N-1}(\L_{1,N-1}\>,\Phi)$ where $\L_{1,N-1}$ is the cube of edge
length $L$ in $(\d_{N-1}\math{Z})^{3}$ centered at the origin. The fundamental
goal in this paper is to control this sequence of transformations when $N$
is indefinitely large in the infinite volume limit (as explained at the end
of section~3). 

\vglue0.3cm
\\1.3. {\bf Polymer gas representation}.
\vglue0.2cm

\\In order to analyze the RG evolution we will write the densities 
${\cal Z}_{n}$ in a {\it polymer gas} representation whose form is preserved
under RG transformations. 

\vglue0.2cm

\\{\it Polymers} : 
\textRed
\\We pave $\math{R}^{3}$ with a disjoint union of open cubes 
$\D\subset\math{R}^{3}$ of edge length 1 called unit cubes or $1$-cubes
defined by

$$\D = (-{1\over 2}+ m_{1},{1\over 2}+ m_{1})
\times (-{1\over 2}+  m_{2},{1\over 2}+ m_{2})
\times (-{1\over 2}+ m_{3},{1\over 2}+ m_{3})\Eq(1.2201) $$

\\where $(m_1,m_2,m_3)\in \math{Z}^{3}$. \textBlack
{\it We say two unit cubes from the paving 
are connected if their closures share at least a vertex in  
common.} If they are not connected (i.e. their closures are disjoint) we say
that they are {\it strictly disjoint}. A continuum (connected) $1$- polymer 
$X$ is a (connected) union of a 
finite subset of unit cubes chosen from the paving and is thus open. 
Henceforth, unless otherwise mentioned, a polymer is connected by default.

\\We will measure distances
in $\math{R}^{3}$ and in all embedded lattices in the norm

$$|x-y|=\max_{1\le j\le 3}|x_{j}-y_{j}|  \Eq(1.220) $$

\\If $\D_1$ and $\D_2$ are two unit cubes from the paving then the distance
between them is

$$d(\D_1,\D_2)=\inf_{x\in \D_1,\>y\in \D_2}|x-y| \Eq(1.220a) $$

\\If $\D_1$ and $\D_2$ are strictly disjoint than $d(\D_1,\D_2)\ge 1$.

\textRed
\\Let $\d_{n}= L^{-n}$ where $L=3^{p}$ is a triadic integer. Let $\d$ be any 
member of the sequence $\{\d_n\}_{n\ge 0}$. \textBlack  Define 
the {\it unit block
or $1$-block} in $(\d\math{Z})^{3}$ by

$$\D_{\d}=\D \cap (\d\math{Z})^{3} \Eq(1.22a)$$

\\and the {\it lattice $1$-polymer $X_{\d}$} by

$$X_{\d}=X \cap (\d\math{Z})^{3} \Eq(1.22b)$$

\\where $X$ is a continuum $1$-polymer. Note that as point sets 
$X_{\d_{n}}\subset X_{\d_{n+1}}$. 

\\We denote by $|X_{\d}|$ the volume of $X_{\d}$ measured 
in accordance with \equ(0.18a). 
The  $1$-blocks are lattice restrictions of the
open continuum unit cubes defined above. Therefore, as is easy to verify,
$|\D_{\d_n}|=1$ and 

$$|X_{\d_n}| = \sharp \{\D_{\d_n}\>:\> \D_{\d_n}\subset X_{\d}\} \Eq(1.22f) $$    
                
\\the total number of $1$-blocks in $X_{\d_n}$. This is equal to $|X|$ the 
total number of $1$-cubes in $X$ by our construction. As a consequence we
have $|X_{\d_n}|=|X_{\d_{n+1}}|$.

\vglue0.2cm

\\{\it We say two $1$-blocks in $X_{\d}$ are connected if the continuum 
$1$-cubes of which they are the lattice restrictions are connected 
( see above).} If the $1$-blocks are not connected we say that they are 
{\it strictly disjoint}. The distance between two strictly disjoint $1$-blocks
is $\ge 1$. The lattice (connected) polymer $X_{\d}$ is a (connected) union of 
a finite subset of disjoint $1$-blocks $\D_{\d}$. Let $X_{\d}$ and $Y_{\d}$
be each a connected polymer. {\it We say that $X_{\d},Y_{\d}$ are strictly
disjoint if they are mutually disconnected i.e.
if every $1$-block from $X_{\d}$ is strictly disjoint from every $1$-block
from $Y_{\d}$ }. Then the distance $d(X_{\d},Y_{\d})\ge 1$.  

\\Given an integer $n\ge 1$ we define the $n$-{\it collar} of 
$X_{\d}$, denoted $\dpr_{n} X_{\d}$ by

$$\dpr_{n} X_{\d}
=\{y\notin X_{\d}\> : |x-y|\le n\d,\> {\rm some}\> x\in X_{\d}\}
\Eq(1.22c)  $$

\\where $|\cdot|$ is the distance function inherited from $\math{R}^{3}$. We
define

$${\tilde X^{(n)}_{\d}}=  X_{\d} \cup \dpr_{n} X_{\d}  \Eq(1.22d) $$

\\Let $f : (\d\math{Z})^{3}\rightarrow \math{C}$. We define the forward 
lattice partial derivative $\dpr_{\d,\m}$ and the backward lattice derivative
 $\dpr_{\d,-\m}$ by

$$\dpr_{\d,\m}f(x)=\d^{-1}(f(x+\d e_{\m})-f(x))
\Eq(1.22e)$$

$$\dpr_{\d,-\m}f(x)=\dpr_{\d,\m}^{*}f(x)=\d^{-1}(f(x-\d e_{\m})-f(x))
\Eq(1.22g)$$

\\where $e_{1},e_{2},e_{3}$ is the standard basis of unit vectors which 
provides the orientation of $\math{R}^{3}$ and thus of all the embedded lattices
we will encounter. $\dpr_{\d,\m}^{*}$ is the $L^{2}((\d\math{Z})^{3})$ adjoint
of $\dpr_{\d,\m}$.

\vglue0.2cm

\\{\it Polymer activity} :

\vglue0.2cm

\\A {\it polymer activity} $K(X_{\d},\Phi)= {\tilde K}(X_{\d},\f, \psi)$, where
it is henceforth understood that it also depends on $\bar\f,\bar\psi$, 
is a map \textRed
$X_{\d},\>\Phi \rightarrow \Omega^{0}_{\tilde X_{\d}^{(2)}}$ where the fields 
$\Phi$ depend only on the points of $\tilde X_{\d}^{(2)}$. \textBlack 

\vglue0.2cm

\\{\it The polymer activities of this paper are of degree $0$, 
gauge invariant and supersymmetric, and invariant under translations,
reflections and
rotations which leave the lattice invariant. In addition they satisfy 
the condition $K(X_{\d},\Phi)= K(X_{\d},-\Phi)$ together with the support
condition : $K(X_{\d},\Phi)=0$ if $X$ is not connected. Furthermore
$K(X_{\d},0)=0.$ }

\vglue0.2cm
\\We write the generic density 
${\cal Z}({\L_{\d}})(\Phi)$ in the form

$${\cal Z}(\L_{\d})=\sum_{N=0}^{\infty}{1\over
N!}e^{-V(X_{\d}^{(c)})} \sum_{X_{\d,1},..,X_{\d,N}} 
\prod_{j=1}^N K(X_{\d,j})
\Eq(1.22)$$ 

\\where the connected polymers $X_{\d,j}\subset\L_{\d}$ are strictly disjoint,
$X_{\d}=\cup_1^N X_{\d,j}$, $X_{\d}^{(c)}=\L_{\d}\backslash X_{\d}$
and $V(Y_{\d})=V(Y_{\d},\Phi,C,g,\mu)$ is given 
by\equ(0.21)
with parameters $g,\mu$ and integration over $Y_{\d}$ with measure $dx$ defined
as the counting measure in  $(\d\math{Z})^{3}$ times $\d^{3}$. The 
Wick ordering covariance $C =C_{n}$ (see\equ(0.25)) if $\d=\d_{n}$. 
We have 
suppressed the field dependence in \equ(1.22). Initially the activities
$K$ vanish but they do arise under RG transformations. The representation 
\equ(1.22) remains stable under RG transformations as we will see in Section~3.

\vglue0.3cm

\textRed

\\Polymer activities $K(X_{\d},\Phi)= K(X_{\d},\f,\psi)
\in \Omega^{0}_{\tilde X_{\d}^{(2)}}$ can be  represented uniquely
as a ( finite) series in the fermionic fields $\psi, \bar\psi$
with coefficients which are functionals of the bosonic fields $\f$ :

$$K(X_{\d},\Phi)= K(X_{\d},\f,\psi)=\sum_{p\ge 0} {1\over (p!)^{2}}
\int_{X_{\d}^{p}\times X_{\d}^{p}} d{\bf x}d{\bf y} \ 
(D_{F}^{2p}K)(X_{\d},\f,{\bf x},{\bf y}) 
\prod_{j=1}^{p}\psi(x_{j})\bar\psi(y_{j}) \Eq(2.1)$$ 

\\where: 
 
\\${\bf x}=(x_{1},...,x_{p})$, ${\bf y}=(y_{1},...,y_{p})$ 
and $d{\bf x}=\prod_{i=1}^{2p}dx_{i}$ where
$dx_{i}$ is the counting measure multiplied by $\d^{3}$ on $(\d\math{Z})^{3}$.
${\bf y}$ and $d{\bf y}$ are similarly defined.
The coefficient $(D_{F}^{2p}K)(X_{\d},\f,{\bf x},{\bf y})$ is defined by

$$(D_{F}^{2p}K)(X_{\d},\f,{\bf x},{\bf y})=\prod_{j=0}^{p-1} 
{\dpr\over{\dpr\bar\psi(y_{p-j})}}{\dpr\over{\dpr\psi(x_{p-j})}}
K(X_{\d},\f,\psi)\Bigr|_{\psi=\bar\psi=0}   \Eq(2.1b)$$

\\{\it This defines a lattice analogue of a distributional kernel which is 
henceforth restricted so as to contain at most (lattice) delta functions and 
their first and second (lattice) derivatives.}
It is clearly antisymmetric in 
$(x_1,...,x_p)$ and in $(y_1,...,y_p)$. \textBlack
It is gauge invariant as is the 
Grassmann monomial of degree $0$.

\\The polymer activities in question also satisfy

$$K(X_{\d},0)=0\Eq(2.111)$$

\vglue0.1cm

\\{\it Remarks} : 
We will see that the representations \equ(1.22), \equ(2.1) are preserved 
by renormalization group transformations. {\it The RG transformations are 
gauge invariant, preserve supersymmetry by virtue of \equ(0.2411),
as well as the vanishing condition \equ(2.111) by virtue of Lemma~1.1. The
RG transformations preserve invariance of the polymer activities under 
translations, reflections and rotations which leave the lattice invariant. 
}

\vglue.5truecm 
\\{\bf 2. REGULATORS, DERIVATIVES AND NORMS } 
\numsec=2\numfor=1 
\vskip.5truecm   
 
In this section we will introduce Banach spaces of polymer activities.
These are lattice analogues of the continuum constructions in [BDH-est, BMS,
A] albeit with changes because of the presence of Grassman variable. 
The Banach space norms that we will presently introduce measure 
differentiability properties of the activities with respect to fields $
\f,\psi$, as well
as the behaviour with respect to large fields $\dpr\f$ and large sets. The 
behaviour for large $\f$ itself will be controlled with the help of
lattice Sobolev inequalities and the local potential.    

\vskip.5truecm 
 
\\2.1 {\bf Regulators }

\vglue0.3cm

\\Let $\dpr_{\d,\m}$ and $\dpr_{\d,-\m}$
be respectively the forward and backward lattice derivatives in 
$(\d\math{Z})^{3}$ along the unit vector $e_{\m}$ defined in \equ(1.22e) 
and \equ(1.22g).
Here as before $\d$ is any member of the sequence $\{\d_{n}\}$
where $\d_{n}=L^{-n}$ and $L=3^{p}$ with integer $p\ge 2$.
Define 

$$\dpr^{j}_{\m_1, \m_2,..,\m_j}= 
\dpr_{\d,\m_1}\dpr_{\d,\m_2}...\dpr_{\d,\m_j}$$

\\Let $X$ be a connected polymer in $\math{R}^3$ and 
$X_{\d}=X\cap (\d\math{Z})^3$. Let 
${\tilde X}^{(n)}_{\d}= X_{\d}\cup \dpr_{n} X_{\d}$ as defined earlier 
(\equ(1.22c) and \equ(1.22d)). Let 
$\f : {\tilde X}^{(5)}_{\d}\rightarrow \math{C}$. We define a norm 
$\Vert\cdot\Vert_{X_{\d},1,5}$ : 

$$\Vert\f\Vert^2_{X_{\d},1,5}=\sum_{j=1}^{5}\> {1\over 2^{j}} 
\sum_{\m_{j}\in S,\forall j}
\int_{X_{\d}}dx\> |\dpr^{j}_{\m_1, \m_2,..,\m_j}\f(x)|^{2} \Eq(2.10reg) $$

\\where $S=\{1,-1,2,-2,3,-3\}$. This is a lattice Sobolev norm of the type
introduced
in Section 5, page 421 of [BGM] but now without the $L^{2}$ piece.

\\We define now the {\it large field regulator}

$$G_{\k}\> :\> X_{\d}\times {\cal F}_{{\tilde X}^{(5)}_{\d}}
\rightarrow \math{R} \Eq(2.11reg) $$ 

\\where ${\cal F}_{{\tilde X}^{(5)}_{\d}}$ is the 
algebra of  
$\bf C$ valued functions on $ {\tilde X}^{(5)}_{\d} $ by

$$G_{\k}(X_{\d},\f)=e^{\kappa\Vert\f\Vert^2_{X_{\d},1,5}} 
\Eq(2.1reg)$$ 

\\$G_{\k}$ satisfies the multiplicative property: If $X_{\d}, Y_{\d}$
are disjoint sets then

$$G_{\k}(X_{\d}\cup Y_{\d} ,\f)= G_{\k}(X_{\d},\f) G_{\k}(Y_{\d},\f)
\Eq(2.110) $$

\\$G_{\k}^{-1}$ will be a weight function in polymer activity norms.
The norm 
$\Vert\cdot\Vert_{X_{\d},1,5}$ can be used in lattice Sobolev inequalities, in 
conjunction with the stability provided by the local potential, to control 
$\f$ and its first two lattice derivatives pointwise. The parameter 
$\k=\k(L) > 0$ is chosen so that for all $L\ge 2$  
the large field regulator satisfies the stability property given in the
following Lemma :

\vglue0.2cm
\\ {\it Lemma~2.1 (stability property)} : There exists a constant 
$\k_{0}=\k_{0}(L) >0$
independent of $n$ such that for all $\k$ with $0<\k\le\k_{0}$ 

$$\int d\mu_{\G_{n}}(\zeta)\> G_{\k}(X_{\d_{n}},\zeta +\f)
\le 2^{|X_{\d}|}G_{2\k} (X_{\d_{n}},\f) \Eq(2.3reg)$$ 

\\where $|X_{\d_{n}}|$ is the number of unit blocks in $X_{\d_{n}}$. 

\\{\it Proof} : \equ(2.3reg)
is proved in  exactly the same way as in the proof of the stability property 
of the continuum large field regulator in Lemma~3 of [BDH-est]. The proof
uses a flow equation for the measure convolution with interpolated covariance
which remains true for the lattice. Another ingredient is Young's 
convolution inequality for functions 
which is also true on the lattice. In the cited proof we replace the covariance
$C$ by $\G_{n}$ and continuum derivatives by lattice derivatives. From
the proof of Lemma~3 of [BDH-est] we see that two conditions have to be 
satisfied by $\k_{0}$, namely : 
1)  $\k_{0} \max_{2\le m\le 10} 
\Vert\dpr_{\d_{n}}^{m}\G_{n}\Vert_{L^{\infty}((\d_{n}\bf Z)^{3})}$ is 
sufficiently small and
2) $\k_{0}\Vert \G_{n}\Vert_{L^{1}((\d_{n}\bf Z)^{3})}$ is sufficiently 
small. 
Parts (5a) of Theorem~1.1 shows that that 1) and 2) above can be 
assured by a $\k_{0}$ independent of $n$. From (5a) we have 

$$\k_{0} \max_{2\le m\le 10}
\Vert\dpr_{\d_{n}}^{m}\G_{n}\Vert_{L^{\infty}((\d_{n}\bf Z)^{3})}
\le \k_{0}c_{L}$$

\\and from (5a) and the finite range property

$$\k_{0}\Vert \G_{n}\Vert_{L^{1}((\d_{n}\bf Z)^{3})} \le
\k_{0} L^{3}c'_{L}$$

\\It is sufficient to choose $\k_{0}$ so that the right hand side of both
inequalities are sufficiently small. This is achieved independent of $n$. 
\bull

\vglue0.1cm

\\Now hold $L=3^{p}$ sufficiently large by taking $p$ large. 
Recall that $\a= {3+\e\over 2}$ where $0<\e<1$ so that $\a<2$. 
Then we get after rescaling 
$$\int d\mu_{\Gamma}(\zeta)\> G_{\k}(X_{\d_{n}},\zeta +S_{L}\f)
\le 2^{|X_{\d_{n}}|} G_{\k}(L^{-1}X_{\d_{n+1}},\f) \Eq(2.3.1reg)$$ 
because  from the scaling property  of the  
fields $\f$ , see \equ(1.41), \equ(1.42)we have 

$$\Vert S_{L}\f\Vert^2_{X_{\d_{n}},1,5}\le L^{-(2-\a)} 
\Vert\f\Vert^2_{L^{-1}X_{\d_{n+1}} ,1,5}  \Eq(2.3.1.sc) $$ 
 
\vglue0.3cm

\\Next we introduce a {\it large set regulator}. Let $X_{\d}$ be a connected 
$1$-polymer in $(\d\math{Z})^{3}$. This is a connected union of $1$-blocks 
defined earlier. We define 
$$\AA_{p} (X_{\d})= 2^{p|X_{\d}|}L^{(D+2)|X_{\d}|} \Eq(2.4reg)$$ 
where for us the dimension of space $D=3$, and $p$ is an integer. 

\vglue0.2cm

\\{\it Small sets} : 
We call a connected polymer $X_{\d}$ {\it small} if $|X_{\d}|\le 2^D$. 
A connected polymer which is not small is called {\it large}. 

\vglue0.2cm
\textRed
\\{\it $L$-polymers and $L$-closure} : Pave 
$\math{R}^{3}$ by a disjoint union of open cubes $L\D$ of edge length $L$, 
called $L$-cubes:

$$L\D== (-{L\over 2}+ m_{1}L,{L\over 2}+ m_{1}L)
\times (-{L\over 2}+  m_{2}L,{L\over 2}+ m_{2}L)
\times (-{L\over 2}+ m_{3}L,{L\over 2}+ m_{3}L) \Eq(2.44reg)$$

\\where $(m_1,m_2,m_3)\in \math{Z}^{3}$. \textBlack
Each $L$-cube is a union of $1$-cubes. Let $\d$ be any member of
the sequence $\{\d_n\}_{n\ge 0}$ where $\d_n=L^{-n}$, $L=3^p$ and $p\ge 2$.
Take the restriction of these $L$-cubes to
$(\d\math{Z})^{3}$ and call the latter cubes $L$-blocks. Each $L$-block is a 
union of $1$-blocks. The paving of $\math{R}^{3}$ by $L$-cubes induces a 
paving of $(\d\math{Z})^{3}$ by $L$-blocks. An $L$-polymer is a union of $L$-
blocks.
{\it We define the $L$-closure of the $1$-polymer $X_{\d}$, denoted 
$\bar{X_{\d}}^{(L)}$, as the $L$-polymer given by the smallest union of  
$L$-blocks containing  $X_{\d}$}. The notions of connectedness and strict
disjointness carry over from the case of $1$ blocks and $1$-polymers. Thus
we say two $L$-blocks from the
$L$-paving are connected if the closures of the corresponding continuum 
$L$-cubes are connected ( i.e. share at least a vertex in common). If they
are not connected we say that they are strictly disjoint. Strictly disjoint
$L$-blocks are separated by a distance $\ge L$. A connected
$L$-polymer is a connected union of $L$-blocks. If two connected $L$-polymers
are not connected to each other we say they are strictly disjoint. Strictly
disjoint $L$-polymers are separated by a distance $\ge L$.

\vglue0.2cm

\\{\it Lemma~2.2} :  Fix any 
integer $p\ge 0$ and let $L$ be sufficiently large depending on $p$. Then 
for any connected $1$-polymer $X_{\d}$   

$$\AA (L^{-1}\bar{X_{\d}}^{(L)}) \le c_{p}\AA_{-p}(X_{\d}) \Eq(2.5reg)$$ 
For $X_{\d}$ a {\it large} connected $1$-polymer, 
$$\AA (L^{-1}\bar{X_{\d}}^{(L)}) \le c_{p}L^{-D-1}\AA_{-p}(X_{\d})\Eq(2.6reg)$$  
Here $c_p =O(1)$ is a constant independent of $L$ and $\d$.  
 
\\{\it Remark} : This is the lattice version of Lemma~1 of [BDH-est]. 
It is purely geometrical and proved in the same way.

\vskip0.3cm 
 
\\2.2 {\bf Field derivatives and norms}: 
The polymer activities in question are
degree $0$ gauge invariant supersymmetric functionals
of the complex bosonic fields $\f, \bar\f$ and the fermionic fields 
$\psi,\bar\psi$. Lattice field derivatives are partial
derivatives with respect to the fields at different points of the lattice.
The fermionic derivative is an antiderivation. 
However in order to measure the size of the lattice
field derivatives it turns out to be useful to generalize the notion of 
field derivatives as directional derivatives (directional in field space).
For the bosonic coefficient this is the lattice transcription
of that given in [BDH].
For the fermionic part there is no clear sense of direction and the definition
we give below suggested to us by David Brydges is both natural and useful.

\textRed

\vglue0.2cm
\\Let $X_{\d}\subset (\d\math{Z})^{3}$ be a connected polymer. 
Let $f_{j}$ for $j=1,....,m$ be $\bf C$ valued functions on  ${\tilde X}^{(2)}
_{\d}$. 
Let $g_{2p}({\bf x},{\bf y})=: g_{2p}(x_1,..,x_p,y_1,..,y_p)$ 
be a $\bf C$ valued function on
$({\tilde X_{\d}}^{(2)})^{p}\times ({\tilde X_{\d}}^{(2)})^{p}$, 
antisymmetric in the $x_j$ and in the $y_j$.
A polymer activity $K(X_{\d},\Phi)$ has the 
representation \equ(2.1) with the coefficients defined in 
\equ(2.1b).
We consider it as a function of $\f,\bar\f,\psi,\bar\psi$ denoted
as ${K}(X_{\d},\f,\psi)$ where we have suppressed
the dependence on $\bar\f,\bar\psi$. We define using the notations of 
\equ(2.1), \equ(2.1b) for the coefficients,

$$D^{2p,m}K(X_{\d},\f,0;f^{\times m},g_{2p})=:  
\int_{X_{\d}^{p}\times X_{\d}^{p}} d{\bf x} d{\bf y}  D_{B}^m  
D_{F}^{2p}(X_{\d},\f,{\bf x},{\bf y};f^{\times m})g_{2p}(x_{1}, 
...x_{p},y_1,..,y_p) \Eq(2.2)$$ 
 
\\where $f^{\times m}=(f_1,..,f_m)$ and

$$D_B^m D_{F}^{2p}K_{2p}(X_{\d},\f,{\bf x},{\bf y};f^{\times m})
=\dpr_{s_{1}}....\dpr_{s_{m}}D_{F}^{2p} K
(X_{\d},\f+s_{1}f_{1},....,\f+s_{m}f_{m},{\bf x},{\bf y})|_{s_1=..=s_m =0}
\Eq(2.211)$$

\\and the $s_j$ are real parameters.

\\Let $\dpr_{\d,\m},\> \dpr_{\d,-\m}$ be the forward and backward
lattice derivative in the direction $ e_{\m}$. Let the index set $S$
be defined as after \equ(2.10reg).
We endow the linear space of $\bf C$ valued functions $f$ as above
with the norm

$$\Vert f\Vert_{C^{2}(X_{\d})}= \sup_{\m,\n \in S}
(\Vert f\Vert_{L^{\infty}(X_{\d})}\Vert,\> 
\Vert \dpr_{\d,\m}f\Vert_{L^{\infty}(X_{\d})},\>
\Vert \dpr_{\d,\m}\dpr_{\d,\n} f\Vert_{L^{\infty}(X_{\d})}) \Eq(2.111a)$$

\\and call the resulting normed space $C^{2}(X_{\d})$.

\\Let $\dpr_{\d,\m_{j}}$, $\m_j\in S$, acting on 
$g_{2p}({\bf x}, {\bf y})$ denote the 
forward or backward lattice derivative with respect to $x_j$ or $y_j$
in the direction $e_{\m_{j}}$. We endow
the linear space  of $\bf C$ valued functions  $g_{2p}({\bf x}, {\bf y})$ on 
$({\tilde X_{\d}}^{(2)})^{p}\times ({\tilde X_{\d}}^{(2)})^{p}$,
antisymmetric in the $x_j$, and in the $y_j$, with 
the norm

$$\Vert g_{2p}\Vert_{C^{2}(X_{\d}^{2p})}= \sup_{\m_{j},\m_{k}\in S
\atop {1\le j,k\le 2p}}
(\Vert g_{2p}\Vert_{L^{\infty}(X_{\d}^{2p})},\>
\Vert \dpr_{\d,\m_{j}} g_{2p}\Vert_{L^{\infty}(X_{\d}^{2p})},\>
\Vert \dpr_{\d,\m_{j}}\dpr_{\d,\m_{k}} g_{2p}\Vert_{L^{\infty}(X_{\d}^{2p})}) 
\Eq(2.112)$$

\\and call the resulting normed space $C^{2}_{a}(X_{\d}^{2p})$.
The above norms always exist for lattice functions since $X_{\d}$ is a 
finite set.

\\ \equ(2.2) then defines a $\bf C$ valued multilinear functional on  
$ C^{2}(X_{\d})^{m} \times C^{2}_{a}(X_{\d}^{2p})$ whose norm is 
defined to be  
 
$$\Vert D^{2p,m}K(X_{\d},\f,0)\Vert = 
\sup_{\Vert f_j\Vert_{C^{2}(X_{\d})}\le 1
\atop{ \Vert g_{2p}\Vert_{C^{2}(X_{\d}^{2p})} 
\le 1\atop \> \forall 1\le j\le m}}\left\vert  
D^{2p,m}K(X_{\d},\f,0;f^{\times m},g_{2p}) \right\vert\Eq(2.3)$$

\\The space of $\bf C$ valued multilinear functionals defined in \equ(2.2) 
which are bounded in the norm \equ(2.3) is complete and thus a Banach space.  

\vglue0.2cm
\\{\it Remarks}: 
It is well known that the space of bounded $\bf C$ valued multilinear 
functionals
on a normed space is complete ( even if the normed space is not).
The completeness follows on using the completeness of the number field 
$\bf C$ by a standard argument.

\vskip0.2cm 
 
\\Let ${\bf h}=(h_F,h_B)$  where $h_F,h_B >0$ are strictly positive real
numbers. We define the following set of norms. The $\bf h$ norm is defined by  
 
$$\Vert K(X_{\d},\f,0)\Vert_{\bf h}= 
\sum_{p=0}^\io\sum_{m=0}^{m_0}{h_F^{2p}\over (p!)^{2}} {h_B^m\over m!} 
\Vert D^{2p,m} K(X_{\d},\f,0)\Vert    \Eq(2.4) $$ 
 
\\In addition we define a {\it kernel} norm with ${\bf h}_*=(h_F,h_{B*})$ 

$$\vert K(X_{\d})\vert_{\bf h_*} = 
\sum_{p=0}^\io\sum_{m=0}^{m_0}{h_F^{2p}\over (p!)^{2}}{h_{B*}^m\over m!} 
\Vert D^{2p,m} K(X_{\d},0,0)\Vert  \Eq(2.5)$$ 
 
\textBlack

\\${\bf h},{\bf h}_*$ will be chosen later in Section~5. We now define
the ${\bf h}, G_{\k}$ norm by 

$$\Vert K(X_{\d})\Vert_{{\bf h},G_{\k}}=
\sup_{\f\in {\cal F}_{\tilde X_{\d}^{(5)}}} 
\Vert K(X_{\d},\f,0)\Vert_{\bf h}G^{-1}_{\k}(X_{\d},\f)\Eq(2.6)$$ 

\\Let $\AA(X_{\d})$ be the large set regulator defined earlier. We then have 
our final set of  norms 
$$\Vert  K\Vert_{{\bf h},G_{\k},\AA,\d}=
\sup_{\Delta_{\d}}\sum_{X_{\d}\supset\Delta_{\d}} 
\Vert (K(X_{\d})\Vert_{{\bf h},G_{\k}}\AA (X_{\d}) \Eq(2.611) $$ 

\\where $\Delta_{\d}=\D\cap (\d\math{Z})^{3}$ and $\D$ is a unit cube in 
$\math{R}^{3}$ as defined earlier, and 

$$\vert K\vert_{{\bf h}_{*},\AA,\d}= 
\sup_{\Delta_{\d}}\sum_{X_{\d}\supset\Delta_{\d}}\vert K(X_{\d})
\vert_{{\bf h}_*}\AA (X_{\d}) \Eq(2.612) $$ 

\\The index $\d$ in our final norms \equ(2.611) and \equ(2.612) indicate that 
the large set 
norm is being taken over polymers in $(\d\math{Z})^{3}$. Under each of above
norms we have Banach spaces. 
Moreover it is easy to verify that the {\it multiplicative (Banach algebra) 
property} holds 
for the polymer activities $\tilde K(X_{\d})$ under the {\bf h}-norm \equ(2.4),
the kernel norm \equ(2.5), and, for activities supported on disjoint
polymers , under the $\bf h, G_{\k} $ norm. 
The multiplicative property
plays a very important role in the estimates in the rest of the paper. We 
therefore state it as {\it Proposition 2.3} below and supply a proof.

\vglue0.2cm
\textRed
\\{\it Proposition 2.3} : Let $X_{\d,1}$, $X_{\d,2}$ denote two connected 
polymers. Let $X_{\d,1}=X_{\d,2}$ or $X_{\d,1}\cap X_{\d,2}=\emptyset$.
$K_j(X_{\d,j},\f,\psi)$, $j=1,2$ are  polymer activities of degree $0$.
Define a new polymer activity

$${\bf K}(X_{\d,1}\cup X_{\d,2},\f,\psi)=
K_1(X_{\d,1},\f,\psi)K_2(X_{\d,2},\f,\psi)$$

\\Then

$$\Vert{\bf K}(X_{\d,1}\cup X_{\d,2},\f, 0)\Vert_{\bf h}\le
\Vert K_1(X_{\d,1},\f, 0)\Vert_{\bf h}
\Vert K_2(X_{\d,2},\f, 0)\Vert_{\bf h}$$

\\The same inequality holds for the ${\bf h}_*$ norm.
If $X_{\d,1}$ and $X_{\d,2}$ are disjoint we have

$$\Vert{\bf K}(X_{\d,1}\cup X_{\d,2})\Vert_{\bf h, G_{\k}}\le
\Vert K_1(X_{\d,1})\Vert_{\bf h, G_{\k}}
\Vert K_2(X_{\d,2})\Vert_{\bf h, G_{\k}}$$

\\{\it Proof}

\\Let $f_j,\ j=1,...,m$ be functions on 
${\tilde X}^{(2)}_{\d,1}\cup {\tilde X}^{(2)}_{\d,2}$ 
and $g_{2p}({\bf x},{\bf y})= g_{2p}(x_1,..,x_p,y_1,..y_p) $ be a function on 
$({\tilde X}^{(2)}_{\d,1}\cup {\tilde X}^{(2)}_{\d,2})^{p}\times
({\tilde X}^{(2)}_{\d,1}\cup {\tilde X}^{(2)}_{\d,2})^{p}=
({\tilde X}^{(2)}_{\d,1}\cup {\tilde X}^{(2)}_{\d,2})^{2p}$. 
$g_{2p}$ is antisymmetric in the $x_j$ and in the $y_j$.
By definition

$$\Vert D^{2p,m}{\bf K}(X_{\d,1}\cup X_{\d,2},\f,0)\Vert=
\sup_{\Vert f_j\Vert_{C^2(X_{\d,1}\cup X_{\d,2})}\le 1
\atop \Vert g_{2p}\Vert_{C^2((X_{\d,1}\cup X_{\d,2})^{2p})}\le 1}
\vert D^{2p,m}{\bf K}(X_{\d,1}\cup X_{\d,2},\f,0;f^{\times m},
g_{2p})\vert$$

\\where $f^{\times m}=(f_1,...,f_m)$,
$f^{\times M}=\{f_i\}_{i\in M}$ and $M\subset \{1,2,.....,m\}$.
We extend the coefficients of 

\\$K_j(X_{\d,j},\f,\psi)$, $j=1,2$ to 
$X_{\d,1}\cup X_{\d,2}$
by declaring that they have support in  $X_{\d,j}$. Now  $D_{F}^{2p}$
is a partial (anti) derivation of order $2p$. $D_{B}^{2m}$ a derivation of
order $m$. Distributing $D_{F}^{2p}$ and $D_{B}^{2m}$ on the product of 
polymer activities gives

$$D_{B}^{m}D_{F}^{2p}{\bf K}(X_{\d,1}\cup X_{\d,2},\f,
x_1,..,x_p,y_1,..,y_p); f^{\times m})=
\sum_{p_1+p_2=p}
\sum_{m_1+m_2=m}\sum_{M_1\cup M_2=\{1,...,m\}\atop
{M_1\cap M_2=\emptyset\atop
|M_1|=m_1, |M_2|=m_2}}\sum_{I, J\subset \{1,...,p\}\atop
|I|=|J|=p_1}$$
$$D_{B}^{m_1}D_{F}^{2p_{1}}K_{1}
(X_{\d,1},\f,{\bf x}_{I},{\bf y}_{J};f^{\times M_1})
D_{B}^{m_2}D_{F}^{2p_{2}}K_{2}
(X_{\d,2},\f,{\bf x}_{I^{c}},{\bf y}_{J^{c}};f^{\times M_2}) \times $$
$$g_{2p}({\bf x}_{I},{\bf x}_{I^{c}},{\bf y}_{J}, {\bf y}_{J^{c}})
\times (-1)^{\sharp} \Eq(2.616) $$

\\where $(-1)^{\sharp}$ is a sign factor which plays no role in the 
norm bounds to
follow, $I^{c},J^{c}$ are repectively the complements of $I,J$ in
$\{1,...,p\}$. We have $|I^{c}|=|J^{c}|=p_2$. We now integrate this
with respect to $x_1,..,x_p,y_1,..y_p$ in 
$(X_{\d,1}\cup X_{\d,2})^{p}\times (X_{\d,1}\cup X_{\d,2})^{p}$. Because of 
the support properties of the coefficients the integral splits over the
products on the right hand side. We get 

$$D^{2p,m}{\bf K}(X_{\d,1}\cup X_{\d,2},\f,0;f^{\times m},g_{2p})=
\sum_{p_1+p_2=p}
\sum_{m_1+m_2=m}\sum_{M_1\cup M_2=\{1,...,m\}\atop
{M_1\cap M_2=\emptyset\atop
|M_1|=m_1, |M_2|=m_2}}\sum_{I, J\subset \{1,...,p\}\atop
|I|=|J|=p_1}  $$
$$\int_{X_{\d,1}^{2p_1}}d{\bf x}_{I}d{\bf y}_{J} 
\int_{X_{\d,2}^{2p_2}}d{\bf x}_{I^{c}}d{\bf y}_{J^{c}} 
D_{B}^{m_1}D_{F}^{2p_{1}}K_{1}
(X_{\d,1},\f,{\bf x}_{I},{\bf y}_{J};f^{\times M_1})
D_{B}^{m_2}D_{F}^{2p_{2}}K_{2}
(X_{\d,2},\f,{\bf x}_{I^{c}},{\bf y}_{J^{c}};f^{\times M_2}) \times $$
$$g_{2p}({\bf x}_{I},{\bf x}_{I^{c}},{\bf y}_{J}, {\bf y}_{J^{c}})
\times (-1)^{\sharp} \Eq(2.616) $$

\\where $X_{\d,j}^{2p_j}= X_{\d,j}^{p_j}\times X_{\d,j}^{p_j} $. Define

$$\eqalign{{\tilde g}_{2p_1}({\bf x}_{I},{\bf y}_{J}) &=
\int_{X_{\d,2}^{2p_2}}d{\bf x}_{I^{c}}d{\bf y}_{J^{c}} 
D_{B}^{m_2}D_{F}^{2p_{2}}K_{2}
(X_{\d,2},\f,{\bf x}_{I^{c}},{\bf y}_{J^{c}};f^{\times M_2})
g_{2p}({\bf x}_{I},{\bf x}_{I^{c}},{\bf y}_{J}, {\bf y}_{J^{c}})\cr
&= D^{2p_2,m_2}K_{2}(X_{\d,2},\f,0;g_{2p}({\bf x}_{I},\cdot,{\bf y}_{J},\cdot 
),f^{\times M_2}) \cr} \Eq(2.617)$$

\\where the dependence of ${\tilde g}_{2p_1}({\bf x}_{I},{\bf y}_{J})$ on
$K_2,X_{\d,2},p_2,f^{\times M_2}$ has been suppressed. Note that
${\tilde g}_{2p_1}({\bf x}_{I},{\bf y}_{J})$ is antisymmetric in the 
$\{x_i:\>i \in I\}$ and in the $\{y_j:\> j \in J\}$ and therefore qualifies as
a test function. 

\\From \equ(2.616) and \equ(2.617) we have

$$D^{2p,m}{\bf K}(X_{\d,1}\cup X_{\d,2},\f,0;f^{\times m},g_{2p})=
\sum_{p_1+p_2=p}
\sum_{m_1+m_2=m}\sum_{M_1\cup M_2=\{1,...,m\}\atop
{M_1\cap M_2=\emptyset\atop
|M_1|=m_1, |M_2|=m_2}}\sum_{I, J\subset \{1,...,p\}\atop
|I|=|J|=p_1}$$
$$D^{2p_1,m_1}K_{1}(X_{1,\d},\f,0;f^{M_1},
{\tilde g}_{2p_1}) \times (-1)^{\sharp}\Eq(2.618)$$

\\Therefore

$$|D^{2p,m}{\bf K}(X_{\d,1}\cup X_{\d,2},\f,0;f^{\times m},g_{2p})|\le
\sum_{p_1+p_2=p}
\sum_{m_1+m_2=m}\sum_{M_1\cup M_2=\{1,...,m\}\atop
{M_1\cap M_2=\emptyset\atop
|M_1|=m_1, |M_2|=m_2}}\sum_{I, J\subset \{1,...,p\}\atop
|I|=|J|=p_1}$$
$$\Vert D^{2p_1,m_1}K_{1}(X_{1,\d},\f,0)\Vert \> 
\prod_{i\in M_1}\Vert f_i\Vert_{C^2(X_{1,\d})}
\Vert{\tilde g}_{2p_1}
\Vert_{C^{2}(X_{1,\d}^{2p_1})}\Eq(2.619) $$

\\From \equ(2.617) we have for $0\le k\le 2$

$$\partial_{\d}^k {\tilde g}_{2p_1}({\bf x}_{I},{\bf y}_{J})=
 D^{2p_2,m_2}K_{2}(X_{\d,2},\f,0;
\partial_{\d}^k g_{2p}({\bf x}_{I},\cdot,{\bf y}_{J},\cdot ),f^{\times M_2})
$$

\\where $\partial_{\d}^k$ is the lattice partial derivative of degree $k$ with 
respect to  ${\bf x}_{I},{\bf y}_{J}$ in multi-index notation. Whence 

$$\vert\partial_{\d}^k {\tilde g}_{2p_1}({\bf x}_{I},{\bf y}_{J})\vert \le
\Vert D^{p_2,m_2}K_{2}(X_{\d,2},\f,0)\Vert
\prod_{j\in M_2}\Vert f_j\Vert_{C^2(X_{\d,2})}
\Vert \partial_{\d}^k g_p({\bf x}_{I},\cdot,{\bf y}_J,\cdot)
\Vert_{C^2( X_{\d,2}^{2p_2})} $$

\\and therefore

$$\Vert {\tilde g}_{2p_1} \Vert_{C^2(X_{\d,1}^{2p_1})}\le 
\Vert D^{p_2,m_2}K_{2}(X_{\d,2},\f,0)\Vert
\prod_{j\in M_2}\Vert f_j\Vert_{C^2(X_{\d,2})}
\Vert  g_{2p}\Vert_{C^2( X_{\d,2}^{2p_2}\times X_{\d,1}^{2p_1})}\Eq(2.620) $$

\\Now $X_{\d,2}^{2p_2}\times X_{\d,1}^{2p_1}\subset 
(X_{\d,2}\cup X_{\d,1})^{2p}$ where $p=p_1+p_2$.
Therefore from \equ(2.619) and \equ(2.620)
we get

$$\Vert D^{2p,m}{\bf K}(X_{\d,1}\cup X_{\d,2})\Vert \le
\sum_{p_1+p_2=p}
\sum_{m_1+m_2=m}\sum_{M_1\cup M_2=\{1,...,m\}\atop
{M_1\cap M_2=\emptyset\atop
|M_1|=m_1, |M_2|=m_2}}\sum_{I, J\subset \{1,...,p\}\atop
|I|=|J|=p_1}$$
$$\Vert D^{2p_1,m_1}K_{1}(X_{1,\d},\f,0)\Vert \> 
\Vert D^{2p_2,m_2}K_{2}(X_{2,\d},\f,0)\Vert $$

\\Now

$$\sum_{M_1\cup M_2=\{1,...,m\}\atop
{M_1\cap M_2=\emptyset\atop
|M_1|=m_1, |M_2|=m_2}}\sum_{I, J\subset \{1,...,p\}\atop
|I|=|J|=p_1}\>1 = {m!\over m_1! m_2!} {(p!)^{2}\over (p_1!)^{2}(p_2!)^{2}}
$$

\\Therefore

$$\Vert D^{2p,m}{\bf K}(X_{\d,1}\cup X_{\d,2},\f,0)\Vert \le
\sum_{p_1+p_2=p}\sum_{m_1+m_2=m} 
{m!\over m_1! m_2!} {(p!)^{2}\over (p_1!)^{2}(p_2!)^{2}}
\Vert D^{2p_1,m_1}K_{1}(X_{1,\d},\f,0)\Vert \times $$ 
$$\Vert D^{2p_2,m_2}K_{2}(X_{2,\d},\f,0)\Vert \Eq(2.621) $$

\\Multiply both sides of the previous inequality by $h_B^m/m!$ and 
$h_F^{2p}/(p!)^{2}$. 
Sum over integers $m$, $0\le m\le m_0$, and over all integers 
$p\ge 0$ to obtain
$$\Vert{\bf K}(X_{\d,1}\cup X_{\d,2},\f,0)\Vert_{\bf h}\le\Vert K_1(X_{\d,1},\f,0)\Vert_{\bf h}
\Vert K_2(X_{\d,2},\f,0)\Vert_{\bf h}$$ 

\\This proves the first inequality of Proposition 2.3. The second inequality
follows from the first because for union of disjoint sets

$$G_{\k}(X_{\d}\cup Y_{\d},\f)=G_{\k}(X_{\d},\f)G_{\k}(Y_{\d},\f)$$
\bull

\textBlack

\vskip0.5cm

\\{\bf 3. THE RG MAP } 
\numsec=3\numfor=1 
\vskip.3truecm 
In this section we describe the RG map applied to the generic density
in the polymer representation given in \equ(1.22). This is a lattice 
transcription of the continuum RG map described in [BMS], ( see also [M]).
This goes in several steps. First we must perform the
fluctuation integration and rescaling (see \equ(1.48))

$${\cal Z}'(L^{-1}\L_{L^{-1}\d},\f)=S_{L}\mu_{\G}*{\cal Z}(\L_{\d},\Phi) 
\Eq(3.0) $$

\\where $\L_{\d}\subset (\d\math{Z})^{3}$ is the volume arrived at after a 
certain number of previous RG steps and $\G$ is the fluctuation covariance for 
the next step. $\G$ is one of the covariances $\G_{n}$ of Theorem~1.1 and
has the finite range property stated in that theorem. Thus after $n$ RG steps
(see \equ(1.45a)-\equ(1.48a))  $\d=\d_{n}$, $\G= \G_{n}$, 
$\L_{\d}=\L_{N-n,n}$ and $L^{-1}\L_{L^{-1}\d}=\L_{N-n-1,n+1}$. 

\\The polymer representation \equ(1.22) for ${\cal Z}(\L_{\d})$
is parametrized by the {\it coordinates}
$(V,K)$ on the scale $\d$ where $V$ is a local functional (potential): 

$$V(X_{\d})=\sum_{\D_{\d}\subset X_{\d}}V(\D_{\d}) \Eq(3.01)$$

\\Let ${\tilde V}(X_{\d},\Phi)$ be an arbitrary local supersymmetric 
functional with ${\tilde V}(X_{\d},0)=0$. We will see
that the polymer representation is preserved under the RG transformation
\equ(3.0) with new coordinates ${\tilde V}_{L},{\cal F}( K)$ on the 
next scale $L^{-1}\d$. ${\cal F}$ depends on ${\tilde V}$. 
The finite range property of $\G$ leads to
a simple description of this map :

$$V\rightarrow {\tilde V}_{L},\quad  
{\tilde V}_{L}(\D_{L^{-1}\d},\Phi)=(S_L V)(\D_{L^{-1}\d},\Phi)
={\tilde V}(L\D_{\d},S_L \Phi) $$

$$K\rightarrow  {\cal F}( K) :\> {\cal F}( K)(X_{L^{-1}\d},\Phi)
=\int d\m_{\G}(\xi){\cal B}K(LX_{\d},\xi, S_{L}\Phi) \Eq(3.02) $$

\\where ${\cal B}K$ is a ${\tilde V}$ dependent nonlinear functional of $K$
to be presently described. We call this map the {\it fluctuation map}.

\\We can take advantage of the arbitrariness of the local potential 
$\tilde V$ in the above map so as to remove the
expanding ( relevant) parts $F$ in the polymer activity ${\cal F}(K)$
and compensate by a change ${\tilde V}_{L}(F)$
in the local potential ${\tilde V}_{L}$ in such a way that the evolved 
density ${\cal Z}'(L^{-1}\L_{L^{-1}\d})$  on the left hand side of \equ(3.0) 
remains unchanged. This operation gives rise to the 
{\it extraction} map, [BDH-est]

$${\tilde V}_{L}\rightarrow V'(F)={\tilde V}_{L}-{\tilde V}_{L}(F),
\quad {\cal F}(K)\rightarrow K'={\cal E}({\cal F}(K),F) 
\Eq(3.03) $$

\\where the image is on the same scale $L^{-1}\d$.
$V'(F)$ and the nonlinear map $\cal E$ have simple expressions which are 
lattice transcriptions of those given in [BDH-est]. 
The composition of the fluctation map \equ(3.02) and 
the extraction map \equ(3.03) gives the {\it RG map} 

$$f : \quad f(V,K)=(f_{V}(V,K), f_{K}(V,K))$$

\\where

$$f_{V} :  V\rightarrow {\tilde V}_{L}\rightarrow V'(F) $$

$$f_{K} :  K\rightarrow {\cal F}(K) \rightarrow K'={\cal E}({\cal F}(K),F) 
\Eq(3.04) $$

\\The operation of
extraction leads in particular to a discrete flow of the coupling
constants in $V$ on scale $L^{-1}\d$  provided we choose $F,{\tilde V_{L}}(F)$ 
appropriately. The expanding functionals will be gathered in the local 
potential $V'(F)$ whereas the polymer activity ${\cal E}({\cal F}(K),F)$ 
will be a contracting (irrelevant) error term.

\vskip.3truecm  
\\3.1 {\bf The Fluctuation Map} 
 
We now construct the map \equ(3.02) starting from \equ(3.0) with the density
in the polymer representation \equ(1.22). In performing the fluctuation 
integration

$$\mu_{\G}*{\cal Z}(\L_{\d},\Phi)=\int d\m_{\G}(\xi) \sum_{N} {1\over N!} 
e^{-V(X_{\d}^{(c)},\Phi +\xi)} \sum_{X_{\d,1},..,X_{\d,N}} 
\prod_{j=1}^N K(X_{\d,j},\Phi+\xi)  \Eq(3.501) $$

\\we will exploit 
the independence of $\xi (x)$ and $\xi (y)$ when $|x-y|\ge L$. To 
this end we construct an $L$-paving of $\L_{\d}$ and the $L$-closure of
$\bar{X_{\d}}^{(L)}$ of a connected $1$-polymer $X_{\d}$ as in the paragraph 
preceding Lemma~2.1. The $1$-polymers will be 
combined into larger connected $L-$polymers which by definition are  
connected unions of $L$-blocks, ( for the relevant definitions intervening here
and in the following see the paragraph on {\it $L$-polymers and $L$-closures}
before Lemma~2.1). The combination is performed in 
such a way that the new polymers are associated to independent 
functionals of $\xi$. This is the lattice adaptation of Section~3.1 of
[BMS]. 
 
\vskip0.2cm
\\Define the polymer activity P, supported on  unit blocks, by: 
 
$$P(\Delta_{\d} ,\xi ,\Phi) =e^{-V(\Delta_{\d} ,\xi + \Phi)} - e^{-\tilde{V} 
(\Delta_{\d},\Phi )}  \Eq(3.5.v)$$ 

\\with $\tilde{V}$ , to be chosen. $\tilde{V}(\Delta_{\d},\Phi )$ is required
to satisfy $\tilde{V}(\Delta_{\d},0)=0$.
{\it In the following $V,K$ has field 
argument $\xi + \Phi $ whereas $\tilde{V}$ depends only on $\Phi$ }. 
The dependence of P on $\xi ,\Phi$ is as defined above. 

\\$X_{\d}^{c}=\L_{\d}\setminus \cup_{j=1}^{N}X_{\d,j}$ is a 
union of disjoint $1$-blocks $\D_{\d}$. Therefore 

$$e^{-V(X_{\d}^{c})}= {\prod}_{{\Delta_{\d}}\subset  
{X}_{\d}^{c}}[e^{-\tilde{V}(\Delta_{\d})} + P(\Delta_{\d})] $$ 
 
\\Expand the product and insert the expansion into the integrand of
in \equ(3.501) which gives 
 
$${\rm integrand} =\sum_{N} {1\over N!}   
{\sum}_{(X_{\d,j}),(\Delta_{\d,i})} 
 e^{-\tilde{V}(X_{\d,0})} 
{\prod}_{j=1}^{N} {K}(X_{\d,j}){\prod}_{i=1}^{M} {P}(\Delta_{\d,i}) 
\Eq(3.4.v)$$  
 
\\where $X_{\d,0} = \Lambda_{\d} \setminus (\cup X_{\d,j}) \cup (\cup 
\Delta_{\d,i})$. Let $Y_{\d}$ be the $L-$closure of $(\cup X_{\d,j}) 
\cup (\cup \Delta_{\d,i})$ and let $Y_{\d,1},\dots , Y_{\d,P}$ be the 
connected components 
of $Y_{\d}$.  These are $L-$ polymers. Let $f$ be the function that maps 
$\pi: = (X_{\d,j}),(\Delta_{\d,i})$ into $\{Y_{\d,1},\dots ,Y_{\d,P} \}$. 
Now we 
perform the sum over $(X_{\d,j}),(\Delta_{\d,i})$ in \equ(3.4.v) by summing 
over $\pi \in f^{-1} (\{Y_{\d,1},\dots ,Y_{\d,P} \})$ and then 
$\{Y_{\d,1},\dots ,Y_{\d,P} \}$.  The result is: 
 
$${\rm integrand} =\sum_{N} {1\over N!}   
{\sum}_{(Y_{\d,j})} 
 e^{-\tilde{V}(Y_{\d}^{c})} 
{\prod}_{j=1}^{N} {\cal B}K(Y_{\d,j}) 
\Eq(3.12.v)$$ 
 
\\where the sum is over strictly disjoint connected $L$ polymers and
 
$$({\cal B}K)(Y_{\d}) =  \sum_{N+M\ge 1}{1\over N !M!} 
{\sum}_{(X_{\d,j}),(\Delta_{\d,i})\rightarrow \{Y \}}  
e^{-\tilde{V}(X_{\d,0})} 
{\prod}_{j=1}^{N}K(X_{\d,j}) 
{\prod}_{i=1}^{M} {P}(\Delta_{\d,i})   \Eq(3.13)$$ 
 
\\where $X_{\d,0}=Y\setminus (\cup X_{\d,j})\cup (\cup\Delta_{\d,i})$  
and the $\rightarrow $ is the map $f$. In other words the sum in \equ(3.13)
is over distinct $\Delta_{\d,i}$ and disjoint $1$-polymers $ X_{\d,j}$
such that their $L$-closure is the connected $L$-polymer $Y_{\d}$.

\\We now perform the fluctuation integration of \equ(3.12.v) over $\xi$ 
followed by rescaling.
Since $\tilde{V}(Y_{\d}^{c})$ is independent of $\xi$ the $\xi$ integration
factors through and acts on the product of polymer activities
$\prod_{j}({\cal B}K)(Y_{\d,j})$. A polymer activity 
$({\cal B}K)(Y_{\d,j})$ belongs to $\Omega^{0}(\tilde Y_{\d,j}^{(2)})$.
The $Y_{\d,j}$ are strictly disjoint 
connected $L$-polymers and thus necessarily separated from each other by a 
distance $\ge L$. The $2$-collar attached $L$-polymers  
$\Omega^{0}(\tilde Y_{\d,j}^{(2)})$ are therefore separated from each other
by a distance $\ge L-4$. The fluctuation covariance $\G$ has finite range
$L/2$ and for $L$ sufficiently large $L-4\ge L/2$. 
Therefore the fluctuation integration over the product of
polymer activities factorizes. We now follow this up
by applying the rescaling operator to both sides. This has the effect of
bringing us back to $1$-polymers but on the scale $L^{-1}\d$.
Therefore we obtain 

$$(S_{L}\mu_{\G}*{\cal Z})(L^{-1}\L_{L^{-1}\d},\Phi)=
\sum_{N} {1\over N!}   
{\sum}_{(X_{L^{-1}\d,j})} 
 e^{-\tilde{V_{L}}(X_{L^{-1}\d}^{c},\Phi)} 
{\prod}_{j=1}^{N} \int d\mu_{\G}(\xi){\cal B}K(LX_{\d,j},S_{L}\Phi,\xi) 
\Eq(3.14)$$ 
 
\\where $X_{L^{-1}\d,j}=X_{j}\cap (L^{-1}\d\math{Z})^{3}$ ( as well as
$X_{\d,j}=X_{j}\cap (\d\math{Z})^{3}$ ) are disjoint $1$-polymers,
$X_{L^{-1}\d}^{c}=L^{-1}\L_{L^{-1}\d}\setminus \cup_{j}X_{L^{-1}\d,j}$.
and $\tilde{V_{L}}(\D_{L^{-1}\d})= S_{L}{\tilde V}(L\D_{\d})$.
This gives the fluctuation map \equ(3.02) : $V\rightarrow {\tilde V}_{L}$,
$K\rightarrow {\cal F}(K)$ with ${\cal B}K$ defined as above. 
At the same time we have
shown that the polymer representation is stable with respect to the RG 
transformation.

\\Consider

$${\cal F}( K)(X_{L^{-1}\d},\Phi)
=\int d\m_{\G}(\xi){\cal B}K(LX_{\d},\xi, S_{L}\Phi)$$

\\By construction ${\cal B}K$ is supersymmetric. Therefore since the 
supersymmetry operator commutes with the measure ${\cal F}( K)$ is also 
supersymmetric. Now since $P(\Delta_{\d} ,\xi ,0)$ and $K(X_{\d},\xi)$
vanish for $\xi=0$ ( the latter by hypothesis, see \equ(2.111)) it follows that
${\cal B}K(LX_{\d},\xi, 0)$ also vanishes for $\xi=0$. Therefore by Lemma~1.1

$${\cal F}( K)(X_{L^{-1}\d},0)
=\int d\m_{\G}(\xi){\cal B}K(LX_{\d},\xi, 0)={\cal B}K(LX_{\d},0,0)=0
\Eq(3.141) $$

\\Thus the condition \equ(2.111) is satisfied by the new polymer activities.
This implies in particular that no field independent relevant parts are 
generated by the fluctuation integration as a consequence of supersymmetry.

\vskip.3truecm 
\\3.2 {\bf  Extraction } 
 
\vskip0.2cm

\\Let $\d'=L^{-1}\d$ and let $\L'=L^{-1}\L_{\d'}$.
The fluctuation map gave us ${\tilde V}_{L}, {\cal F}(K)$ as the coordinates
of the evolved density ${\cal Z}(\L')$. We want to change the
local potential ${\tilde V}_{L}$ and the polymer activity ${\cal F}(K)$
simultaneously such that ${\cal Z}(\L')$ remains invariant.
To this end let $P (\Phi (x))$ be a {\it local} polynomial, which means 
that it is a polynomial in $\Phi (x)$ for $x\in \L'$. 
Furthemore we require that $P(0)=0$, i.e. $P$ has no field independent part. 

\\Given $\D_{\d'}$ a unit block in $\L'$
we consider a change in ${\tilde V}_{L}(\D_{\d'})$ of the form 

$${\tilde V}_{L}(F)(\D_{\d'})= \sum_{P}\int_{\D_{\d'}} dx \  \alpha_{P} (x) 
P(\Phi (x)) \Eq(33.2d)  $$ 

\\where the sum ranges over finitely many local polynomials and, 
for each such $P$, $\alpha _{P}(x)$ has the form 

$$ \alpha _{P}(x)  = \sum_{X_{\d'}\supset x}\alpha _{P}(X_{\d'},x) 
\Eq(33.2cc) $$ 

\\such that $\alpha _{P}(X_{\d'},x)=0$ if $x\notin X_{\d'}$, 
$\alpha_{P} (X_{\d'},x)=0$ if $X_{\d'} \not \subset \Lambda' $ and 
$\alpha_{P} (X_{\d'},x)=0$ if $X_{\d'}$ is not a small set (see definition
after \equ(2.4reg)).
The corresponding change in ${\cal F}(K)$ is given in terms of the 
{\it relevant parts} 
$$ F(X_{\d'},\Phi)= \sum_{P}\int_{X_{\d'}} dx \  \alpha _{P} (X_{\d'},x) 
P(\Phi (x)), \quad 
F(X_{\d'},\D_{\d'})=  
\sum_{P} \int_{\D_{\d'}}dx \  \alpha _{P} (X_{\d'},x) 
P(\Phi (x)) \Eq(33.2aa) $$ 
 
\\Note that $F(X_{\d'},0)=0$.

\vskip0.2cm
\\{\it Extraction Map: } 

\vskip0.2cm

\\{\it Theorem 3.1 ( after Brydges,Dimock and Hurd, [BDH-est])} : 
Given $F,\> {\tilde V}_{L}(F)$ as above 
there exists a polymer activity which is a non-linear  
functional  ${\cal E}({\cal F}(K),F)$ of ${\cal F}(K),\> F$ such that 

$$V_{L}\rightarrow V'(F)={\tilde V}_{L}-{\tilde V}_{L}(F), 
\quad \quad {\cal F}(K)\rightarrow K'={\cal E}({\cal F}(K),F) \Eq(33.17b)$$

\\preserves the polymer representation for the density ${\cal Z}(\L')$ 
with new coordinates coordinates $(V',\> K')$ satisfing 
$V'(F)(\D_{\d'},0)=K'(X_{\d'},0)=0$. Let
${\cal E}_{1}$ denote the linearization of ${\cal E}$. Then the linearization 
of the extraction map is given by

$${\cal E}_{1}({\cal F}(K),F) = {\cal F}(K)-Fe^{-{\tilde V}_{L}},  
\quad  \quad  V'(F) ={\tilde V}_{L}- {\tilde V}_{L}(F) 
\Eq(33.17c)$$ 

\\We say that ${\tilde V}_{L}$ is stable with respect to perturbation $F$ 
if there are positive numbers $f(X)$ such that 

$$\Vert e^{-{\tilde V}_{L}(\D_{\d'})-
\sum_{X_{\d'}\supset \D_{\d'}}z(X)F(X_{\d'},\D_{\d'})}
\Vert_{{\bf h},G_{\k}} \le 2  \Eq(33.17d) $$

\\for all complex numbers $z(X_{\d'})$ with $|z(X_{\d'})|f(X_{\d'})\le 2$. 
Assume that ${\tilde V}_{L}$ is stable.
Then ${\cal E}({\cal F}(K),F)$ is norm analytic and satisfies the bounds

$$\Vert {\cal E}({\cal F}(K),F)\Vert_{{\bf h},G_{\k},\AA,\d'}
\le O(1)(\Vert {\cal F}(K) \Vert_{{\bf h},G_{\k},\AA_{1},\d'}
+\Vert f\Vert_{\AA_{3},\d'}) \Eq(33.17e) $$

$$\vert {\cal E}({\cal F}(K),F)\vert_{{\bf h},\AA,\d'}
\le O(1)(\Vert {\cal F}(K) \Vert_{{\bf h},\AA_{1},\d'}
+\Vert f\Vert_{\AA_{3},\d'}) \Eq(33.17f) $$

\\ {\it Proof} : 
This is a restatement of Theorem 5 in Sec. 4.2 of [BDH-est] with the 
substitution $(\tilde V_{L},{\cal F}(K))$ for $(V,K)$, adapted to the 
lattice. The proof of Theorem 5 exploited Lemmas~10, 11, 12, 13 the last of 
them
providing the extraction formula in equation (121), page 781 of [BDH-est].
In [BDH-est] the continuum unit blocks are open. Our lattice unit
blocks are lattice restrictions of continuum  open unit cubes. Overlap 
connectedness is replaced by connectedness. With this in mind
the proofs of Lemmas~10, 11, 12, 13 go through intact on the lattice providing
the extraction map above. 
The estimates in Theorem 5 on the norms of 
${\cal E}(K,F)$ together with norm analyticity remain valid on the
lattice. \bull. 

\vskip0.2cm
\\{\it Remark} :  The stability property \equ(33.17d) is proved in Section 5 
once we have chosen $\tilde V$ appropriately. The estimate \equ(33.17e) on the 
extraction operator $\cal E$ plays an essential role and is exploited in 
Section 5. 
 
\vskip0.2cm
\\{\it Formal infinite volume limit}:
We reestablish the notations leading to \equ(1.48a). Choose $\d=\d_{n}$,
$\G=\G_{n}$, $\d'=L^{-1}\d=\d_{n+1}$. $\L_{\d}= \L_{N-n,n},\>
\L_{\d'}= \L_{N-n-1,n+1}$ and ${\cal F}= {\cal F}_{n+1}$ in \equ(3.02).
The RG transformation $T_{N-n-1,n+1}$ of \equ(1.48a)
induces the RG map $f_{N-n-1,n+1}(V,K)$ of \equ(3.04) for the 
coordinates of the density ${\cal Z}_{n-1}(\L_{N-n,n})$
in the polymer representation.
$\alpha _{P}(X_{\d_{n+1}},x)$ in \equ({33.2cc}) is chosen later in 
Section 4. This choice will be local, in the sense that it is determined 
by ${\tilde V}_{L}(\D_{\d_{n+1}}), \ \Delta_{\d_{n+1}} \subset X_{\d_{n+1}}$ 
and by 
${\cal F}_{n+1}(K) (X_{\d_{n+1}})$.  Lemma~13 and 
equation (112) of [BDH-est] imply that 
${\cal E}({\cal F}_{n+1}(K),F))(X_{\d_{n+1}})$ 
also is local: it is determined by ${\cal F}_{n+1}(K) (Y_{\d_{n+1}}), 
\ Y_{\d_{n+1}} \subset X_{\d_{n+1}}$ and ${\tilde V}_{L}(\D_{\d_{n+1}}), 
\ \D_{\d_{n+1}} \subset \tilde{X_{\d_{n+1}}}$, where $\tilde{X_{\d_{n+1}}}$ 
is a neighbourhood of 
$X_{\d_{n+1}}$, namely the union of $X_{\d_{n+1}}$ with all small sets that 
intersect $X_{\d_{n}}$. Therefore the $K$ component of the map $f_{N-n-1,n+1}$
representing the action of the $n+1$th step of RG, namely
$f_{N-n-1,n+1,K}(K,V)(X_{\d_{n+1}},\Phi )$ is 
independent of $N$ for all $N$ large enough so that $\L_{N-n-1,n+1}$ 
contains $\tilde{X_{\d_{n+1}}}$.  Thus $\lim_{N\rightarrow 
\infty}f_{N-n-1,n+1,K}(K,V)(X_{\d_{n+1}},\Phi )$ exists pointwise in 
$X_{\d_{n+1}}$.  
In this paper we are studying the action of this pointwise infinite volume 
limit called the {\it formal infinite volume limit}.

\vskip.3truecm \\{\it 3.3   Appendix}: 
\textRed

\\We record here some definitions which have either already been used or will 
be used later. The object
is to be able to move scaling past fluctuation integration. \textBlack 

\vglue0.2cm
\\We define for $x,y\in (L^{-1}\d\math{Z})^{3}$ and any covariance $u$ on
$(\d\math{Z})^{3}$ 

$$u_{L}(x-y)= S_{L^{-1}}u(x-y)= L^{2d_{s}}u(L(x-y)) \Eq(3.151)$$ 

\\Since the fluctuation covariance $\G$ defined on $(\d\math{Z})^{3}$
has finite range $L/2$ we have that $\G_{L}$ defined on 
$(L^{-1}\d\math{Z})^{3}$
has finite range $1/2$. We recall from section~1.3 that a polymer $X_{\d}$
is defined by $X_{\d}=X\cap (\d\math{Z})^{3}$ where $X$ is a continuum polymer.
We define the rescaling of polymer activities by 

$$ S_{L}K(X_{L^{-1}\d},\Phi)=K_{L}(X_{L^{-1}\d},\Phi)
=K(LX_{\d},S_{L}\Phi) \Eq(3.152) $$
 
\\We write the fluctuation integration of the polymer activity  
$K(X_{\d},\Phi,\xi)$ with respect to $\mu_{\G}$    as

$$K^{\sharp}(X_{\d},\Phi) = \int d\mu_{\G}(\xi)
K(X_{\d},\Phi,\xi) \Eq(3.153)    $$.

\\We write the fluctuation integration of the polymer activity 
$K(X_{L^{-1}\d},\Phi,\xi)$ with respect to $\mu_{\G_{L}}$ as

$$K^{\natural}(X_{L^{-1}\d},\Phi) = \int d\mu_{\G_{L}}(\xi)
K(X_{L^{-1}\d},\Phi,\xi) \Eq(3.154)    $$.  

\\We define 

$$\SS_{L}=S_{L}{\cal B} \Eq(3.155)   $$. 

\\With these notations it is easy to see that
the fluctuation map can be written as

$$ {\cal F}(K)(X_{L^{-1}\d},\Phi) = ({\cal B}K)^{\sharp}(LX_{\d},S_L\Phi )
=(\SS_{L}K)^{\natural}(X_{L^{-1}\d},\Phi) \Eq(3.15)$$

\vskip.5truecm 
\\{\bf 4. THE RENORMALIZATION GROUP MAP APPLIED  } 
\numsec=4\numfor=1 
\vskip.5truecm 
 
\\In this section we specify the RG map of Section~3 by making choices for 
the local potential $\tilde{V}$, and relevant parts $ F$. $\tilde{V}$ is chosen
via first order perturbation theory. $F$ is chosen so as to remove the 
expanding part of the fluctuation map. This is the extraction step. This will
be done in second order perturbation theory as well as in the error term.
We will follow closely the strategy in Section~4 of [BMS]. We will use the
notations established in the Appendix to section~3.3, \equ(3.151)-\equ(3.15).
We take $\d=\d_{n}$, $\G=\G_n$. Recall that, see \equ(1.41a),
$\a= {3+\e\over 2}$ where we take $0<\e<1$. The field scaling dimension 
is $d_s={3-\e\over 4}$, see \equ(1.41),\equ(1.42).

\vglue0.2cm
\\We assume that starting
from the unit lattice where only the local potential \equ(0.21) is present
$n$ steps of the renormalization group map has been carried out. This produces
a new local potential

$$V_{n}(\D_{\d_n},\Phi)=V(\D_{\d_n},\Phi,C_{n},g_{n},\m_{n})=
g_{n}\int_{\D_{\d_n}} dx:(\Phi\bar\Phi)^2(x):_{C_{n}}+ 
\m_{n}\int_{\D_{\d_n}} dx(\Phi\bar\Phi)(x)\Eq(4.1)$$ 

\\together with a polymer activity $K_{n}$ supported on connected polymers in 
$(\d_{n}\math{Z})^3$. Note that $(\Phi\bar\Phi)(x)=:\Phi\bar\Phi:_{C_{n}}(x):$
by virtue of \equ(0.19). 
We write $K_{n}$ in the form

$$K_{n}=Q_{n}e^{-V_{n}}+R_{n}   \Eq(4.12.5)$$ 

\\where $Q_{n}$ is a polymer activity which is given  by
second order perturbation theory in $g$ assuming that $\m$ is $O(g^2)$. 
$Q_{n}$ is specified below. $R_{n}$ is the remainder which is formally of 
$O(g^3)$. $Q_{n},R_{n}$ vanish when $\Phi=0$ by hypothesis. The RG map will
preserve this property.

In order to carry through the next step of the RG map as described
in Section~3 we must also specify $ \tilde V(\D_{\d_n},\Phi)$. We define

$$\tilde V_n(\D_{\d_n},\Phi)=V(\D_{\d_n},\Phi,C_{n+1,L^{-1}},g_{n},\m_{n})= 
g_{n}\int_{\D_{\d_n}} dx:(\Phi\bar\Phi)^2(x):_{C_{n+1,L^{-1}}}+ 
\m_{n}\int_{\D_{\d_n}} d^3x(\Phi\bar\Phi)(x)\Eq(4.1.1)$$

\\where we have used the notation $C_{n+1,L^{-1}}=S_{L}C_{n+1}$. 
Here and in what
follows we adopt the notations introduced in the Appendix of Section~3.3. Thus
$\sharp$ denotes fluctuation integration with
respect to the measure $d\m_{\G_n}(\xi)$ and $\natural$ denotes
fluctuation integration with
respect to the measure $d\m_{\G_{n,L}}(\xi)$, with $\G_{n,L}=S_{L^{-1}}\G_n$.
We recall (see section 3.1) that when we perform the 
fluctuation integration the fluctuation field $\xi$ enters $V$ through
$V(\D_{\d_n},\Phi + \xi)$ but $\tilde V$ will remain independent of $\xi$.

\\We now define $Q_n$: $Q_n$ is  supported on connected polymers 
$X_{\d_{n}} $ such that
$|X_{\d_{n}} |\le 2$. We assume it can be written in the form

$$Q_{n}(X_{\d_n},\Phi)=Q(X_{\d_n},\Phi;C_{n},{\bf w}_{n},g_n) =  
g_{n}^{2}\sum_{j=1}^{3}Q^{(j,j)}(\hat X_{\d_n},\Phi;C_n,w_{n}^{(4-j)})
\Eq(4.19.1)$$ 
 
\\where 
${\bf w}_{n}=(w_{n}^{(1)},w_{n}^{(2)},w_{n}^{(3)})$ is a triple of 
integral kernels to be obtained inductively and 

$$\hat X_{\d_n}=\cases{\D_{\d_n}\times\D_{\d_n}&if $X_{\d_n}=\D_{\d_n}$\cr 
(\D_{\d_n,1}\times\D_{\d_n,2})\cup(\D_{\d_n,2}\times\D_{\d_n,1})
&if $X_{\d_n}=\D_{\d_n,1}\cup\D_{\d_n,2}$\cr 
0&otherwise}\Eq(4.17) 
$$ 

$$\eqalign{Q^{(1,1)}(\hat X_{\d_n},\Phi;C_n,w_{n}^{(3)})&=-2
\int_{\hat X_{\d_n}}dxdy(\Phi(x)-\Phi(y))(\bar\Phi(x)-\bar\Phi(y)) 
w_{n}^{(3)}(x-y)\cr 
Q^{(2,2)}(\hat X_{\d_n},\Phi;C_n,w_{n}^{(2)})&= 
-\int_{\hat  
X_{\d_n}}dxdy\left[:(\Phi(x)-\Phi(y))(\bar\Phi(x)-\bar\Phi(y)) 
(\Phi(x)+\Phi(y))(\bar\Phi(x)+\bar\Phi(y)):_{C_n} 
+\right.\cr 
&\left.+3:[(\Phi\bar\Phi)(x)-(\Phi\bar\Phi)(y)]^2:_{C_n} 
\right]w_{n}^{(2)}(x-y)\cr 
Q^{(3,3)}(\hat X_{\d_n},\Phi;C_n,w_{n}^{(1)})&=4 
\int_{\hat X_{\d_n}}dxdy\  
:\Phi(x)\bar\Phi(x)\Phi(x)\bar\Phi(y)\Phi(y)\bar\Phi(y):_{C_n} 
w_{n}^{(1)}(x-y)\cr}\Eq(4.18)$$ 
 
\\Note that in the expression for $Q^{(1,1)}$ is equal to its $C_n$ Wick
ordered form because of \equ(0.19).
 
\\Next we define the second order approximation to the RG map. 
Let $p_n$ be the activity supported on unit blocks 
defined by  
$$p_n(\D_{\d_n},\xi ,\Phi)=V_n(\D_{\d_n},\xi + \Phi)-\tilde V_n 
(\D_{\d_n}, \Phi )  
 = p_{n,g} + p_{n,\mu} 
\Eq(4.8a)$$

\\where  
$$p_{n,g}=g\int_{\D_{\d_n}} dx\Bigl(:(\xi\bar\xi)^2:_{\G_n}(x)+ 
2\sum_\a\bigl[\Phi_\a(x):\bar\x_\a(\x\bar\x):_{\G_ n }(x) +
:(\x\bar\x)\x_\a:_{\G_n}(x)\Phi_a(x)\bigr]  + $$
$$+2(\Phi\bar\Phi)(x)(\x\bar\x)(x)+ 
(\Phi\bar\x)^2(x)+ 
(\x\bar\Phi)^2(x)+ 2\sum_{\a,\b}:(\x_\a\bar\x_\b):_{\G_n} (x)
:(\bar\Phi_\a\Phi_\b)(x):_{C_{n+1,L^{-1}}}+ $$ 
$$ +2\sum_\a\bigl[\x_\a(x):\bar\Phi_\a(\Phi\bar\Phi):_{C_{n+1,L^{-1}}}(x)+ 
:(\Phi\bar\Phi)\Phi_\a:_{C_{n+1,L^{-1}}}(x)\bar\x_\a(x)\bigr] 
\Bigr) 
$$ 
$$p_{n,\mu}=\m\int_{\D_{\d_n}} dx\left((\x\bar\Phi)(x)+ 
(\Phi\bar\x)(x)+(\x\bar\x)(x)\right) 
\Eq(4.8b)$$ 
 
\\In \equ(4.8b) we have used a component notation. Thus $\Phi_1=\f,\>
\Phi_2=\psi$. Similarly for
the fluctuation field $\xi$, $\xi_1=\zeta,\> \xi_2=\eta$. 
$\zeta$ is bosonic (degree $0$)and $\eta$ fermionic (degree $1$). In deriving
\equ(4.8b) from \equ(4.8a) have used $C_n=\G_n + C_{n+1,L^{-1}}$ ( see
\equ(0.28)), the independence of $\Phi,\> \xi$ in the sense that their 
components are independent and distributed 
with covariances $ C_{n+1,L^{-1}},\> \G_n$ respectively. The unordered objects
in \equ(4.8b) are equal to their Wick ordered form.

\vglue0.1cm

\\We will effectuate the RG map of Section~3 following closely the strategy 
in [BMS]. 
Namely, we insert a complex parameter $\lambda$ into our previous definitions 
in 
such a way that (i) at $\lambda =1$ our $\lambda$ dependent objects 
correspond with the previous definitions. (ii) The expansion through 
order $\lambda^{2}$ is second order perturbation theory in $g_n$ 
counting $\mu_n =O (g_n^{2})$. (iii) Powers of $\lambda$ are determined so 
as to correspond with leading powers of $g_n$ buried inside polymer 
activities. (iv) All functions will turn out to be norm analytic in 
$\lambda$ and this will enable us in section~5 to profit from Cauchy estimates.

\\We define

$$ 
P_n (\lambda) =  e^{-\tilde{V_n}} \big( 
-\lambda p_{n,g}-\lambda^{2}p_{n,\mu}+{1\over 2}\lambda^{2}p_{n,g}^{2} 
\big) + \lambda^{3}r_{n,1} 
\Eq(4.9)$$ 
 
\\where $r_{n,1}$ is defined by the condition $P_n (\lambda = 1) = P_n = 
e^{-V_n} - e^{-\tilde{V_n}}$.  Similarly, we define 
$$ 
K_n (\lambda) =  
\lambda^{2}e^{-\tilde{V_n}}Q_n  +\lambda^{3} \big( 
[e^{-V_n} - e^{-\tilde{V_n}}]Q_n + R_n  
\big) 
\Eq(4.10)$$ 
 
\\which, for $\lambda =1$ coincides with $K_n = e^{-V_n}Q_n+R_n$. 
Corresponding to \equ({3.13}) we define 
$${\cal B}(\lambda ,K_n) (Y_{\d_n}) =  \sum_{N+M\ge 1}{1\over N !M!} 
{\sum}_{(X_{\d_n,j}),(\Delta_{\d_n,i})\rightarrow \{Y_{\d_n} \}}  
e^{-\tilde{V_n}(X_{\d_n,0})} 
{\prod}_{j=1}^{N} K_n (\lambda ,X_{\d_n,j})  
{\prod}_{i=1}^{M}P_n (\lambda ,\Delta_{\d_n,i})   \Eq(4.13)$$ 
where $X_{\d_n,0}=Y_{\d_n}\setminus (\cup X_{\d_n,j})
\cup (\cup\Delta_{\d_n,i})$. 
Let $\SS(\lambda,K_n) = S_{L}{\cal B} (\lambda,K_n)$, 
where $S_{L}$ is the rescaling defined in the last section.  

\\The RG map ( see section~3) 
for $K_n$ with parameter $\lambda$ is $K_n\mapsto f_{n+1,K}(\l,K_n)=
{\cal E} (\SS (\lambda,K_n)^{\natural},F_{n} (\lambda ))$, where the 
superscript
$\natural$ denotes integration over the fluctuation field $\xi=(\z,\eta)$ 
with the measure $d\m_{\G_{n,L}}$ and $\G_{n,L}$ is the rescaled covariance
$S_{L^{-1}}\G_n$ as in the Appendix to Section~3. The relevant part 
$F_{n} (\lambda ))$ is defined on polymers in $(\d_{n+1}\math{Z})^{3}$  
and will be written as

$$F_{n} (\lambda) = \lambda^{2}F_{Q_n}+ \lambda^{3}F_{R_n}   \Eq(4.12)$$ 
 
\\and $F_{n} (\lambda)= F_{n}$, when $\lambda =1$. 

\vglue0.2cm

\\{\it Perturbative contribution to $f_{n+1}$. }

\\Given a function $f (\lambda)$ let 
 
$$T_{\lambda}f = f (0)+f' (0)+{1\over2}f'' (0)  \Eq(4.13.5)$$  
 
\\be the Taylor expansion to second order evaluated at $\lambda 
=1$. Then the second order approximation to the RG map is  
$f_{n+1}^{(\le 2)} = 
(f^{(\le 2)}_{n+1,K},f^{(\le 2)}_{n+1,V})$ with 
 
$$ 
f^{(\le 2)}_{n+1,K} (K_n,V_n)  
= T_{\lambda} {\cal E} (\SS(\lambda,K_n)^{\natural},F_{n}(\lambda )) 
= {\cal E}_{1} ( T_{\lambda} \SS(\lambda,K_n)^{\natural},F_{Q_n}), \>
\qquad f^{(\le 2)}_{n+1,V}(K_n,V_n) = V_{n+1}^{(\le 2)}\Eq(4.14.1)$$ 
 
\\where 

$$V_{n+1}^{(\le 2)}= {\tilde V}_{n,L}-{\tilde V}_{n,L}(F_{Q_n})$$

\\Note also that only the linearized ${\cal E}_{1}$ intervenes, because 
it will turn out that the nonlinear part of extraction generates terms 
only at order $\lambda ^{3}$ or higher. 

\vskip.5truecm 

\\{\it Proposition 4.1: 
\\There is a choice of $F_{Q}$ such that 
the form of $Q$ remains invariant under the RG evolution at second 
order. In more detail, $f_{n+1}^{(\le 2)}(V_n,Q_n e^{-V_n}) = 
(V_{n+1}^{(\le 2)},Q_{n+1}^{(\le 2)} e^{-\tilde V_{n,L}})$ where the 
parameters in 

$$V_{n+1}^{(\le 2)}(\D_{\d_{n+1}})
=V(\D_{\d_{n+1}},C_{n+1},g'_{n+1,(\le 2)},\m'_{n+1,(\le 2)})$$ 
\\evolved according to 
$$g_{n+1,(\le 2)}=L^\e g_{n}(1-L^\e a_{n}g_{n})\qquad\qquad 
\m'_{n+1,(\le 2)}=L^{3+\e\over 2}\m_{n}-L^{2\e}b_{n}g_{n}^{2} 
\Eq(4.47)$$ 
 
\\The parameters in $Q_{n+1}^{(\le 2)}=Q(C_{n+1},{\bf w_{n+1}},g_{n,L})$, where
$g_{n,L}=L^{\e}g_{n}$, evolved 
according to 
$${\bf w_{n+1}}={\bf v_{n+1}}+{\bf w}_{n,L}\qquad v_{n+1}^{(1)}=\G_{n,L} 
\qquad v_{n+1}^{(p)}=(C_{n,L})^p-(C_{n+1})^p\qquad p=2,3\Eq(4.54) 
$$ 
The constants $a_n, b_n$ are given by

$$a_n = 4\int_{(\d_{n+1}\bf Z)^3} dy\> v_{n+1}^{(2)}(y), \quad\quad
b_n = 2\int_{(\d_{n+1}\bf Z)^3} dy \> v_{n+1}^{(3)}(y) \Eq(4.54.1)   $$
}

\\{\it Proof:} We define a polymer activity $\hat Q_{n,L}$ supported on 
connected polymers $X_{\d_{n+1}}$ with $|X_{\d_{n+1}}|\le 2$ as follows:
if $|X_{\d_{n+1}}|=1$, say  $X_{\d_{n+1}}=\D_{\d_{n+1}}$, then

$$\hat Q_{n,L}(\D_{\d_{n+1}},\xi,\Phi)=
{1\over 2}(p_{n,L }(\D_{\d_{n+1}},\xi,\Phi))^2 $$

\\If $|X_{\d_{n+1}}|=2$ then

$$\hat Q_{n,L}(X_{\d_{n+1}},\xi,\Phi)=
{1\over 2}
\sum_{\D_{n+1,1},\D_{n+1,2}\atop\D_{n+1,1}\cup\D_{n+1,2}=X_{\d_{n+1}}} 
p_{n,L,g}(\D_{n+1,1},\xi,\Phi)p_{n,L,g}(\D_{n+1,2},\xi,\Phi) \Eq(4.9.1)  $$

\\where $p_{n,L,g}$ is defined by replacing in \equ(4.8a) and \equ(4.8b)
$(g_n,\m_n,\G_n,C_{n+1,L^{-1}})$ by  $(g_{n,L},\m_{n,L},\G_{n,L},C_{n+1})$
with $g_{n,L}=L^{\e}g_n$ and $\m_{n,L}=L^{3+\e\over 2}\m_n$.
 
\\It is easy to check that 

$$T_{\lambda }\SS (K_n,\lambda ) =  - p_{n,L}e^{-\tilde{V}_{n,L}} + (
e^{-{\tilde V}_{n,L}}\hat Q_{n,L} 
+e^{-{\tilde V}_{n,L}}Q_{n,L} )   \Eq(4.21.1)$$ 
 
\\where 

$$Q_{n,L}(X_{\d_{n+1}},\xi+\Phi)= Q(X_{\d_{n+1}},\xi+\Phi, C_{n,L},
{\bf w_{n,L}},g_{n,L})$$    

\\Using $C_{n,L}=\G_{n,L} + C_{n+1}$ and remembering that $\tilde V$ 
depends only on $\Phi$ we get

$$(e^{-\tilde{V}_{n,L}}Q_{n,L})^{\natural}=e^{-{\tilde V}_{n,L}}
Q(X_{\d_{n+1}},\Phi, C_{n+1},{\bf w_{n,L}},g_{n,L})$$    

\\Therefore

$$T_{\lambda }\SS (K_n,\lambda )^{\natural}(X_{\d_{n+1}},\Phi)=
e^{-{\tilde V}_{n,L}}\Bigl(
Q(X_{\d_{n+1}},\Phi, C_{n+1},{\bf w_{n,L}},g_{n,L}) + 
{\tilde Q}_n(X_{\d_{n+1}},\Phi,{\bf v}_{n+1}, C_{n+1},g_{n,L} )\Bigr) 
\Eq(4.30)  $$

\\where ${\tilde Q}_n= \hat Q_{n,L}^{\natural}$ is given after a 
straightforward but lengthy computation by

$${\tilde Q}_n(X_{\d_{n+1}},\Phi,C_{n+1},{\bf v}_{n+1},g_{n,L} )
= g_{n,L}^{2}\sum_{j=1}^{3}
\tilde Q^{(j,j)}(\hat X_{\d_{n+1}} ,\Phi;C_{n+1},v^{(4-j)}_{n+1}) 
\Eq(4.14.2)$$ 
 
\\where  
 
$$\eqalign{\tilde Q^{(1,1)}(\hat X_{\d_{n+1}},\Phi;C_{n+1},u)=&2
\int_{\hat X_{\d_{n+1}}}dxdy[\Phi(x)\bar\Phi(y)+\Phi(y)\bar\Phi(x)] 
u(x-y)\cr 
\tilde Q^{(2,2)}(\hat X_{\d_{n+1}},\Phi;C_{n+1},u)=&
\int_{\hat  
X_{\d_{n+1}}}dxdy\ 
\Bigl \{:[\Phi(x)\bar\Phi(y)+\Phi(y)\bar\Phi(x) ]^2:_{C_{n+1}}\cr
&+4:(\Phi(x)\bar\Phi(x))(\Phi(y)\bar\Phi(y)):_{C_{n+1}}\Bigr \}
u(x-y)\cr 
\tilde Q^{(3,3)}(\hat X_{\d_{n+1}},\Phi;C_{n+1},u)=&4 
\int_{\hat X_{\d_{n+1}}}dxdy 
:\Phi(x)\bar\Phi(x)\Phi(x)\bar\Phi(y)\Phi(y)\bar\Phi(y):_{C_{n+1}} 
u(x-y)\cr}\Eq(4.25)$$ 

\\Define 

$$F_{Q_n} =  
\tilde Q(C_{n+1},{\bf v}_{n+1},g_{n,L}) - Q(C_{n+1},{\bf v}_{n+1},g_{n,L}) 
\Eq(4.52) $$ 

\\evaluated on $X_{\d_{n+1}},\Phi$.

\\Then we have from \equ({4.30}) and \equ({4.52}) 

$$ {\cal E}_{1} \bigg(T_{\lambda }\SS 
(\lambda,K_n)^{\natural},F_n\bigg) 
= T_{\lambda }\SS(\lambda,K_n)^{\natural} - F_{Q_n}e^{-\tilde{V}_{n,L}} = 
e^{-\tilde{V}_{n,L}}Q(C_{n+1},{\bf w}_{n+1},g_{n,L})  
\Eq(4.15)$$ 
  
\\which shows that $Q$ is stable under RG evolution and verifies 
\equ(4.54). It remains to show that the chosen perturbative relevant part
$F_{Q_n}$ given by \equ(4.52) is of the form \equ(33.2aa) and thus suitable
for extraction.

\\To compute the difference in \equ(4.52) we will make use of the 
following {\it localization formulae} 
$$\Phi(x)\bar\Phi(y)+\Phi(y)\bar\Phi(x)=\Phi(x)\bar\Phi(x)+\Phi(y)\bar\Phi(y) 
-(\Phi(x)-\Phi(y))(\bar\Phi(x)-\bar\Phi(y))\Eq(4.52.1)$$ 
$$(\Phi(x)\bar\Phi(y)+\Phi(y)\bar\Phi(x))^2+4(\Phi(x)\bar\Phi(x))(\Phi(y)\bar\Phi(y))= 
4[(\Phi\bar\Phi)^2(x)+(\Phi\bar\Phi)^2(y)] -$$
$$-(\Phi(x)-\Phi(y))(\bar\Phi(x)-\bar\Phi(y)) 
(\Phi(x)+\Phi(y))(\bar\Phi(x)+\bar\Phi(y))-3[(\Phi\bar\Phi)(x)-(\Phi\bar\Phi)(y)]^2 
\Eq(4.52.2)$$ 
that are immediate to check. We get 

\eject

$$F_{Q_n}(X_{\d_{n+1}})=2g_{n,L}^{2} 
\int_{\hat X_{\d_{n+1}}}dxdy\Bigl[(\Phi\bar\Phi)(x)+(\Phi\bar\Phi)(y)\Bigr]
v^{(3)}(x-y)+4g_{n,L}^{2}
\int_{\hat X_{\d_{n+1}}} dxdy\Bigl[:(\Phi\bar\Phi)^2:_{C_{n+1}}(x)+$$
$$:(\Phi\bar\Phi)^2:_{C_{n+1}}(y)\Bigr]v^{(2)}(x-y) \Eq(4.33)$$ 
 
\\Note that due to supersymmetry 
there is no field independent part in $F_{Q_n}$. We can write
$F_{Q_n}(X_{\d_{n+1}})$  as: 

$$F_{Q_n}(X_{\d_{n+1}})=\sum_{\D_{\d_{n+1}}\subset X_{\d_{n+1}}}F_{Q_n}(X_{\d_{n+1}},\D_{\d_{n+1}})\Eq(4.34)$$ 

\\where 

$$F_{Q_n}(X_{\d_{n+1}},\D_{\d_{n+1}})=
4g_{n,L}^2F^{(2)}_{Q_n}(X_{\d_{n+1}},\D_{\d_{n+1}}) 
+2g_{n,L}^2F^{(1)}_{Q_n}(X_{\d_{n+1}},\D_{\d_{n+1}})\Eq(4.35)$$ 

\\and 

$$F^{(m)}_{Q_n}(X_{\d_{n+1}},\D_{\d_{n+1}})=
\int_{\D_{\d_{n+1}}} dx:(\Phi\bar\Phi)^m(x):_{C_{n+1}} 
f^{(m)}_{Q_n}(x,X_{\d_{n+1}},\D_{\d_{n+1}}) 
\Eq(4.36)$$ 

\\with 

$$ f^{(m)}_{Q_n}(x,X_{\d_{n+1}},\D_{\d_{n+1}})=\cases{\int_{\D_{\d_{n+1}}} dy v^{(m')}(x-y)&$X_{\d_{n+1}}=\D_{\d_{n+1}}$\cr 
\cr 
\int_{\D_{\d_{n+1}}'}dy v^{(m')}(x-y)&$X_{\d_{n+1}}=\D_{\d_{n+1}}\cup\D_{\d_{n+1}}'$, connected\cr}\Eq(4.36a)$$ 
and $m' = 4- m$.

$$V(F_{Q_n},\D_{\d_{n+1}})= 
\sum_{X_{\d_{n+1}}\supset\D_{\d_{n+1}}}F_{Q_n}(X_{\d_{n+1}},\D_{\d_{n+1}})
=4g_{n,L}^2\sum_{X_{\d_{n+1}}\supset\D_{\d_{n+1}}} 
F^{(2)}_{Q_n}(X_{\d_{n+1}},\D_{\d_{n+1}}) +$$ 
$$+2g_{n,L}^2\sum_{X_{\d_{n+1}}\supset\D_{\d_{n+1}}}
F^{(1)}_{Q_n}(X_{\d_{n+1}},\D_{\d_{n+1}})\Eq(4.37)$$ 

\textRed

\\where we have used \equ(33.2d), \equ(33.2cc) and \equ(33.2aa) for the
first equality.

\textBlack

\\In the following we will use the fact that the 
$v^{(j)}_{n+1}(x-y),\ 1\le j\le 3$ vanish for $|x-y|\ge 1$. This follows 
from the fact that $\G_{n,L}(x-y)$ appears as a factor in the expression 
\equ(4.54) for  $v^{(j)}_{\d_{n+1}}(x-y)$ and  $\G_{n,L}$ has range 
$1$. Returning to \equ(4.37) we have 

$$\sum_{X_{\d_{n+1}}\supset\D_{\d_{n+1}}}F^{(m)}_{Q_n}(X_{\d_{n+1}},\D_{\d_{n+1}})= 
\int_{\D_{\d_{n+1}}} dx 
:(\Phi\bar\Phi)^m(x):_{C_{n+1}} 
\Big[\int_{\D_{\d_{n+1}}} dy v^{(m')}_{n+1}(x-y) +$$ 
$$+\sum_{\D_{\d_{n+1}}'\ne\D_{\d_{n+1}}\atop(\D_{\d_{n+1}},\D_{\d_{n+1}}')\ {\rm connected}} 
\int_{\D_{\d_{n+1}}'} dy v^{(m')}_{n+1}(x-y)\Big]$$ 

\\On the r.h.s. use $v^{(m')}_{n+1}(x-y)=0$ for $|x-y|\ge 1/2$ 
to extend the sum 
on $\D_{\d_{n+1}}'$ to {\it all} the $\D_{\d_{n+1}}'\ne\D_{\d_{n+1}}$. 
We then get 

$$\sum_{X_{\d_{n+1}}\supset\D_{\d_{n+1}}}F^{(m)}_{Q_n}(X_{\d_{n+1}},\D_{\d_{n+1}})= 
\int_{\D_{\d_{n+1}}} dx 
:(\Phi\bar\Phi)^m(x):_{C_{n+1}} \int dy v^{(m')}(x-y)$$ 

\\Hence from \equ(4.37) and above we get 

$$V(F_{Q_n},\D_{\d_{n+1}})=
a_n g_{n,L}^2\int_{\D_{\d_{n+1}}} dx:(\Phi\bar\Phi)^2(x):_{C_{n+1}} 
+b_n g_{n,L}^2\int_{\D_{\d_{n+1}}} dx(\Phi\bar\Phi)(x)\Eq(4.38)$$ 

\\where 

$$a_n=4\int_{(\d_{n+1}\math{Z})^{3}} dy\ v^{(2)}_{n+1}(y)\qquad\qquad 
b=2\int_{(\d_{n+1}\math{Z})^{3}} dy\ v^{(3)}_{n+1}(y)\Eq(4.39)$$ 

\bull

\\{\it Remark} :  $a_n$ and $b_N$ are well defined since the 
$v^{j}_{n+1}$ have compact support. They are positive and their properties  
are discussed in Lemma~5.12 of Section~5.

\vskip.5truecm 

\\{\it The exact RG map $f_{n+1}$ for $K_n=Q_ne^{-V_n}+R_n$}. 

\vskip.2truecm 

$$ K_n\mapsto K_{n+1} = f_{n+1,K} (\lambda,K_n,V_n)\vert_{\lambda =1}  
=  {\cal E} (\SS(\lambda,K_n)^{\natural},F_n (\lambda ))\vert_{\lambda =1}
\Eq(4.16)$$ 

\\induces an evolution of the remainder $R_n$ which is studied by Taylor 
series around $\lambda =0$ with remainder written using the Cauchy 
formula: 

$$ f_{n+1,K} (\lambda =1) = \sum_{j=0}^{3} {f_{n+1,K}^{(j)} (0)\over j!} +  
{1\over 2\pi i}  
\oint_{\gamma} {d\lambda \over \lambda^{4} (\lambda -1)} 
f_{n+1,K} (\lambda) 
$$ 
 
\\The terms $j=0,1,2$ are the second order part $f^{(\le 2)}_{n+1,K}$.  In 
the $j=3$ term there are no terms mixing $R_n$ with $Q_n,P_n$ because of the 
$\lambda^{3}$ in front of $R_n$.  Therefore it splits 
$ 
{f_{K}^{(3)} (0)\over 3!} = R_{n+1,1} + R_{n+1,2}  
$ into the third order derivative at $R_n=0$, which we write 
using the Cauchy formula as 

$$ 
R_{n+1,1} \equiv R_{n+1,\rm main}  
= {1\over 2\pi i}  
\oint_{\gamma} {d\lambda \over \lambda^{4}} 
{\cal E} \bigg(  
\SS(\lambda ,Q_n e^{-V_n})^{\natural},F_{Q_n} (\lambda )  
\bigg) 
\Eq(4.188)$$ 
 
\\and terms linear in $R_n$: 
$$\eqalign{ 
&R_{n+1,2} \equiv R_{n+1,\rm linear} 
= \big(\SS_1 R_n\big)^{\natural} - F_{R_n}e^{-\tilde V_{L,n}}\cr 
&\SS_1 R_n(Z_{\d_{n+1}}) = 
\sum_{X_{\d_{n+1}}:L^{-1}{\bar X}_{\d_{n+1}}^{L}=Z_{\d_{n+1}} } 
e^{-\tilde V_{n,L}(Z_{\d_{n+1}}\backslash 
L^{-1}X_{\d_{n+1}})} R_{n,L}(L^{-1}X_{\d_{n+1}})\cr 
}       \Eq(4.19)$$

\\The remainder term in the Taylor expansion is 
 
$$ 
R_{n+1,3} =  
{1\over 2\pi i}  
\oint_{\gamma} {d\lambda \over \lambda^{4} (\lambda -1)} 
{\cal E} (\SS (\lambda,K_n)^{\natural},F_n (\lambda 
)) 
\Eq(4.21)$$ 
 
In Proposition~4.1 the coupling constant in $Q_{n+1}^{(\le 2)}$ is not the 
same as 
the coupling constant in $V_{n+1}^{(\le 2)}$.  Furthermore, the coupling 
constant in $V_{n+1}^{(\le 2)}$ will further change because of the 
contribution from $F_{R}$.  To take this into account we introduce 
 
$$\eqalign{ 
&V_{n+1}(\D_{\d_{n+1}})=V(\D_{\d_{n+1}} ,C_{n+1} ,g_{n+1} ,\m_{n+1})\cr 
&g_{n+1} =L^\e g_n(1-L^\e a_n g_n)+\xi_n(u_n)\cr 
&\m_{n+1} =L^{3+\e\over 2}\m_n-L^{2\e}b_n g_n^2+\r_n(u_n)\cr} 
\Eq(4.47A)$$ 
 
\\where $u_n=(g_n,\m_n,R_n)$ and
the remainders $\xi_n(u_n), \r_n(u_n)$ anticipate the effects of a 
yet-to-be-specified $F_{R_n}$.  Then we set 
 
$$ 
R_{n+1,4} = e^{-V_{n+1}}Q(C_{n+1},{\bf w}_{n+1},g_{n+1})
- e^{-\tilde{V}_{n,L}}Q(C_{n+1},{\bf w}_{n+1},g_{n,L}) 
\Eq(4.22) 
$$ 
 
\\and define

$$\eqalign{ Q_{n+1}&=Q(C_{n+1},{\bf w}_{n+1},g_{n+1})\cr 
R_{n+1}&=R_{n+1,{\rm main}}+R_{n+1,{\rm linear}}+R_{n+1,3}+R_{n+1,4} \cr
K_{n+1}&=Q_{n+1} e^{-V_{n+1}}+R_{n+1} \cr }
\Eq(4.221)$$

\\With these definitions we have obtained the RG map 
 
$$ f_{n+1,V}(V_n,K_n)=V_{n+1},\quad f_{n+1,K}(V_n,K_n) = K_{n+1} \Eq(4.222)$$

\vskip.2truecm 
 
\\{\it Definition of $F_{R_n}$} 
\vglue.3truecm 
 
\\To complete the RG step we must specify the relevant part $F_{R_n}$
from the remainder $R_n$.   
The goal is to choose $F_{R_n}$ so that the map $R_n 
\rightarrow R_{n+1,\rm linear}$ will be contractive in the following sense.
$R_n$ is measured in the norm \equ(2.611), and the kernel norm \equ(2.612),
with $\d=\d_{n}$, with a choice of $\bf h$ and $\bf h'$ 
(to be made in section~5). $R_{n+1}$ is measured in the same norms but on the
lattice scale $\d_{n+1}$. We will say that the map is contractive if the
size of $R_{n+1,\rm linear}$ is less than the size of $R_n$. 

\vglue0.1cm

\\$F_{R_n}$ will have the form given in \equ(33.2aa) with $P$ a 
supersymmetric polynomial vanishing at $\Phi=0$. The coefficients $\a_P$
will be
identified via {\it normalization conditions} on the small set part of 
$R_{n+1,\rm linear}$. This means that certain derivatives with respect to
$\Phi=(\f,\psi)$ vanish when $\Phi=0$. That the map in question is contractive
when $R_{n+1,\rm linear}$ is suitably normalized is proven in Section~5.

\vglue0.2cm 
 
\\For given coefficients ${\tilde \alpha}_{n,P} (X)$, 
we define 

$$\tilde F_{R_n}(X_{\d_n},\Phi)= \sum_{P} \int_{X_{\d_n}} dx\  
{\tilde \alpha}_{n,P} (X_{\d_n})P (\Phi (x),\dpr_{\d_n} \Phi (x)) 
\Eq(4.57) $$ 
 
$$\tilde F_{R_n}(X_{\d_n},\Phi)=0\> : X_{\d_n} {\rm is\ not\ a\ small\ set}
\Eq(4.571)   $$

\\$P$ runs over the {\it relevant} monomials which in this 
model are  
$P = \Phi\bar\Phi,(\Phi\bar\Phi)^{2},\Phi \dpr_{\d_n,\m}\bar\Phi, 
\dpr_{\d_n,\m}\Phi \bar\Phi $, $\mu \in S$,
 with the corresponding coefficients  
${\tilde\alpha}_{P}(X_{\d_n})={\tilde\alpha}_{n,2,0}(X_{\d_n}), 
{\tilde\alpha}_{n,4}(X_{\d_n}), 
{\tilde\alpha}_{n,2,\bar 1}(X_{\d_n},\m),
{\tilde\alpha}_{n,2,1}(X_{\d_n},\m) $. 
The index set $S$ was defined in section~2.1 after \equ(2.10reg).
Note that $P=1$ is not a relevant monomial in this model: $R_n$ vanishes when
$\Phi=0$ vanishes by hypothesis. Then 
$R_n^{\sharp}(X_{\d_n},\Phi)$ vanishes when $\Phi=0$ by supersymmetry,
(Lemma~1.1) so that no subtraction is necessary at $\Phi=0$.
 
\\Choose the coefficients ${\tilde \alpha}_{n,P}$ so that 

$$J_n= R_n^{\sharp}-\tilde F_{R_n}e^{-\tilde{V_n}} 
\Eq(4.59) $$ 

\\is normalized (details are given below). Note that $J(X_{\d_n},0)=0$.
We define the relevant 
part, supported on small sets, by  

$$F_{R_n}(Z_{\d_{n+1}},\Phi)=
\sum_{X_{\d_{n+1}}: {\rm small\ sets}\atop L^{-1}{\bar X}_{\d_{n+1}}^{L}
=Z_{\d_{n+1}}} \tilde F_{R_n,L}(L^{-1}X_{\d_{n+1}} ,\Phi) 
=\sum_{X_{\d_{n}}: {\rm small\ sets}\atop L^{-1}{\bar X}_{\d_{n}}^{L}
=Z_{\d_{n}}}\tilde F_{R_n}(X_{\d_{n}} ,S_L\Phi) 
\Eq(4.58)$$ 
 
\\$F_{R_n}$ is supported on small sets by construction.
From the definition of $R_{n+1,\rm linear}$ in \equ(4.19) we get 

$$ R_{n+1,\rm linear} (Z_{\d_{n+1}}) 
= \sum_{X_{\d_{n+1}}: {\rm small\ sets}
\atop L^{-1}{\bar X_{\d_{n+1}}}^{L}=Z_{\d_{n+1}}} 
e^{-\tilde{V}_{L} (Z_{\d_{n+1}}\setminus L^{-1}X_{\d_{n+1}})}
J_{n,L} (L^{-1}X_{\d_{n+1}}) + $$
$$+\sum_{X_{\d_{n+1}}: {\rm large\ sets}
\atop L^{-1}{\bar X_{\d_{n+1}}}^{L}=Z_{\d_{n+1}}} 
e^{-\tilde{V}_{L} (Z_{\d_{n+1}}\setminus L^{-1}X_{\d_{n+1}})}
J_{n,L} (L^{-1}X_{\d_{n+1}}) \Eq(4.59b)$$ 
 
\\Therefore the first sum in $R_{\rm linear}$ is also normalized 
because normalization as defined below is preserved under multiplication by 
smooth functionals of $\Phi$ and rescaling. 

\\Substitution of \equ(4.57) in \equ(4.58) shows that $F_{R_n}$ is of 
the form required in \equ({33.2aa}). We have

$$F_{R_n}(Z_{\d_{n+1}},\Phi)= \sum_{P}\int_{Z_{\d_{n+1}}}  dx\  
\alpha_{n,P} (Z_{\d_{n+1}},x)P (\Phi (x),\dpr_{\d_{n+1}}\Phi (x)) 
\Eq(4.57b) 
$$ 
 
\\where

$$ \alpha_{n,P}(Z_{\d_{n+1}},x) =  
\sum_{X_{\d_{n+1}}\ {\rm small\ set}\atop L^{-1}{\bar X}_{\d_{n+1}}^{L}
=Z_{\d_{n+1}}} {\tilde \alpha}_{n,P}(X_{\d_{n}} ) 
L^{-[P]+3}1_{L^{-1}X_{\d_{n+1}} } (x) \Eq(4.67a)$$ 
 
\\In \equ(4.67a)
$1_{X}$ is the characteristic function of the set $X$. Note that 
$X_{\d_{n+1}}$ fixes $X_{\d_{n}}$ by restriction by our construction of
polymers in section~1.3. $[P]$ is the dimension of the monomial 
$P$, ($2kd_s$ for 
$(\Phi\bar\Phi)^{k}$ and $2d_s+1$ for $\Phi \dpr_{\d_{n+1}}\bar\Phi $).  
$\alpha_{n,P}(Z_{\d_{n+1}},x)$  is supported on small sets $Z_{\d_{n+1}}$
and vanishes if $x\notin Z_{\d_{n+1}}$.

\vglue0.1cm

\\We now compute $V_{F_{R}}$ following \equ({33.2d}). Define

$$ 
\alpha_{n,P} : = \sum_{Z_{\d_{n+1}}\supset x}\alpha_{n,P}(Z_{\d_{n+1}},x)  
\Eq(4.57c)$$ 
 
\\This is independent of $x$ by translation invariance. In fact
given an $x$ it belongs uniquely to a block $\D_{\d_{n+1}}$, since our blocks 
which are restrictions of half open continuum cubes are always disjoint 
(see section~1.3). The sum over all polymers containing a block  
$\D_{\d_{n+1}}$ is independent of $\D_{\d_{n+1}}$ 
by translation invariance.

\\From \equ(4.67a) and \equ(4.57c) we get 

$$\alpha_{n,P} = L^{-[P]+3} \sum_{X_{\d_{n+1}}  \ {\rm small\ set}: 
L^{-1}X_{\d_{n+1}}\supset x }  
\tilde{\alpha}_{n,P}(X_{\d_{n}} ) \Eq({4.70})$$ 
 
\\$\alpha_{n,P}=0$ for $P=\Phi\dpr_{\d_{n+1}}\bar\Phi\> {\rm or}\>
\dpr_{\d_{n+1}}\Phi\bar\Phi$ by reflection invariance of polymer activities.

\\Therefore 
$$\eqalign{ \tilde V_{L}(F_{R_n},\D_{\d_{n+1}}) 
&=\int_{\D_{\d_{n+1}}} dx\bigg\{\alpha_{n,2,0} \Phi\bar\Phi + 
\alpha_{n,4,0} (\Phi\bar\Phi)^{2} 
\bigg\}\cr  
&=\int_{\D_{\d_{n+1}}} dx\bigg\{\r_n(u_n):\Phi\bar\Phi:_{C_{n+1}} + 
\xi_n(u_n):(\Phi\bar\Phi)^{2}:_{C_{n+1}}\bigg\}\cr}\Eq(4.71) $$

\\where $u_n=(g_n,\m_n,R_n)$ and 
 
$$ 
\r_{n}=\alpha_{n,2,0}+ 2C_{n+1}(0)\a_{n,4,0}\qquad\qquad 
\xi_{n}=\alpha_{4,0} 
\Eq(4.64)$$

\\which are formulas for the error terms in \equ(4.47A).

\vglue0.3cm 
\\{\it Normalization conditions} 
\vglue0.3cm 

\\By an abuse of notation let $1$ denote the constant function in 
$C^{2}(X_{\d_n})$  
equal to  $1$. Similarly let $1^{2p}$ denote the constant  
function in $C^{2}(X_{\d_n}^{2p})$ equal to $1$. We will identify the 
$C^{2}(X_{\d_n})$  function 
$f(x)=x_{\m}$ with $x_{\m}$. \textRed Note that
$x_{\m}$ is defined with respect to an origin which belongs to 
$X_{\d_n}$. \textBlack
Similarly we will identify $C^{2}(X_{\d_n}^{2})$ 
functions $g_2(x_{1}, x_{2})=x_{1,\m}$, $g_2(x_{1}, x_{2})= x_{2,\m}$ with 
$ x_{1,\m},\>x_{2,\m}$ respectively. 

\\Suppose the polymer activity $J(X_{\d_n},\Phi)= J(X_{\d_n},\f,\psi)$ is of
degree $0$, gauge invariant and supersymmetric. We have the following 
identities:

$$D^{2,0}J(X_{\d_n},0,0;1^{2})=D^{0,2}J(X_{\d_n},0,0,;1,1) \Eq(4.40)$$     

$$D^{2,0}J(X_{\d_n},0,0;x_{1,\m})=D^{0,2}J(X_{\d_n},0,0,;x_{\m},1) 
\Eq(4.402)$$     

$$D^{2,0}J(X_{\d_n},0,0;x_{2,\m})=D^{0,2}J(X_{\d_n},0,0,;1,x_{\m}) 
\Eq(4.403)$$

$$D^{2,2}J(X_{\d_n},0,0;1,1,1^{2})=2D^{0,4}J(X_{\d_n},0,0;1,1,1,1) \Eq(4.41)$$

$$D^{4,0}J(X_{\d_n},0,0;1^{4})=0 \Eq(4.401)$$

\\where the field derivatives are taken according to \equ(2.2). The 
identities \equ(4.40)-\equ(4.41)
follow by expanding $J(X_{\d_n},\Phi)$ in the fields, retaining a degree 4 
supersymmetric polynomial in $\Phi$ and $\dpr_{\d_n}\Phi$ which is all that
enters into the computation. Then express it in the Grassmann 
representation \equ(2.1).
\equ(4.401) is trivial. Because $J$ is of degree $0$ the only term that 
survives for the computation of \equ(4.401) is of the form 
$\int_{X_{\d_n}^{4}}dx\>a(x_1,x_2,x_3,x_4)\psi(x_1)\bar\psi(x_2)\psi(x_3)
\bar\psi(x_4)$ where the kernel $a$ is antisymmetric in $x_1,x_3$ and in
$x_2,x_4$. The integral vanishes if we replace the grassmann piece by $1^4$.
Derivatives on the grassmann fields annihilate $1^4$. 

\vglue0.2cm

\\We say that a degree $0$, gauge invariant, supersymmetric polymer activity  
$J(X_{\d_n},\Phi)= J(X_{\d_n},\f,\psi)$ with $J(X_{\d_n},0)=0$
is {\it normalized} if, for all small sets $X_{\d_n}$, 

$$\eqalign{D^{2,0}J(X_{\d_n},0,0;1^{2})=& 
D^{0,2}J(X_{\d_n},0,0,;1,1)=0\cr 
D^{2,0}J(X_{\d_n},0,0;x_{1,\m})=&  D^{2,0}J(X_{\d_n},0,0;x_{2,\m})=0\cr 
D^{0,2}J(X_{\d_n},0,0;1,x_\m)=& 
D^{0,2}J(X_{\d_n},0,0;x_\m,1)=0\cr 
2D^{0,4}J(X_{\d_n},0,0;1,1,1,1)=& 
D^{2,2}J(X_{\d_n},0,0;1,1,1^{2})=0\cr} 
\Eq(4.42)$$ 
 
\vskip.5truecm 
\\{\it Determining coefficients from \equ({4.42})} 
\vskip.3truecm 
 
\\We will apply the normalization conditions to $J=J_n$ 
defined in \equ(4.59).
This will determine the dependence of the error terms 
$\xi_n,\>\r_n$ on $R_n$. Lemma~5.17 will show that these terms are bounded
by the kernel norm of $R_n$. 

\\In doing the following computations
note that $J_n(X_{\d_n},0,0)=0$ as shown earlier.
Moreover
the odd derivatives $D^{0,j}J_n(X_{\d_n},0;f^{\times j})$,  
$j$=odd integer, vanish identically by gauge invariance. It is enough 
to take derivatives with respect to the bosonic fields $\f$ because of the
identities stated above, \equ(4.40) et seq.
Taking derivatives of \equ(4.59) and  
remembering that $\tilde F_{R_n}(X_{\d_n},0)=0,\> {\tilde V_n}(X_{\d_n},0)=0 $ 
we get 

$$\eqalign{ 
D^{0,2}J_n(X_{\d_n},0,0;f,\bar f)&=D^{0,2}R_n^{\sharp}(X_{\d_n},0,0;f,\bar f)- 
D^{0,2}\tilde F_{R_n}(X_{\d_n},0,0;f,\bar f)\cr 
D^{0,4}J_n(X_{\d_n},0,0;f_1,\bar f_1,f_2,\bar f_2)&=D^{0,4}R_n^{\sharp}(X_{\d_n},0,0;f_1,\bar f_1,f_2,\bar f_2)+ 
D^{0,4}\tilde F_{R_n}(X_{\d_n},0,0;f_1,\bar f_1,f_2,\bar f_2)+\cr 
&+4D^{0,2}\tilde F_{R_n}(X_{\d_n},0,0;f,\bar f)D^{0,2}{\tilde V_n}(X_{\d_n},0,0;f,\bar f) 
}\Eq(4.64.1)$$ 

\\where the $f$ are complex valued functions in $C^{2}(X_{\d_n})$. A variation
of $\f$ along $f$ implies that we vary $\bar\f$ along $\bar f$.
Note that from \equ(4.57) 

$$\eqalign{ 
D^{0,2}F_{R_n}(X_{\d_n},0,0;1,1)=&\ |X_{\d_n}| 
{\tilde \a}_{n,2,0}(X_{\d_n})\cr 
D^{0,2}F_{R_n}(X_{\d_n},0,0;1,x_\m)=&\ |X_{\d_n}| 
{\tilde \a}_{n,2,\bar 1}(X_{\d_n},\m)+  {\tilde \a}_{n,2,0}(X_{\d_n})\int_{X_{\d_n}} dx\> x_{\m} \cr 
D^{0,2}F_{R_n}(X_{\d_n},0,0;x_{\m},1)=&\ |X_{\d_n}| 
{\tilde \a}_{n,2,1}(X_{\d_n},\m)+  {\tilde \a}_{n,2,0}(X_{\d_n})\int_{X_{\d_n}} dx\> x_{\m} \cr 
D^{0,4}F_{R_n}(X_{\d_n},0,0;1,1,1,1)=&\ 4|X_{\d_n}| 
{\tilde \a}_{n,4}(X_{\d_n})\cr}$$ 
 
\\Now imposing successively the conditions \equ(4.42) we get  

$$\eqalign{ 
{\tilde \a}_{n,2,0}(X_{\d_n})&= {1\over |X_{\d_n}|}D^{0,2}R_n^{\sharp}(X_{\d_n},0,0;1,1)\cr 
{\tilde \a}_{n,2,\bar 1}(X_{\d_n},\m)&={1\over |X_{\d_n}|} 
D^{0,2}R_n^{\sharp}(X_{\d_n},0,0;1,x_\m) - 
{1\over |X_{\d_n}|}{\tilde \a}_{2,0}(X_{\d_n})\int_{X_{\d_n}} dx\> x_{\m} \cr 
{\tilde \a}_{n,2,1}(X_{\d_n},\m)&={1\over |X_{\d_n}|} 
D^{0,2}R_n^{\sharp}(X_{\d_n},0,0;,x_{\m},1) - 
{1\over |X_{\d_n}|}{\tilde \a}_{n,2,0}(X_{\d_n})\int_{X_{\d_n}} dx\> x_{\m} \cr 
{\tilde \a}_{n,4}(X_{\d_n})&={1\over 4}{1\over |X_{\d_n}|}\Bigl( 
D^{0,4}R_n^{\sharp}(X_{\d_n},0,0;1,1,1,1) \cr 
&+D^{0,2}{\tilde V_n}(X_{\d_n},0,0;1,1)D^{0,2} 
R_n^{\sharp}(X_{\d_n},0,0;1,1)\Bigr)\cr} \Eq(4.62)$$ 
\textBlack 
 
\\We remind the reader that RG transformations preserve the invariance of 
polymer activities under translations, reflections, and rotations which 
leave the lattice invariant. 

\vskip.5truecm 
\\{\bf 5. ESTIMATES} 
\numsec=5\numfor=1 
\vskip.5truecm 
 
\\Let $u_{n}=(g_{n},\m_{n},R_{n})$. Then $({\bf w}_{n},u_n)$
are the coordinates of the  
measure density in the polymer representation after $n$ successive applications
of the RG map $f_j$, $1\le j\le n$, of  Section~4. The ${\bf w}_{n}$ evolve
according to ${\bf w}_{n+1}=f_{n+1,{\bf w}}({\bf w}_{n})
={\bf v}_{n+1}+{\bf w}_{n,L}$
as given in \equ(4.54). This evolution is independent of
$u_n$ and is solved in Lemma~5.9 below.
The sequence $\{{\bf w}_{n},u_{n}\}$ with $u_{n+1}=   f_{n+1}(u_{n})$, 
where the solution for ${\bf w}_{n}$ is incorporated in the map $f_n$, is the 
RG trajectory. The index $n$ in $R_{n}$ also indicates that $R_{n}$ is 
supported on polymers in $(\d_{n}\math{Z})^{3}$.
Correspondingly the norms for Banach spaces of polymer activities given in
Section~2 are indexed
by the lattice spacing $\d_{n}$.  In this section we first set up a uniformly
bounded domain ${\cal D}_{n}$ for $u_{n}$. The rest of this section
is then devoted to the proof of Theorem~5.1 below. This theorem
controls the remainders  $(\xi_n,\r_n)$ in the flow equations \equ(4.47A)  
together with $R_{n+1}$ in \equ(4.221) when $u_{n}$ 
belongs $ {\cal D}_n$. It also gives bounds on $g_{n+1}$ and 
$\m_{n+1}$.  
Theorem~5.1 will provide essential ingredients for
the proof (in Section~6,) of existence of an initial choice of the 
mass parameter such that there is a uniformly bounded RG trajectory 
at all scales labelled by $n$. 

\vglue0.1cm
The aforementioned domain will be a ball defined with Banach space norms
with the
center of the ball fixed i.e. independent of $n$. To this end we first 
obtain an approximate discrete flow of the coupling constant $g_{n}$ from 
the first equation in \equ(4.47A) by ignoring the remainder 
$\xi_{n}(g_n,\m_n,R_n)$. The approximate flow equation has $n$-dependent
coefficients. However 
we show below (Lemma~5.12), with no assumption about the domain 
${\cal D}_{n}$
given below, that the the positive coefficients $a_{n}$
converge geometrically as $n\rightarrow \infty$  to a constant $a_{c,*}>0$.
This leads us to set up a reference approximate
discrete flow of the coupling constant

$$ g_{c,n+1}= L^{\e}g_{c,n}(1-L^{\e}a_{c,*}g_{c,n})\Eq(55.0)$$
 
\\This may be thought of as an approximate flow in an underlying continuum 
theory. This approximate flow has a nontrivial fixed point, namely 

$${\bar g}= {L^{\e}-1\over L^{2\e}a_{c,*}}>0 \Eq(55.4) $$ 

\\The constant $a_{c,*}=a_{c,*} (L,\e)$
depends on $L,\e$ in such a way that when $\e\rightarrow 0$ with $L$ fixed
$a_{c,*} (L,\e) \rightarrow  \bar a_{c,*}(L)$ which depends only on $L$.
We will assume $L$ large but fixed for the rest of the paper. We then choose
$\e$ sufficiently small depending on $L$.

\\We have

$$0< {\bar g}< C_{L}\e \Eq(55.01)$$

\\where $C_{L}$ is a constant which depends only on $L$. $\e$ is then a measure
of smallness of $\bar g$. 

\\In the following $O(1)$ denotes a constant {\it independent of $L$, $\e$ and 
$n$}. Constants C are {\it independent of $\e$ and $n$ but may 
depend on $L$}. These constants may change from line to line. It will not be 
necessary to keep track of these changes. 

\vskip0.3cm

\\{\it The Domain ${\cal D}_{n}$} :

\vskip0.1cm

\\{\it We will say that $u_{n}= (g_n,\m_n,R_n) $  belongs to the 
domain ${\cal D}_{n}$ if 
 
$$\vert g_n -{\bar g}\vert < \n{\bar g},
\ \ \vert\mu_n\vert < {\bar g}^{2-\delta}   
\Eq(55.1)$$ 

$$||| R_n |||_{n}< {\bar g}^{11/4-\eta}  \Eq(55.02)$$

\\where the constant $\n$ is held in the range $0<\n<{1\over 2}$, and

$$||| R_n |||_{n}=\max \{\vert R_n\vert_{{\bf h}_{*},\AA,\d_{n}},\>
{\bar g}^{2}\Vert R_n\Vert_{{\bf h},G_{\k},\AA,\d_{n}}\} \Eq(55.03)$$

\\We choose $\k=\k(L)$ as in Lemma~2.1 and $\r=\r(L)$ as in 
Lemma~5.3 (independent of the domain hypothesis). $\k,\r$ will be held
fixed after $L$ has been chosen sufficiently large.
$\delta,\eta =O(1)> 0$ are very small fixed numbers, say $1/64$, and 
$h_{B}=c{\bar g}^{-1/4}$ with $c=O(1) > 0$ a very small number. 
Furthermore we take \textRed
$h_{B*}=\r^{-1/2}+\k^{-1/2}$ \textBlack
and choose $m_{0}=9$. $h_{F}=h_{F}(L)$ is an $\e$ 
independent constant which depends on $L$ and is taken to be sufficiently 
large. (The dependence of $h_{F}$ on $L$ enters in the proofs of Lemma~5.15 and
Lemma~5.16 below). We recall that 
${\bf h} = (h_{B}, h_{F})$ , ${\bf h_*} = (h_{B*}, h_{F})$.}

\vskip0.1cm

\\{\it Remark} :1. Note that condition \equ(55.02) is equivalent to having both

$$\Vert R_n\Vert_{{\bf h},G_{\k},\AA,\d_{n}}< {\bar g}^{3/4-\eta} 
\Eq(55.2)$$ 
$$\vert R_n\vert_{{\bf h}_{*},\AA,\d_{n}} < {\bar g}^{11/4-\eta} 
\Eq(55.3)$$  
\textRed 

\\2. In [BMS], see equations (5.1)-(5.3)therein, 
the domain was specified using $\e$. In contrast here we specify
the domain as in [A] by using $\bar g$ instead of $\e$ and moreover 
we enlarge the domain of $g_n$
slightly for technical reasons. \textBlack   

\vglue0.2cm
\\Recall the definitions of $\rho_{n}(g_n,\m_n, R_n)$ and 
$\xi_{n}(g_n,\m_n, R_n)$ from 
\equ({4.64}). We will prove in this section  
 
\vglue.3truecm

\\{\it Theorem~5.1
 
\\Let $u_{n}=(g_n,\m_n,R_n)\in {\cal D}_{n}$. Let $L$ be  
large but fixed followed by $\e$ sufficiently small depending on $L$. 
${\bar g}$ is thus sufficiently small. Let $u_{n+1}=f_{n+1}(u_n)$ where
$f_{n+1}$ is the RG map of section~4. Then there exist constants $C_{L}$ 
independent of $n$ and $\e$ such that}

$$\vert\xi_{n} \vert \le C_L{\bar g}^{11/4-\eta} \Eq(555.3)$$ 

$$\vert\rho_{n}\vert \le C_L{\bar g}^{11/4-\eta} \Eq(555.4)$$ 

$$|||R_{n+1}|||_{n+1}\le L^{-1/4}{\bar g}^{11/4-\eta} \Eq(55.7)$$

$$|g_{n+1}-\bar g| < 2\n{\bar g}^{3/2}, \quad 
|\m_{n+1}|< O(1)L^{3+\e\over 2} {\bar g}^{2-\d}
\Eq(55.5) $$

\vskip.2truecm 
 
\\{\it Remark } :
The lemmas which follow will serve to prove Theorem~5.1.  They are 
organized as in Section~5 of [BMS]. We 
remark that Lemmas~5.1-5.4, 5.9, and Lemma~ 5.12 are independent of the domain
hypothesis. {\it All the other lemmas are under the assumption that
$(g_n,\m_n,R_n)$ belong to the domain ${\cal D}_{n}$}.
Lemmas~5.21, 5.22, 5.23 and 5.27 are the major 
parts of the program. $R_{n+1,\rm main}$ is bounded in Lemma~5.21 and 
this result determines the qualitative form of the bound on the 
remainder.  $R_{n+1,3}$ and $R_{n+1,4}$ are seen, in Lemmas~5.22, 5.23 to be 
negligible in comparison.  $R_{n+1,\rm linear}$ is the crux of the program 
and it is bounded in Lemma~5.27.  The remaining Lemmas are auxiliary 
results on the way to these Lemmas. 
These auxiliary lemmas implement some of the following principles: Wick 
constants $C_{n}(0)$ are uniformly bounded by constants $C=C_{L}$. In
bounds by $G_{\k},{\bf h},{\cal A}$ norms, a fluctuation field $\zeta$ 
contributes a constant $C= O(1)(\r(L)\k(L))^{-1/2}$ and a field 
$\f$ contributes a constant $O (1){\bar g}^{-1/4}$.  
The contributions of these fields as well as of the Grassmann fields 
$\psi, \h$ are controlled by the structure  
of the norms defined in section 2  (with above choice of ${\bf h},{\bf h}_*$)
and later in this section  ((5.20) et seq). 
Integration over the Grassman fluctuation fields $\h$ is controlled by the 
Gramm inequality. 
In bounds by the ${\bf h_{*}},{\cal A}$ norms, fluctuation fields $\z$ 
contribute a constant $C= O(1)(\r(L)\k(L))^{-1/2}$ and 
fields $\f$ contribute $O (1)$. $h_{B*}$ has been adjusted to take care
of the constant $C$ above in the contribution of the fluctuation field.

\vskip0.3cm

\\{\it Lattice Taylor expansions}

\\In the following we will have occasion to estimate the difference of lattice
fields at two different points of a hypercubic lattice $(\d_n\math{Z})^{d}$. 
Let $f$ be a lattice function and $x,y$ be two points in the lattice. We
write $y-x=\sum_{j=1}^{d}\d_n \e_j h_j e_j$ where $h_j\in \math{Z}_{+}$, 
$\e_j={\rm sign}(y_j-x_j)$ and the $e_j$ are
the unit vectors of the lattice. We will express the difference 
$f(y)-f(x)$ as a sum of forward and backward lattice derivatives of $f$ along 
segments of a specified lattice path. As usual a forward derivative
in the direction $e_j$ is denoted $\dpr_{\d_n,e_j}$ and the backward derivative
is denoted $\dpr_{\d_n,-e_j}$. Given $j\in \{1,2,..,d\}$, 
$s\in \math{Z}_{+}$ we define $p_j(x-y,s)\in (\d_n\math{Z})^{d}$ by

$$p_j(y-x,s)=\sum_{i=1}^{j-1}(y-x,e_i)e_i +\d_n \e_j s e_j \Eq(5.001) $$

\\with the convention that if $j=1$ the sum is empty.
Then it is a simple matter to verify that

$$f(y)-f(x)=\d_n\sum_{j=1}^{d}\sum_{s_j=0}^{h_j-1}
\dpr_{\d_n,\e_j e_j}f(x+p_j(y-x,s_j))\Eq(5.002)  $$

\\By iterating \equ(5.002) we get the second order lattice Taylor expansion

$$f(y)-f(x)=\sum_{j=1}^{d} ((y-x),e_{j})
\dpr_{\d_n,\e_j e_j}f(x) +\d_n^{2}\sum_{j,k=1}^{d}\sum_{s_j=0}^{h_j-1}
\sum_{s_k=0}^{h_k-1}\dpr_{\d_n,\e_j e_j}\dpr_{\d_n,\e_k e_k} 
f(x+ p_k(p_j(y-x,s_j),s_k)) \Eq(5.003) $$

\vskip.3truecm 
 
\\{\it Lemma 5.1 
 
\\Let $Z_{\d_{n}},X_{\d_{n}}$ be connected 1-polymers in
$(\d_{n}\math{Z})^{3}$. Let $Y_{\d_{n}}=\emptyset$ or 
$ Y_{\d_{n}} =L^{-1}X_{\d_{n}} \subset Z_{\d_{n}}$ such that 
vol$(Z_{\d_{n}}\setminus Y_{\d_{n}})$

\\$\ge {1\over 2}$. Choose 
any $\g=O(1)>0$ and $\k=\k(L)>0$ as in Lemma~2.1. \textRed
Let $\bar g$ be sufficiently
small so that $0\le {\bar g}\le \k^{2}$. \textBlack    Let 
$\f :{\tilde Z}_{\d_{n}}^{(5)}\rightarrow \math{C}$ where 
${\tilde Z}_{\d_{n}}^{(5)}$ is $Z_{\d_{n}}$ with $5$-collar attached 
(see \equ(1.22c), \equ(1.22d)). Then there exists an $O(1)$ constant 
which depends on $j$ such that

$$\Vert\f\Vert_{C^{2}(Z_{\d_{n}})}^j\le  O(1)2^{|Z|}{\bar g}^{-{j\over 4}} 
e^{\g {\bar g}\int_{Z_{\d_{n}}\backslash Y_{\d_{n}}}dy
|\f(y)|^4}G_\k(Z_{\d_{n}},\f)\Eq(5.1)$$ 

\\For $Y_{\d_{n}}=\emptyset$ the above bound holds without the factor
$2^{|Z|}$. 
}
 
\\{\it Proof}: This is the lattice analogue of Lemma~5.1 in [BMS].
The proof reposes on the H\"older inequality and
the lattice Sobolev inequality ( see [BGM], Appendix B for an
elementary proof). Let $x\in Z_{\d_{n}}$. Write

$$\f(x)= {1\over vol( Z_{\d_{n}}\backslash Y_{\d_{n}})}
\int_{ Z_{\d_{n}}\backslash Y_{\d_{n}} }dy (\f(y)+\f(x) -\f(y))$$

\\and bound

$$|\f(x)|\le {1\over vol(Z_{\d_{n}}\backslash Y_{\d_{n}} )}
\int_{Z_{\d_{n}}\backslash Y_{\d_{n}} }dy |\f(y)|+ 
{1\over vol( Z_{\d_{n}}\backslash Y_{\d_{n}} )}
\int_{Z_{\d_{n}}\backslash Y_{\d_{n}} }dy | \f(x) -\f(y)| $$

\\We bound the first term on the right hand side by 
$ O(1)\Vert \f\Vert_{L^{4}(Z_{\d_{n}}\backslash Y_{\d_{n}} )}$. 
To bound the second term we
write the difference $\f(y) -\f(x)$ as a sum of lattice derivatives along the
segments of the path as in \equ(5.002). Any connected polymer 
$Z_{\d_n}$ as defined in 
section~1.3 can be represented as $Z_{\d_n}=Z\cap (\d_n\math{Z})^{3}$ where 
$Z$ is a connected continuum polymer. If $Z_{\d_n}$ is a \textRed
block (unit cube)  \textBlack
then the path $p_j(y-x,s_j)$  in \equ(5.002)
lies entirely in $Z_{\d_n}$. If  $Z_{\d_n}$ is not a \textRed block \textBlack
then it decomposes
as a connected union of \textRed blocks. \textBlack If $x,y$ are not in the 
same \textRed block \textBlack 
then it suffices to consider the case when they are in 
adjacent components. We pick a point $z_0$ in the intersection of the closures,
write $f(x)-f(y)=(f(x)-f(z_0))+(f(z_0)-f(y))$ and use the above first order
taylor expansion for each summand. The estimates below remain valid. Therefore
it is sufficient to consider the case  $Z_{\d_n}$ is a block. From

$$\f(y)-\f(x)=\d_n\sum_{j=1}^{3}\sum_{s_j=0}^{h_j-1}
\dpr_{\d_n,\e_j e_j}\f(x+p_j(y-x,s_j))\Eq(5.1a) $$

\\we get the bound

$$|\f(y) -\f(x)|\le \sum_{j=1}^{3}\d_n|h_j|
\sup_{z\in Z_{\d_{n}}}|\dpr_{\d_{n},\e_j e_j}\f(z)| 
\le 3\d_n(\max_{j}|h_j|) \max_{j}\sup_{z\in Z_{\d_{n}}}
|\dpr_{\d_{n},\e_j e_j}\f(z)| $$
$$\le O(1)|y-x|\> \Vert \f \Vert_{Z_{\d_{n}},1,5}\Eq(5.1b)$$

\\where in the last step we have used the lattice Sobolev embedding theorem.
We have $|y-x|\le |Z|$. Putting the above bounds together we get

$$|\f(x)|\le 0(1)|Z|(\Vert \f\Vert_{L^{4}(Z_{\d_{n}}\setminus Y_{\d_{n}})} +
\Vert \f \Vert_{Z_{\d_{n}},1,5})$$
 
\\We also have for $k=1,2$, 
$|\dpr_{\d_{n},\m_1,..,\m_k}^{k}\f(x)|\le \Vert \f \Vert_{Z_{\d_{n}},1,5}$
by Sobolev embedding. Therefore combining with the previous inequality we get

$$\Vert\f\Vert_{C^{2}(Z_{\d_{n}})}\le 
0(1)|Z|(\Vert \f\Vert_{L^{4}(Z_{\d_{n}}\setminus Y_{\d_{n}})} +
\Vert \f\Vert_{Z_{\d_{n}},1,5})$$
 
\\Hence

$$\Vert\f\Vert_{C^{2}(Z_{\d_{n}})}^{j}\le 
0(1)^{j}|Z|^{j}(\Vert \f\Vert_{L^{4}(Z_{\d_{n}}\setminus Y_{\d_{n}})}^{j} +
\Vert \f\Vert_{Z_{\d_{n}},1,5}^{j})\le C(j) \> 2^{|Z|}
{\bar g}^{-j/4} e^{\g {\bar g}\int_{Z_{\d_{n}}\setminus Y_{\d_{n}}}dy\> 
\f^{4}(y)}G_{\k}(Z_{\d_{n}},\f) $$

\\where $C(j)$ is an O(1) constant that depends on $j$.   
We have used the hypothesis
that $\bar g$ is sufficiently small so that $0\le\bar g\le \k^{2}$.
This proves the 
bound \equ(5.1). We now
prove the statement following \equ(5.1). Let 
$Y_{\d_{n}}=\emptyset$. For $x\in Z_{\d_{n}}$ pick the unit 
block $\D_{\d_{n}}\subset Z_{\d_{n}},\> \D_{\d_{n}}\ni x $. We have

$$|\f(x)|\le \int_{\D_{\d_{n}}}dy\>|f(y)|+\int_{\D_{\d_{n}}}dy\>|\f(x)-\f(y|$$

\\Proceeding as before the first term is bounded by the $L^{4}(\D_{\d_{n}})$
norm which is less than the $L^{4}(Z_{\d_{n}})$ norm. The second term is 
bounded as before except that since $x,y\in \D_{\d_{n}}$ we have 
$|x-y| \le O(1)$. The rest is as before. \bull

\vglue.3truecm 
 
\\In effecting the fluctuation map in Section~3.1 we created polymer activities
which depended separately on $\Phi$ and the fluctuation field $\xi$. The 
following lemmas will enable us to  estimate the contributions of the bosonic 
fluctuation field $\z$ at various steps.  Define a large field regulator for
the bosonic fluctuation field $\z : \tilde {X}_{\d_{n}}^{5}\rightarrow \bf C$

$$\tilde G_{\k,\r}(X_{\d_{n}},\z)=e^{\r\Vert\z\Vert^{2}_{L^{2}(X_{\d_{n}})}}
G_{\k}(X_{\d_{n}} ,\z) ,\quad \r,\k>0\Eq(5.2)$$ 

\\$\k$ is chosen as in Lemma~2.1 and is held sufficiently small. The choice 
of $\r >0$ is dictated by Lemma 5.3 below. 
\vglue.3truecm 
 
\\{\it Lemma 5.2 
 
\\For any $x\in X_{\d_{n}}$ 

$$|\z(x)|^j\le C_{\r,j,\k}\tilde G_{\k,\r}(X_{\d_{n}},\z)\Eq(5.3)$$ 
 
\\where \textRed   $C_{\r,j}=(\r^{-1/2}+ \k^{-1/2})^{j}O(1) $ \textBlack 
and $O(1)$ depends  on  $j$. We have isolated out the $\r,\k$ 
dependence in the bound. 
} 

\\{\it Proof} : The proof follows the lines of the proof of Lemma~5.1 for the
case $Y_{\d_{n}}=\emptyset$. Take the unit block $\D_{\d_{n}}
\subset X_{\d_{n}}$ such that $\D_{\d_{n}}\ni x$. We replace
the $L^{4}$ norm by the $L^{2}$ norm in the appropriate place and estimate 
$|\z(x)-\z(y)|$ with 
$x,y\in \D_{\d_{n}}$ as before now using the regulator $ \tilde G_{\k,\r}$.
\bull
 
\vglue.3truecm 
\\The parameter $\r > 0$ is chosen such that the following
Lemma~5.3 holds. $\r$ depends on $L$.

\vglue.1truecm
\\{\it Lemma 5.3 : Let $\k >0$ be chosen as in Lemma~2.1. Then there exists
$\r_{0}=\r_{0}(L) >0$ independent of $n$ such that for all 
$\r$, $0 <\r\le \r_{0}$
 
$$\int d\m_{\G_{n}}(\z)\tilde G_{\k,\r}(X_{\d_{n}},\z)\le 2^{|X_{\d_{n}}|}
\Eq(5.4)$$ 
 
} 
 
\\Lemma~5.3 is proved in the same way as Lemma~2.1. 
 
\vglue.3truecm 
\\We introduce a new  
intermediate large field regulator which combines the ones introduced earlier
 
$$\hat G_{\k,\r}(X_{\d_{n}},\z,\f)
=G_\k(X_{\d_{n}},\z+\f)G_\k(X_{\d_{n}},\f)\tilde G_{\k,\r}(X_{\d_{n}},\z) 
\Eq(5.9)$$ 
  
\vglue.2truecm 
 
\\{\it Lemma 5.4 : Let $\k,\>\r$ be chosen as in Lemma~2.1 and Lemma~5.3
respectively. Then we have
 
$$\int d\m_{\G_{n}}(\z)\hat G_{\k,\r}(X_{\d_{n}},\z,\f)\le 
2^{|X_{\d_{n}}|}G_{3\k}(X_{\d_{n}},\f) 
\Eq(5.12)$$ 

} 
 
\vglue.3truecm 
 
\\{\it Proof}: The proof follows from an application of the \textRed
H$\ddot{\rm o}$lder \textBlack
inequality and Lemmas~2.1, 5.3. 
 
\vglue.3truecm 
\textRed
\\{\it Intermediate norms}:

\vglue.2truecm

\\We will set up some additional norms to help us control intermediate steps
where we encounter polymer activities which are functions  
of the  four separate fields $\f,\z,\psi,\h$. These norms supplement the  
basic norms defined in section~2, \equ(2.2)-\equ(2.6).
 
\vglue0.2cm

\\Let $\tilde\Omega(X_{\d_{n}})$ be the Grassmann algebra with (bosonic) 
coefficients in ${\cal F}(X_{\d_{n}})$ generated by
$\psi(x)$, $\bar\psi(x)$, $\h(x)$, $\bar\h(x)$ for all $x\in X_{\d_{n}}$. 
We assign to $\h,\bar \h$ the same degrees as for $\psi,\bar\psi$. This 
is a graded algebra and $\tilde\Omega^{0}(X_{\d_{n}})$ denotes the subalgebra
of degree $0$ elements. Note that 
$\Omega^{0}(X_{\d_{n}})\subset \tilde\Omega^{0}(X_{\d_{n}})$. 
Consider any polymer activity 
${\tilde K}(X_{\d_{n}},\f,\z,\psi,\h) \in \tilde\Omega^{0}(X_{\d_{n}})$.
Let $I\subset \{1,2,...,p\}$ and $J\subset \{1,2,...,p\}$. Define
$I^{c}=\{1,2,...,p\}\setminus I$. We introduce the
abbreviated notation ${\bf x}_{I}:=(x_{i_1},...,x_{i_|I|})$ where for 
$i_j \in I$ for $j=1,..,|I|$. We will refer to the $x_{i_j}$ as the members
of ${\bf x}_{I}$. Define
$\psi({\bf x}_I):=\psi(x_{i_1}).....\psi(x_{i_|I|})$ and
${\dpr\over \dpr\psi({\bf x}_I)}:= \prod_{j=0}^{|I|-1}
{\dpr\over \dpr\psi(x_{|I|-j})}$. We now define

$$D_{F}^{2p,IJ}
{\tilde K}(X_{\d_{n}},\f,\z,{\bf x}_{I},{\bf x}_{I^{c}},{\bf y}_{J},
{\bf y}_{J^{c}}):={\dpr\over \dpr\bar\h({\bf y}_{J^{c}})}
{\dpr\over \dpr\h({\bf x}_{I^{c}})}{\dpr\over \dpr\bar\psi({\bf y}_J)}
{\dpr\over \dpr\psi({\bf x}_I)}{\tilde K}(X_{\d_{n}},\f,\z,\psi,\h)
\Bigl|_{\psi=\h=0} \Eq(2.11.22)$$

\\Note that the left hand side is antisymmetric respectively in the members
of ${\bf x}_{I},{\bf x}_{I^{c}},{\bf y}_{J},{\bf y}_{J^{c}}$.
Let ${\bf x}:=({\bf x}_{I},{\bf x}_{I^c})$ and 
${\bf y}:=({\bf y}_{J},{\bf y}_{J^c})$.
Then ${\tilde K}(X_{\d_{n}},\f,\z,\psi,\h)$ can be represented uniquely as

$${\tilde K}(X_{\d_{n}},\f,\z,\psi,\h)=\sum_{p\ge 0} 
\sum_{{I\subset\{1,...,p\}}\atop {J\subset\{1,...,p\}}} 
{1\over |I|!\>|I^{c}|!\>|J|!\>|J^c|!}
\int_{X_{\d_{n}}^{2p}} d{\bf x}d{\bf y} \>
D_{F}^{2p,IJ}{\tilde K}(X_{\d_{n}},\f,\z,{\bf x}_{I},{\bf x}_{I^{c}},{\bf y}_{J},
{\bf y}_{J^{c}}) \times $$
$$\psi({\bf x}_I)\bar\psi({\bf y}_J)\h({\bf x}_{I^c})\bar\h({\bf y}_{J^c})
\Eq(2.1.2)$$ 

\\Let $g_{IJ}:(\tilde X_{\d_n}^{(2)})^{|I|}\times (\tilde X_{\d_n}^{(2)})^{|J|}
\rightarrow {\bf C}$ such that  $g_{IJ}({\bf x}_{I},{\bf y}_{J})$ is
antisymmetric respectively in the members of ${\bf x}_{I}$ and those of
${\bf y}_{J}$. 
Let $h_{I^{c}J^{c}}:(\tilde X_{\d_n}^{(2)})^{|I^{c}|}
\times (\tilde X_{\d_n}^{(2)})^{|J^{c}|}
\rightarrow {\bf C}$ such that  
$h_{I^{c}J^{c}}({\bf x}_{I^{c}},{\bf y}_{J^{c}})$ is
antisymmetric respectively in the members of ${\bf x}_{I^{c}}$ and those of
${\bf y}_{J^{c}}$. The tensor product $g_{IJ}\otimes h_{I^{c}J^{c}}$ maps
$(\tilde X_{\d_n}^{(2)})^{|I|}\times (\tilde X_{\d_n}^{(2)})^{|J|}\times
(\tilde X_{\d_n}^{(2)})^{|I^c|}\times (\tilde X_{\d_n}^{(2)})^{|J^c|}
=(\tilde X_{\d_n}^{(2)})^{2p} \rightarrow {\bf C} $. By definition
$(g_{IJ}\otimes h_{I^{c}J^{c}})
({\bf x}_{I},{\bf x}_{I^{c}},{\bf y}_{J},{\bf y}_{J^{c}})=
g_{IJ}({\bf x}_{I},{\bf y}_{J})h_{I^{c}J^{c}}({\bf x}_{I^{c}},{\bf y}_{J^{c}})$
. We consider the space of functions $g_{IJ}$ endowed  with the 
$C^{2}((X_{\d_{n}})^{|I|}\times (X_{\d_{n}})^{|J|})$ norm. Similarly we 
consider the space of functions $h_{I^{c}J^{c}} $ endowed  with the 
$C^{2}((X_{\d_{n}})^{|I^{c}|}\times (X_{\d_{n}})^{|J^{c}|})$ norm. Define

$$D^{2p,IJ,m}{\tilde K}(X_{\d_{n}},\f,\z,0,0; f^{\times m},
g_{IJ}\otimes h_{I^{c}J^{c}} )=
\int_{X_{\d_{n}}^{2p}} d{\bf x} d{\bf y} \>D_{B}^m D_{F}^{2p,IJ}
{\tilde K}(X_{\d_{n}},\f,\z,{\bf x}_{I},{\bf x}_{I^{c}},{\bf y}_{J},
{\bf y}_{J^{c}};f^{\times m} )\times $$
$$(g_{IJ}\otimes h_{I^{c}J^{c}})
({\bf x}_{I},{\bf x}_{I^{c}},{\bf y}_{J},{\bf y}_{J^{c}})
\Eq(2.2.2) $$

\\where the derivative $D_{B}^m$ of the bosonic coefficient
is with respect to the field $\f$ (and not the fluctuation field $\z$). 
This defines a multilinear functional on the normed subspace of antisymmetric 
functions in $C^{2}((X_{\d_{n}})^{|I|}\times (X_{\d_{n}})^{|J|})\times
C^{2}((X_{\d_{n}})^{|I^{c}|}\times (X_{\d_{n}})^{|J^{c}|})$. 

\\The norm of the multilinear functional \equ(2.2.2) is defined analogously to 
\equ(2.3), namely
 
$$\Vert D^{2p,IJ,m}{\tilde K}(X_{\d_{n}},\f,\z,0,0)\Vert = 
\sup_{\Vert f_j\Vert_{C^{2}(X_{\d_{n}})}\le 1,\> \forall 1\le j\le m
\atop{ \Vert g_{IJ}\Vert_{C^{2}(X_{\d_{n}}^{|I|}\times (X_{\d_{n}}^{|J|})}
\le 1
\atop  \Vert h_{I^{c}J^{c}}\Vert_{C^{2}(X_{\d_{n}}^{|I^{c}|}\times 
(X_{\d_{n}}^{|J^{c}|})}\le 1 }}\Bigl| 
\int_{X_{\d_{n}}^{2p}} d{\bf x}d{\bf y} $$
$$D_{B}^{m}D_{F}^{2p,IJ}
{\tilde K}(X_{\d_{n}},\f,\z,{\bf x}_{I},{\bf x}_{I^{c}},{\bf y}_{J},
{\bf y}_{J^{c}}; f^{\times m})(g_{IJ}\otimes h_{I^{c}J^{c}})
({\bf x}_{I},{\bf x}_{I^{c}},{\bf y}_{J},{\bf y}_{J^{c}})\Bigr|
\Eq(2.3.2)$$ 

\\In the beginning of this section we specified 
${\bf h}=(h_F,h_B)$ and ${\bf h}_{*}=(h_F,h_{B*})$.

\\We define the norms 
 
$$\Vert {\tilde K}(X_{\d_{n}},\f,\z,0,0)\Vert_{\bf h}= 
\sum_{p=0}^\io\sum_{m=0}^{m_0} 
\sum_{{I\subset\{1,...,p\}}\atop {J\subset\{1,...,p\}}} 
{h_B^m\over m!}{h_F^{2p} \over |I|!\>|I^{c}|!\>|J|!\>|J^c|!} 
\Vert D^{2p,IJ,m} {\tilde K}(X_{\d_{n}},\f,\z,0,0)\Vert   \Eq(2.4.2) $$ 
 
$$\Vert {\tilde K}(X_{\d_{n}})\Vert_{{\bf h},\hat G_{\k,\r}}= 
\sup_{\f,\z\in {\cal F}_{\tilde X_{\d}^{(5)}} }
\Vert {\tilde K}(X_{\d_{n}},\f,\z,0,0) 
\Vert_{\bf  h}\hat G^{-1}_{\k,\r}(X_{\d_{n}},\f,\z)\Eq(2.6.2)$$ 
 
$$\Vert {\tilde K}(X_{\d_{n}},0,\z,0,0)\Vert_{\bf h_{*}}= 
\sum_{p=0}^\io\sum_{m=0}^{m_0}
\sum_{{I\subset\{1,...,p\}}\atop {J\subset\{1,...,p\}}} 
{h_{B*}^m\over m!}{h_F^{2p} \over |I|!\>|I^{c}|!\>|J|!\>|J^c|!} 
\Vert D^{2p,IJ,m}\tilde {\tilde K}(X_{\d_{n}},0,\z,0,0)\Vert  \Eq(2.6.2.1)$$ 
 
$$\Vert {\tilde K}(X_{\d_{n}})\Vert_{\bf h_{*},\tilde G_{\k,\r}}     
= \sup_{\z\in {\cal F}_{\tilde X_{\d}^{(5)}}} 
\Vert {\tilde K}(X_{\d_{n}},0,\z,0,0)\Vert_{\bf h_{*}}  
{\tilde G}^{-1}_{\k,\r}(X_{\d_{n}},\z)   \Eq(2.6.2.2)    $$

$$\vert {\tilde K}(X_{\d_{n}})\vert_{\bf {h_{*}}}= 
\sum_{p=0}^\io\sum_{m=0}^{m_0}
\sum_{{I\subset\{1,...,p\}}\atop {J\subset\{1,...,p\}}} 
{h_{B*}^m\over m!}{h_F^{2p} \over |I|!\>|I^{c}|!\>|J|!\>|J^c|!} 
\Vert D^{2p,IJ,m} {\tilde K}(X_{\d_{n}},0,0,0,0)\Vert\Eq(2.5.2)$$

\\It is straightforward to prove that the above $\bf h$ and ${\bf h}_{*}$ 
norms satisfy the multiplicative property of Proposition~2.3. 

\vglue0.2cm
\\{\it Special case}: Consider the map  
$\tilde\Omega^{0}(X_{\d_{n}})\rightarrow \Omega^{0}(X_{\d_{n}})$
given by 

$${\tilde K}(X_{\d_{n}},\f,\z,\psi,\h)=K(X_{\d_{n}},\f +\z, \psi +\h)
\Eq(2.6.2.3b) $$. 

\\Norms for $K$ were defined earlier 
earlier in section~2, see \equ(2.2)-\equ(2.5). On the other hand 
the $\bf h$ and ${\bf h}_*$ norms of $\tilde K$ are defined in \equ(2.4.2),
\equ(2.6.2.1) above. We have 

\vglue0.1cm
\\{\it Lemma~5.4A: 
\vglue0.1cm

\\Define $\hat{\bf h}:=({h_F\over 2}, h_B)$ and
$\hat{\bf h}_{*}:=({h_F\over 2}, h_{B*})$. Then we have for the polymer 
activity defined in \equ(2.6.2.3b) 

$$\Vert {\tilde K}(X_{\d_{n}},\f,\z,0,0)\Vert_{\hat{\bf h}}\le
\Vert K(X_{\d_{n}},\f +\z, 0)\Vert_{\bf h} \Eq(2.6.2.3) $$

$$\Vert {\tilde K}(X_{\d_{n}},0,\z,0,0)\Vert_{\hat{\bf h}_{*}}\le
\Vert K(X_{\d_{n}},\z, 0)\Vert_{{\bf h}_{*}} \Eq(2.6.2.3a) $$

}

\\{\it Proof}: Let $I\subset \{1,..,p\}$ and $J\subset \{1,..,p\}$.
From the definitions \equ(2.11.22) and \equ(2.1b) we have upto a sign factor

$$D_{B}^{m}D_{F}^{2p,IJ}
{\tilde K}(X_{\d_{n}},\f,\z,{\bf x}_{I},{\bf x}_{I^{c}},{\bf y}_{J},
{\bf y}_{J^{c}};f^{\times m} )
=(-1)^{\sharp}D_{B}^{m}D_{F}^{2p}
K(X_{\d_{n}},\f+\z,{\bf x}_{I},{\bf x}_{I^{c}},{\bf y}_{J},
{\bf y}_{J^{c}};f^{\times m} ) \Eq(2.6.2.3d) $$

\\Let $(g_{IJ}\otimes h_{I^{c}J^{c}})({\bf x}_{I},{\bf x}_{I^{c}},{\bf y}_{J},
{\bf y}_{J^{c}})$ be a test function as defined after \equ(2.1.2). It is
antisymmetric respectively in the 
members of ${\bf x}_{I},{\bf x}_{I^{c}},{\bf y}_{J},
{\bf y}_{J^{c}}$. From \equ(2.2.2) and \equ(2.6.2.3d)  we get

$$D^{2p,IJ,m}{\tilde K}(X_{\d_{n}},\f,\z,0,0; f^{\times m}, g_{2p,IJ})=
(-1)^{\sharp}\int_{X_{\d_{n}}^{2p}} d{\bf x} d{\bf y} \>D_{B}^m D_{F}^{2p}
K(X_{\d_{n}},\f+\z,{\bf x}_{I},{\bf x}_{I^{c}},{\bf y}_{J},
{\bf y}_{J^{c}};f^{\times m} )\times $$
$$(g_{IJ}\otimes h_{I^{c}J^{c}}) 
({\bf x}_{I},{\bf x}_{I^{c}},{\bf y}_{J},{\bf y}_{J^{c}})
\Eq(2.6.2.3e)$$

\\Let $({\bf x}_{I},{\bf x}_{I^{c}})=(x_1,...,x_p)$ and 
$({\bf y}_{J},{\bf y}_{J^{c}})=(y_1,..,y_p)$. Write
$(g_{IJ}\otimes h_{I^{c}J^{c}}) 
({\bf x}_{I},{\bf x}_{I^{c}},{\bf y}_{J},{\bf y}_{J^{c}})=
(g_{IJ}\otimes h_{I^{c}J^{c}})(x_1,..,x_p,y_1,..y_p)$. 
Let $S_p$ be the permutation 
group of $\{1,..,p\}$. Define 

$${\cal A}(g_{IJ}\otimes h_{I^{c}J^{c}})(x_1,...,x_p,y_1,..y_p)=
{1\over (p!)^{2}}\sum_{\s,\s'\in S_p}
(g_{IJ}\otimes h_{I^{c}J^{c}})
(x_{\s(1)},...x_{\s(p)}, y_{\s'(1)},...y_{\s'(p)})$$

\\Now the coefficient function $D_{B}^m D_{F}^{2p}
K(X_{\d_{n}},\f+\z,{\bf x}_{I},{\bf x}_{I^{c}},{\bf y}_{J},
{\bf y}_{J^{c}};f^{\times m} )$ is  antisymmetric in  
$({\bf x}_{I},{\bf x}_{I^{c}})=$

\\$(x_1,...,x_p)$ and in $({\bf y}_{J},{\bf y}_{J^{c}})=(y_1,..,y_p)$.
Therefore we can replace $g_{IJ}\otimes h_{I^{c}J^{c}}$ in \equ(2.6.2.3e) 
by ${\cal A}(g_{IJ}\otimes h_{I^{c}J^{c}})$ and hence

$$\Bigl|D^{2p,IJ,m}{\tilde K}(X_{\d_{n}},\f,\z,0,0; f^{\times m}, g_{2p,IJ})
\Bigr|\le \Vert D^{2p,m}K(X_{\d_{n}},\f+\z,0\Vert 
\prod_{j=1}^{m}\Vert f_j\Vert_{C^{2}(X_{\d_{n}})}
\Vert {\cal A}(g_{IJ}\otimes h_{I^{c}J^{c}})\Vert_{C^{2}(X_{\d_{n}}^{2p})}$$

\\Now 
$\Vert {\cal A}(g_{IJ}\otimes h_{I^{c}J^{c}})\Vert_{C^{2}(X_{\d_{n}}^{2p})}\le
\Vert g_{IJ}\Vert_{C^{2}((X_{\d_{n}})^{|I|}\times (X_{\d_{n}})^{|J|})} \>
\Vert h_{I^{c}J^{c}}\Vert_{C^{2}((X_{\d_{n}})^{|I^{c}|}\times 
(X_{\d_{n}})^{|J^{c}|})}$. 
Using this in the previous inequality gives 

$$\Vert D^{2p,IJ,m}{\tilde K}(X_{\d_{n}},\f,\z,0,0)\Vert
      \le \Vert D^{2p,m}K(X_{\d_{n}},\f+\z,0\Vert  \Eq(2.6.2.3f) $$

\\Therefore

$$\Vert {\tilde K}(X_{\d_{n}},\f,\z,0,0)\Vert_{\hat{\bf h}}\le
\sum_{p=0}^\io\sum_{m=0}^{m_0} 
\sum_{{I\subset\{1,...,p\}}\atop {J\subset\{1,...,p\}}} 
{h_B^m\over m!}{2^{-2p}h_F^{2p}\over |I|!\>|I^{c}|!\>|J|!\>|J^c|!} 
\Vert D^{2p,m}K(X_{\d_{n}},\f,\z,0)\Vert  \Eq(2.6.2.3g) $$ 

\\Now

$$\sum_{{I\subset\{1,...,p\}}\atop {J\subset\{1,...,p\}}} 
{1\over |I|!\>|I^{c}|!\>|J|!\>|J^c|!}={1\over (p!)^{2}}2^{2p}$$

\\Substituting this in the previous inequlity and using the definition
\equ(2.4) gives

$$\Vert {\tilde K}(X_{\d_{n}},\f,\z,0,0)\Vert_{\hat{\bf h}}\le
\Vert K(X_{\d_{n}},\f +\z, 0)\Vert_{\bf h} $$

\\which proves \equ(2.6.2.3). The proof of \equ(2.6.2.3a) is the same. \bull

\textBlack

\vglue.3truecm 

\\{\it Lemma 5.5 
 
\\Let $g_n, \m_{n}$ belong to ${\cal D}_{n}$. Let $Y_{\d_{n}}$ be a 
1-polymer. Then
for $V_{n} (Y_{\d_{n}},\Phi,\xi)= 
V(Y_{\d_{n}},\Phi+\xi,C_{n},g_n,\m_n)$ or 
$V(Y_{\d_{n}},\Phi,S_{L}C_{n+1} ,g_n,\m_n)$, we have 
 
$$\Vert e^{- V_{n}(Y_{\d_{n}} ,\Phi,\xi )}\Vert_{\bf h}\le 2^{|Y_{\d_{n}}|} 
e^{-{\bar g\over 8}\int_{Y_{\d_{n}}} dx |\f +\z|^4(x)}\Eq(5.5)$$ 
 
$$ 
\Vert e^{-V_{n}(Y_{\d_{n}},\Phi,\xi)}\Vert_{{\bf h}_{*}}
\le  2^{|Y_{\d_{n}}|}\Eq(5.lem5.1) $$ 
 
\\for $\e >0$ sufficiently small depending 
on $L$. In the above norms ${\bf h}=(h_{B},h_{F})$ and  
${{\bf h}_{*}}=(h_{B*},h_{F})$ are chosen as in the hypothesis for the
domain $\cal D_{\d_{n}}$. Thus $h_{F}=h_{F}(L)$, 
$h_{B}=c{\bar g}^{-{1\over 4}}$ with $c=O(1)$ sufficiently small, and  
$h_{B*}=\r^{-1/2} + \k^{-1/2}$. Note that $ h_{B*}$ depends on $L$ via 
$\r$ and $\k$.
} 
 
\vglue.3truecm 
 
\\{\it Proof} :  It is sufficient to prove this when $Y_{\d_n}$ is a 1-block
$\D_{\d_{n}}$. Because otherwise we can write $Y_{\D_n}$ as a disjoint union
of 1-blocks and write the left hand side as a product over 1-block 
contributions. Then the multiplicative property of the $\bf h$ norm (
Proposition~2.3) gives the lemma.

\\From the definition of $V$ in \equ(4.1) we get on undoing
the Wick ordering , (see \equ(0.20)),

$$V_{n}(\Delta_{\d_{n}},\Phi)=V_{n,u}(\Delta_{\d_{n}},\f)+
2g_{n}\int_{\D_{\d_{n}}}dx\f\bar\f(x)\psi\bar\psi(x)+ 
{\tilde\m_{n}}\int_{\D_{\d_{n}}
}dx\psi\bar\psi(x) \Eq(5.12.1)$$ 

\\where  

$$V_{u,n}(\D_{\d_{n}},\f)=
\int_{\D_{\d_{n}}}dx [g_{n}(\f\bar\f(x))^{2}+{\tilde\m_{n}}\f\bar\f(x)] 
\Eq(5.12.2)$$ 

\\and ${\tilde\m_{n}}=\m_{n} -2g_{n} C_{n}(0)$.

\\By the multiplicative property of the $\bf h$ norm, 
$$\Vert e^{- V(\D_{\d_{n}},\Phi )}\Vert_{\bf h}\le 
\Vert e^{- V_{u,n}(\D_{\d_{n}},\f )}\Vert_{h_{B}}
\Vert e^{-2g_{n}\int_{\D_{\d_{n}}}dx\f\bar\f(x)\psi\bar\psi(x)}\Vert_{\bf h}
\Vert e^{-{\tilde\m_{n}}\int_{\D_{\d_{n}}}dx\psi\bar\psi(x)} 
\Vert_{h_{F}} \Eq(5.12.3)$$ 

\\We estimate each of the factors on the right hand side in turn. We observe 
that by taking $\e$ sufficiently small we can make $\bar g$ as small as 
necessary since $0<{\bar g}\le C\e$. Since \textRed 
$g_{n}, \m_{n}$ belong to $ {\cal D}_{n}$ and $0< \n < {1\over 2}$,
we have  ${\bar g\over 2}\le g_{n}\le {3\over 2}\bar g $ 
and $\m =O({\bar g}^{2-\d})$. \textBlack
Moreover from from (5a) of Theorem~1.1 and \equ(0.25) we have the uniform bound
$|C_{n}(0)|\le C_{L}$.  Therefore $|\tilde\m_n|\le C_L \bar g$ with a new
constant $C_L$.

\\For the first factor on the right hand side of \equ(5.12.3) 
we have the bound  
 
$$\Vert e^{- V_{u,n}(\D_{\d_{n}},\f )}\Vert_{h_{B}}\le  
2^{{1\over 2}|\D_{\d_{n}}|}
e^{-{g_n\over 2}\int_{\D_{\d_{n}}}dx (\f\bar\f)^2(x)}
\Eq(5.12.4) $$ 

\textRed
\\where $|\D_{\d_{n}}|=1$. This can be proved on the lines of the proof of  
Lemma~ 5.5 of [BMS]
by substituting there $\bar g$ for $\e$ and
taking account of the previous observations. Thus 

$$ V_{u,n}(\D_{\d_{n}},\f )-{g_n \over 2}\int_{\D_n}dx |\f|^{4}
\ge {\bar g\over 4}\int_{\D_n}dx(|\f|^{4}-C_L |\f|^{2})$$

\\Now $C_L |\f|^{2}\le {1\over 2}(|\f|^{4}+ {C_L}^{2})$. {\it Let $\bar g$ be 
sufficiently small so that ${\bar g}{C_L}^{2}\le {\bar g}^{1\over 2}$. } 
Using these two observations we get from the previous inequality 

$$e^{-V_{u,n}(\D_{\d_{n}},\f )}\le (1+O({\bar g}^{1\over 2}))
e^{-{g_n \over 2}\int_{\D_n}dx |\f|^{4}} $$

\\The rest of the proof which consists of estimating, for $k\ge 1$,
${h_B^{k}\over k!}\Vert D^{k}e^{-V_{u,n}}\Vert $ goes through as in the proof
of Lemma~ 5.5 of [BMS] on replacing $\e$ by $\bar g$.

\vglue0.2cm
\\Now consider the second factor in the right hand side of \equ(5.12.3).
From the multiplicative property of the $\bf h$ norm applied to the series
expansion of the exponential we get
 
$$\Vert e^{-2g_{n}\int_{\D_{\d_{n}}}dx\f\bar\f(x)\psi\bar\psi(x)}\Vert_{\bf h} 
\le  e^{2g_{n} h_{F}^{2}\int_{\D}dx\Vert\f\bar\f(x)\Vert_{h_B}} $$ 

\\Now $g_n\le {3\over 2}\bar g$ and ${\bar g}^{1\over 2}= O(1) h_B^{-2}$.
Let $t={|\f(x)|\over h_B}$. Then 

$$2g_{n} h_{F}^{2}\Vert\f\bar\f(x)\Vert_{h_B}\le 
O(1){h_F^{2}\over h_B^{2}}(t^2 +t +1)\le  
O(1){h_F^{2}\over h_B^{2}}(t^4 +1)  $$ 
 
\\which can be proved by two applications of H$\ddot o$lder's inequality.
and therefore for the second factor we have the bound 
 
$$\Vert e^{2g_{n}\int_{\D_{\d_{n}}}dx\f\bar\f(x)\psi\bar\psi(x)}\Vert_{\bf h}
\le 
2^{O(1)(h_F/h_B)^{2}|\D_{\d_n}|}  
e^{O(1)(h_F/ h_B)^{2}{\bar g} \int_{\D_{\d_n|}}dx (\f\bar\f (x))^{2}} 
\Eq(5.12.5) $$ 
 
\\Finally for the third factor we have straightforwardly the bound 
 
$$\Vert e^{{\tilde\m_{n}}\int_{\D_{\d_{n}}}dx\psi\bar\psi(x)} 
\Vert_{h_{F}} \le 2^{h_F^{2}C_{L}{\bar g}|\D_{\d_n}|} \Eq(5.12.6) $$ 
 
\\where we have used the bound $|\tilde\m_n|\le C_L {\bar g}$ ( see above).

\\Put together the bounds for the three factors. In the bound \equ(5.12.4)
use $g_n\ge {{\bar g}\over 2}$. Let $\bar g$ be 
sufficiently small ( thus making $h_B$ sufficiently large) so that
$\max (O(1),C_L) (h_F/ h_B)^{2}\le {1\over 16}$
where $O(1)$ is the constant in \equ(5.12.5)
and $C_L$ is the constant encountered
above. This ensures that in the bound \equ(5.12.5) 
the exponent $O(1)(h_F/ h_B)^{2}\le {1\over 16}$. It also ensures that
in the bound \equ(5.12.6) the exponent 
$h_F^{2}C_{L}{\bar g}\le {1\over 16}$. Thus we have obtained for $\bar g$
sufficiently small 
 
$$\Vert e^{- V(\D_{\d_{n}},\Phi )}\Vert_{\bf h} \le 
2^{|\D_{\d_n}|}e^{-{\bar g}/8\int_{\D}dx (\f{\bar \f})^2(x)}$$ 
 
\\and the first part of the lemma now follows on invoking the argument in the
beginning of  the proof.
The proof of the second part follows the same lines. \bull  

\textBlack 
\vglue0.3cm 
\\{\it Lemma 5.6 
 
\\Let $p_{n,g}(\D_{\d_n},\xi,\Phi)$ and $p_{n,\m} (\D_{d_n}, \xi, \Phi)$ 
be as given in 
\equ(4.8b). Let $g_n, \m_n$ belong to ${\cal D}_{n}$. Let
$h_{B}=c{\bar g}^{-1/4}$ and $h_{B*}$ be as in the definition of 
${\cal D}_{n}$. Recall that ${\bf h}=(h_{B},h_{F})$,  
and ${\bf h_{*}}=(h_{B*},h_{F})$, where $h_{F}=h_{F}(L)$.  
Let $\k=\k(L)>0$ and  $\r=\r(L)>0$, be as specified in Lemmas~2.1 and 5.3. 
Then for
any $\g=O(1)>0$, $0\le s < 1$ we have constants $C_{L}$ independent of $n$ and
$\e$ but depending on $L$ such that  

$$\Vert p_{n,g}(\D_{\d_n},\f,\z,0,0)\Vert_{\bf h}\le C_{L} {\bar g}^{1/4}(1-s)^{-3/4} 
\tilde G_{\k,\r}(\D_{\d_n},\z)G_{\k}(\D_{\d_n},\f)e^{{\bar g}(1-s)
\g\int_{\D_{\d_n}} 
dx\ (\f{\bar \f})^2(x)}\Eq(5.6)$$ 

$$\Vert p_{n,\m}(\D_{\d_n},\f,\z,0,0)\Vert_{\bf h}\le C_{L} {\bar g}^{7/4-\delta 
}(1-s)^{-1/2} \tilde G_{\k,\r}(\D_{\d_n},\z)G_{\k}(\D_{\d_n},\f)
e^{{\bar g}(1-s)\g\int_{\D_{\d_n}} 
dx\ (\f{\bar \f})^2(x)}\Eq(5.6b)$$ 

$$\Vert p_{n,g}(\D_{\d_n},0,\z,0,0)\Vert_{{\bf h}_{*}}\le C_{L}\> {\bar g}
\tilde G_{\k,\r}(\D_{\d_n},\z) 
\Eq(5.6.1) 
$$ 

$$\Vert p_{n,\m}(\D_{\d_n},0,\z,0,0)\Vert_{{\bf h}_{*}}\le C_{L} 
{\bar g}^{2-\delta }\tilde G_{\k,\r}(\D_{\d_n},\z) \Eq(5.6.2) 
$$ 
} 
 
\\{\it Proof} :  
$p_{n,g}(\D_{\d_n},\xi,\Phi)$ is given in  \equ(4.8a). We undo the Wick ordering  
which produces constants $C_{n}(0)$ uniformly bounded by constant
$C_{L}$ from Corollary~1.2. We can then write it in the form   
\equ(2.1.2) by expanding out in the Grassmann fields. Since it is a local 
polynomial of degree four we get,  
 
$$p_{n,g}(\D_{\d_n},\f,\z,\psi,\h)=\sum_{p=0}^{2} 
\sum_{{\bf a}}\sum_{I\subset\{1,...,2p\}}\int_{\D_{\d_n}} 
dx{\tilde p}_{n,g,2p}^{{\bf 0},{\bf a},I}(\D_{\d_n},\f,\z,x) 
\prod_{i\in I}\psi_{a_i}(x) 
\prod_{i\in I^c}\h_{a_i}(x)\Eq(5.6.3)$$ 
 
\\where ${\bf 0}$ means that 
$l_{i}=0\  \forall i$. We have following the definition of the norm
in \equ(2.4.2) with $\hat{\bf h}$ replaced by ${\bf h}$ 

$$\Vert p_{n,g}(\D_{\d_n},\f,\z,0,0)\Vert_{\bf h} \le 
\sum_{p=0}^{2} h_{F}^{2p} \sup_{||g_{2p}||\le 1}
\sum_{{\bf a}}\sum_{I\subset\{1,...,2p\}} 
\int_{\D_{\d_n}}dx\Vert {\tilde p}_{n,g,2p}^{{\bf 0},{\bf a},I}(\D_{\d_n},\f,\z,x)
\Vert_{h_{B}} 
|g_{2p}({\bf x})|$$ 
$$\le \sum_{p=0}^{2} h_{F}^{2p} 
\sum_{{\bf a}}\sum_{I\subset\{1,...,2n\}\atop |I|\ {\rm even}} 
\int_{\D_{\d_n}}dx\Vert {\tilde p}_{n,g,2p}^{{\bf 0},{\bf a},I}(\D_{\d_n},\f,\z,x)
\Vert_{h_{B}} 
\Eq(5.6.4)$$ 
 
\\where $g_{2p}({\bf x})=g_{2p}(x,x,...,x)$ and 
$\Vert g_{2p}\Vert$ is the $C^2(\D_{\d_{n}}^{2p})$ norm of $g_{2p}$. 
${\tilde p}_{n,g,2p}^{{\bf 0},{\bf a},I}(\D_{\d_n},\f,\z,x)$ is a polynomial 
in  $\f,\z$ 
and every term in the $h_B$ norm of ${\tilde p}_{n,g,2p}$ 
can be estimated as in the proof of Lemma 5.6  
of [BMS]. Each term carries a factor $g_{n}=O({\bar g})$. 
The fluctuation fields $\z$ are estimated via Lemma 5.2 and the 
fields $\f$ via Lemma 5.1. For each field $\f$ we lose 
${\bar g}^{-{1\over 4}}$. 
In the $p=0$ term the maximum power of $\f$ in ${\tilde p}_{n,g,p}$ 
is $3$, for $p=1$ the maximum power is $2$, and for $p=2$ it is $0$. 
The bound \equ(5.6) now follows as in Lemma 5.6, [BMS]. The bound \equ(5.6b) 
for $p_{n,\mu}(\D_{\d_n},\xi,\Phi)$  is proved in the same way. We just have 
to  
remember that $\m_{n} =O({\bar g}^{2-\d})$ from the domain hypothesis, and 
that the 
maximum power of the field $\f$ in the ${\tilde p}_{n,\m,p}$ is $1$. The  
remaining parts are proved in the same way. \bull
 
\vglue0.3cm 
 
Define $p_n (s) = p_n (s,\D_{\d_{n}} ,\Phi,\xi )$ by  
$p_n (s) = sp_{n,g}+s^{2}p_{n,\mu} $. 
Then $r_{n,1} = r_{n,1} (\D_{\d_{n}} ,\Phi ,\xi)$ defined by \equ({4.9}) is 
given by 
$$r_{n,1} = 
{1\over 2}\int_0^1ds(1-s)^{2}e^{-p_n (s)-\tilde{V_n}} 
\big(-p'_n (s)^{3}+6p'_n (s)p_{n,\mu }\big)\Eq(5.7)$$ 
 
\\with $p'_n (s)={d\over ds}p_n (s)=p_{n,g}+2sp_{n,\mu }$ and 
$p''_n (s)=2p_{n,\mu}$.

\vglue.3truecm 
 
\\{\it Lemma 5.7 

\\Under the conditions of the domain ${\cal D}_{n}$ there exists
a constant $C_L$ independent of $n$ and $\e$ but dependent on $L$ such that
 
$$\Vert r_{n,1}(\D_{\d_n})\Vert_{h,\hat G_{\k,\r}} \le C_{L}{\bar g}^{3/4}
\Eq(5.8)$$ 
 
$$\Vert r_{n,1}(\D_{\d_n})\Vert_{h_{*},\tilde G_{\k,\r}}\le 
C_{L}{\bar g}^{3-\delta } \Eq(5.8.1)$$ 
} 
 
\\{\it Proof} :  Follow the proof of the corresponding Lemma 5.7 of [BMS]. 
Write ${\tilde V}_{n} +p_{n}(s)= V_{n,1}(s) + V_{n,2}(s)$ where

$$ V_{n,1}(s)=V(\D_{\d_n}, \Phi +\xi, C_{n}, sg_{n}, s^{2}\m_{n}),\quad
V_{n,2}(s)=V(\D_{\d_n},\Phi,S_{L}C_{n},(1- s)g_{n}, (1-s^{2})\m_{n})$$

\\We have

$$\Vert r_{n,1}(\D_{\d_n},\f,\z,0,0)\Vert_{\bf h}
\le {1\over 2} \int_{0}^{1} ds(1-s)^{2}\Vert e^{-V_{n,1}(s)}\Vert_{\bf h}
\Vert e^{-V_{n,2}(s)}\Vert_{\bf h}\Bigl(\Vert p'(s)\Vert_{\bf h}^{3} +
6\Vert p'(s)\Vert_{\bf h} \Vert p_{\m}\Vert_{\bf h}\Bigr) $$

\\$g_n,\m_n$ belong to ${\cal D}_n$. Lemmas~ 5.5 and 5.6 continue to hold
with $g_n,\m_n$ replaced by $sg_n,s^{2}\m_n$ or $(1-s)g_n,(1-s^{2})\m_n$.
We bound 
$\Vert e^{-V_{n,1}(s)}\Vert_{\bf h}\le 2$ and 
$\Vert e^{-V_{n,2}(s)}\Vert_{\bf h} \le 2e^{-(1-s){{\bar g}\over 4}
\int_{\D_{\d_n}}dx |\f|^{4}(x)} $. We bound the remaining factor (using 
Lemma~5.6) by $C_{L}{\bar g}^{3\over 4}{\hat G}_{\r,\k}(\D_{\d_n},\f,\z)
e^{(1-s){\bar g}3\g \int_{\D_{\d_n}}dx |\f|^{4}(x)}$. We put the three bounds 
together and choose $0<\g< {1\over 12}$. This gives the bound \equ(5.8).
The proof of \equ(5.8.1) is similar. \bull

\vglue0.3cm 
 
\\{\it Lemma 5.8 
 
\\Under the conditions for the domain ${\cal D}_{\d_n}$ there exists a 
constant $C_L$, independent of $n$ and $\e$ but dependent on $L$ such that  
 
$$\Vert P_{n}(\l)\Vert_{{\bf h},\hat G_{\k,\r},\AA,\d_n}\le  
C_{L}\vert\l {\bar g}^{1/4}\vert 
\qquad {\rm for } \ |\l {\bar g}^{1/4}| \le 1 
\Eq(5.13) 
$$ 
 
$$\Vert P_{n}(\l)\Vert_{{\bf h}_{*},\tilde G_{\k,\r},\AA,\d_n} 
\le C_{L}\vert\l {\bar g}^{1-\delta /2}\vert    
\qquad {\rm for } \ |\l {\bar g}^{1-\delta /2}| \le 1 
\Eq(5.13.1)$$ 
} 
 
\vglue.3truecm 
\\{\it Proof}: 
This follows on applying  Lemmas 5.5, 5.6 and 5.7 to $ P_{n}(\l)$ defined in 
\equ(4.9). \bull

\vglue0.3cm

\\{\it Estimates for $Q_ne^{-V_n}$} 
 
\\We now turn to the estimate of $Q_n e^{-V_n}$. From \equ(4.19.1) 

$$ Q_{n}(X_{\d_n},\Phi)=Q(X_{\d_n},\Phi;C_{n},{\bf w}_{n},g) =  
g_{n}^{2}\sum_{j=1}^{3}Q^{(j,j)}(\hat X_{\d_n},\Phi;C_n,w_{n}^{(4-j)}) 
\Eq(5.18)$$ 
where the $Q^{(m,m)}$ are given 
in \equ(4.18). Under an iteration, see Proposition~4.1, we have 

$$w_{n}^{(p)}\rightarrow w_{n+1}^{(p)}=v_{n+1}^{(p)}+w_{n,L}^{p}$$, 

\\where $p=1,2,3$ and the $v_{n}^{(p)}$ are given in Proposition~4.1.  
Starting with $w_0^{(p)}=0$ we get by iterating

$$w_{n}^{(p)}=\sum_{j=0}^{n-1}v_{n-j,L^{j}}^{(p)}\Eq(5.19)$$ 

\\For every integer $n\ge 0$ 
we consider the Banach spaces ${\cal W}_{p,\d_n}$ of functions 
$f:\> (\d_n\math{Z})^3\mapsto \bf R$ with norms 
$\Vert\cdot\Vert_{p,n}$, $p=1,2,3$: 

$$\Vert f\Vert_{p,n}= \sup_{x\in (\d_n\math{Z})^3} 
\left((|x|+\d_n)^{6p+1\over 
4}|f(x)|\right)\Eq(5.20)$$  

\\We define the Banach space ${\cal W}_{n}= {\cal 
W}_{1,n}\times {\cal W}_{2,n}\times{\cal W}_{3,n}$ consisting of 
vectors 
${\bf f}=(f^{(1)},f^{(2)},f^{(3)})$, $f^{(p)}:\>(\d_n\math{Z})^3\mapsto \bf R$  
with the norm 

$$\Vert{\bf f}\Vert_{n} = \max_{p} \Vert f^{(p)}\Vert_{n}\Eq(5.20a) $$ 

\\Let ${\bf w}_n=(w_n^{(1)},w_n^{(2)},w_n^{(3)})$ as above.

\vglue.3truecm 
 
\\{\it Lemma~5.9 
 
\\1. For $L$ sufficiently large and $\e>0$ sufficiently small there 
exists a constant $k_{L}$ independent of $n$ and $\e$ such that 
for all $n\ge 1$,

$$\Vert {\bf w}_n\Vert_{n}\le k_{L}/2  \Eq(5.20b)  $$ 

\\If we start the sequence $\{{\bf w}_n\}_{n\ge 0}$ with 
$ {\bf w}_0\not = 0$, with $\Vert {\bf w}_n\Vert_{\d_0}\le k_{L}/2 $, then

$$\Vert {\bf w}_n\Vert_{n}\le k_{L},\> \forall\> n\ge 0  \Eq(5.20c)  $$

\\2. There exists  a function  ${\bf w}_{*}$ defined on 
$\cup_{n\ge 0}(\d_n\math{Z})^{3}\subset \math{Q}^{3}$ 
such that for every integer $l\ge 0$ held fixed , the sequence  
$\{{\bf w}_n\}_{l\le n} $ converges to ${\bf w}_{*}$ in the 
norm $\Vert\cdot\Vert_{l}$ as $n\rightarrow \io$. The convergence rate
is given by

$$\Vert {\bf w}_n-{\bf w}_{*}\Vert_{l}\le \tilde c_L L^{-qn}  \Eq(5.20d) $$

\\where $q>0$ is the constant in Theorem~1.1 and Corollary~1.2..   
We have ${\bf w}_{*}={\bf v}_{c*}+{\bf w}_{*,L}$ in ${\cal W}_l$ 
for every $l\ge 0$.

} 
 
\vglue.3truecm 
\\{\it Proof}: 

\\1. Let $m=n-j$ with $0\le j\le n-1$. By definition
$v_{m}^{(p)}=C_{m,L}^{p}- C_{m+1}^{p}$ with pointwise multiplication. Since
$C_{m,L}=\G_{m,L}+ C_{m+1}$, it follows that $v_{m}^{(p)}$ has $\G_{m,L}$ as 
factor. From the finite range property of $\G_{m,L}$ it follows that  

$$v_{m}^{(p)}(x)=0:\> |x|\ge 1$$

\\Theorem~1.1, part (5a), and Corollary~1.2
give uniform bounds on the $\G_{m}$ and $C_{m}$. Therefore there exists a 
constant $c_{L,p}$ independent of $n$ such that 

$$\Vert v_{m}^{(p)}\Vert_{L^{\infty}((\d_m\math{Z})^{3})} \le c_{L,p}$$
 
\\2. By definition 

$$\eqalign{\Vert v_{n-j,L^{j}}^{(p)}\Vert_{p,n}& = 
\sup_{x\in (\d_n\math{Z})^{3}}
\left(L^{2d_{s}j}(|x|+\d_n)^{6p+1\over 4}|v_{n-j}^{(p)}(L^{j}x)|\right)\cr 
&= L^{-j({6p+1\over 4}- 2d_{s})}\sup_{y\in (\d_{n-j}\math{Z})^{3}} 
\left((|y|+\d_{n-j})^{6p+1\over 4}|v_{n-j}^{(p)}(y)|\right) \cr} 
$$

\\Because of the finite range property of $v_{n-j}^{(p)}$ of paragraph 1,
we can bound $|y|\le 1$ in the weight factor on the right. Because 
$n-j\ge 1$ we can bound in the weight factor $\d_{n-j}\le \d_1=L^{-1}$. 
Therefore on using the bound
on $v_{n-j}^{(p)}$ of paragraph 1 we get

$$\Vert v_{n-j,L^{j}}^{(p)}\Vert_{p,n}
\le  L^{-j({6p+1\over 4}- 2d_{s})}(1+L^{-1})^{6p+1\over 4}c_{L,p}$$

\\We bound the first geometric factor by taking $p=1$ and 
$\e>0$ very small in $d_{s}=(3-\e)/4$. This gives $L^{-j/5}$.
This gives the bound

$$\Vert v_{n-j,L^{j}}^{(p)}\Vert_{p,n}\le L^{-j/5}\>c_{L,p} $$

\\with a new constant $c_{L,p}$ independent of $n$. Using the above bound
we get from \equ(5.19) the bound

$$\Vert w_n^{(p)}\Vert_{p,n} \le c_{L,p}
\sum_{j=0}^{\infty}L^{-j/5}\le 2c_{L,p} $$

\\for $L$ sufficiently large. Therefore setting $k_{L}=4\max_{p}c_{L,p}$ we get

$$\Vert {\bf w}_n\Vert_{n} \le k_{L}/2$$

\\which proves \equ(5.20b). \equ(5.20c) is a trivial consequence of the above.
This proves the first part of the lemma.

\\3. Let $v_{c*}^{(p)}=C_{c*,L}^p-C_{c*}^p$, with pointwise multiplication,
where $C_{c*}$ is the smooth continuum covariance in $\math{R}^3$ of 
Corollary~1.2. By factoring out $\G_{c*,l}$, Theorem~1.1 and Corollary~1.2
we have that $v_{c*}^{(p)}$ exists in $L^{\io}(\math{R}^3)$ and has finite
range: $v_{c*}^{(p)}(x)=0 : |x|\ge 1$. Moreover by Theorem~1.1 and 
Corollary~1.2 we have, see the proof of Lemma~5.12 for the detailed argument,

$$\Vert v_{m}^{(p)}- v_{c*}^{(p)}\Vert_{L^{\infty}((\d_{m}\math{Z})^3)}
\le c_{L,p}L^{-q(m-1)} $$
  
\\Define

$$w_{*}^{(p)}=\sum_{j=0}^{n-1}v_{c*,L^{j}}^{(p)} $$

\\Fix any integer $l\ge 0$. Then for $n\ge l$ the $(p,l)$ norm is dominated by 
the $(p,n)$ norm. Then proceeding as in the first part and using the previous 
inequality we get

$$\eqalign{\Vert w_n^{p}-w_{*}^{p}\Vert_{p,l} &\le
\sum_{j=0}^{n-1}\Vert v_{n-j,L^j}^{p}-v_{c*,L^j}^{p}\Vert_{p,n}\le
c_{L,p} \sum_{j=0}^{n-1}L^{-j({6p+1\over 4}- 2d_{s})}   
\Vert v_{n-j}^{p}-v_{c*}^{p}\Vert_{L^{\infty}((\d_{n-j}\math{Z})^3)} \cr
&\le c_{L,p}\sum_{j=0}^{n-1} L^{-j/5}L^{-q(n-j-1)} \le c'_{L,p} L^{-qn}
}$$ 

\\Now take the maximum over $p$. This proves \equ(5.20) and at the same time
the convergence statement of part 2 of the lemma. The last statement of 
part 2 is trivial to prove. This completes the proof of the second part of 
the lemma. \bull

\vglue.3cm 
 
\\{\it Lemma 5.10 
 
\\Under the conditions of the domain ${\cal D}_n$ there exists 
constants $C_{p,L}$ independent of $n$ such that 

$$\Vert Q_{n}e^{-V_{n}}\Vert_{{\bf h},G_{\k},\AA_p,\d_n}\le  
C_{p,L}\>{\bar g}^{1/2} 
\Eq(5.55.1) 
$$ 
 
$$\vert Q_{n}e^{-V_{n}}\vert_{{\bf h}_{*},\AA_p,\d_n} 
\le C_{p,L}\>{\bar g}^{2} 
\Eq(5.55.2)$$ 

}

\vglue.3truecm 
\\{\it Proof} 
 
$$Q_{n}(X_{\d_n})e^{-V_{n}(X_{\d_n})}=
g_{n}^2\sum_{m=1}^3 Q^{(m,m)}(\hat X_{\d_n},\Phi;C_{n},w_{n}^{(4-m)}) 
e^{-V_{n}(X_{\d_n})}\Eq(5.55.3)$$ 
 
$$\left\Vert Q_{n}(X_{\d_n},\f)e^{-V_{n}(X_{\d_n},\f)}\right\Vert_{\bf h} \le 
g_{n}^{2}\sum_{m=1}^3
\left\Vert  Q^{(m,m)}(\hat X_{\d_n},\Phi;C_{n},w_{n}^{(4-m)})\right\Vert_{\bf h} 
\left\Vert e^{-V_{n}(X_{\d_n})}\right\Vert_{\bf h}\Eq(5.55.4)$$ 
Here $X_{\d_n}$ is a small set because of the support properties of $Q_{n}$. The last 
factor will be estimated by Lemma 5.5. From \equ(4.18) we have 
$$Q^{(3,3)}(\hat X_{\d_n},\Phi;w_{n}^{(1)})=4 
\int_{\hat X_{\d_n}}dxdy\  
:\Phi(x)\bar\Phi(x)\Phi(x)\bar\Phi(y)\Phi(y)\bar\Phi(y):_{C_{n}} 
w_{n}^{(1)}(x-y)\Eq(5.55.5)$$ 

\\We exhibit \equ(5.55.5) as an element of the Grassmann algebra:

$$Q^{(3,3)}(\hat X_{\d_n},\Phi;w_{n}^{(1)})= 
Q_0^{(3,3)}(\hat X_{\d_n},\f;w_{n}^{(1)})+ 
\int_{\hat X_{\d_n}}dxdy\  
Q_1^{(3,3)}(\hat X_{\d_n},\f,x,y;w_{n}^{(1)}):\psi(x)\bar\psi(y):_{C_n}+$$ 
$$+\int_{X_{\d_n}}dx\  
Q_2^{(3,3)}(\hat X_{\d_n},\f,x;w_{n}^{(1)}):\psi(x)\bar\psi(x):_{C_n}+$$ 
$$+\int_{\hat X_{\d_n}}dxdy\  
Q_3^{(3,3)}(\hat X_{\d_n},\f,x,y;w_{n}^{(1)}):\psi(x)\bar\psi(x)\psi(y)\bar\psi(vy):_{C_n} 
\Eq(5.55.6)$$ 

\\where 

$$\eqalign{Q_0^{(3,3)}(\hat X_{\d_n},\f;w_{n}^{(1)})&= 
4\int_{\hat X_{\d_n}}dxdy\  
:\f(x)\bar\f(x)\f(x)\bar\f(y)\f(y)\bar\f(y):_{C_{n}} 
w_{n}^{(1)}(x-y)\cr 
Q_1^{(3,3)}(\hat X_{\d_n},\f,x,y;w_{n}^{(1)})&= 
4:\f(x)\bar\f(x)\f(y)\bar\f(y):_{C_{n}} 
w_{n}^{(1)}(x-y)\cr 
Q_2^{(3,3)}(\hat X_{\d_n},\f,x;w_{n}^{(1)})&= 
4\int_{X_{\d_n}^{\prime}}dy:(\f(x)\bar\f(y)+\f(y)\bar\f(x))
\f(y)\bar\f(y):_{C_{n}} 
w_{n}^{(1)}(x-y)\cr 
Q_3^{(3,3)}(\hat X_{\d_n},\f,x,y;w_{n}^{(1)})&= 
4:\f(x)\bar\f(y):_{C_{n}} 
w_{n}^{(1)}(x-y)\cr}\Eq(5.55.7)$$ 
where, denoting with $\D(x)$ the block $\D$ such that 
$x\in\D$, we have $X_{\d_n}^{\prime}=\cases{\D&if $X_{\d_n}=\D$\cr X_{\d_n}\setminus\D(x)&if  
$X_{\d_n}=\D_1\cup\D_2$\cr}$. 
 
\\Undo the Wick ordering, which produces lower order terms with coefficients
which are uniformly bounded independent of $n$ by Corollary~1.2. It is 
therefore enough to estimate with the Wick ordering taken off. We get 

$$\Vert Q^{(3,3)}(\hat X_{\d_n},\Phi;w_{n}^{(1)})\Vert_{\bf h}\le 
\Vert Q_0^{(3,3)}(\hat X_{\d_n},\f;w_{n}^{(1)})\Vert_{h_B}+$$ 
$$+ 
h_F^2\sup_{\Vert g_2\Vert_{C^{2}(\hat X_{\d_n}^2)} \le 1}
\int_{\hat X_{\d_n}}dxdy\  
\Vert Q_1^{(3,3)}(\hat X_{\d_n},\f,x,y;w_{n}^{(1)})\Vert_{h_B} 
|g_2(x,y)|+$$ 
$$+h_F^2\sup_{\Vert g_2\Vert_{C^{2}(\hat X_{\d_n}^2)}\le 1}\int_{X_{\d_n}}dx\  
\Vert Q_2^{(3,3)}(\hat X_{\d_n},\f,x;w_{n}^{(1)})\Vert_{h_B} 
|g_2(x,x)|+$$ 
$$+h_F^4\sup_{\Vert g_4 \Vert_{C^{2}(\hat X_{\d_n}^4)}\le 1}
\int_{\hat X_{\d_n}}dxdy\  
\Vert Q_3^{(3,3)}(\hat X_{\d_n},\f,x,y;w_{n}^{(1)})\Vert_{h_B} 
|g_4(x,x,y,y)|$$ 

\\To estimate the $h_B$ norm of the $Q_j^{(3,3)}$ we apply to \equ(5.55.7) 
$h_B^k D^{k}$, with $D$ the bosonic field derivative, $h_B=c{\bar g}^{-1/4}$ 
and use Lemma~5.1. Contributions for  $k>4$ vanish. 
We use $g_{n}=O({\bar g})$ from the domain hypothesis of
Theorem~5.1. We estimate the kernel $w_{n}^{(1)}(x-y)$ using Lemma~5.9.
As a result we get

$$\Vert Q^{(3,3)}(\hat X_{\d_n},\Phi;w_{n}^{(1)})\Vert_{\bf h}\le 
C_L {\bar g}^{-3/2}\int_{\hat X_{\d_n}}dxdy\  
{1\over(|x-y| +\d_n)^{7/4}}
e^{ {\bar g}/4\int_{X_{\d_n}}dx|\f\bar\f(x)|^2}G_\k(X_{\d_n},\f) 
\Eq(5.55.8)$$ 
 
\\The integral over $\hat X_{\d_n}$ exists and is of $O(1)$ since 
$X_{\d_n}$ is a small set in $(\d_n\math{Z})^{3}$. Therefore

$$\Vert Q^{(3,3)}(\hat X_{\d_n},\Phi;w_{n}^{(1)})\Vert_{\bf h}\le 
C_L {\bar g}^{-3/2}
e^{ {\bar g}/4\int_{X_{\d_n}}dx|\f\bar\f(x)|^2}G_\k(X_{\d_n},\f) 
\Eq(5.55.8a)$$

\\Next turn to $Q^{(m,m)}$, $m=1,2$. From \equ(4.18) 
$$Q^{(2,2)}(\hat X_{\d_n},\Phi;C,w_{n}^{(2)})= 
-\int_{\hat  
X_{\d_n}}dxdy\left[:(\Phi(x)-\Phi(y))(\bar\Phi(x)-\bar\Phi(y)) 
(\Phi(x)+\Phi(y))(\bar\Phi(x)+\bar\Phi(y)):_{C_{n}}\right. 
+$$ 
$$\left.+:[(\Phi\bar\Phi)(x)-(\Phi\bar\Phi)(y)]^2:_{C_{n}} 
\right]w_{n}^{(2)}(x-y)$$

\\We exhibit this as an element of the Grassmann algebra. This gives

$$Q^{(2,2)}(\hat X_{\d_n},\Phi;C_n,w_{n}^{(2)})= 
-Q^{(2,2)}_{0}(\hat X_{\d_n},\f;C_n,w_{n}^{(2)})-\int_{\hat  
X_{\d_n}}dxdy $$ 
$$\left[Q^{(2,2)}_{1}(\hat X_{\d_n},\f,x,y;C_n,w_{n}^{(2)}) 
:(\psi(x)+\psi(y))(\bar\psi(x)+\bar\psi(y)):_{C_{n}}+\right.$$ 
$$Q^{(2,2)}_{2}(\hat X_{\d_n},\f,x,y;C_n,w_{n}^{(2)},) 
:(\psi(x)-\psi(y))(\bar\psi(x)-\bar\psi(y)):_{C_{n}}+$$ 
$$+Q^{(2,2)}_{3}(\hat X_{\d_n},\f,x,y;C_n,w_{n}^{(2)}) 
:(\psi\bar\psi(x)-\psi\bar\psi(y)):_{C_{n}}+$$ 
$$+Q^{(2,2)}_{4}(\hat X_{\d_n},x,y,w_{n}^{(2)}) 
\{(:\psi(x)-\psi(y))(\bar\psi(x)-\bar\psi(y)) 
(\psi(x)+\psi(y))(\bar\psi(x)+\bar\psi(y)):_{C_{n}}+$$ 
$$\left. :(\psi\bar\psi(x)-\psi\bar\psi(y))^{2}):_{C_{n}})\} 
\right]\Eq(5.55.10) 
$$ 

\\where 

$$\eqalign{Q^{(2,2)}_{0}(\hat X_{\d_n},\f;C_n,w_{n}^{(2)})=&\int_{\hat  
X_{\d_n}}dxdy\>w_{n}^{(2)}(x-y)(:|\f(x)-\f(y)|^{2}|\f(x)+\f(y)|^{2}:_{C_{n}}\cr
&+:(|\f|^{2}(x)-|\f|^{2}(y))^{2}:_{C_{n}})\cr 
Q^{(2,2)}_{1}(\hat X_{\d_n},\f,x,y;C_n,w_{n}^{(2)})=& 
w_{n}^{(2)}(x-y):|\f(x)- \f(y)|^{2}_{C_n} \cr 
Q^{(2,2)}_{2}(\hat X_{\d_n},\f,x,y;C_n,w_{n}^{(2)})=& 
w_{n}^{(2)}(x-y) :|\f(x)+\f(y)|^{2}:_{C_{n}} \cr 
Q^{(2,2)}_{3}(\hat X_{\d_n},\f,x,y;C_n,w_{n}^{(2)} )=& 
2w_{n}^{(2)}(x-y):(|\f|^{2}(x)-|\f|^{2}(y)) :_{C_{n}}\cr
Q^{(2,2)}_{4}(\hat X_{\d_n},x,y,w_{n}^{(2)} )=&
w_{n}^{(2)}(x-y) \cr}\Eq(5.55.11)$$

\\The $\Vert \cdot\Vert_{{\bf h},G_{\k},\AA_{p}}$ norm
estimate for $Q^{(2,2)}(\hat X_{\d_n},\Phi;C,w_{n}^{(2)})$ reposes on
the following principles: 

\\1. Undoing the Wick ordering produces lower order terms  
with Wick coefficients which \textRed together with their derivatives 
\textBlack
are uniformly bounded independent of $n$ by 
Corollary~1.2. Moreover by the domain hypothesis $g_{n}=O({\bar g})$.

\\2. By Lemma~5.9, the kernel $w_{n}^{(2)}$ has the bound
$|w_{n}^{(2)}(x-y)|\le k_{L}(|x-y|+\d_n)^{-13/4}$ where the constant $k_L$ is 
independent of $n$.  

\\3. The fields $\f(x)$ are estimated by Lemma~5.1. Differences of fields
$|\f(x)-\f(y)|$ are estimated by \equ(5.1b). This produces a factor $|x-y|$
which we retain, and majorise the Sobolev factor by the large field regulator.
Diffferences of fields $\f\bar\f(x)-\f\bar\f(y)$ can also be expressed as 
in \equ(5.1a), substituting $\f\bar\f$ for $\f$. This requires estimating
$(\dpr_{\d_n,e_{j}}\f\bar\f)(x+\cdot\cdot)$. We apply the lattice Leibniz which 
modifies
the continuum rule by producing an extra term 
$\d_{n} |\dpr_{\d_n,e_{j}}\f(x+\cdot\cdot)|^{2}$, 
(see equation (5.2),page 432 of [BGM]).
We estimate the $\f$ by Lemma~5.1, with $\k/2$ in the large field regulator.
We estimate the gradient pieces by the Sobolev inequality as in 
\equ(5.1b), and then by the large field regulator with $\k/2$. We have also
produced a factor $|x-y|$ as in \equ(5.1b).

\\Invoking the above principles we get the  following bounds for the bosonic
coefficients:

$${h_{B}^{k}\over k!}
\Vert D^{k}Q^{(2,2)}_{0}(\hat X_{\d_n},\f;C_n,w_{n}^{(2)})\Vert
\le c_{L}{\bar g}^{-1/2}\int_{\hat X_{\d_n}}dxdy\>
(|x-y|+\d_n) ^{-({13\over 4} +2)}
e^{ {\bar g}/4\int_{X_{\d_n}}dx|\f\bar\f(x)|^2}G_\k(X_{\d_n},\f) \Eq(5.55.81)$$

\\and for $j=1,2,3$

$${h_{B}^{k}\over k!}
\Vert D^{k}Q^{(2,2)}_{j}(\hat X_{\d_n},\f;x,y,C_n,w_{n}^{(2)})\Vert
\le c_{L}{\bar g}^{-1/2}\>(|x-y|+\d_n)^{-13/4}f_{j}(x-y)
e^{ {\bar g}/4\int_{X_{\d_n}}dx|\f\bar\f(x)|^2}G_\k(X_{\d_n},\f) \Eq(5.55.82)$$

\\where the maximum value of $k$ which gives a nonvanising contribution 
is $4$ and 

$$f_{1}(x-y)=|x-y|^{2},\>f_{2}(x-y)=1, \>
f_{3}(x-y)=|x-y|,f_{4}(x-y)=1 \Eq(5.55.83)$$

\\4. We must estimate the contribution of the fermionic pieces to the 
$\bf h$ norm. To this end 
denote by $F_j(\psi)$ the fermionic factor multiplying $Q^{(2,2)}_{j}$
in \equ(5.55.10). Express the differences $\psi(x)-\psi(y)$ by the fermionic
analogue of \equ(5.002). We do the same also for
$\psi\bar\psi(x)-\psi\bar\psi(y)$ and then apply the lattice Leibnitz rule
to $\dpr_{\d_{n},\e_j}\psi\bar\psi(x+\cdot)$. We replace the fermionic
pieces by the functions $g_{2p}$ on $\hat X_{\d_n}\cup\dpr_{2}\hat X_{\d_n}$. 
and their lattice derivatives. Corresponding to $F_j(\psi)$ we get the
contribution $G_{ j}$ which is a linear form on $g_{2p_j}$, where
$p_j=1$ for $j=1,2,3$ and $p_4=2$. 
Let $\d_n h_j$, $h_j\in \bf Z$  be the component of $y-x$ along the unit vector
$e_j$. We have

$$G_{1}=g_2(x,x)+g_2(x,y)+g_2(y,x)+g_2(y,y)$$

$$G_{2}=\d_n^{2}\sum_{i_1,i_2=1}^{3}\sum_{0\le s_{i_l}\le h_{i_l}-1,\>l=1,2}
\dpr_{\d_n,e_{i_1}}^{(1)}\dpr_{\d_n,e_{i_2}}^{(2)}
g_2(x+ p_{i_1}(y-x, s_{i_1}),x+  p_{i_2}(y-x, s_{i_2})  )$$

$$G_{3}=\d_n\sum_{i=1}^{3}\sum_{s_i=0}^{h_i-1}
\Bigl[\dpr_{\d_n,e_{i}}^{(1)}g_2(x+ p_{i}(y-x,s_{i}),
x+p_{i}(y-x,s_{i})) +\dpr_{\d_n,e_{i}}^{(2)}g_2(x+p_{i}(y-x,s_{i}),
x+p_{i}(y-x,s_{i}))+$$
$$+\d_n\dpr_{\d_n,e_{i}}^{(1)}\dpr_{\d_n,e_{i}}^{(2)}
g_2(x+ p_{i}(y-x,s_{i}),x+p_{i}(y-x,s_{i}))\Bigr] $$

$$G_{4}=\d_n^{2}\sum_{i_1,i_2=1}^{3}\sum_{0\le s_{i_l}\le h_{i_l}-1,
\>l=1,2}\dpr_{\d_n,e_{i_1}}^{(1)}\dpr_{\d_n,e_{i_2}}^{(2)}
\Bigl(g_4(x+p_{i_1}(y-x,s_{i_i}),x+ p_{i_2}(y-x,s_{i_2}),x,x)+$$
$$+g_4(x+\cdot\cdot,x+\cdot\cdot,x,y)
+g_4(x+p_{i_1}(y-x,s_{i_1}),x+p_{i_2}(y-x,s_{i_2}),y,x)+$$
$$+g_4(x+p_{i_1}(y-x,s_{i_1}),x+ p_{i_2}(y-x,s_{i_2}) ,y,y)\Bigr) 
  +\cdot\cdot $$

\\where the superscript on the lattice derivative denotes the argument on which
it acts. The omitted terms $\cdot\cdot$ in $G_4$ comes from the square of the
(first order) lattice Taylor expansion of $\psi\bar\psi(x)-\psi\bar\psi(y)$
and then replacing the product of $4$ Grassmann fields by the test function 
$g_4$. For $j=1,2,3$ we have the bounds

$$|G_{j}|
\le O(1){\tilde f}_j(x-y)\Vert g_2\Vert_{C^{2}({\hat X}_{\d_n}^{2}) } 
\Eq(5.55.84)  $$

\\and for $j=4$ we have

$$|G_{4}|
\le O(1){\tilde f}_4(x-y)\Vert g_4\Vert_{C^{2}({\hat X}_{\d_n}^{4}) }
\Eq(5.55.85)   $$

\\where 

$$\tilde f_1=1,\>\tilde f_2=|x-y|^2,\>\tilde f_3=|x-y|,
\>\tilde f_4=|x-y|^2  \Eq(5.55.86)  $$.

\\On using the bounds \equ(5.55.81)-\equ(5.55.86) we get for $0\le k\le 4$
and $0\le p\le 2$

$$h_{F}^{2p}{h_B^k\over k!}
\vert D^{2p,n}Q^{(2,2)}({\hat X}_{\d_n},\f,0;f^{\times k},g_{2p})\vert
\le c_{L}{\bar g}^{-1/2}\int_{\hat X_{\d_n}}dxdy\>
(|x-y|+\d_n)^{-({13\over 4} -2)}$$
$$\times e^{ {\bar g}/4\int_{\hat X_{\d_n}}dx|\f\bar\f(x)|^2}G_\k(X_{\d_n},\f) 
\Vert f\Vert_{C^{2}({\hat X}_{\d_n})}^{\times k}
\Vert g_{2p}\Vert_{C^{2}({\hat X}_{\d_n}^{2p}) } \Eq(5.55.87)$$ 

\\For $k>4$ or $p>2$ we have vanishing contribution. The integral over
$\hat X_{\d_n}$ exists and gives a contribution of $O(1)$ since 
$X_{\d_n}$ is a small set in $ (\d_n\math{Z})^3$. Therefore we obtain
from the previous inequality

$$\Vert Q^{(2,2)}({\hat X}_{\d_n},\f,0)\Vert_{\bf h}
\le c_{L}{\bar g}^{-1/2}
e^{ {\bar g}/4\int_{\hat X_{\d_n}}dx|\f\bar\f(x)|^2}G_\k(X_{\d_n},\f) 
\Eq(5.55.88) $$   

\\We can estimate in the same way the case $m=1$.  
We have 
$$\Vert Q^{(1,1)}(\hat X_{\d_n},\Phi;C,w^{(3)})\Vert_{\bf h}\le  
c_{L}{\bar g}^{-1/2} \int_{\hat X_{\d_n}}dxdy\  
(|x-y|+\d_n)^{-({19\over 4}-2)}
e^{ {\bar g}/4\int_{X_{\d_n}}dx|\f\bar\f(x)|^2}G_\k(X_{\d_n},\f) \Eq(5.55.13) 
$$ 

\\The integral over $\hat X_{\d_n}$ exists and gives a contribution of 
$O(1)$. Therefore 

$$\Vert Q^{(1,1)}(\hat X_{\d_n},\Phi;C,w^{(3)})\Vert_{\bf h}\le  
c_{L}{\bar g}^{-1/2}
e^{{\bar g}/4\int_{X_{\d_n}}dx|\f\bar\f(x)|^2}G_\k(X_{\d_n},\f) \Eq(5.55.13a) 
$$ 

\\Therefore from \equ(5.55.4), Lemma 5.5 and the bounds \equ(5.55.8a),
\equ(5.55.88),\equ(5.55.13a) we get 

$$\left\Vert Q_n(X_{\d_n})e^{-V_n(X_{\d_n})}\right\Vert_{\bf h,G_\k}
\le c_{L}{\bar g}^{1/2}$$  
and since $Q_n$ is supported on small sets we get 
$$\Vert Q_ne^{-V_n}\Vert_{{\bf h},G_{\k},\AA_p}\le  
C_{L,p}{\bar g}^{1/2} 
\Eq(5.55.14) 
$$ 
which is \equ(5.55.1). To prove \equ(5.55.2) we estimate the r.h.s of 
\equ(5.55.4) at $\Phi=0$ after undoing the Wick ordering, set 
${\bf h}={\bf h}_*$, and use Lemma~5.5.  \bull

\vglue0.1cm 
\\In the following lemma we consider $Q_n(\Phi+\xi)e^{-V_n(\Phi+\xi)}$ as 
a function of $\f,\z,\psi,\eta$.   

\vglue.3truecm 
\\{\it Lemma 5.11 
 
\\Under the conditions of the domain ${\cal D}_n$ there exists constants 
$C_{L,p}$ independent of $n$ such that 

$$\Vert Q_ne^{-V_n}\Vert_{{\bf h},\hat G_{\k,\r},\AA_p,\d_n}\le  
C_{L,p}{\bar g}^{1/2} 
\Eq(5.55.15) 
$$ 
 
$$\Vert Q_ne^{-V_n}\Vert_{{\bf h}_{*},\tilde G_{\k,\r},\AA_p,\d_n} 
\le C_{L,p}{\bar g}^{2} 
\Eq(5.55.16)$$ 
} \vglue.3truecm 
\\{\it Proof} 
 
\\The bound \equ(5.55.15) follows  from \equ(5.55.1) since
$\hat G_{\k,\r}>G_{\k}$. To prove \equ(5.55.16) we first express 
$Q_n^{(m,m)}(X_{\d_n},\f+\z,\psi+\eta) $ in the Grassmann representation as in the proof of Lemma~5.10, substituting in the expressions there
$\f\rightarrow\f+\z$, $\psi\rightarrow\psi+\h$. Field derivatives are 
defined as in \equ(2.2.2). For the bosonic 
coefficients we take derivatives at $\f=0$.
The resulting dependence on $\z$ is estimated by Lemma~5.2. The rest of the
proof follows that of Lemma~5.10. We use Lemma~5.5 which implies that 
$\Vert e^{-V_n(X_{\d_n},\z,\eta)}\Vert_{{\bf h}_*}\le 2^{|X_{\d_n}|}$,
and $X_{\d_n}$ is a small set.  \bull

\vglue.3truecm 

\\We now prove a lemma to 
control the perturbative flow coefficients $a_n,b_n$ given in \equ(4.47) and
\equ(4.54.1). This lemma is independent of the domain ${\cal D}_n$.

\vglue.3truecm 

\\{\it Lemma~5.12 

\\Let $v_{c*}^{(p)}=C_{c*,L}^p-C_{c*}^p$, with pointwise multiplication,
where $C_{c*}$ is the smooth continuum covariance in $\math{R}^3$ of 
Corollary~1.2. Define 

$$a_{c*}=2\int_{\math{R}^3}dy\> v_{c*}^{(2)}(y),\quad  
b_{c*}=4\int_{\math{R}^3}dy\> v_{c*}^{(3)}(y)$$
 
\\We have that $a_n,b_n,a_{c*},b_{c*}$ are strictly positive. Moreover
there exist constants $c_{L}$ independent of $n$ such that

$$|a_n|\le c_L, \quad |b_n|\le c_L , \quad |a_{c*}|\le c_L, 
\quad |b_{c*}| \le c_L \Eq(5.361)  $$

\\and

$$|a_n-a_{c*}|\le c_{L}L^{-qn},\quad |b_n-b_{c*}|\le c_{L}L^{-qn} \Eq(5.362) $$

\\where $q>0$ is as in Therem~1.1.
}

\\{\it Remark} : The convergence rate estimates \equ(5.362) play no role
in the estimates of the present section. They are used in Section~6 for the
existence proof of a global renormalization group trajectory.

\vglue.3truecm 
\\{\it Proof} 
 
\\From \equ(4.54), for $p=2,3$, using $C_{n,L}=\G_{n,L}+C_{n+1}$ 
 
$$v_{n+1}^{(p)}=C_{n,L}^p-C_{n+1}^p
=\G_{n,L}(\G_{n,L}^{p-1}+p\G_{n,L}^{p-2}C_{n+1}+\d_{p,3}3C_{n+1}^2)
\Eq(5.36)$$ 
 
\\with pointwise multiplication. The positivity in Fourier space of the
integral kernels on the right hand side implies that $a_n>0,\> b_n>0$ 
as claimed.  The common factor of $\G_{n,L}(x)$ which has finite range $1$
implies that $v_{n+1}^{(p)}(x)$ has support in the unit ball in 
$(\d_{n+1}\math{Z})^3$. From Theorem.1.1 and Corollary~1.2 we have that 
$v_{n+1}^{p}$ above are uniformly bounded in 
$L^{\infty}((\d_{n+1}\bf Z)^{3})$ by constants $c_L$. By the same arguments
$v_{c*}$ has finite range and belongs to $L^{\infty}(\math{R}^{3})$.
The uniform bounds in the first part of the lemma now follow. 

\\By the same arguments using 
$C_{*,L}=\G_{c*}+C_{c*}$ we have that $a_{c*}>0,\>b_{c*}>0$ and 
$v_{c*}^{(p)}(x)$ has support in the unit ball in $\math{R}^3$. Moreover
using Corollary~1.2 we have $\Vert v_{c*}^{(p)}\Vert_{C^{k}(\math{R}^{3})} 
\le c_{k,L}$ for all $\k\ge 0$.

\\Define 

$$a_n^{(p)}=\int_{(\d_{n+1}\math{Z})^3}dy\>v_{n+1}^{(p)}(y),\quad 
a_{c*}^{(p)}=\int_{\math{R}^3}dy\>v_{c*}^{(p)}(y) \Eq(5.36.1)$$

\\Then using the compact support property of $v_{n+1}^{(p)}$ and 
$v_{c*}^{(p)}$ we get

$$|a_n^{(p)}-a_{c*}^{(p)}|\le \Vert v_{n+1}^{(p)}- v_{c*}^{(p)}
\Vert_{L^{\infty}((\d_{n+1}\math{Z})^3)} 
+\left\vert \int_{(\d_{n+1}\math{Z})^3}dy\>v_{c*}^{(p)}(y)-
\int_{\math{R}^3}dy\>v_{c*}^{(p)}(y)\right\vert \Eq(5.36.2)
$$

\\We estimate the first term on the right hand side of \equ(5.36.2). We have

$$\Vert v_{n+1}^{(p)}- v_{c*}^{(p)}\Vert_{L^{\infty}((\d_{n+1}\math{Z})^3)}
\le \Vert C_{c*,L}^{p}-C_{n,L}^{p}\Vert_{L^{\infty}((\d_{n+1}\math{Z})^3)}+
\Vert C_{c*}^{p}-C_{n+1}^{p}\Vert_{L^{\infty}((\d_{n+1}\math{Z})^3)}\Eq(5.36.3) 
$$

\\In the first term on the right in \equ(5.36.3)
 we factor out $C_{c*,L}-C_{n,L}$ and
in the second term we factor out $C_{c*}-C_{n+1}$. Then use of the bounds
in Corollary~1.2 gives

$$\Vert v_{n+1}^{(p)}- v_{c*}^{(p)}\Vert_{L^{\infty}((\d_{n+1}\math{Z})^3)}
\le c_{L,p}L^{-qn} \Eq(5.36.4)   $$

\\We estimate the second term on the right in \equ(5.36.2) using Lemma~6.6
of [BGM] and the compact support of $v_{c*}$. This gives

$$\left\vert \int_{(\d_{n+1}\math{Z})^3}dy\>v_{c*}^{(p)}(y)-
\int_{\math{R}^3}dy\>v_{c*}^{(p)}(y)\right\vert \le O(1)\d_{n+1}
\Vert v_{c*}^{(p)}\Vert_{C^{1}(\math{R}^3)} \le c_{L,p}L^{-(n+1)}\Eq(5.36.5)$$

\\From \equ(5.36.2), \equ(5.36.4) and \equ(5.36.5) we get with $q$ that of
Theorem~1.1

$$|a_n^{(p)}-a_{c*}^{(p)}|\le c_{L,p}L^{-qn} \Eq(5.36.6)$$

\\which completes the proof of the lemma. \bull

\vglue.3truecm 
\\{\it Lemma 5.13 

\\Under the conditions of the domain ${\cal D}_n$ there exist constants 
$C_{p,L}$ independent of $n$ such that 

$$\Vert Q_n(e^{-V_n}-e^{-\tilde V_n})\Vert_{{\bf h},\hat G_{\k,\r},\AA_p,\d_n}
\le  C_{L,p}{\bar g}^{3/4} 
\Eq(5.55.18) 
$$ 
 
$$\Vert Q_n(e^{-V_n}-e^{-\tilde V_n})\Vert_{{\bf h}_{*},\tilde G_{\k,\r},\AA_p
,\d_n} 
\le C_{L,p}{\bar g}^{3} 
\Eq(5.55.19)$$ 
} 
 
\\{\it Proof}: The proof is on the same lines as that of the corresponding 
Lemma 5.13 in [BMS]. It follows from Lemmas 5.11, 5.6, 5.10, and 5.5 which
are lattice equivalents of the corresponding lemmas in [BMS]. \bull

\vglue.3truecm 
\\{\it Lemma 5.14 
 
\\Under the conditions of the domain ${\cal D}_n$ there exists constants
$C_L$ independent of $n$ such that 
$K_n(\l)$ given by \equ(4.10) satisfies the bounds 

$$\Vert K_n(\l)\Vert_{{\bf h},\hat G_{\k,\r},\AA,\d_n}\le  
C_{L}|\l {\bar g}^{1/4-\h/3}|^2\quad {\rm for}\ |\l{\bar g}^{1/4-\h/3}|<1 
\Eq(5.55.20) 
$$ 
 
$$\Vert K(\l)\Vert_{{\bf h}_{*},\tilde G_{\k,\r},\AA,\d_n } 
\le C_{L}|\l{\bar g}^{11/12-\h/3}|^2\quad {\rm for}\ 
|\l{\bar g}^{11/12-\h/3}|<1 
\Eq(5.55.21)$$ 
}

\\{\it Proof}: This follows from Lemmas 5.11 and 5.13 and the hypothesis 
\equ(55.02) on $R_n$. \bull
  
\vglue.1truecm

\\The following proposition shows how fluctuation integration of polymer 
activities passes through $\bf h$ and ${\bf h}_{*}$ norms. It will be put
to use in the subsequent lemmas. 

\vglue0.2cm

\textRed

\\{\it Lemma~5.14A 

\\1. Let ${\tilde K}(X_{\d_n},\f,\z,\psi,\h)$ be a polymer activity  
(see\equ(2.11.22) and \equ(2.1.2)) with norms defined as in 
\equ(2.2.2)-\equ(2.5.2). Let ${\bf h}=(h_B, h_F)$ and 
${\bf h}_{*}=(h_{B*}, h_F)$. Let 
${\tilde K}^{\sharp}(X_{\d_n},\f,\psi)=\int d\m_{\G_n}(\z)\>d\m_{\G_n}(\h)
\tilde K(X_{\d_n},\f,\z,\psi,\h)$.
Then for $h_F$ sufficiently large depending on
$L$ we have

$$\vert {\tilde K}^{\sharp}(X_{\d_n})\vert_{{\bf h}_{*}}\le 
\int d\m_{\G_{n}}(\z)
\Vert {\tilde K}(X_{\d_n},0,\z,0,0)\Vert_{{\bf h}_{*}} \Eq(5.366)$$

$$\Vert {\tilde K}^{\sharp}(X_{\d_n},\f,0)\Vert_{{\bf h}}\le 
\int d\m_{\G_{n}}(\z)
\Vert {\tilde K}(X_{\d_n},\f,\z,0,0)\Vert_{{\bf h}} \Eq(5.367)$$

\\where the norms on the left hand side are as in \equ(2.3)-\equ(2.5).

\\2. Let $K(X_{\d_n},\f,\psi)$ be a polymer activity in $\Omega^{0}(X_{\d_n})$.
and let $K^{\sharp}(X_{\d_n},\f,\psi)=\int d\m_{\G_n}(\z)\>d\m_{\G_n}(\h)
K(X_{\d_n},\f+\z,\psi+\h)$. Let $\hat{\bf h}=(h_B, {h_F\over 2})$
and $\hat{\bf h}_{*}=({h_B}_{*}, {h_F\over 2})$
Then for $h_F$ sufficiently large depending on $L$ we have

$$\vert K^{\sharp}(X_{\d_n})\vert_{\hat{\bf h}_{*}}\le 
\int d\m_{\G_{n}}(\z)
\Vert K(X_{\d_n},\z,0)\Vert_{{\bf h}_{*}} \Eq(5.368)$$

$$\Vert K^{\sharp}(X_{\d_n},\f,0)\Vert_{\hat{\bf h}}\le 
\int d\m_{\G_{n}}(\z)
\Vert K(X_{\d_n},\f+\z,0)\Vert_{{\bf h}} \Eq(5.369)$$

\\where the norms on both sides are as in \equ(2.3)-\equ(2.5).

}

\vglue0.2cm

\\{\it Proof}: We get from the representation \equ(2.1.2) and using the 
notations introduced there (\equ(2.11.22),\equ(2.1.2))

$${\tilde K}^{\sharp}(X_{\d_{n}},\f,\psi)=
\int d\m_{\G_n}(\z)\Bigl[ \sum_{p\ge 0} 
\sum_{{I\subset\{1,...,p\}}\atop {J\subset\{1,...,p\}}} \d_{|I|,|J|} 
{1\over |I|!\>|I^{c}|!\>|J|!\>|J^c|!}\times $$
$$\int_{X_{\d_{n}}^{2p}} d{\bf x}d{\bf y} \>
D_{F}^{2p,IJ}{\tilde K}
(X_{\d_{n}},\f,\z,{\bf x}_{I},{\bf x}_{I^{c}},{\bf y}_{J},
{\bf y}_{J^{c}}) 
\psi(x_I)\bar\psi(y_J){\det}_{\G_n,I^c,J^c}({\bf x}_{I^{c}},{\bf y}_{J^{c}})
\times (-1)^{\sharp} \Bigr]\Eq(5.4660) $$ 

\\Note that $|I^{c}|=|J^{c}|$ since $|I|=|J|$.
The matrix $ {\det\G_n}_{I^c,J^c} $ is an $|I^c|\times |I^c| $   
square matrix whose entry  $({{\G_n}_{I^c,J^c}})_{rs}$  is given by 
 
$$({{\G_n}_{I^c,J^c}})_{rs}= \G_{n}(x_{r}-y_{s})  \Eq(5.4661)  $$  

\\for $r\in I^c$, $s\in J^{c}$. $(-1)^{\sharp}$ is a sign factor which it
is not necessary to specify. It may change from line to line.

\\We have

$${h_F^{2j}\over (j!)^{2}}  
D^{2j,m}{\tilde K}^{\sharp}(X_{\d_{n}},\f,0,f^{\times m},g_{2j})
=\int d\m_{\G_n}(\z)\Bigl[ \sum_{p\ge 0} 
\sum_{{I\subset\{1,...,p\}}\atop {J\subset\{1,...,p\}}} \d_{|I|,j}\d_{|J|,j}
{h_F^{2|I|}\over |I|!\>|I^{c}|!\>|J|!\>|J^c|!}\times $$
$$\int_{X_{\d_{n}}^{2p}} d{\bf x}d{\bf y} \>
D_B^{m}D_{F}^{2p,IJ}{\tilde K}(X_{\d_{n}},\f,\z,{\bf x}_{I},{\bf x}_{I^{c}},{\bf y}_{J},
{\bf y}_{J^{c}};f^{\times m})g_{2j}({\bf x}_{I},{\bf y}_{J}) 
{\det}_{\G_n,I^c,J^c}({\bf x}_{I^{c}},{\bf y}_{J^{c}})
\times (-1)^{\sharp}   \Bigr] \Eq(4.6662)  $$ 

\\By definition $g_{2j}({\bf x}_{I},{\bf y}_{J})$, (note that $|I|=|J|=j$),
is antisymmetric in the members of ${\bf x}_{I}$ and in the members of
${\bf y}_{J}$. The determinant is antisymmetric in the members of 
${\bf x}_{I^{c}}$ and in the members of ${\bf y}_{J^{c}}$. 
Therefore the function  

$$(g_{2j}\otimes {\det}_{\G_n,I^c,J^c}) 
({\bf x}_{I},{\bf x}_{I^{c}},{\bf y}_{J},{\bf y}_{J^{c}})=
g_{2j}({\bf x}_{I},{\bf y}_{J}) 
{\det}_{\G_n,I^c,J^c}({\bf x}_{I^{c}},{\bf y}_{J^{c}}) \Eq(4.6663)$$

\\on $\tilde X_{\d_n}^{|I|}\times \tilde X_{\d_n}^{|I^{c}|}
\times \tilde X_{\d_n}^{|J|}\times \tilde X_{\d_n}^{|J^c|}
=\tilde X_{\d_n}^{2p}$
is an admissable test function for the norm defined in \equ(2.2.2),
\equ(2.3.2). Hence we get from \equ(4.6662) and \equ(4.6663)

$${h_F^{2j}\over (j!)^{2}}  
\Bigl|D^{2j,m}{\tilde K}^{\sharp}(X_{\d_{n}},\f,0,f^{\times m},g_{2j})\Bigr|
\le\int d\m_{\G_n}(\z)\Bigl[ \sum_{p\ge 0} 
\sum_{{I\subset\{1,...,p\}}\atop {J\subset\{1,...,p\}}} \d_{|I|,j}\d_{|J|,j}
{h_F^{2|I|}\over |I|!\>|I^{c}|!\>|J|!\>|J^c|!}\times $$
$$\Vert D^{2p,IJ,m}{\tilde K}(X_{\d_{n}},\f,\z,,0,0)\Vert \>
\prod_{j=1}^{m} \Vert f_j\Vert_{C^{2}(X_{\d_n})}\> 
\Vert g_{2j}\otimes {\det}_{\G_n,I^c,J^c}\Vert_{C^{2}(X_{\d_n}^{2p})}\Bigr] 
\Eq(4.6664) $$ 

\\We have 

$$\Vert g_{2j}\otimes {\det}_{\G_n,I^c,J^c}\Vert_{C^{2}(X_{\d_n}^{2p})}\le 
\Vert g_{2j}\Vert_{C^{2}(X_{\d_n}^{2j})}
\Vert {\det}_{\G_n,I^c,J^c}\Vert_{C^{2}(X_{\d_n}^{2(p-j)})} \Eq(4.6665)$$

\\where $X_{\d_n}^{2j}=X_{\d_n}^{|I|}\times X_{\d_n}^{|J|}$ and
$X_{\d_n}^{2(p-j)}=X_{\d_n}^{|I^{c}|}\times X_{\d_n}^{|J^{c}|}$, since
$|I|=|J|=j$ and $|I^{c}|=|J^{c}|=p-j$.

\\Let $\dpr_{\d_n}^{k}$ be the lattice forward derivative of order $k$ in 
multi-index notation. By part (5a) of 
Theorem~1.1 we have 

$$\max_{0\le k\le 4}\Vert\dpr_{\d_n}^{k}\G_{n}
\Vert_{L^{\infty}((\d_n\math{Z})^{3})}\le C_{L} \Eq(5.61.11) $$ 

\\where $C_{L}$ is a constant independent of $n$. Relabel
$(\xi_1,...,\xi_{2(p-j)})=(x_1,..,x_{|I^c|},y_1,..,y_{|J|^c })$ where the
$x_i\in I^c$ and the $y_j\in J^c$.
Let now $\dpr_{\d_n}^{k_r}$ be the forward lattice derivative 
of order $k_r$, $0\le k_r\le 2$ 
with respect to the points $\xi_{r}$. 
Let ${\bf k}=(k_1,...,k_{2(p-j)})$  and define 
$\dpr_{\d_n}^{\bf k}=\prod_{r=1}^{2(p-j)}\dpr_{\d_n}^{k_r}$. 
Let  $\dpr_{\d_n}^{\bf k}$  act on the determinant. This 
produces another determinant with derivatives acting on the  
matrix elements $\G_{n} (x_{r}-y_{s})$. Since $\G_n$ is positive definite 
these matrices can be written
as Gram matrices by a standard argument. Thus the matrix
$a_{rs}=\dpr_{\d_n}^{k_r}\dpr_{\d_n}^{k_s}\G_n(x_r-y_s)$ with 
$(x_r,y_s)\in X_{\d_n}\times X_{\d_n}$ 
can be written as $a_{rs}=(f_r,g_s)_{L^{2}((\d_n\math{Z})^3)}$
where $f_r(\cdot)=\dpr_{\d_n}^{k_r}\G_n^{1\over 2}(x_r,\cdot)$ and
$g_s(\cdot)=\dpr_{\d_n}^{k_s}\G_n^{1\over 2}(\cdot,y_s)$.
Gram's inequality says $|{\det}\>a_{rs}|\le 
\prod_{r=1}^{p-j}\Vert f_r\Vert_{L^{2}((\d_n\math{Z})^3)}
\prod_{s=1}^{p-j}\Vert g_s\Vert_{L^{2}((\d_n\math{Z})^3)}$. 
We have
$\Vert f_r\Vert_{L^{2}((\d_n\math{Z})^3)}^{2}=\dpr_{\d_n}^{k_r}
\dpr_{\d_n}^{k_s}\G_n(x_r-x_s)|_{r=s}$. Similarly,
$\Vert g_s\Vert_{L^{2}((\d_n\math{Z})^3)}^{2}=\dpr_{\d_n}^{k_s}
\dpr_{\d_n}^{k_s}\G_n(y_r-y_s)|_{r=s}$. Since 
$2k_r\le 4$, we have by \equ(5.61.11) the bound 
$\Vert f_r\Vert_{L^{2}((\d_n\math{Z})^3)}^{2}\le C_{L}$. Similarly
$\Vert g_s\Vert_{L^{2}((\d_n\math{Z})^3)}^{2}\le C_{L}$. 
We therefore get  
 
$$|\dpr_{\d_n}^{\bf k}\det {\G_{n}}_{I^c,J^c}({\bf x}_{I^c},{\bf y}_{J^c} )|
\le C_{L}^{p-j} 
\Eq(5.61.1)$$ 

\\Hence 

$$\Vert {\det\G_n}_{I^c,J^c}\Vert_{C^{2}(X_{\d_n}^{2(p-j)})} 
\le  C_{L}^{p-j} \Eq(5.61.2)   $$ 
 
\\From \equ(5.61.1), \equ(4.6665) and \equ(4.6664) we get

$${h_F^{2j}\over (j!)^{2}}  
\Vert D^{2j,m}{\tilde K}^{\sharp}(X_{\d_{n}},\f,0)\Vert
\le\int d\m_{\G_n}(\z)\Bigl[\> \sum_{p\ge 0} 
\sum_{{I\subset\{1,...,p\}}\atop {J\subset\{1,...,p\}}} \d_{|I|,j}\d_{|J|,j}
\Bigl({C_L\over h_F^2})^{p-j}
{h_F^{2p}\over |I|!\>|I^{c}|!\>|J|!\>|J^c|!}\times $$
$$\Vert D^{2p,IJ,m}{\tilde K}(X_{\d_{n}},\f,\z,,0,0)\Vert \> ] \Eq(4.6666) $$ 

\\Now {\it choose $h_F$ sufficiently large depending on $L$ such that
$h_F^2\ge C_L$} which implies that $({C_L\over h_F^2})^{p-j}\le 1$. 
Putting in this bound and then summing over $j$ we get

$$\sum_{j\ge 0}{h_F^{2j}\over (j!)^{2}}  
\Vert D^{2j,m}{\tilde K}^{\sharp}(X_{\d_{n}},\f,0)\Vert
\le\int d\m_{\G_n}(\z)\Bigl[\> \sum_{p\ge 0} 
\sum_{{I\subset\{1,...,p\}}\atop {J\subset\{1,...,p\}}} \d_{|I|,|J|}
{h_F^{2p}\over |I|!\>|I^{c}|!\>|J|!\>|J^c|!}\times $$
$$\Vert D^{2p,IJ,m}{\tilde K}(X_{\d_{n}},\f,\z,,0,0)\Vert \> ] \Eq(4.6667) $$ 

\\Set $\f=0$ in \equ(4.6667). 
Multiply both sides by ${h_{B*}^{m}\over m!}$ and sum over $m$,
$0\le m\le m_0$. Majorize by dropping the constraint $\d_{|I|,|J|}$
on the right hand side. This gives

$$\vert {\tilde K}^{\sharp}(X_{\d_{n}})\vert_{{\bf h}_*} \le 
\int d\m_{\G_n}(\z)\> \Vert {\tilde K}(X_{\d_{n}},0,\z,,0,0)\Vert_{{\bf h}_*} $$

\\On the other hand multiplying both sides of  \equ(4.6667) by 
${h_{B}^{m}\over m!}$ and  summming over $m$, $0\le m\le m_0$ gives

$$\Vert {\tilde K}^{\sharp}(X_{\d_{n}},\f,0)\Vert_{{\bf h}} \le 
\int d\m_{\G_n}(\z)\> \Vert {\tilde K}(X_{\d_{n}},\f,\z,,0,0)\Vert_{{\bf h}} $$

\\This proves the first part of the Lemma.

\vglue0.2cm

\\We next turn to the second part of the Lemma. This is a consequence of the 
first part and Lemma~5.4A. Define 
${\tilde K}(X_{\d_n},\f,\z,\psi,\h)=K(X_{\d_n},\f+\z,\psi+\h)$. Then 
from \equ(5.367) of the first part and \equ(2.6.2.3) of Lemma~5.4A  we have

$$\eqalign{\Vert K^{\sharp}(X_{\d_n},\f,0)\Vert_{\hat{\bf h}} 
=\Vert {\tilde K}^{\sharp}(X_{\d_n},\f,0)\Vert_{\hat{\bf h}} &\le
\int d\m_{\G_{n}}(\z)\Vert {\tilde K}(X_{\d_n},\f,\z,0,0)\Vert_{\hat{\bf h}}\cr
&\le \int d\m_{\G_{n}}(\z)\Vert K(X_{\d_n},\f+\z,0)\Vert_{{\bf h}}\cr} $$

\\which proves \equ(5.369). \equ(5.368) follows similarly by using \equ(5.366)
of the first part followed by \equ(2.6.2.3a) of Lemma~5.4A. \bull 

\textBlack

\vglue.3truecm 

\\The following lemma generalizes Lemma~5.15 of [BMS] to the lattice and
the additional presence of Grassmann fields. It will play a key role later 
in obtaining contractive estimates.

\vglue0.2cm
\\{\it Lemma 5.15 
 
\\For any polymer activity ${\tilde K}(X_{\d_n},\phi+\z,\psi+\h)$: 
 
$$\Vert \tilde K(X_{\d_n},\z,0)\Vert_{{\bf h}_{*}}\le  
O(1) \tilde G_{\r,\k}(X_{\d_n},\z)\left[ 
\vert \tilde K(X_{\d_n})\vert_{{\bf h}_{*}}+h_B^{-m_0}h_{B*}^{m_{0}}\Vert 
\tilde K(X_{\d_n})\Vert_{{\bf h},G_\k}\right]\Eq(5.45)$$ 
 
$$\Vert \tilde K(Y_{\d_n},\phi,0)\Vert_{{\bf h}}\le O(1)  
e^{\g {\bar g}\int_{Z_{\d_n}\backslash Y_{\d_n}}dy(\phi\bar\phi(y))^2}G_\k(Z_{\d_n},\phi) 
\left[ 
\vert \tilde K(Y_{\d_n})\vert_{{\bf h}}+L^{-m_0 d_s}\Vert 
\tilde K(Y_{\d_n})\Vert_{h_F,L^{d_s}h_B,G_\k}\right]\Eq(5.45b)$$ 
 
$$\vert \tilde K^\sharp(X_{\d_n})\vert_{\hat {\bf h}_{*}}\le O(1) 2^{|X_{\d_n}|}\left[ \vert 
\tilde K(X_{\d_n})\vert_{{\bf h}_{*}}+h_B^{-m_0}h_{B*}^{m_{0}}\Vert 
\tilde K(X_{\d_n})\Vert_{{\bf h},G_\k}\right]\Eq(5.46)$$ 
 
$$\Vert \tilde K^\sharp(X_{\d_n})\Vert_{\hat {\bf h}, G_{2\k}} \le 2^{|X_{\d_n}|} 
\Vert \tilde K(X_{\d_n})\Vert_{{\bf h}, G_{\k}} \Eq(5.4666)    $$

\\where $\tilde G_{\r,\k}$ is as defined in \equ(5.2), and $m_0=9$ is 
the maximum number of derivatives appearing in the definition of 
Kernel and $h$ norms. In \equ({5.45b}), $Y_{\d_n},Z_{\d_n},\gamma$ are as described 
in Lemma~5.1. Moreover in the above norms  ${\bf h}_{*}=(h_{F},h_{B*})$, 
and ${\hat {\bf h}}_{*}=({h_{F}\over 2},h_{B*})$ where  
$h_{B*}=(\r\k)^{-1/2}$, ${\bf h}=(h_F,h_B)$,
${\hat {\bf h}}=({h_{F}\over 2},h_{B})$, $h_B=c{\bar g}^{-1/4}$, c=O(1) very 
small, and $h_{F}$ is taken to be 
sufficiently large depending on $L$.} 
 
\vglue.2truecm 
\\The superscript $\sharp$ stands for 
$d\m_{\G_n}(\z)$ integration. $\r$ is chosen as in Lemma~5.3, and $\k$ as in
Lemma~2.1. 
Note that we have that the constant $C(\r,\k,j)$ appearing in 
Lemma~5.2 ( this bounds $\z^{j}$)   satisfies 

$$C(\r,\k,j)=h_{B*}^{j}O(1)^{j}  \Eq(5.45a)   $$.

\vglue.3truecm 

\\{\it Proof}

\\We will first prove \equ(5.45) 
following the lines of the proof of lemma 5.15 of [BMS] where the Grassman  
fields were absent. Recall from the definition in \equ(2.6.2.1) that 
 
$$\Vert\tilde K(X_{\d_n},0,0,\z,0,0)\Vert_{\bf h_{*}} 
=\sum_{m=0}^{m_0}{h_{B*}^m\over m!} A_m\Eq(2.6.2.3)$$

\\where

$$A_m =\sum_{p\ge 0} h_{F}^{2p}{h_{B*}^{m}\over m!}
\Vert D^{2p,m}\tilde K(X_{\d_n},0,0,\z,0,0)\Vert \Eq(2.6.2.3a)$$ 
 
\\First conside the case $m=m_0$. Then

$$A_{m_{0}}\le h_{B*}^{m_0}h_{B}^{-m_0}\Vert\tilde K(X_{\d_n}\Vert_{{\bf h},G_{\k}}
\tilde G_{\r,\k} \Eq(5.46.2) $$

\\since $G_\k \le \tilde G_{\r,\k}$. Now let $m<m_0$.  
 
\\We expand in $\z$ in Taylor series with remainder 
 
$$(D^{2p,m}\tilde K)(X_{\d_n},\z,0;f^{\times m},g_{2n}  )= 
\sum_{j=0}^{m_0-m-1}{1\over j!} 
(D^{2p,j+m}\tilde K)(X_{\d_n},0,0;f^{\times m},\z^{\times j}, g_{2p})+$$ 
$$+{1\over (m_0-m-1)!}\int_0^1ds(1-s)^{m_0-m-1} 
(D^{2p,m}\tilde K)(X_{\d_n},s\z,0;f^{\times m},\z^{\times m_0-m},g_{2p}) 
\Eq(5.46.22)  $$ 
 
\\Therefore 
 
$$\Vert (D^{2p,m}\tilde K)(X_{\d_n},\z,0)\Vert\le 
\sum_{j=0}^{m_0-m-1}{1\over j!} 
\Vert (D^{2p,j+m}\tilde K)(X_{\d_n},0,0)\Vert\Vert\z\Vert_{C^{2}(X_{\d_n})}^{j} +$$ 
$$+{1\over (m_0-m-1)!}\int_0^1ds(1-s)^{m_0-m-1}  
\Vert\z\Vert_{C^{2}(X_{\d_n})}^{m_0 -m}
\Vert(D^{2p,m_0}\tilde K)(X_{\d_n},s\z,0)\Vert $$   
 
\\Hence 
 
$$A_m \le \sum_{j=0}^{m_0-m-1}{(j+m)!\over j!m!} h_{B*}^{-j}
\Vert\z\Vert_{C^{2}(X_{\d_n})}^{j}
\Vert\tilde K(X_{\d_n})\Vert_{{\bf h}_*}  +$$  
$$+{m_{0}!h_{B*}^{m_0}h_{B}^{-m_0}\over m!(m_0-m-1)!}
\int_0^1ds(1-s)^{m_0-m-1} h_{B*}^{-(m_0 -m)}
\Vert\z\Vert_{C^{2}(X_{\d_n})}^{m_0 -m} 
\Vert\tilde K(X_{\d_n})\Vert_{{\bf h},G_{\k}}\tilde G_{\r,\k}(X_{\d_n},s\z) 
$$   
 
\\By Lemma~5.2, with $\z$ replaced by $\sqrt{1-s^{2}}\z$, and  
\equ(5.45a), 
 
$$h_{B*}^{-j}\Vert\z\Vert^j_{C^2(X_{\d_n})}\le O(1)^{j}  
\tilde G_{\r,\k}(X_{\d_n},\sqrt{1-s^2}\,\z) 
{1\over(1-s^2)^{j/2}} \Eq(5.466)$$ 
 
\\where $0\le s<1$.  With $s=0$ this 
bound is applied to the terms in the sum over $j$.  For the Taylor 
remainder term take $j=m_0 -m$ and
note that $(1-s)^{(m_0-m)/2-1}$ is integrable since $m_0>m$.  Hence: 
 
$$\eqalign{A_m &\le O(1)\tilde G_{\r,\k}(X,\z)\Bigr[
\sum_{j=0}^{m_0-m-1}{(j+m)!\over j!m!}
\Vert\tilde K(X_{\d_n})\Vert_{{\bf h}_*}  +
{m_{0}!h_{B*}^{m_0}h_{B}^{-m_0}\over m!(m_0-m-1)!}
\Vert \tilde K(X_{\d_n})\Vert_{{\bf h},G_{\k}}\Bigr] \cr 
&\le O(1)\tilde G_{\r,\k}(X,\z)\Bigr[\Vert\tilde K(X_{\d_n})\Vert_{{\bf h}_*}
+h_{B*}^{m_0}h_{B}^{-m_0}\Vert\tilde K(X_{\d_n})\Vert_{{\bf h},G_{\k}}\Bigr] 
\cr } \Eq(5.46.3) $$   

\\Summing \equ(5.46.3) over $0\le m\le m_0-1$ and adding 
\equ(5.46.2) proves \equ(5.45).

\vglue0.2truecm 

\\Inequality \equ(5.45b) is also proved in the same way as \equ(5.45). 
We are estimating the $\bf h$ norm which is given by \equ(2.6.2.3) with 
$\bf h_{*}$ replaced by $\bf h$,  $h_B*$ by $h_B$ and $\z$ by $\f$. We replace
$\tilde G_{\r,\k}$ by $G_\k$. Then 
\equ(5.46.2) remains true with $\z$ replaced by $\f$ and $\tilde G_{\r,\k}$ 
replaced by $G_k$. Subsequently for $m < m_0$ we expand in Taylor series  
as above but now in $\f$. We  do the norm estimate as above but now using  
Lemma 5.1 in place of Lemma 5.2. For $\e$ sufficiently small depending on
$L$ we have $\bar g$ sufficiently small and therefore $h_{B}^{-1}$ is 
sufficiently small. Hence $h_{B}^{-j}C\le O(1)$ where $C=\k^{-j/2}O(1)$ is the 
constant appearing in Lemma~5.1. In the Taylor remainder term we replace $h_B$
by $L^{d_s}h_B$. which leads to the factor $L^{-m_0 d_s}$. 
 
\vglue.2cm 

\textRed 

\\Finally note that the inequality \equ(5.46) now follows on using 
\equ(5.368) of Lemma~5.14A, followed by \equ(5.45) and then Lemma~5.3.
\equ(5.4666) follows from \equ(5.369) of Lemma~5.14A on using the 
stability of the large field regulator $G_\k$.  \bull  

\textBlack

\vglue0.2truecm 
 
\\The next lemma extends lemma~5.16 of [BMS] to the case when
Grassmann fields are also present.
 
\vglue0.2truecm 

\\{\it Lemma 5.16 
 
\\For any $q>0$, there exists constants $c_L$ independent of $n$
such that for $L$ sufficiently large, $\e$ sufficiently small and
$h_F$ sufficiently large depending on $L$,
 
$$\Vert\SS(\l, K_n)^\natural\Vert_{{\bf h},G_{\k},\AA_{p},\d_{n+1}}\le  q 
\qquad {\rm when} \ |\l {\bar g}^{1/4-\eta/3} |\le c_L
\Eq(5.57) $$ 
 
$$\vert\SS(\l, K_n)^\natural\vert_{{\bf h}_{*},\AA_{p},\d_{n+1}}\le q 
\qquad {\rm when} \ |\l {\bar g}^{11/12-\eta/3} |\le c_L
\Eq(5.58)$$ 
where ${\bf h}_{*}=(h_{B*},h_{F})$ and $\natural$ denotes integration with
respect to $d\m_{\G_{n,L}}(\xi)$, $\G_{n,L}=S_{L^{-1}}\G_n$ being the 
rescaled fluctuation covariance.   

\\When $R_n=0$ we may set $\eta =0$ in \equ(5.57) and replace 
$\l {\bar g}^{11/12-\eta/3}$ by $\l {\bar g}^{1/2-\delta/2}$ in \equ(5.58)
} 
 
\vglue.3truecm 
\\{\it Proof} 
\textRed

\\We suppress the dependence on $\l$ which plays a passive role in most of the
following and make the dependence explicit towards the end when necessary. 
\textRed We will apply the first part of Lemma~5.14A to the the reblocked 
polymer activity ${\cal B}K_n(LZ_{\d_n},S_L\Phi,\xi)= 
{\cal B}K_n(LZ_{\d_n},S_L\f,\z,S_L\psi,\h)$ which is a functional of 
$\tilde K_n$ and $P_n$ where $\tilde K_n(X_{\d_n},\f,\z,\psi,\h)
=K_n(X_{\d_n},\f+\z,\psi+\h)$ 
( see the definition of reblocking in section~3.1). Recall the definition of
rescaled polymer activities and rescaled covariances given by \equ(3.152) 
and \equ(3.151) in the Appendix to section~3. The rescaled, reblocked 
activities are defined by \equ(3.155). We get by virtue of
\equ(5.367) and \equ(5.366)

$$\Vert \SS(K_n))^{\natural}(Z_{\d_{n+1}},\f,0)\Vert_{\bf h}
\le \int d\m_{\G_{n,L}}(\z)\Vert (S_{L}{\cal B}K_n)
(Z_{\d_n},\f,\z,0)\Vert_{\bf h}
\Eq(5.58.4.4) $$

$$\vert\SS(K_n))^{\natural}(Z_{\d_{n+1}})\vert_{\bf h_*} \le
\int d\m_{\G_{n,L}}(\z)\Vert (S_{L}{\cal B}K_n)(Z_{\d_n},0,\z,0)\Vert_{\bf h_*}
\Eq(5.58.4.5) $$

\\We now prove \equ(5.57) starting from \equ(5.58.4.4). This
follows the lines of the proof of Lemma~5.16, [BMS]. 
Unfortunately, in the latter proof a minor error crept in  
\footnote{$^{(1)}$}{A. Abdesselam, private communication } 
and we take this opportunity to correct it.  
Inserting the definition \equ(4.13) in \equ(5.58.4.4) and 
using the multiplicative property of the $\bf h$ norm we get 

$$ \Vert (\SS (K_n)^{\natural})(Z_{\d_{n+1}},\f,0)\Vert_{\bf h} 
\le \sum_{N+M\ge 1}{1\over N !M!}
{\sum}_{(X_{\d_n,j}),(\Delta_{\d_n,i})\rightarrow LZ_{\d_{n+1}}}
\Vert e^{-\tilde V_{n,L}(Z_{\d_{n+1}}\setminus L^{-1}({\bf X}_{\d_{n+1}}
\cup {\bf \Delta_{\d_{n+1}}}),\f)}\Vert_{\bf h}\times $$
$$\times \int d\m_{\G_n,L}(\z)  
{\prod}_{j=1}^{N}\Vert \tilde K_{n,L}(L^{-1}X_{\d_n,j},\f,\z,0)\Vert_{\bf h}  
{\prod}_{i=1}^{M}\Vert P_{n,L}(L^{-1}\Delta_{\d_n,i},\f,\z,0)\Vert_{\bf h}$$

\\whence

$$\Vert (\SS (K_n)^{\natural})(Z_{\d_{n+1}},\f,0)\Vert_{\bf h} 
\le 2^{|Z_{\d_{n+1}}|}\sum_{N+M\ge 1}{1\over N !M!}
{\sum}_{(X_{j}),(\Delta_{\d_n,i})\rightarrow LZ_{\d_n}}\int d\m_{\G_n}(\z)
{\hat G}_{\k,\r}({\bf X}_{\d_n}\cup {\bf \Delta_{\d_n}}, S_L\f,\z)\times  $$
$${\prod}_{j=1}^{N}\Vert\tilde K_n(X_{j})\Vert_{{\bf h}_{L},{\hat G}_{\k,\r}}
{\prod}_{i=1}^{M}\Vert P_n(\Delta_{\d_n,i})\Vert_{{\bf h}_{L},{\hat G}_{\k,\r}}
\Eq(5.58.9)   $$

\\where ${\bf h}_{L}=(L^{-d_s}h_B,L^{-d_s}h_F)$,        
${\bf X}_{\d_n}=\cup X_{\d_n,j}$, 
${\bf\Delta}_{\d_n}=\cup\Delta_{\d_n,i}$ and we have bounded 
$e^{-\tilde V_{n,L}}$ using lemma~5.5 which continues to apply. 

\\Lemma~5.4 bounds the $\z$ integral by 

$$ 2^{|{\bf X}_{\d_n}\cup{\bf\Delta_{\d_n}}|} 
G_{3\k}({\bf X}_{\d_n}\cup {\bf\Delta_{\d_n}},S_L\f)
\le 2^{|{\bf X}_{\d_n}\cup{\bf\Delta_{\d_n}}|}
G_{\k}(L^{-1}({\bf X}_{\d_{n+1}}\cup{\bf\Delta}_{\d_{n+1}}), \phi) $$
$$\le \prod_{j=1}^{M} 2^{|X_{\d_n,j}|}\prod_{i=1}^{N} 2^{|\Delta_{\d_n,i}|} 
G_{\k}(Z_{\d_{n+1}},\phi)$$

\\since $L^{-1}({\bf X}_{\d_{n+1}}\cup{\bf\Delta_{\d_{n+1}}})
\subset Z_{\d_{n+1}}$. Moreover for $L$ sufficiently large

$$\Vert\tilde K_n(X_{j})\Vert_{{\bf h}_{L},{\hat G}_{\k,\r}}
\le \Vert\tilde K_n(X_{j})\Vert_{\hat{\bf h},{\hat G}_{\k,\r}}
\le \Vert K_n(X_{j})\Vert_{{\bf h},{\hat G}_{\k,\r}} $$

\\where we have used Lemma~5.4A, \equ(2.6.2.3) in the last step.
Therefore

$$\Vert (\SS (K_n)^{\natural})(Z_{\d_{n+1}})\Vert_{\bf h, G_{\k}}
\le 2^{|Z_{\d_{n+1}}|}\sum_{N+M\ge 1}{1\over N !M!} 
{\sum}_{(X_{\d_n,j}),(\Delta_{\d_n,i})\rightarrow LZ_{\d_n}}
{\prod}_{j=1}^{M}2^{|X_{\d_n,j}|}\Vert K_n(X_{j})\Vert_{{\bf h},{\hat G}_{\k,\r}} 
\times $$
$${\prod}_{i=1}^{N}2^{|\Delta_{\d_n,i}|}
\Vert P_n(\Delta_{\d_n,i})\Vert_{{\bf h},{\hat G}_{\k,\r}}
$$

\\From this point on we proceed as in the proof of Lemma~5.16, [BMS], the
only difference being is that now we are on a lattice. \textBlack     
The condition on the sum over the polymers above implies that
$Z_{\d_n}=(\cup L^{-1}\bar X_{\d_n,j}^{L})
\cup (\cup L^{-1}\bar \Delta_{\d_n,i}^{L})$.
This also implies that
$Z_{\d_{n+1}}=(\cup L^{-1}\bar X_{\d_{n+1},j}^{L})
\cup (\cup L^{-1}\bar\Delta_{\d_{n+1},i}^{L})$. 

\\Multiply both sides by $\AA_p(Z_{\d_{n+1}})$ and observe on the right hand 
side 

$$\AA_{p+1}(Z_{\d_{n+1}})
\le  {\prod}_{j=1}^{M} \AA_{p+1}(L^{-1}\bar X_{j,\d_{n+1}})
{\prod}_{i=1}^{N} \AA_{p+1}(L^{-1}\bar \Delta_{\d_{n+1},i})$$ 
$$\le O(1)^{N+M} {\prod}_{j=1}^{M} \AA_{-2}(X_{j,\d_{n+1}})          
{\prod}_{i=1}^{N} \AA_{-2}(\Delta_{\d_{n+1},i})$$

\\where we have first used the fact that the $L$-closures of the polymers 
are connected by definition of the reblocking operation,
then Lemma~2.2 together with $|X_{j,\d_{n+1}}|=|X_{j,\d_{n}}|$ and 
$|\Delta_{\d_{n+1},i}|=|\Delta_{\d_{n},i}|$ . The last observation follows
from our definition of polymers in section~1.3 and \equ(1.22f). Therefore 

$$\Vert (\SS (K_n)^{\natural})(Z_{\d_{n+1}})\Vert_{\bf h, G_{\k}}
\AA_{p}(Z_{\d_{n+1}})
\le \sum_{N+M\ge 1}{1\over N !M!} O(1)^{N+M} 
{\sum}_{(X_{j}),(\Delta_{\d_n,i})\rightarrow LZ } $$
$${\prod}_{j=1}^{M}\Vert K_n(X_{j,\d_n})\Vert_{{\bf h},{\hat G}_{\k,\r}}
\AA_{-1}(X_{j,\d_n}){\prod}_{i=1}^{N}
\Vert P_n(\Delta_{\d_n,i})\Vert_{{\bf h},{\hat G}_{\k,\r}}
\AA_{-1}(\Delta_{\d_n,i}) $$

\\Fix any $\D_{\d_{n+1}}$ and sum over $Z_{\d_{n+1}}\ni \Delta_{\d_{n+1}}$. 
This fixes on the right hand side the sum over $Z_{\d_{n}}\ni \Delta_{\d_{n}}$
with $\D_{\d_{n}}$ fixed by restriction. The spanning tree 
argument of Lemma~7.1 of [BY] controls the sums over 
$N, M, Z_{\d_n}, (X_{j,\d_n} ), (\Delta_{i,\d_n})\rightarrow LZ_{\d_n} $ 
with the result ( we have now made
explicit the dependence on $\l$)

$$\Vert (\SS (\l, K_n)^{\natural})\Vert_{\bf h, G_{\k},\AA_{p},\d_{n+1}}
\le O(1)\sum_{N\ge 1} O(1)^{N} L^{3N}
\Bigl (\Vert K_n(\l)\Vert_{{\bf h},{\hat G}_{\k,\r},\AA,\d_n} +
\Vert P_n(\l)\Vert_{{\bf h},{\hat G}_{\k,\r},\AA,\d_n} \Bigr )^{N} $$

\\The proof of \equ(5.57) is completed by Lemmas~5.8 and 5.14. When $R=0$
we can use Lemma~5.8 and replace Lemma~5.14 by Lemma~5.11.

\\To prove \equ(5.58) we start from \equ(5.58.4.5) and proceed as before. We
replace ${\hat G}_{\k,\r}$ by 
${\tilde G}_{\k,\r}$ and then  use Lemma~5.3 to estimate the $\z$ integral.
We use \equ(2.6.2.3a) of Lemma~5.4A. Proceeding as before now leads to

$$\vert (\SS (\l, K_n)^{\natural})\vert_{{\bf h}_{*}, \AA_{p},\d_{n+1}}
\le O(1)\sum_{N\ge 1} O(1)^{N} L^{3N}
\Bigl (\Vert K_n(\l)\Vert_{{\bf h}_{*},{\tilde G}_{\k,\r},\AA,\d_n} +
\Vert P(\l)\Vert_{{\bf h}_{*},{\tilde G}_{\k,\r},\AA,\d_n} \Bigr )^{N} $$

\\Now use  Lemmas~5.8 and 5.14 to complete the proof of \equ(5.58). Finally
when $R = 0$ use Lemmas~5.8 and 5.11 as before. \bull

\vglue.3truecm 
\\{\it Estimates on relevant parts and flow coefficients  
from the remainder} 
 
\vglue.3truecm 
 
\\Let $({\tilde \a}_{n,P})$ be the coefficients $({\tilde 
\a}_{n,2,0}, {\tilde \a}_{n,2,1}, {\tilde \a}_{n,2,\bar 1}
{\tilde \a}_{n,4})$ defined in \equ(4.57) 
and \equ({4.62}).  The flow coefficients $\xi_{n, R}$, $\r_{n, R}$  
are given in \equ({4.64}). 
 
\vglue.3truecm 
\\{\it Lemma 5.17 : Under the conditions of the domain ${\cal D}_n$ we have  

$$\Vert  R_n^\sharp\Vert_{{\bf \hat h},G_{3\k},\AA_{-1},\d_n}\le 
{\bar g}^{3/4-\eta}  \Eq(5.65)$$ 
 
$$\vert  R_n^\sharp\vert_{{\bf {\hat h}}_*,\AA_{-1},\d_n}
\le O(1){\bar g}^{11/4-\eta} \Eq(5.66)$$ 
 
$$\vert{\tilde \a}_{n,P}\vert_{\AA,\d_n}
\le O(1){\bar g}^{11/4-\eta} \Eq(5.67)$$ 

$$|\xi_{n}|\le C_L{\bar g}^{11/4-\eta} 
\Eq(5.69)$$ 
$$|\r_{n}|\le C_L{\bar g}^{11/4-\eta} \Eq(5.70)$$ 
where the constants $C_L$ are independent of $n$ and $\e$ }

\vglue.3truecm 
\\{\it Proof} 
 
\\\equ(5.65) follows from \equ({55.2}) and Lemma~5.15, 
\equ(5.4666). \equ(5.66) follow from  
\equ({55.3}) and lemma 5.15 with  $m_0 =9$  
and $\e$ sufficiently small depending on L so that $\bar g$ is 
sufficiently small. In fact in lemma~5.15 (with $\tilde K=R_n$) the first term
has the desired bound by \equ(55.3). By \equ({55.2}) rogether with 
$h_B^{-1}=c{\bar g}^{1\over 4}$ and $h_{B*}=h_{B*}(L)$ we see that
the second term is bounded by 
$O(1){\bar g}^{1\over 4}h_{B*}^{9}{\bar g}^{11/4 -\eta}
\le {\bar g}^{11/4 -\eta}$ for $\bar g$ sufficiently small.  

\\Recall that ${\tilde \a}_{n,P}(X_{\d_n})$,  are supported on small 
sets. Then \equ(5.67) follows from \equ(4.62) and
\equ(5.66). In fact the dominant contribution 
comes by setting ${\tilde V_n}=0$ because the difference gives 
additional powers of $\bar g$. Then  we have   

$$\Vert {\tilde \a}_{n,P}\Vert_{\AA,\d_n} \le O(1) n(P)!  
{\bf {\hat h}}_{*}^{-n(P)} 
\vert 1_{S}R_n^\sharp\vert_{{\bf {\hat h}}_{*},\AA,\d_n}$$  
where $n(P)$ is the number of fields in the monomial $P$, 
we have used the shorthand notation  
${\bf {\hat h}}_{*}^{-n(P)}=\max_{n(P)_{F}+n(P)_{B}=n(P)}(h_{*B}^{-n(P)_{B}} 
{\hat h}_F^{-n(P)_{F}})$  
and $1_{S}$ is the 
indicator function on small sets. Now use \equ(5.66) to get \equ(5.67).  
\equ(5.69), \equ(5.70) follow from \equ(5.67), the 
definitions \equ(4.64), \equ(4.70) and Wick 
coefficients $C_n(0)$ are uniformly bounded by a $L$ dependent constant by
Corollary~1.2. \bull

\vglue.3truecm 
\\{\it Lemma~5.18 
 
\\Under the conditions of ${\cal D}_n$ and
$\e$ sufficiently small depending on $L$, there exists a constant
$C_L$ independent of $\e$ and $n$ such that

$$\vert g_{n+1}-\bar g\vert < 2\n{\bar g}^{3/2}, 
\quad \vert\mu_{n+1}\vert < C_L {\bar g}^{2-\d}  \Eq(55.9) $$

} 
 
\\{\it Proof} : 
 
\\It is convenient to define

$$\tilde g_n = g_n -\bar g $$

\\Then from the flow equation \equ(4.47A) for
$g_n$ and the definition of ${\bar g}$ in \equ(55.4) we get

$$\tilde g_{n+1}=(2-L^{\e})\tilde g_{n} + \tilde\xi_n \Eq(55.9a) $$

\\where

$$\tilde\xi_n=-L^{2\e}a_{c*}{\tilde g_n}^{2}-L^{2\e}(a_n-a_{c*}) g_n^{2}+ 
\xi_{n} \Eq(55.9b)  $$

\\From Lemma~5.12, Lemma~5.17 and $g_n\in {\cal D}_n$
we get for  $\e$ sufficiently small depending on $L$
the bound 

$$|\tilde\xi_n|\le C_L{\bar g}^{2} \Eq(55.9c)  $$ 

\\Therefore

$$|\tilde g_{n+1}|< \n{\bar g}((2-L^{\e}) + 
  {C_L\over \n} {\bar g}) \Eq(55.9d)  $$

\\For $\e$ sufficiently small depending on $L$ we get 

$$|(1-L^{\e}) + {C_L\over \n} {\bar g}|\le 1 \Eq(55.9e)  $$ 
\\Therefore $|\tilde g_{n+1}|\le 2\n{\bar g}$ which proves the first
inequality of \equ(55.9).

\\The bound on $\mu_{n+1}$ 
follows from the second of the flow equations \equ({4.47}), on using
$\mu_{n}$, $g_n$ belong to ${\cal D}_n$, Lemma~5.12 
and the bound \equ(5.70) on $\rho_{n}$. \bull

\vglue0.2cm

\\As stated in Theorem~3.1 borrowed from [BDH] the assumption
of stability of the local potential with respect to perturbation by 
relevant parts (see \equ(33.17d) ensures the extraction estimate of 
\equ(33.17e). The following lemma proves the stability for the case at hand,
namely that of $\tilde V_{n,L}(\D_{n+1})$ with respect to the relevant
part $F_n$ defined in section~4.  
 
\vglue0.2cm
\\Recall from \equ(4.12) that $F_n (\lambda) = \lambda^{2}F_{Q_n}+ 
\lambda^{3}F_{R_n}$ and from \equ({33.2aa}) that (each part of) $F_n$ 
decomposes: $F_n (X_{\d_{n+1}})=\sum_{\Delta_{\d_{n+1}} \subset X_{\d_{n+1}}}
F_n(X_{\d_{n+1}},\Delta_{\d_{n+1}})$. 
 
\vglue.3truecm 
 
\\{\it Lemma 5.19 
 
\\For any $R>0$ and $\xi :=R \max(|\lambda^{2}|\bar g, 
|\lambda^{3}|\bar g^{7/4-\eta})$ sufficiently small, 
 
$$ 
\Vert e^{-\tilde V_{n,L}(\D_{n+1})-\sum_{X_{\d_{n+1}}\supset\D_{n+1}} z(X_{\d_{n+1}})F(\lambda, 
X_{\d_{n+1}},\D_{n+1})}\Vert_{{\bf h},G_\k}\le 2^2 \Eq(5.88)$$ 
 
\\where $z(X_{\d_{n+1}})$ are complex parameters with $|z(X_{\d_{n+1}})|
\le R$. 
} 
 
\vglue.3truecm 
 
\\{\it Proof} 
 
\\It is easy to see that Lemma~5.5 still holds if we replace $\tilde V_n$ by 
$\tilde V_{n,L}$ provided $\e$ is sufficiently small. This implies
that $\bar g$ is sufficiently small. We then have 
 
$$ 
\Vert e^{-\tilde V_{n,L}(\D_{n+1})-\sum_{X_{\d_{n+1}}\supset\D_{n+1}}
z(X_{\d_{n+1}})F(\lambda, X_{\d_{n+1}},\D_{n+1})}\Vert_{\bf h}\le $$

$$ 2\ e^{-{\bar g}/4\int_{\D_{n+1}} 
dx(|\f(x)|^{2})^2+\sum_{X_{\d_{n+1}}\supset\D_{n+1}}R\Vert F(\lambda 
,X_{\d_{n+1}},\D_{n+1})\Vert_{\bf h}} \Eq(55.88) $$ 
 
\\Recall that the relevant parts $F(X_{\d_{n+1}},\D_{n+1})$ are supported on 
small sets $X_{\d_{n+1}}$. The proof now  follows easily from the following 
 
\vglue.3truecm 
 
\\{\it Claim:} For $\e$ sufficiently small 
 
$$|z(X_{\d_{n+1}})|\Vert F(\lambda,X_{\d_{n+1}},\D_{n+1})\Vert_{\bf h} 
\le C_{L}\xi\left(\bar g \int_{\D_{n+1}} 
d^{3}x(|\f(x)|^{2})^{2} + \bar g^{1/2}\Vert|\f|^{2} 
\Vert_{\D_{n+1},1,5}+1\right)\Eq(5.91) 
$$ 
 
\\where $\Vert\phi\Vert^{2}_{\D_{n+1},1,5}$ is the square of the lattice 
Sobolev norm defined in \equ(2.10reg).

\vglue0.1cm 
\\{\it Proof of the Claim:} 

\\We have 
$\Vert F(\lambda,X_{\d_{n+1}},\D_{n+1})\Vert_{\bf h}\le |\lambda |^{2}\Vert 
F_Q(X_{\d_{n+1}},\D_{n+1})\Vert_{\bf h}+|\lambda |^{3}
\Vert F_{ R}(X_{\d_{n+1}},\D_{n+1})\Vert_{\bf h}$. 
 
\\Consider \equ(4.33)-\equ(4.36a). Undo the Wick ordering on the superfield
field 
and note that, by virtue of supersymmetry, no field independent terms arise. 
The $m=1$ 
term in \equ(4.36) remains unchanged and in the $m=2$ case there results 
an additional contribution $-2C_{n+1}(0)\Phi\bar\Phi$. The Wick
constant $C_{n+1}(0)$ has a uniform bound $C$ which depends only on $L$
by Corollary~1.2. We write this in the Grassmann representation and notice that
for the $m=2$ case the $\psi\bar\psi(x)^{2}$ contribution vanishes by 
statistics. From the definition of the $\bf h$ norm with 
$h_B=c{\bar g}^{-1/4}$ and $h_F=h_F(L)$ we get the bound 

$$\Vert F_Q(X_{\d_{n+1}},\D_{n+1})\Vert_{\bf h}\le C_{L}{\bar g}^2\left( 
\int_{\D_{n+1}} d^{3}x\Vert(|\f|^{2})^2(x)\Vert_{{h_{B}}} 
\sup_{x\in \D_{n+1}}|f^{(2)}_{Q}(X_{\d_{n+1}},x,\D_{n+1})|+\right.$$ 
$$\left.\left(\int_{\D_{n+1}} d^{3}x\Vert(|\f|^{2})(x)\Vert_{h_{B}}+1\right) 
\sum_{m=1}^{2} \sup_{x\in \D_{n+1}}|f^{(m)}_{Q}(X_{\d_{n+1}},x,\D_{n+1})|
\right)$$ 
 
\\Now for $m=1,2$ 
$$ 
\bar g\int_{\D_{n+1}} d^{3}x\Vert(|\f|^{2})^m(x)\Vert_{h_{B}}\le O(1)\left( 
\bar g\int_{\D_{n+1}} d^{3}x(|\f|^{2})^{2}(x)+1\right)$$ 
 
\\From the definition $\equ(4.36a)$ and the estimates obtained  
in the course of proving Lemma~5.12, we have 
 $$\sup_{x\in \D_{n+1}}|f^{(m)}_{Q}(X_{\d_{n+1}},x,\D_{n+1})|\le C_{L}$$ 
Therefore 
$$|\lambda^{2}||z(X_{\d_{n+1}})| \ \Vert F_Q(X_{\d_{n+1}},\D_{n+1})\Vert_h 
\le C_{L} R |\lambda |^{2}\bar g \left(\bar g\int_{\D_{n+1}} dx(|\f|^{2})^2(x) 
+1\right)\Eq(5.96)$$ 
 
\\Next consider $F_{ R_n}$, supported on small sets, defined in 
\equ(4.57), \equ(4.58). Recall \equ({4.57b}), 
$$F_{R}(X_{\d_{n+1}},{\bf\Phi})= \sum_{P}\int_{\D_{n+1} } dx\  
\alpha_{P} (X_{\d_{n+1}},x)P ({\bf\Phi} (x),\dpr_{\d_{n+1}}{\bf\Phi} (x)) 
$$ 
 
\\By Lemma~5.17 and \equ(4.67a) we have
$|\a_P(X_{\d_{n+1}},x)|\le C_{L} \bar g^{11/4-\eta}$, 
so that 
$$\eqalign{ 
|\lambda |^{3}\vert z(X_{\d_{n+1}})\vert \Vert F_{ R}(X_{\d_{n+1}},\D_{n+1})\Vert_{\bf h} 
\le &C_{L} R|\lambda |^{3}\bar g^{11/4-\eta}\sum_{P} 
\int_{\D_{n+1}} dx\ \Vert 
P(\Phi (x),\dpr_{\d_{n+1}}\Phi (x)) 
\Vert_{\bf h}\cr 
\le &C_{L} R|\lambda |^{3}\bar g^{7/4-\eta}\bigg(  
\bar g\int_{\D_{n+1}} dx \, (|\f|^{2})^2(x)+ 
\bar g^{1/2}\Vert|\f|^{2}\Vert_{\D_{n+1},1,5}+1 
\bigg)\cr}$$ 
 
\\The claim follows by combining this with \equ({5.96}). In 
the above inequality the Sobolev norm  when estimating the 
term giving arise to $\phi\dpr_{\d_n,\m}\bar\phi$. We bound 
$\vert\phi\dpr_{\d_n,\m}\bar\phi\vert \le 1/2(\vert|\f|^{2}\vert 
+\vert \dpr_{\d_n,\m}\phi\vert^{2})$  
and then use the lattice Sobolev embedding inequality.  \bull  
 
\vglue.3truecm 
 
\\{\it Lemma 5.20 
 
\\For any $R>0$ and $\xi :=R \max(|\lambda^{2}|{\bar g}^{2}, 
|\lambda^{3}|{\bar g}^{11/4-\eta})$ sufficiently small, 
 
$$ 
\vert e^{-\tilde V_{n,L}(\D_{n+1})-\sum_{X_{\d_{n+1}}\supset\D_{n+1}} 
z(X_{\d_{n+1}})F(\lambda,X_{\d_{n+1}},\D_{n+1})}
\vert_{{\bf h}_{*}}\le 2^2 
\Eq(5.88.1)$$ 
 
\\where $z(X)$ are complex parameters with $|z(X)|\le R$. 
} 
 
\vglue.3truecm 
 
\\{\it Proof} 
 
\\The proof is similar to the previous one except that we can use the estimate 
$\vert F(\lambda,X,\D)\vert_{{\bf h}_{*}} 
\le C_{L}\xi$ in place of \equ({5.91}) since the ${\bf h}_{*}$ norm is computed
with field derivatives at $\Phi=0$. \bull  
 
\vglue0.2cm

We will now bound the remainder $R_{n+1}$ given in \equ(4.221). It consist 
of a sum of four contributions, namely $R_{n+1,\rm main}$, $R_{n+1,linear}$, 
$R_{n+1,3}$ and $R_{n+1,4}$ which we will estimate in turn. These estimates
parallel those obtained for the continuum bosonic theory in [BMS].

\vglue0.2cm
Recall from \equ({4.188}) that 
 
$$ 
R_{n+1,\rm main}  
= {1\over 2\pi i}  
\oint_{\gamma} {d\lambda \over \lambda^{4}} 
{\cal E} \bigg(  
\SS(\lambda ,Q_ne^{-V_n})^{\natural},F_{Q_n} (\lambda )  
\bigg) 
\Eq(5.lem35-0) 
$$ 
 
\vglue.2truecm 
\\{\it Lemma 5.21 
 
$$\Vert R_{n+1,\rm main}\Vert_{{\bf h},G_{\k},\AA,\d_{n+1}}
\le C_{L}{\bar g}^{3/4} \Eq(5.lem35-1)$$ 
 
$$\vert R_{n+1,\rm main}\vert_{{\bf h}_{*},\AA,\d_{n+1}} 
\le C_{L}{\bar g}^{3-3\d/2} \Eq(5.lem35-2)$$ 

}  
 
\\{\it Proof}
 
\textRed

\\The proof is identical to that of Lemma~5.21 of [BMS] except that we replace
$\e$ by $\bar g$, put in lattice subscript $n$ where appropriate, and 
note that the field independant piece $F_0$ is now absent. We apply 
Theorem~3.1 (which is a restatement of Theorem~5 in section~4.2 of [BDH-est])
instead of Theorem~6 of [BDH-est]. \bull  

\textBlack

Recall from \equ({4.21}) that 
 
$$ 
R_{n+1,3} =  
{1\over 2\pi i}  
\oint_{\gamma} {d\lambda \over \lambda^{4} (\lambda -1)} 
{\cal E}\Bigl (\SS (\lambda,K_n)^{\natural},F_n(\lambda )\Bigr) 
\Eq(5.lem36-3)$$

\vglue.3truecm 
\\{\it Lemma 5.22 
 
$$\Vert R_{n+1,3}\Vert_{{\bf h},G_{\k},\AA,\d_{n+1}}\le C_{L}
{\bar g}^{1-4\eta /3}  \Eq(5.lem36-4)$$

$$\vert R_{n+1,3}\vert_{{\bf h}_{*},\AA,\d_{n+1}} 
\le C_{L}{\bar g}^{10/3-4\eta /3}  \Eq(5.lem36-2)$$ 

}

\\{\it Proof} 
 
\\The proof follows the lines of that of Lemma~5.21. To prove \equ(5.lem36-4)
we take the contour $\g$ to of radius 
$|\lambda|= c_{L}{\bar g}^{-(1/4-\eta/3)}$.
This ensures that the hypothesis of Lemma~5.16, \equ(5.58) is satisfied
and $\xi$ of Lemma~5.19 is sufficiently small so that stability holds.
\equ(5.lem36-4) now follows from the extraction estimate \equ(33.17e)
and the Cauchy bound as before. To prove \equ(5.lem36-4) we take
$\g$ to be of radius $|\lambda|= c_{L}{\bar g}^{-(5/6-\eta/3)}$. Then for
$\bar g$ sufficiently small the hypothesis for \equ(5.58) of Lemma~5.16
is satisfied. Moreover then $\xi$ of Lemma~5.20 is sufficiently small and
and Lemma~5.20 holds. \equ(5.lem36-4) now follows from the extraction
estimate \equ(33.17f) and the Cauchy bound as before. \bull  

\vglue0.2cm

From the definition of $R_{n+1,4}$ in \equ(4.22) we have

$$R_{n+1,4} = \bigg(e^{-V_{n+1}} - e^{-\tilde{V}_{n,L}}\bigg) 
Q(C_{n+1},{\bf w}_{n+1},g_{n+1}) +  e^{-\tilde{V}_{n,L}} 
Q(C_{n+1},{\bf w}_{n+1},(g_{n+1}^{2}-g_{n,L}^{2})) 
$$ 

\\{\it Lemma 5.23 
 
$$ 
\Vert R_{n+1,4}  \Vert_{{\bf h},G_{\kappa},\AA,\d_{n+1}}  
\le C_L {\bar g}^{3/2}
\qquad\qquad\vert R_{n+1,4}\vert_{{\bf h}_{*},\AA,\d_{n+1}} 
\le C_L {\bar g}^{3} 
$$ 
} 
 
\vglue.2truecm 
 
\\{\it Proof} \textRed 
 
\\The proof is the same as that of Lemma~5.23 of [BMS] except that
we replace $\e$ by $\bar g$, insert lattice subscript $n$ as appropriate
and use the lattice counterparts (that we have already established) of
the lemmas exploited in [BMS] for the proof. \bull \textBlack

\vglue.3truecm 
 
\\{\it Lemma 5.24  
 
\\
Let $X_{\d_n}$ be a small set and let $J(X_{\d_n},\Phi)$ be normalized as in 
\equ({4.42}). Recall that the rescaled activity $J_{L}$ is defined by 
$J_{L}(L^{-1}X_{\d_{n+1}},\f,\psi)=J(X_{\d_n},\f_{L^{-1}},\psi_{L^{-1}})$
where $\f_{L^{-1}}=S_L\f$ and $\psi_{L^{-1}}=S_L\psi$.
Then we have  
 
\\1. For  $ 2p+m=2 $  so that  $(p,m)=(1,0),\> (0,2)$ 
$$\Vert D^{2p,m}J_{L}(L^{-1}X_{\d_{n+1}},0,0)\Vert   
\le O(1)L^{-(7-\e)/2}\Vert 
D^{2p,m}J(X_{\d_n},0,0)\Vert  \Eq(5.79)   $$ 
 
\\2. For $ 2p+m=4 $ so that $(p,m)=(2,0),(1,2),(0,4)$  
 
$$\Vert D^{2p,m}J_{L}(L^{-1}X_{\d_{n+1}},0,0)\Vert \le O(1)L^{-(4-\e)}\Vert 
D^{2p,m}J(X_{\d_n},0,0)\Vert 
\Eq(5.80)$$ 
}

\vglue.2truecm 
 
\\{\it Proof} :
\\In the proof of this lemma we will need to use the lattice Taylor expansion
introduced in \equ(5.001),\equ(5.002) and \equ(5.003) with a particular 
choice of 
a lattice path joining two points. The polymer $X_{\d_n}$ being a small set
is connected. It can be represented
as $X_{\d_n}=X\cap (\d_n\math{Z})^{3}$ where $X$ is a continuum connected 
polymer which is a small set. \textRed
By the argument in the proof of
Lemma~5.1 it suffices to consider the case when $X_{\d_n}$ is a block. Then
the lattice path lies entirely in $X_{\d_n}$. \textBlack

\\Let $u_{k}$ be a function defined on $(\tilde X_{\d_n}^{(2)})^{k}$ 
for $k\ge 2$. In the following $u$ is one of the test functions of 
subsection 2.2. Thus $u$  will represent either 
one of the functions $f_j$ defined on $\tilde X_{\d_n}^{(2)}$ 
giving a direction for a bosonic derivative or a function $g_{2p}$ 
defined on $(\tilde X_{\d_n}^{(2)})^{2p}= 
(\tilde X_{\d_n}^{(2)})^{p}\times (\tilde X_{\d_n}^{(2)})^{p}$ 
associated with a fermionic derivative of order $2p$. Note that $g_{2p}$
is restricted to be antisymmetric in the sense explained in the lines preceding
equation \equ(2.2). The $C^{2}(X_{\d_n}^{k})$ norms of these functions for
$k=1,\> 2p$ are 
defined as in \equ(2.111a) and \equ(2.112) of subsection 2.2. 

\\We recall the definition of the rescaled function  
 
$$S_{L}u_{k}(x)= u_{k,L^{-1}}(x)= L^{-kd_{s}}u_{k}({x\over L})$$              
 
\\Observe that

$$\Vert u_{k,L^{-1}}\Vert_{C^{2}(X_{\d_n}^{k})} \le L^{-kd_{s}} 
\Vert u_{k}\Vert_{C^{2}(L^{-1}X_{\d_{n+1}}^{k})} \Eq(8.303.5) $$

\\Let $e_1,e_2,.....,e_{3k}$ be the basis
vectors of $(\d_n\math{Z})^{3k}$. Let $x=\sum_{i=1}^{3k}(x,e_i)e_i $ 
denote a point in $X_{\d_n}^{k}$. Fix a point $x_0\in X_{\d_n}^{k}$. We write 
$x-x_0=\d_n\sum_{i=1}^{3k}h_i\e_i e_i$  where $h_i$ are non-negative 
integers and $\e_i={\rm sign}(x-x_0)_{i}$.

\\From \equ(5.003) we have 

$$u_{k,L^{-1}}(x) = u_{k,L^{-1}}(x_0)+  \sum_{i=1}^{3k} ((x-x_0),e_{i})
\dpr_{\d_n,\e_i e_i}u_{k,L^{-1}}(x_0) +
\d_n^{2}\sum_{i,j=1}^{3k}\sum_{s_i=0}^{h_i-1}
\sum_{s_j=0}^{h_j-1}$$
$$\dpr_{\d_n,\e_i e_i}\dpr_{\d_n,\e_j e_j} 
u_{k,L^{-1}}(x_0+ p_j(p_i(x-x_0,s_i),s_j)) $$

\\The argument of $u_{k,L^{-1}}$ in the last term lies entirely in 
$X_{\d_n}^{k}$ since $X_{\d_n}$ is a block.

\\Define 
$$ \d u_{k,L^{-1}}(x)=u_{k,L^{-1}}(x)- 
u_{k,L^{-1}}(x_0)= \d_n\sum_{j=1}^{3k}\sum_{s_j=0}^{h_j-1}
\dpr_{\d_n,\e_j e_j}u_{k,L^{-1}}(x_0+p_j(x-x_0,s_j)) \Eq(5.803) $$

\\and

$$\d^2u_{k,L^{-1}}( x)=h_{k,L^{-1}}(x)- u_{k,L^{-1}}(x_0)-\sum_{i=1}^{3k}
(( x- x_0),e_i)\dpr_{\d_n,\e_i e_i}u_{k,L^{-1}}(x_0)$$
$$=\d_n^{2}\sum_{i,j=1}^{3k}\sum_{s_i=0}^{h_i-1}
\sum_{s_j=0}^{h_j-1}\dpr_{\d_n,\e_i e_i}\dpr_{\d_n,\e_j e_j} 
u_{k,L^{-1}}(x_0+ p_j(p_i(x-x_0,s_i),s_j))\Eq(5.803.2)  $$

\\Now from \equ(5.803), \equ(5.803.2) we have 
using the definition of the rescaled function $u_{k,L^{-1}}$ 
$$\d u_{k,L^{-1}}(x)=L^{-(1+k{3-\e\over 4})} 
\d_n\sum_{j=1}^{3k}\sum_{s_j=0}^{h_j-1}
(\dpr_{\d_n,\e_j e_j}u_k)({L^{-1}}(x_0+p_j(x-x_0,s_j))) $$
 
\\and for $l\ge 1$ 
$${\dpr_{\d_n}}^{l}\d u_{k,L^{-1}}(x)=L^{-(l+k{3-\e\over 4})} 
({\dpr_{\d_n}}^{l}u_{k})(L^{-1}x)$$ 

\\where for $ {\dpr_{\d_n}}^{l}$ a multi-index convention is implicit. 
This implies that 
$$\Vert \d u_{k,L^{-1}}\Vert_{C^{2}(X_{\d_n}^{k})} 
\le c_{1} L^{-(1+k{3-\e\over 4})} 
\Vert u_k \Vert_{C^{2}(L^{-1}X_{\d_{n+1}} ^{k})} \Eq(8.303.3)$$ 
where $c_{1}=O(1)$ since $X$ is a small set. 
In the same way starting from \equ(5.803.2) 
a little bit of work shows that  
$$\Vert \d^{2}u_{k,L^{-1}}\Vert_{C^{2}(X_{\d_n}^{k})}
\le c_{2} L^{-(2+k{3-\e\over 4})} 
\Vert u_k\Vert_{C^{2}(L^{-1}X_{\d_{n+1}} ^{k})}   \Eq(8.303.4)    $$ 
where $c_{2}= O(1)$. 
 
\vskip0.2cm

\vskip0.2cm 
 
\\We will first prove the bounds of \equ(5.79).  
 
\\Consider first the case $(p,m)=(1,0)$. We have  
 
$$ D^{2,0}J_{L}(L^{-1}X_{\d_{n+1}},0,0;g_{2}) = 
D^{2,0}J(X_{\d_{n}},0,0;g_{2,L^{-1}}) 
= D^{2,0}J(X_{\d_{n}},0,0;\d^2 g_{2,L^{-1}}) \Eq(5.809.1)  $$ 
where we have Taylor expanded the function $ g_{2,L^{-1}}$ as in \equ(5.803.2) 
and then used the first and  second normalization conditions in \equ(4.42). 
Therefore  
 
$$\vert D^{2,0}J_{L}(L^{-1}X_{\d_{n+1}},0,0;g_2 )\vert 
\le \Vert D^{2,0}J(X_{\d_{n}},,0,0)\Vert \>  
\Vert \d^{2}g_{2,L^{-1}}\Vert_{C^{2}(X_{\d_{n}}^{2})}$$ 
$$\le O(1)L^{-(7-\e)/2}\Vert D^{2,0}J(X_{\d_{n}},0)\Vert\> 
\Vert g_2\Vert_{C^{2}(L^{-1}X_{\d_{n+1}}^{2})}          $$ 
where in the last step we have used the bound in \equ(8.303.4) for $k=2$ 
This proves the case $(p,m)=(1,0)$ of the lemma. 
 
\\To prove the case $(p,m)=(0,2)$ we Taylor expand the function $f_{L^{-1}}(x)$ 
to second order, then use the first and the third 
conditions in \equ(4.42) to get 
$$\eqalign{ D^{0,2}J(X_{\d_{n}},0,0;f_{L^{-1}}^{\times 2}) =  
D^{0,2}J(X_{\d_{n}},0,0;f_{1,L^{-1}}(x_0),\d^2 f_{2,L^{-1}}) 
+& D^{0,2}J(X_{\d_{n}},0,0;\d^2 f_{1,L^{-1}}, f_{2,L^{-1}}(x_0))\cr 
+& D^{0,2}J(X_{\d_{n}},0,0;\d f_{1,L^{-1}},\d f_{2,L^{-1}}) \cr }\Eq(5.809)   $$ 
 
\\Therefore 
 
$$\eqalign{ 
\vert D^{0,2}J(X_{\d_{n}},0,0;f_{L^{-1}}^{\times 2})\vert 
&\le \Vert D^{0,2}J(X_{\d_{n}},0,0)\Vert  
\Bigl (\Vert f_{1,L^{-1}}\Vert_{C^{2}(X_{\d_{n}})} 
\Vert \d^{2}f_{2,L^{-1}}\Vert_{C^{2}(X_{\d_{n}})} \cr   
&+\Vert f_{2,L^{-1}}\Vert_{C^{2}(X_{\d_{n}})}\Vert \d^{2}f_{1,L^{-1}}\Vert_{C^{2}(X_{\d_{n}})}\cr 
&+\Vert \d f_{1,L^{-1}}\Vert_{C^{2}(X_{\d_{n}})}\Vert \d f_{2,L^{-1}}\Vert_{C^{2}(X_{\d_{n}})}) 
\Bigr ) \cr 
&\le O(1)L^{-(7-\e)/2}\Vert D^{0,2}J(X_{\d_{n}},0,0)\Vert \prod_{j=1}^{2}
\Vert f_{j}\Vert_{C^{2}(L^{-1}X_{\d_{n+1}})} \cr } $$ 
 
\\where we have used the bounds \equ(8.303.5), 
\equ(8.303.3) and \equ(8.303.4) for the case 
$k=1$.  This proves the case $(p,m)=(0,2)$.
 
\\ Next we prove the bounds  \equ(5.80) . For this case  
$(p,m)=(2,0),(1,2),(0,4)$ so that $2p+m=4$. Taylor expand test functions 
around the fixed point $x_{0} \in X_{\d_{n}}$ to first order with remainder.  
We get for $(p,m)=(2,0)$ 
$$ D^{4,0}J_{L}(L^{-1}X_{\d_{n+1}},0;g_{4}) = D^{4,0}J(X_{\d_{n}},0;g_{4,L^{-1}}) 
= D^{4,0}J(X_{\d_{n}},0;\d g_{4,L^{-1}}) \Eq(5.810)  $$ 
where we have Taylor expanded the function $ g_{4,L^{-1}}$ as in \equ(5.803) 
and then used  \equ(4.401). 
 
\\Therefore exploiting the bound \equ(8.303.3) for $k=4$  we get 
$$\vert D^{4,0}J_{L}(L^{-1}X_{\d_{n+1}},0,0;g_{4} )\vert 
\le O(1)L^{-(4-\e)}\Vert D^{4,0}J(X_{\d_{n}},0)\Vert 
\Vert g_{4}\Vert_{C^{2}(L^{-1}X_{\d_{n+1}}^4)}  $$ 
which proves the case $(p,m)=(2,0)$. 

\\Next we turn to the case $(p,m)=(1,2)$. We have 
$$ D^{2,2}J_{L}(L^{-1}X_{\d_{n+1}},0;f^{\times 2},g_{2})=D^{2,2}J(X_{\d_{n}},0;f_{L^{-1}}^{\times 2},g_{2,L^{-1}})  
= D^{2,2}J(X_{\d_{n}},0;\d f_{L^{-1}}^{(1)},f_{L^{-1}}^{(2)},g_{2,L^{-1}})+$$ 
$$+ D^{2,2}J(X_{\d_{n}},0;f_{L^{-1}}^{(1)}(x_0),\d f_{L^{-1}}^{(2)},g_{2,L^{-1}})+ 
D^{2,2}J(X_{\d_{n}},0;f_{L^{-1}}^{(1)}(x_0),f_{L^{-1}}^{(2)}(x_0),\d g_{2,L^{-1}})\Eq(5.812)$$ 
where we have used the fourth condition in \equ(4.42). Therefore  
 
$$\vert D^{2,2}J_{L}(L^{-1}X_{\d_{n+1}},0,0;f^{\times 2},g_{2} )\vert 
\le \Vert D^{2,2}J(X_{\d_{n}},0,0) \Vert   
\Bigl (\Vert \d f^{(1)}_{L^{-1}} 
\Vert_{C^{2}(X_{\d_{n}})}\Vert f^{(2)}_{L^{-1}}\Vert_{C^{2}(X_{\d_{n}})} 
\Vert g_{2,L^{-1}}\Vert_{C^{2}(X_{\d_{n}}^{2})} +  $$ 
$$+\Vert \d f^{(2)}_{L^{-1}}\Vert_{C^{2}(X_{\d_{n}})}
\Vert f^{(1)}_{L^{-1}}\Vert_{C^{2}(X_{\d_{n}})} 
\Vert g_{2,L^{-1}}\Vert_{C^{2}(X_{\d_{n}}^{2})} + 
\Vert \d g_{2,L^{-1}}\Vert_{C^{2}(X_{\d_{n}}^{2})}  
\Vert f^{(1)}_{L^{-1}}\Vert_{C^{2}(X_{\d_{n}})} 
\Vert f^{(2)}_{L^{-1}}\Vert_{C^{2}(X_{\d_{n}})} \Bigr )  $$ 
 
Then using the bounds \equ(8.303.5), \equ(8.303.3) for $k=1,2$ and 
\equ(8.303.4) for $k=1$ we get 
  
$$\vert D^{2,2}J_{L}(L^{-1}X_{\d_{n+1}},0,0;f^{\times 2},g_{2} )\vert 
\le O(1)L^{-(4-\e)}\Vert D^{2,2}J(X_{\d_{n}},0)\Vert 
\prod_{j=1}^{2}\Vert f^{(j)}\Vert_{C^{2}(L^{-1}X_{\d_{n}})}
\Vert g_2\Vert_{C^{2}(L^{-1}X_{\d_{n+1}}^{2})}
$$ 

\\which proves the case $(p,m)=(1,2)$.

\\Finally we treat the case $(n,m)=(0,4)$. 
Let ${\cal N}_{4} =(1,2,3,4)$. Then  
using the fourth condition of \equ(4.42) we get

$$D^{0,4}J_{L}(L^{-1}X_{\d_{n+1}},0,0;f^{\times 4})= 
D^{0,4}J(X_{\d_{n}},0;f_{L^{-1}}^{\times 4})=  
\sum_{I\subset {\cal N}_{4}, |I|\not= 4 }  
D^{0,4}J(X_{\d_{n}},0;f_{L^{-1}}(x_0)^{\times |I|}, 
\d f_{L^{-1}}^{\times |I_{c}|})$$ 
where $f^{\times |I|}=(f_{j})|_{j\in I}$ . 
 
Therefore 
 
$$\vert D^{0,4}J_{L}(L^{-1}X_{\d_{n}},0,0;f^{\times 4} )\vert 
\le \Vert D^{0,4}J(X_{\d_{n}},0,0)\Vert 
\sum_{I\subset {\cal N}_{4}, |I|\not= 4 } 
\Vert f_{L^{-1}}\Vert_{C^{2}(X_{\d_{n}})}^{|I|} 
\Vert \d f_{L^{-1}}\Vert_{C^{2}(X_{\d_{n}})}^{|I_{c}|} $$ 
 
\\Because of the condition on $I$ in the above sum $|I_{c}|\ge 1$.
Therefore using the bounds \equ(8.303.5) and \equ(8.303.3) 
we get 
 
$$\vert D^{0,4}J_{L}(L^{-1}X_{\d_{n}},0,0;f^{\times 4} )\vert 
\le O(1) L^{-(4-\e)}\Vert D^{0,4}J(X_{\d_{n}},0)\Vert \prod_{j=1}^{4} 
\Vert f_j\Vert_{C^{2}(L^{-1}X_{\d_{n+1}})} $$ 

\\which proves the case $(p,m)=(0,4)$ and thus completes the proof of  
Lemma~5.24. \bull

\vglue.3truecm

\\{\it Corollary 5.25 
 
\\Let $Y_{\d_{n+1}}=L^{-1}X_{\d_{n+1}}$ where $X$ is a small set, 
$Z_{\d_{n+1}}=L^{-1}{{\bar X}_{\d_{n+1}}}^{L}$ 
and let $J(X_{\d_n},\Phi)$ be normalized as in \equ({4.42}). By definition
$J_{L}(Y_{\d_{n+1}},\Phi)=J(X_{\d_n},S_L\Phi)$. 
Then  
 
$$ 
|J_{L} (Y_{\d_{n+1}})|_{\bf h} 
\le O (1) L^{-(7-\e)/2}|J(X_{\d_n})|_{\bf \hat h} 
\Eq(5.lem39-1) 
$$ 
 
$$ 
\Vert  
J_{L} (Y_{\d_{n+1}}) e^{-\tilde{V}_{L} (Z_{\d_{n+1}}\setminus Y_{\d_{n+1}})} 
\Vert_{{\bf h},G_{\k}} \le O (1) L^{-(7-\e)/2} 
\bigg[ 
|J(X_{\d_n})|_{\bf \hat h} + \Vert J(X_{\d_n})\Vert_{{\bf \hat h},G_{3\k}}  
\bigg] 
\Eq(5.lem39-2) 
$$ 
 
$$ 
|J_{L}(Y_{\d_{n+1}})|_{\bf h_*} 
\le O (1) L^{-(7-\e)/2}|J(X_{\d_n})|_{\bf {\hat h*}} 
\Eq(5.lem39-3)  
$$ 
\textRed
\\where (see Lemma~5.15) $\hat{\bf h}=(\hat h_F, h_B)$,  
$\hat{\bf h}_* =(\hat h_F, h_{B*})$ and $\hat h_F = h_F/2$.
} \textBlack
 
\vglue.2truecm 
 
\\{\it Proof}

\\\equ({5.lem39-1}), \equ({5.lem39-3}) 
follow  easily from Lemma~5.24
taking advantage of the scaling present to shift 
from $\bf h$ to $\bf \hat h$. To see this observe that 
since $ h_{F}= 2 \hat h_{F}$ we have from the definition of the $\bf h$ norm
 
$$|J_{L} (Y_{\d_{n+1}})|_{\bf h}= \sum_{n=0}^{\infty}\sum_{m=0}^{m_0}  
{\hat h_{F}}^{2n} {h_{B}^{m}\over m !} 2^{2n} 
\Vert D^{2n,m}J_{L}(Y_{\d_{n+1}},0)\Vert  
\Eq(5.995)$$ 
 
\\Only terms with $2n+m$ even contribute. For $2n+m \le 4 $ use Lemma~5.24 and
observe that $2^{2n} \le O(1)$. For $2n+m \ge 6 $, we have for $L$ 
sufficiently large and $\e$ sufficiently small depending on $L$  
 
$$\eqalign{ 
2^{2n} \Vert D^{2n,m}J_{L} (Y_{\d_{n+1}},0)\Vert 
\le 2^{2n} L^{-(2n+m){(3-\e)\over 4}} 
\Vert D^{2n,m}J (X_{\d_n},0)\Vert \cr 
\le L^{-(2n+m){(3-{1\over 3} -\e)\over 4}}
\Vert D^{2n,m}J (X_{\d_n},0)\Vert \cr  
\le O(1) L^{-(7-\e)/2}\Vert D^{2n,m}J (X_{\d_n},0)\Vert 
}$$ 
 
\\Putting the two case together in \equ(5.995) gives \equ({5.lem39-1}). The  
proof of \equ({5.lem39-3}) is the same on replacing $h_B$ by $h_{B*}$.

\\For \equ({5.lem39-2}) we write 
 
$$ 
\Vert J_{L} (Y_{\d_{n+1}},{\bf \Phi}) 
e^{-\tilde{V}_{L} (Z_{\d_{n+1}}\setminus Y_{\d_{n+1}},{\bf \Phi})} 
\Vert_{\bf h} 
\le 
\Vert J_{L} (Y_{\d_{n+1}},{\bf \Phi}) \Vert_{\bf h} 
\Vert e^{-\tilde{V}_{L} (Z_{\d_{n+1}}\setminus Y_{\d_{n+1}},{\bf \Phi})}\Vert_{\bf h}\le$$ 
$$\le O (1)G_\k(Z_{\d_{n+1}},{\phi})\bigg[ 
|J_{L}(Y_{\d_{n+1}})|_{\bf h} +  
L^{-m_{0}d_s}\Vert J_{L} (Y_{\d_{n+1}})\Vert_{h_F,L^{[\phi]}h_B,G_{\kappa}} 
\bigg] 
$$ 
where we used Lemmas~5.5 and Lemma~5.15. By  
\equ({5.lem39-1}), and rewriting the second term by 
moving the scaling from $J$ to the norm,  
$$ 
\Vert J_{L} (Y_{\d_{n+1}},{\bf \Phi}) 
e^{-\tilde{V}_{L} (Z_{\d_{n+1}}\setminus Y_{\d_{n+1}},{\bf \Phi})} 
\Vert_{\bf h}\le 
O (1)G_\k(Z_{\d_{n+1}},\f)\bigg[  
L^{-(7-\e)/2}|J(X_{\d_{n}})|_{\bf \hat h} +  
L^{-m_{0}d_s}\Vert J(X_{\d_{n}})\Vert_{\bf \hat h ,G_{3\kappa}}   
\bigg] 
$$ 
 
\\Recall that $m_0=9$ and the scaling dimension $d_s=(3-\e)/4$.
\equ({5.lem39-2}) now follows by multiplying both sides by 
$G_\k^{-1}(Z_{\d_{n+1}},\f)$, and taking the supremum over $\f$. \bull 
\vglue.2truecm

\\{\it Lemma 5.26 
 
$$\Vert\tilde F_{R_n}e^{-\tilde V_n}\Vert 
_{{\bf h},G_{\k},\AA,\d_n}\le O(1)\e^{3/4-\eta} 
\Eq(5.71)$$ 
 
$$\vert\tilde F_{R_n}e^{-\tilde V_n}\vert 
_{{\bf {\hat h_{*}}},\AA,\d_n}\le O (1)\e^{11/4-\eta} 
\Eq(5.72)$$ 
 
\\and $J_n= R_n^{\sharp}-\tilde F_{R_n}e^{-\tilde{V_n}}$ satisfies on small 
sets the bounds}  

$$\Vert J_n\Vert_{{\bf{\hat h}},G_{3\k},\AA,\d_n}
\le O(1){\bar g}^{{3\over 4}-\eta}  
\qquad \vert J_n \vert_{{\bf{\hat h}},\AA,\d_n } 
\le O(1){\bar g}^{{3\over 4}-\eta} 
\qquad\qquad\vert J_n\vert_{{\bf{\hat h_{*}}},\AA,\d_n}  
\le O(1){\bar g}^{{11\over 4}-\eta} \Eq(5.71a)   $$

\vglue.3truecm 
\\{\it Proof} 
 
\\First we prove \equ({5.71}). $\tilde F_{R_n}$ is defined 
in \equ(4.57) and \equ({4.59}), and is supported on 
small sets. We estimate its $\bf h$ norm as in the proof of Lemma~5.19.
 
$$\eqalign{ \Vert {\tilde F}_{ R_n}(X_{\d_n},{\bf \Phi})\Vert_{\bf h} 
&\le \sum_{P} \vert {\tilde \a}_{n,P}(X_{\d_n})\vert 
\int_{X_{\d_n} } dx\ \Vert 
P ({\bf \Phi}(x),\dpr{\bf \Phi}(x))\Vert_{\bf h} \cr 
&\le C_L\sum_{P}\vert {\tilde\a}_{n,P}(X_{\d_n})\vert {\bar g}^{-1} 
\bigg(  
{\bar g}\int_{X_{\d_n}}dx\> (|\f|^{2})^2(x)
+{\bar g}^{1/2}\Vert\f\Vert^2_{X_{\d_n},1,\s}+1 
\bigg) \cr 
&\le C_L\sum_{P}\vert {\tilde\a}_{n,P}(X_{\d_n})\vert  
{\bar g}^{-1}G_\k(X_{\d_n},\f)e^{\g{\bar g}\int_{X_{\d_n}}dy(|\f|^2)^2(y)} 
\cr } 
$$ 

\\for any $\g =O(1)>0$. Hence, using Lemma 5.5 

$$\Vert\tilde F_{R_n}(X_{\d_n},{\bf \Phi})e^{-\tilde V_n(X_{\d_n},{\bf \Phi})}\Vert_{\bf h} 
\le\Vert\tilde F_{R_n}(X_{\d_n},{\bf \Phi})\Vert_{\bf h} 
\Vert e^{-\tilde V_n(X_{\d_n},{\bf \Phi})}\Vert_{\bf h}  
\le C_L\sum_{P}2^{|X_{\d_n}|}\vert{\tilde\a}_{n,P} 
(X_{\d_n})\vert {\bar g}^{-1}G_\k(X_{\d_n},\f)    $$ 

\\We thus obtain ( remembering that ${\tilde\a}_{n,P}$ are supported on small 
sets) on using \equ(5.67) of Lemma~5.17 for $\e$ sufficiently small depending
on $L$, implying $\bar g$ sufficiently small,  

$$\Vert\tilde F_{R_n}(X_{\d_n},{\bf \Phi})e^{-\tilde V_n(X_{\d_n},{\bf \Phi})} 
\Vert_{{\bf h},G_{\k},\AA,\d_n} 
\le C_L{\bar g}^{-1}\sum_{P}\Vert \tilde{\a}_{n,P}\Vert_{\AA,\d_n} 
\le O(1){\bar g}^{3/4 -\eta}$$ 

\\This proves \equ({5.71}). 
Now we turn to the proof of \equ({5.72}). 
As observed in the proof of Lemma~5.17, 
${\bf {\hat h_{*}}}^{n_P}\vert{\tilde\a}_{n,P}\vert_{\AA,\d_n} \le n(P)!\> 
\vert 1_{S}R_n^\sharp\vert_{\bf {\hat h_{*}},\AA,\d_n}  
\le O(1) {\bar g}^{11/4-\eta}$. 
We have from  the definition of $\tilde F_{R_n}$ given in \equ(4.57)  
$\vert \tilde F_{R_n}(X_{\d_n})\vert_{\bf {\hat h_{*}}} \le O(1)\sum_{P}  
\vert {\tilde\a}_{n,P}(X_{\d_n})\vert {\bf {\hat h_{*}}}^{n_P}$, whence 
$$\vert \tilde F_{R_n}\vert_{{\bf {\hat h_{*}}},\AA,\d_n} \le  
\sum_{P}\vert{\tilde\a}_{n,P}\vert_{\AA,\d_n} 
{\bf {\hat h_{*}}}^{n_P}\le O(1){\bar g}^{11/4-\eta}$$  
which proves \equ({5.72}).   
 
\\To get these bounds for $J_n= R_n^{\sharp}-\tilde F_{R_n}e^{-\tilde{V_n}}$ 
we use \equ({5.71}) and \equ({5.72}) to bound $\tilde 
F_{R_n}e^{-\tilde{V_n}}$ part. We can substitute ${\bf{\hat h}}$ for $\bf h$
in \equ({5.71}) since the ${\bf{\hat h}}$ norm is smaller than the $\bf h$
norm. We bound $R_n^{\sharp}$ by 
Lemma~5.17. We have also used the trivial bound  
$\vert J \vert_{{\bf{\hat h}},\AA,\d_n }
\le\Vert J\Vert_{{\bf{\hat h}},G_{3\k},\AA,\d_n}$ to obtain the second 
inequality for $J$ from the first.   \bull

\vglue.3truecm 
 
\\{\it Lemma 5.27} 
$$\Vert R_{n+1,\rm linear}\Vert 
_{{\bf h},G_{\k},\AA, \d_{n+1}}\le O(1)L^{- (1-\e)/2}{\bar g}^{3/4-\eta} 
\Eq(5.73)$$ 
 
$$\vert R_{n+1,\rm linear}\vert 
_{{\bf h}_{*},\AA, \d_{n+1}}\le O(1)L^{- (1-\e)/2}{\bar g}^{11/4-\eta} 
\Eq(5.74)$$ 
 
\\{\it Remark}: This is a crucial lemma which also figures as Lemma~5.27 in
[BMS]  and its proof is the same. For the readers benefit we give the details
below. The proof is based on the
principle that the contribution to the linearized part of the remainder
from large sets is very small. For small sets the expanding contributions 
have been subtracted out leading to normalized polymer activities and this
is sufficient to provide a contracting factor. 
 
\\{\it Proof} 
 
\\Let $R_{n+1,\rm linear}$, given in \equ(4.59b) is the sum of two
terms which represent contributions from small/large sets respectively. Let
$R_{n+1,\rm linear, s.s}$ denote the first term:

$$R_{n+1,\rm linear, s.s}(Z_{\d_{n+1}}) =\sum_{X_{\d_{n+1}}: {\rm small\ sets}
\atop L^{-1}{\bar X_{\d_{n+1}}}^{L}=Z_{\d_{n+1}}} 
e^{-\tilde{V}_{L} (Z_{\d_{n+1}}\setminus Y_{\d_{n+1}})}
J_{n,L} (Y_{\d_{n+1}}) \Eq(4.59b3) $$

\\where $Y_{\d_{n+1}}=L^{-1}X_{\d_{n+1}}$ and 
$J_n= R_n^{\sharp}-\tilde F_{R_n}e^{-\tilde{V_n}}$. By Corollary~5.25 we get 
 
$$ \Vert R_{n+1,\rm linear, s.s}(Z_{\d_{n+1}})\Vert_{{\bf h},G_{\k}}\le
O (1)L^{-(7-\e)/2}\sum_{X_{\d_{n+1}}: {\rm small\ sets}
\atop L^{-1}{\bar X_{\d_{n+1}}}^{L}=Z_{\d_{n+1}}}       
\bigg[ 
|J_n(X_{\d_n})|_{\hat{\bf h}} + 
\Vert J(X_{\d_{n}})\Vert_{\hat{\bf h},G_{3\k}}  \bigg] 
\Eq(4.59b1) 
$$ 
 
\\Note that $Z_{\d_{n+1}}$ fixes $Z_{\d_{n}}$ by restriction and the sum
on the right hand side is the same as the sum over $X_{\d_{n}}$ such that
$L^{-1}{\bar X_{\d_{n}}}^{L}=Z_{\d_{n}}$. We 
multiply both sides by $\AA(Z_{\d_{n+1}})$. On the right hand side
we have $\AA(Z_{\d_{n+1}})=\AA(Z_{\d_{n}})=\AA({\bar X_{\d_{n}}}^{L})
\le O(1)\AA(X_{\d_{n}})$ by \equ(2.5reg). We fix a unit block $\D_{n+1}$ 
and sum over $Z_{\d_{n+1}}\supset\D_{n+1}$. This fixes by restriction to the
over $Z_{\d_{n}}\supset\D_{n}$ on the right hand side. The argument on 
p.790 of [BDH-est] controls the constrained sum on $X_{\d_{n}}$ such that
$L^{-1}{\bar X_{\d_{n}}}^{L}=Z_{\d_{n}}\supset\D_{n} $ by $L^{3}$ times the 
sum over $X_{\d_{n}}\supset\D_{n}$. Taking then the supremum over the 
fixed unit block gives 

$$\eqalign{\Vert R_{n+1,\rm linear, s.s}\Vert_{{\bf h},G_{\k},\AA,\d_{n+1}}
&\le
O (1)L^{-(7-\e)/2}L^{3} \bigg[ 
|J_n|_{\hat{\bf h},\AA,\d_{n}} + 
\Vert J_n\Vert_{\hat{\bf h},G_{3\k},\AA,\d_{n}}  
\bigg]\cr
&\le O(1)L^{- (1-\e)/2}{\bar g}^{3/4-\eta}} 
\Eq(4.59b3) $$

\\where for the second inequality we used Lemma~5.26.

\\The second term in \equ(4.59b)for $R_{n+1,\rm linear}$ which  gets 
contributions only from large sets is
 
$$R_{n+1,\rm linear,l.s.}(Z_{\d_{n+1}}) =\sum_{X_{\d_{n+1}}: {\rm large\ sets}
\atop L^{-1}{\bar X_{\d_{n+1}}}^{L}=Z_{\d_{n+1}}} 
e^{-\tilde{V}_{L} (Z_{\d_{n+1}}\setminus Y_{\d_{n+1}})}
R_{n,L}^{\natural} (L^{-1}X_{\d_{n+1}}) \Eq(4.59b2) $$

\\where we have used $J_n=R_n^{\sharp}$ since the relevant part 
${\tilde F}_{R_n}$ is supported on small sets. We first bound
in  the ${\bf h}$ norm and observe that because of the rescaling
involved and Lemma~5.17
$$\Vert R_{n,L}^{\natural} (L^{-1}X_{\d_{n+1}})\Vert_{{\bf h}}
\le \Vert R_{n}^{\sharp} (X_{\d_{n}})\Vert_{\hat{\bf h},G_{3\k}}
G_{\k}(X_{\d_{n+1}}) 
\le O(1) 2^{|X_{\d_n}|} {\bar g}^{3/4-\eta} G_{\k}(Z_{\d_{n+1}}) $$

\\so that on using Lemma~5.5 for $e^{-\tilde{V}_{L}}$

$$\Vert R_{n+1,\rm linear,l.s.}(Z_{\d_{n+1}})\Vert_{{\bf h}, G_{\k}}
\le O(1){\bar g}^{3/4-\eta} \sum_{X_{\d_{n+1}}: {\rm large\ sets}
\atop L^{-1}{\bar X_{\d_{n+1}}}^{L}=Z_{\d_{n+1}}} 2^{2|X_{\d_{n+1}}|} $$ 

\\We estimate the $\AA$ norm as before except that for large sets we use
from \equ(2.6reg)

\\$\AA(L^{-1}{\bar X_{\d_{n+1}}}^{L})
\le c_{p}L^{-4}\AA_{-p}(X_{\d_{n+1}})  $ for any positive integer 
$p$ with $c_p =O(1)$. Choose $p=2$. Therefore

$$\Vert R_{n+1,\rm linear,l.s.}\Vert_{{\bf h}, G_{\k}, \AA, \d_{n+1}}
\le O(1) L^{-1}{\bar g}^{3/4-\eta} \Eq(4.59b4) $$

\\Adding the contributions \equ(4.59b3) and \equ(4.59b4) we get \equ(5.73).
\equ(5.74) can be proved in the same way.
For the small set contribution we use the kernel bounds in Corollary~5.25
and Lemma~5.26.  For the large set contribution we first use the rescaling
involved to shift on the right hand side the ${\bf h}_*$ norm to the 
$\hat{\bf h}_*$ norm followed by the kernel bound in Lemma~5.17.  \bull

\vglue.3truecm 
 
\\{\it Proof of Theorem~5.1 } 
 
\vglue.2truecm 
 
\\From \equ(4.221),  $R_{n+1}$ is the sum of $R_{n+1,\rm main}$,
$R_{n+1,\rm linear}$, $R_{n+1,3}$ and $R_{n+1,4}$. $R_{n+1,\rm main}$ 
satisfies the bound given in Lemmas~5.21. For $L$ large and 
$\e$ small depending on $L$ implying $\bar g$ sufficiently small
$C_L{\bar g}^{3/4}\le L^{-1/2}{\bar g}^{3/4 -\eta}$ with $\eta=1/64$. Similarly
$C_L{\bar g}^{3-3\d/2}\le L^{-1/2}{\bar g}^{11/4 -\eta}$ for $\d=\eta$.
Therefore
$|||R_{n+1,\rm main}|||_{n+1}\le $

\\$L^{-1/2}{\bar g}^{11/4 -\eta}$. Similarly from Lemmas~5.22 and 5.23 we 
get $ |||R_{n+1,j}|||_{n+1}\le L^{-1/2}{\bar g}^{11/4 -\eta}$, for
$j=3,4$. Adding these bounds to that provided for $R_{n+1,\rm main}$
by Lemma~5.27 we have that the
sum satisfies the bound \equ(55.7) for $L$ sufficiently large. The bounds
\equ(555.3), \equ(555.4) and \equ(55.5) have been proved in Lemmas~5.17 and 
5.18.  \bull

\vglue0.5truecm

\\{\bf 6.  EXISTENCE OF THE GLOBAL RENORMALIZATION GROUP TRAJECTORY
AND THE STABLE MANIFOLD } 

\numsec=6\numfor=1

\vskip.5cm
\textRed

This section is devoted to the proof of existence of the stable manifold
starting from the unit lattice. Namely, 
there exists an initial critical mass $\m_0$ which is a Lipshitz 
continuous function of the coupling constant $g_0$  such that RG trajectory 
is bounded uniformly on all scales. The proof is complicated because of
the presence of lattice artifacts which become inocuous if we advance
sufficiently on the RG trajectory. We therefore prove the result
by a combination of three theorems, namely 
Theorems~6.2, 6.4, and 6.6.  We first iterate the RG map a finite number
$n_0$ of times and then restart the trajectory. Theorem~6.2 says that
there exists a critical mass such that the RG trajectory is uniformly bounded 
at all scales $n\ge n_0\ge 0$. Theorem~6.4 says that if $n_0$ is 
sufficiently large then
the critical mass $\mu_{n_0}$ at this scale is a Lipshitz continuous functon
continuous function of the contracting variables. The stable ( critical)
manifold at scale $n_0$, appropriately interpreted for a sequence of 
non-autonomous maps, is constructed. Finally Theorem~6.6
says that there exists an initial critical mass $\mu_0$ which is a $C^1$
function of $g_0$ such that after $n_0$ applications of the RG map
we arrive at the critical mass $\mu_{n_0}$ of Theorem~6.4. Combining 
Theorem~6.4 with Theorem~6.6 proves the existence of the stable manifold
starting from the unit lattice. \textBlack   
One consequence is that the coupling constant $g_n$ is bounded away from
$0$ uniformly in $n$.

\vskip0.5truecm 

\\{\it 6.1}. Define $\tilde g_{n}= g_{n}-\bar g $ with $\bar g $ defined by 
\equ(55.4) and $\tilde{\bf w}_n= {\bf w}_n- {\bf w}_*$. Here ${\bf w}_*$ is
the function on $\cup_{n\ge 0}(\d_n\math{Z})^{3}$ defined in Lemma~5.9. 
According to Lemma~5.9, $\tilde{\bf w}_n \rightarrow 0$ geometrically fast
in ${\cal W}_l$ for all $l\ge 0$.

\\We will use as coordinates of the RG trajectory  

$$\y_n=(\tilde g_{n},\m_{n},R_{n},\tilde{\bf w }_n)\Eq(6.2) $$

\\The RG map 

$$\y_{n+1}=f_{n+1}(\y_n) \Eq(6.3)$$

\\can be written in components 
(see \equ(4.47A), \equ({4.222}), \equ(4.54) and Lemma~5.9)

$$\tilde g_{n+1}=f_{n+1,g}(\y_n)=\alpha(\epsilon)\tilde g_n +\tilde\xi_n(\y_n)
\Eq(6.4)$$

$$\m_{n+1}=f_{n+1,\m}(\y_n)=L^{3+\e\over 2}\m_n+\tilde\r_n(\y_n)\Eq(6.5)$$

$$R_{n+1}=f_{n+1,R}(\y_n)=: U_{n+1}(\y_n)\Eq(6.6)$$

$$\tilde{\bf w}_{n+1}= f_{n+1,w}(\y_n)= ({\bf v}_{n+1}-{\bf v}_{c,*})
 +\tilde{\bf w}_{n,L} \Eq(6.7)$$

\\with initial ${\bf w}_0=0$ and $R_0=0$. This implies that 
$\tilde{\bf w}_0=-{\bf w}_{*}$ and our initial condition is

$$\y_0=(\tilde g_{0},\m_{0}, 0, \tilde{\bf w}_0) \Eq(6.8)$$ 

\\$\a(\e),\> \xi_n,\> \r_n$ are defined by  

$$\alpha(\epsilon)=2 - L^{\e} = 1 - O (\log L)\e$$

\\for $\e$ sufficiently small depending on $L$, and

$$\tilde\xi_n(\y_n)=-L^{2\e}a_{c*}\tilde g_n^2 
-L^{2\e}(a_n-a_c*)(\tilde g_n +\bar g)^{2}
+\xi_{n}(\y_n)\Eq(6.9)$$

$$\tilde \r_n(\y_n)=-L^{2\e}b_n(\bar g+\tilde g_n)^2 
+\r_{n}(\y_n)\Eq(6.10)$$

\\Let $E_n$ be the Banach space consisting of elements $\y_n$ with the norm
$$\Vert \y_n\Vert_{n} =\max((\n{\bar g})^{-1}|\tilde g_n|,\> 
{\bar g}^{-(2-\d)}|\m_n|,\> {\bar g}^{-(11/4-\eta)}|||R_n|||_{n}, 
{\tilde c}_L^{-1}\Vert \tilde{\bf w}_n\Vert_{n}) \Eq(6.11)$$

\\where the norm $|||R|||_{\d_n}$ of $R_n$ is as defined in \equ(55.03). $\n$
is the $O(1)$ constant which figures in the specification of the domain 
${\cal D}_n$, \equ(55.1). We have $0<\n<1$. We will take $\n>0$ sufficiently 
small depending on $L$ and this will be specified in the proofs of 
Lemmas~6.3 and 6.4 below. The norm $\Vert \tilde{\bf w}_n\Vert_{\d_n}$ and the
constant $\tilde c_L$ are as specified in Lemma~5.9.

\vglue0.3cm
\\{\it 6.2. Domains and bounds}

\vglue0.2cm

Let $E_n(r)\subset E_n$ be the open ball of radius $r$,
centered at the origin:

$$E_n(r)=\{\y_n\in E_n: \Vert \y_n\Vert_n < r\}\Eq(6.13)$$

\\Let ${\cal D}_n$ be the domain of $(g_n,\m_n,R_n)$ defined in
\equ(55.1) and \equ(55.02).

$$\y_n\in E_n(1)\Rightarrow (g_n,\m_n,R_n)\in {\cal D}_n\Eq(6.14)$$

\\and then Theorem 5.1 holds.

\\Let $\y_n\in E_n(1)$. Let $\e>0$ be sufficiently small (depending on $L$).
Then from Theorem~5.1, Lemma~5.12 , \equ(6.9) and \equ(6.10) we get the bounds
   
$$\eqalign{
|\tilde\xi_{n}(\y_n)|\le &  C_L\Bigl((\n^2 + L^{-nq}){\bar g}^{2} 
+{\bar g}^{11/4-\eta}\Bigr) \cr
|\tilde\r_{n}(\y_n)|\le & C_L {\bar g}^{2}\cr
|||U_{n+1}(\y_n)|||_{n+1}\le &L^{-1/4}{\bar g}^{11/4-\eta} \cr}\Eq(6.15)$$

\\We have the following Lipshitz bounds :

\vglue.3truecm
\\{\it Lemma 6.1}

\\Let $\y_n,\y_n'\in E_n(1/4)$. Then we have:

$$\eqalign{
{\rm (i) }&\quad 
|\tilde\xi_n(\y_n)-\tilde\xi_n(\y_n')|\le C_L\Bigl((\n^2 + L^{-nq}){\bar g}^{2} 
+{\bar g}^{11/4-\eta}\Bigr) \Vert \y_n-\y_n'\Vert_n \cr
{\rm (ii) }&\quad
|\tilde\r_n(\y_n)-\tilde\r_n(\y_n')|\le 
C_L {\bar g}^{2}\Vert \y_n-\y_n'\Vert_n \cr
{\rm (iii) }&\quad
|||U_{n+1}(\y_n)-U_{n+1}(\y_n')|||_{n+1}
\le O(1)L^{-1/4}{\bar g}^{11/4-\eta}\Vert \y_n-\y_n'\Vert_n \cr
 {\rm (iv) } &\quad
{\tilde c}_L^{-1}
\Vert f_{n+1,\bf w}(\y_n)- f_{n+1,\bf w}(\y_n')\Vert_{n+1}
\le L^{-1/5}\Vert \y_n-\y_n'\Vert_n  \cr}
\Eq(6.17)$$

\\{\it Proof}:

\\$\tilde\xi_n,\tilde\r_n,U_{n+1}$ are (norm) analytic functions in 
${\cal D}_n$
and thus in $E_n(1)$. The analyticity follows from the algebraic 
operations in Section 4, the norm analyticity of the reblocking map
together with the norm analyticity of the extraction map 
(Theorem 5 [BDH-est]). Therefore we can use Cauchy estimates exactly as in 
in the proof of Lemma~6.1 of [BMS] together with
the bounds \equ(6.15) to get (i), (ii) and (iii). To get (iv) note that 
from \equ(6.7) and the definition of the norms in \equ(5.20) we have

$$\eqalign{\Vert f_{n+1,\bf w}(\y_n)- f_{n+1,\bf w}(\y_n')\Vert_{n+1}
&\le L^{2d_s} \max_{1\le p\le 3}
\sup_{x\in(\d_{n+1}{\bf Z})^{3}}  
\left((|x|+\d_{n+1}) ^{6p+1\over 4}|{\tilde w}_{n}^{(p)}(Lx)-
{\tilde w}^{\prime (p)}_{n}(Lx)|
\right) \cr
&\le L^{2d_s} \max_{1\le p\le 3}
\sup_{y\in(\d_{n}{\bf Z})^{3}}  
\left(({|y|\over L}+\d_{n+1}) ^{6p+1\over 4}
|\tilde w_{n}^{(p)}(y)-\tilde w_{n}^{\prime (p)}(y)|
\right) \cr
&\le L^{2d_s -7/4} \max_{1\le p\le 3}
\sup_{y\in(\d_{n}{\bf Z})^{3}}  
\left((|y|+\d_{n}) ^{6p+1\over 4}
|\tilde w_{n}^{(p)}(y)-\tilde w_{n}^{\prime (p)}(y)|
\right) \cr
&\le L^{-1/5}\Vert \tilde{\bf w}_n-\tilde{\bf w}_{n}^{\prime}\Vert_{n}\cr }
$$

\\where we have used $d_s={3-\e\over 4}$ and then $L$ large and $\e$ 
sufficiently small. Now dividing both sides by $\tilde c_L$ we get (iv). 
$\bull$

\vglue0.2cm
\\{\it 6.3. Existence of the global RG trajectory}

\vglue0.1cm

Let ${\cal D}_{n}$ be the domain specified by \equ(55.1),\equ(55.02),
and \equ(55.03). 
Let $(\tilde g_0,\m_0, 0)$ belong to 
$\tilde{\cal D}_{0}\subset {\cal D}_{0}$ 
where $\tilde{\cal D}_{0}$ is specified by
 
$$\vert \tilde g_0 \vert < 2^{-(n_0 +5)}\n {\bar g},
\ \ \vert\mu_0\vert < 2^{-(n_0 +5)}L^{-{3+e\over 2}n_0} {\bar g}^{2-\delta}$$

\\Let $n_0$ be a positive integer. By iterating the RG map $n_0$ times
using Theorem~5.1 and the flow equation \equ(4.47A) recursively we obtain for 
$\e$ sufficiently small depending on $L$ and $n_0$, 
$(g_{n_0},\m_{n_0},R_{n_0}) \in {\cal D}_{n_0}(1/32)$ where 
${\cal D}_{n_0}(1/32)$ is specified by

$$\vert \tilde g_{n_0}\vert < {1\over 32}\n {\bar g},
\ \ \vert \m_{n_0}\vert <  {1\over 32} {\bar g}^{2-\delta}  $$ 

$$||| R_{n_0} |||_{n_0}<{1\over 32}{\bar g}^{11/4-\eta}  $$

\\We will now prove the existence of a global solution to
the discrete flow map \equ(6.3):

$$\y_{n+1}=f_{n+1}(\y_{n}),\> \forall \> n\ge n_0    $$
with initial condition

$$\y_{n_0}=(\tilde g_{n_0},\m_{n_0}, \tilde{\bf w}_{n_0},R_{n_0} )$$
 
\\in a bounded domain. 
We will say that $\{\y_{n}:\> \y_{n+1}=f_{n+1}(\y_{n}),\> n\ge n_0\}$ is 
the RG trajectory {\it restarted} at scale $n_0$.  
To this end we consider the Banach space ${\bf E}_{n_0}$ of 
sequences ${\bf s}_{n_0}=\{\y_n\}_{n\ge n_{0}} $, each $\y_n \in E_n$ ,
with the norm

$$\Vert {\bf s}_{n_0}\Vert = \sup_{n\ge n_{0}}\Vert \y_n \Vert_{n} \Eq(6.111)$$

\\and the open ball ${\bf E}_{n_0}(r)\subset \bf E$

$${\bf E}_{n_0}(r)=\{{\bf s}_{n_0} : \Vert{\bf s}_{n_0}\Vert < r \} 
\Eq(6.131)$$

\\We will derive on the space of sequences ${\bf E}_{n_0} $ an equation that a 
global RG trajectory must solve and then prove for  
$\y_{n_0}\in E_{n_0}(1/32)$ 
the existence of a unique solution in the ball 
${\bf E}_{n_0}(1/4)$, 
for a suitable choice of $r$, by the contraction mapping principle. 
This adapts a standard method from the theory of hyperbolic dynamical systems
in Banach spaces due to Irwin in [I]. Irwin's analysis is explained by Shub
in Appendix 2, Chapter 5 of [S]. For earlier applications see 
section 5 of [BDH-eps] and section 6 of [BMS].

\vglue.3truecm
\\{\it Theorem 6.2

\\Let $L$ be large, $\n$ be sufficiently small depending on $L$, then
$\e$ sufficiently small depending on $L$. Let 
$\y_{n_0}\in E_{n_0}(1/32)$ for any integer $n_{0}\ge 0$. Let 
$(\tilde g_{n_0}, R_{n_0}, \tilde{\bf w}_{n_0})$ be held fixed. 
Then there is a  
$\m_{n_0}$ such that there exists a sequence 
${\bf s}_{n_0}=\{\y_n\}_{n\ge n_0} $ in ${\bf E}_{n_0}(1/4)$  satisfying  
$\y_{n+1}=f_{n+1}(\y_{n})$ for all $n\ge n_0$.} 

\vglue.2cm

\\{\it Remark}: The $\m_{n_0}$ of the theorem is called a {\it critical mass}.
We write $\m_{n_0}=\m_{n_0,c}$
\vglue.2truecm
 
\\{\it Proof}

\\Our initial data will be at scale $n_0$. Let $n_0\le n\le N-1$. We iterate 
the map \equ(6.4) forwards $N$ times. We iterate the 
map \equ(6.5) backwards $N-n_0$ times starting from a given $\m_{N}$. 
We then easily derive

$$\tilde g_{n+1}=\alpha(\epsilon)^{n+1-n_0}\tilde g_{n_0} +
\sum_{j=n_0}^{n}\alpha(\epsilon)^{n-j}\tilde\xi_j(\y_j),
\quad n_0\le n\le N-1$$

$$\m_{n+1}=L^{-{3+\e\over 2}(N-n-1)}\m_{N}-
\sum_{j=n+1}^{N-1}L^{-{3+\e\over 2}(j-n)}\tilde\r_j(\y_j),
\quad n_0-1\le n\le N-2 $$

\\Let us fix $\m_{N}=\m_f$ and take $N\rightarrow\io$. In other words we
assume the $\m_n$ flow is bounded and then must show that such a flow exists.
We have

$$\tilde g_{n+1}= \alpha(\epsilon)^{n+1-n_0}\tilde g_{n_0} +
\sum_{j=n_0}^{n} \alpha(\epsilon)^{n-j}\tilde\xi_j(\y_j),\quad n\ge n_0
\Eq(6.18)$$

$$\m_{n+1}=-
\sum_{j=n+1}^{\io}L^{-{3+\e\over 2}(j-n)}\tilde\r_j(\y_j),
\quad n\ge n_0-1\Eq(6.19)$$

\\together with

$$R_{n+1}=U_{n+1}(\y_{n}), \quad n\ge n_0\Eq(6.20)$$

$$\tilde{\bf w}_{n+1}= ({\bf v}_{n+1}-{\bf v}_{c*}) 
+\tilde{\bf w}_{n,L} \Eq(6.201) $$ 

\textRed
\\The flow $\tilde{\bf w}_n$ is independent of that of $g_n, \m_n, R_n$,
is solved by \equ(5.19) and satisfies the bounds of Lemma~5.9. \textBlack
This solution can be incorporated
in the $\y_n$ and $\tilde{\bf w}_n$ is then no longer a flow variable.

\\For $\e$ sufficiently small (depending on $L$)

$$0< \alpha(\epsilon) <1 \Eq(6.21)$$

\\Note that ${\bf s}_{n_0}\in {\bf E}_{n_0}(1/4)$ implies  $\y_j\in E_j(1/4)$ 
for all $j\ge n_0$.
Then the infinite sum of \equ(6.19)
converges by \equ(6.21) and \equ(6.15). So $\m_{n_0}$ has now
been determined provided \equ(6.18)-\equ(6.20) has a solution in the afore
mentioned ball. It is easy to verify that any solution of \equ(6.18)-
\equ(6.19), together with the $\tilde{\bf w}$ flow, is a solution 
of the  RG flow $\y_{n+1}=f_{n+1}(\y_{n})$ for $n\ge n_0$.

\vglue0.2cm

\\We write \equ(6.18)-\equ(6.20) in the form

$$\y_{n+1}=F_{n+1}({\bf s}_{n_0}), \quad n\ge n_0 \Eq(6.22)$$

\\where ${\bf s}_{n_0}=(\y_{n_0}, \y_{n_0+1}, \y_{n_0+2},...)$ and 
$F_{n+1}$ has components
$(F_{n+1}^{(g)},F_{n+1}^{(\m)},F_{n+1}^{(R)})$ given by the r.h.s. of
\equ(6.18), \equ(6.19), \equ(6.20) respectively.

\\If we write

$${\bf F}({\bf s}_{n_0})=
(F_{n_0}({\bf s}_{n_0}), F_{n_0+1}({\bf s}_{n_0}),...)$$

\\then \equ(6.22) can be  written as a fixed point equation

$${\bf s}_{n_0}={\bf F}({\bf s}_{n_0})\Eq(6.23)$$

\\We seek a solution of \equ(6.23) in the open ball ${\bf E}_{n_0}(1/4)$ with
initial data $\y_{n_0}=(\tilde g_{n_0},\m_{n_0},R_{n_0},\tilde{\bf w}_{n_0})$ 
in $E_{n_0}(1/32)$ with $(\tilde g_{n_0},R_{n_0},\tilde{\bf w}_{n_0})$ held 
fixed. The existence of a unique solution follows by the standard
contraction mapping principle and the next Lemma. \bull 

\vglue.3truecm
\\{\it Lemma 6.3}

$${\bf s}_{n_0}\in {\bf E}_{n_0}(1/32)\Rightarrow{\bf F}({\bf s}_{n_0})
\in {\bf E}_{n_0}(1/16)\Eq(6.26)$$
Moreover, for ${\bf s}_{n_0},{\bf s}_{n_0}'\in {\bf E}_{n_0}(1/4)$
$$\Vert{\bf F}({\bf s}_{n_0})-{\bf F}({\bf s}_{n_0}')\Vert\le
{1\over 2}\Vert {\bf s}_{n_0}-{\bf s}_{n_0}'\Vert\Eq(6.27)$$

\vglue.3truecm

\vglue.3truecm
\\{\it Proof}

\\First we prove \equ(6.26), and thus take 
${\bf s}_{n_0}\in {\bf E}_{n_0}(1/32)$.
Then $\y_n\in E_n(1/32)$ for every $n\ge n_0$ and we can use the estimates 
\equ(6.15). From \equ(6.18) and the estimates in \equ(6.15) we have

$$(\n{\bar g})^{-1}|F_{n+1}^{(g)}({\bf s}_{n_0})| < 
\alpha(\epsilon){1\over 32}+ (\n{\bar g})^{-1}
\sum_{j=n_0}^{n} \alpha(\epsilon)^{n-j} C_L\Bigl((\n^2 + L^{-jq}){\bar g}^{2} 
+{\bar g}^{11/4-\eta}\Bigr) $$
$$< {1\over 32}+C_L {{\bar g}^{7/4-\eta}\over \n(1- \alpha(\epsilon)})+
C_L{{\bar g}\n\over 1- \alpha(\epsilon)}
+C_L {{\bar g}\over \n(1-\alpha(\e)^{-1}L^{-q})}$$
$$< {1\over 32}+{C_L \over \n\log L}\e^{3/4-\eta}+ {C_L \over \log L}\n + 
{C_L \over \n(1-L^{-q})}\e < 
{1\over 16}$$

\\for  $L$ sufficiently large, $\n$ sufficiently small depending on $L$ so that
$ {C_L\over \log L}\n \le 1/96$ and then $\e$ sufficiently small depending on 
$L$ so that $\bar g \le C_L\e$ is sufficiently small,   
${C_L \over \n(1-L^{-q})}\e \le 1/96$ and 
${C_L \over \n\log L}\e^{1/4-\eta} \le 1/96$

\\Similarly from \equ(6.19) and \equ(6.15) we have
$${\bar g}^{-(2-\d)}|F_{n+1}^{(\m)}({\bf s}_{n_0})|\le c_L{\bar g}^{\d} 
\sum_{j=n+1}^{\io} L^{-{3+\e\over 2}(j-n)}
\le L^{-{3+\e\over 2}}(1-L^{-{3+\e\over 2}})^{-1}< 
{1\over 16}$$
since $\d=1/64$, $L$ sufficiently large, and $\e$ sufficiently small 
depending on $L$.

\\Finally from \equ(6.20) and \equ(6.15)

$${\bar g}^{-(11/4-\eta)}|||F_{n+1}^{(R)}({\bf s}_{n_0})|||_{n+1}
\le L^{-1/4}< {1\over 16}$$

\\for $L$ sufficiently large. This proves \equ(6.26).

\\To prove \equ(6.27), take ${\bf s}_{n_0},{\bf s}_{n_0}'\in {\bf E}_{n_0}(1/4)$.
This implies that $\y_n, \y'_{n}\in E_n(1/4)$ for every $n\ge n_0$ and we can
use the Lipshitz estimates of lemma 6.1. 
Note that the initial coupling $g_{n_0}$ is held fixed. Then we have

$$(\n{\bar g})^{-1}|F_{n+1}^{(g)}({\bf s}_{n_0})-F_{n+1}^{(g)}
({\bf s}_{n_0}')|\le
\sum_{j=n_0}^{n} \alpha(\epsilon)^{n-j}(\n{\bar g})^{-1}
|\tilde \xi_j(\y_j)-\tilde\xi_j(\y'_j)|$$
$$\le (\n{\bar g})^{-1}\sum_{j=n_0}^{n} \alpha(\epsilon)^{n-j}
C_L\Bigl((\n^2 + L^{-jq}){\bar g}^{2} 
+{\bar g}^{11/4-\eta}\Bigr) \Vert {\bf s}_{n_0}-{\bf s}_{n_0}'\Vert $$ 
$$\le \Bigl({C_L \over \n\log L}\e^{1/4-\eta}+  
C_L{{\bar g}\n\over 1- \alpha(\epsilon)}
+C_L {{\bar g}\over \n(1-\alpha(\e)^{-1}L^{-q}})\Bigr) 
\Vert {\bf s}_{n_0}-{\bf s}_{n_0}'\Vert
\le
{1\over 2}\Vert {\bf s}_{n_0}-{\bf s}_{n_0}'\Vert$$

\\by estimating as above in the bound for $F_{n+1}^{(g)}({\bf s}_{n_0})$
with $L$ sufficiently large, $\n$ sufficiently small
depending on $L$ and $\e$ sufficiently small depending on $L$. Similarly,

$${\bar g}^{-(2-\d)}|F_{n+1}^{(\m)}({\bf s}_{n_0})-F_{n+1}^{(\m)}({\bf s}_{n_0}')|\le 
\sum_{j=n+1}^{\io}
L^{-{3+\e\over 2}(j-n)}{\bar g}^{-(2-\d)}|\tilde \r_j(\y_j)-\tilde\r_j(\y'_j)| $$
$$\le L^{-{3+\e\over 2}}(1-L^{-{3+\e\over 2}})^{-1} 
C_L{\bar g}^{\d}\Vert {\bf s}_{n_0}-{\bf s}_{n_0}'\Vert\le
{1\over 2}\Vert {\bf s}_{n_0}-{\bf s}_{n_0}'\Vert$$

\\for $L$ sufficiently large and $\e$ sufficiently small depending on $L$.  
Finally

$${\bar g}^{-(11/4-\eta)}|||F_{n+1}^{(R)}({\bf s}_{n_0})-F_{n+1}^{(R)}({\bf s}_{n_0}')|||_{n+1} =
{\bar g}^{-(11/4-\eta)}|||U_{n+1}(\y_{n})-U_{n+1}(\y'_{n})|||_{n+1} $$
$$\le O(1)L^{-1/4}\Vert {\bf s}_{n_0}-{\bf s}_{n_0}'\Vert
\le {1\over 2}\Vert {\bf s}_{n_0}-{\bf s}_{n_0}'\Vert$$

\\for $L$ sufficiently large. Thus \equ(6.27) has been proved. 
This completes the proof of Theorem~6.2.   $\bull$

\vglue0.3cm

\\{\it 6.4}. Theorem~6.2 says that 
if $\y_{n_0}=(\tilde g_{n_0}, \m_{n_0}, R_{n_0},\tilde{\bf w}_{n_0})
\in E_{n_0}(1/32)$ for any $n_0\ge 0$
then there is a critical mass $\m_{n_0}=\m_{n_0,c} $ 
such that a uniformly bounded RG trajectory exists.
The Theorem~6.4 below proves the uniqueness of $\m_{n_0,c}$ for $n_0$ 
sufficiently large : $\m_{n_0,c}$ is a Lipshitz continuous function of 
$(\tilde g_{n_0},R_{n_0},\tilde{\bf w}_{n_0})$. \textRed In Theorem~6.6 below
we prove that given $\m_{n_0}$ as above there is a $\m_0$ given by a
$C^{1}$ function of $\tilde g_0$ such that after $n_0$ applications of
the RG map we arrive at $\m_{n_0}$. \textBlack

\vglue0.3cm

\\To this end we represent the Banach space $E_n$ as a 
product of two Banach spaces $E_n= E_{n,1}\times E_{n,2}$. We write 
$\y_n\in E_n$ as $\y_n=(\y_{n,1},\y_{n,2})$ where 
$\y_{n,1}=(\tilde g_n, R_n,\tilde{\bf w}_n )$ and 
$\y_{n,2}=\m_n$. $\y_{n,2}$ is the expanding (relevant)variable.
Let $p_i,\ i=1,2$, denote the projector onto $E_{n,i}$
and $f_{n,i}=p_i\circ f_n$.  The norm $\Vert\cdot\Vert_n$ on $E_n$ being a 
box norm we have

$$\Vert \y_n\Vert_n =\max(\Vert \y_{n,1}\Vert_n,\Vert \y_{n,2}\Vert_n)$$

\\where $\Vert \y_{n,2}\Vert_n={\bar g}^{-(2-\d)}|\m_n|$ and 
$\Vert \y_{n,1}\Vert_n=\max((\n{\bar g})^{-1}|\tilde g_n|,\> 
{\bar g}^{-(11/4-\eta)}|||R_n|||_{\d_n}, 
{\tilde c}_L^{-1}\Vert \tilde{\bf w}_n\Vert_{\d_n})$.

\vglue0.2cm
\\In the following we continue to assume
that $L$ is sufficiently large, followed by $\n$ sufficiently small depending
on $L$, then $\e$ sufficiently small depending on $L$. The last condition also
implies that ${\bar g} \le C_L \e$.  

\vglue0.3cm

\\{\it Theorem~6.4 : Let
${\bf s}_{n_0}= \{\y_n : \y_{n+1}=f_{n+1}(\y_n)\}_{n\ge n_0}\in {\bf E}(1/4)$ 
be the global RG trajectory of Theorem~6.2. Then for $n_0$ sufficiently large
there exists a Lipshitz continuous function 
$h\>:\> E_{n_0,1}\rightarrow \math{R}$   
with Lipshitz constant $1$  such that the stable manifold of the sequence
of maps $\{f_n\}_{n\ge n_0+1}$, 
$W_{n_0}^{s}=\{\y_{n_0}\in E_{n_0}(1/32) :\>{\bf s}_{n_0} \in {\bf E}(1/4)\}$
is the graph 

$$W_{n_0}^{s}= \{\y_{n_0,1}, h(\y_{n_0,1}\} $$

}

\vglue0.2cm

\\We will prove the theorem following the analysis of Shub in 
[S, Section~5]. The Schub analysis has been employed earlier in the context
of continuum models, (see [Section~5.3, BDH-eps] and [Section~6, BMS]). 
\textRed  Here
we have to take account of additional features stemming from the lattice 
which results in $n_0$ having to be taken sufficiently large (sufficiently
fine lattice) for the argument to work. This will be clear from
the proof of the following lemma from which Theorem~6.4 follows. \textBlack 

\vglue.3truecm
\\{\it Lemma 6.5}

\\Let $\y_n,\y_n'\in E_n(1/4)$. Then for $n\ge n_0$, $n_0$ sufficiently large
\textRed depending on $\n$ and $L$ \textBlack 

$$\Vert f_{n+1,1}(\y_n)-f_{n+1,1}(\y_n')\Vert_{n+1}
\le (1-\e)\Vert \y_n-\y_n'\Vert_{n} \Eq(6.30)$$

\\and, if 
$\Vert \y_{n,2}-\y_{n,2}'\Vert_{n}\ge \Vert \y_{n,1}-\y_{n,1}'\Vert_{n}$
then

$$\Vert f_{n+1,2}(\y_n)-f_{n+1,2}(\y_n')\Vert_{n+1}
\ge (1+\e)\Vert \y_n-\y_n'\Vert_{n}\Eq(6.31)$$

\vglue.3truecm
\\{\it Proof}

\\First we prove 
\equ(6.30). $f_{n+1,1}$ has components $f_{n+1,g}$, $f_{n+1,R}$ and
$f_{n+1,w}$. From \equ(6.4)

$$f_{n+1,g}(\y_n)= \alpha(\epsilon) \tilde g_n +\tilde\xi_n(\y_n)$$

\\Since $\y_n,\y_n'\in E_n(1/4)$ we can use lemma 6.1. Therefore for $n\ge n_0$

$$\eqalign{(\n{\bar g})^{-1}|f_{n+1,g}(\y_n)-f_{n+1,g}(\y_n')|
&\le \alpha(\epsilon) \Vert \y_n-\y_n'\Vert_{n} +
(\n{\bar g})^{-1}|\tilde\xi_n(\y_n)-\tilde\xi_n(\y_n')|\cr 
&\le \Bigl(1-\e\log(L) +C_L\e(\n + {1\over \n}L^{-n_{0}q}+ 
{1\over \n} \e^{3/4-\eta}\Bigr)\Vert \y_n-\y_n'\Vert_{n}\cr} $$

\\Let $L$ be large. Let $\n$ be sufficiently small and \textRed {\it
$n_0$ sufficiently large
so that $C_L L^{-n_{0}q/2}\le \n\le {1\over C_L}$ } \textBlack
and 
$\e$ sufficiently small  so that $\n\ge C_L\e^{3/8 -\eta}$. 
Then we have

$$\eqalign{1-\e\log(L) +\e C_L(\n + {1\over \n}L^{-n_0 q}+ 
{1\over \n} \e^{3/4-\eta}) &\le 1+\e(-\log(L) +1+ L^{-qn_0 /2} + \e^{3/8})\cr
&\le 1+\e(-\log(L) +3) \le 1-\e \cr } $$

\\Therefore

$$(\n{\bar g})^{-1}|f_{n+1,g}(\y_n)-f_{n+1,g}(\y_n')|
\le (1-\e)\Vert \y_n-\y_n'\Vert_{n}$$

\\Since $f_{n+1,R}(\y_n)=U_{n+1}(\y_n)$, we have from lemma 6.1 
for $L$ sufficiently large

$${\bar g}^{-(11/4-\eta)}|||f_{n+1,R}(\y_n)-f_{n+1,R}(\y_n')|||_{n+1}
\le(1-\e)\Vert \y_n-\y_n'\Vert_{n}$$

\\as well as

$$\tilde c_L^{-1}
\Vert f_{n+1,\bf w}(\y_n)- f_{n+1,\bf w}(\y_n')\Vert_{n+1}\le (1-\e)
\Vert \y_n-\y_n'\Vert_{n}$$

\\These three inequalities prove \equ(6.30). 

\vglue 0.1cm  \textRed

\\{\it Remark: $n_0$ had to be chosen sufficiently large because the
$\tilde g_n$ flow coefficient $a_n$, see \equ(6.8), depends on the lattice 
scale $n$. $a_n$ converges geometrically (Lemma~5.12)
to a constant $a_{c*}$. As a result we have to wait sufficiently long before 
$\tilde g_n$ becomes irrelevant. A consequence of this is the presence of the
$L^{-qn}{\bar g}^{2}$ term in the first inequality of Lemma~6.1. It has to
be sufficiently small to ensure the validity of the 
first of three inqualities above.} \textBlack 

\vglue 0.1cm  

\\Next we turn to \equ(6.31). In this case by assumption
$\Vert \y_{n,2}-\y_{n,2}'\Vert\ge \Vert \y_{n,1}-\y_{n,1}'\Vert$ and hence,
since our norms are box norms, we have

$$\Vert \y_n-\y_n'\Vert_n=\Vert \y_{n,2}-\y_{n,2}'\Vert_n
={\bar g}^{-(2-\d)}|\m_n-\m'_n|$$

\\From \equ(6.5)

$$f_{n+1,\m}(\y_n)=L^{3+\e\over 2}\m_n +\tilde\r_n(\y_n)$$

\\Then, using lemma 6.1, we have 

$$\eqalign{{\bar g}^{-(2-\d)}|f_\m(\y_n)-f_\m(\y_n')|\ge
& L^{3+\e\over 2}\Vert \y_n-\y_n'\Vert_{n} -
{\bar g}^{-(2-\d)}|\tilde\r(\y_n)-\tilde\r(\y_n')| \cr 
& \ge(L^{3+\e\over 2}-C_L{\bar g}^{\d})\Vert \y_n-\y_n'\Vert_{n} \cr
& \ge (1+\e)\Vert \y_n-\y_n'\Vert}_{n}$$

\\for $\e$ sufficiently small depending on $L$. This proves \equ(6.31). $\bull$
 
\vglue.3truecm

\\{\it Proof of Theorem~6.4 }

\\Given Lemma~6.5, the proof follows the Schub argument as in [BMS]. Namely,
to prove that $W_{n_0}^{s}$ is given by a graph of a function  
$\y_{n_0,2}=h(y_{n_0,1}) $ it is enough to prove that 
if in $W_{n_0}^{s}$ we take two points $\y_{n_0}=(\y_{n_0,1},\y_{n_0,2})$ and 
$\y_{n_0}'=(\y_{n_0,1}',\y_{n_0,2}')$ then

$$\Vert \y_{n_0,2}-\y_{n_0,2}'\Vert_{n_0}\le
\Vert \y_{n_0,1}-\y_{n_0,1}'\Vert_{n_0}\Eq(6.33)$$

\\because then for a given $\y_{n_0,1}$ we would have at most one
$\y_{n_0,2}$, and by theorem 6.2 there exists such a $\y_{n_0,2}$. This means
that $W_{n_0}^{s}$ is the graph of a function $h$, $\y_{n_0,2}=h(\y_{n_0,1})$,
and moreover 
$$\Vert h(\y_{n_0,1})-h(\y_{n_0,1}')\Vert_{n_0}\le\Vert \y_{n_0,1}-\y_{n_0,1}'\Vert_{n_0}$$

\\Suppose \equ(6.33) is not true. Then

$$\Vert \y_{n_0,2}-\y_{n_0,2}'\Vert_{n_0}>
\Vert \y_{n_0,1}-\y_{n_0,1}'\Vert_{n_0}\Eq(6.33.1)$$

\\\equ(6.33.1) implies that \equ(6.31) holds. The latter followed by 
\equ(6.30) gives

$$\Vert f_{n_0+1,2}(\y_{n_0})-f_{n_0+1,2}(\y_{n_0}')\Vert_{n_0+1}
\ge (1+\e)\Vert \y_{n_0}-\y_{n_0}'\Vert_{n_0}   >
(1-\e)\Vert \y_{n_0}-\y_{n_0}'\Vert_{n_0}$$
$$\ge\Vert f_{n_0+1,1}(\y_{n_0})-f_{n_0+1,1}(\y_{n_0}')\Vert_{n_0+1}
\Eq(6.33.2)  $$

\\and hence

$$\Vert f_{n_0+1}(\y_{n_0})-f_{n_0+1}(\y_{n_0}')\Vert_{n_0+1}= 
\Vert f_{n_0+1,2}(\y_{n_0})-f_{n_0+1,2}(\y_{n_0}')\Vert_{n_0+1}
 \ge (1+\e)\Vert \y_{n_0}-\y_{n_0}'\Vert_{n_0} \Eq(6.33.3) $$

\\Define the composition of maps 

$${\cal P}_{n}^{k}\doteq f_{n+k}\circ .....\circ f_{n+2}\circ f_{n+1}$$ 

\\Now

$$\Vert {\cal P}_{n_0}^{2}(\y_{n_0})-{\cal P}_{n_0}^{2}(\y_{n_0}')\Vert_{n_0+2}
=\Vert f_{n_0+2}(f_{n_0+1}(\y_{n_0}))-f_{n_0+2}(f_{n_0+1}(\y_{n_0}'))
\Vert_{n_0+2} $$

\\By \equ(6.33.2) and the second part of Lemma 6.4 followed by 
\equ(6.33.3) we get

$$\Vert{\cal P}_{n_0}^{2}(\y_{n_0})-{\cal P}_{n_0}^{2} (\y_{n_0}')\Vert_{n_0+2}
\ge(1+\e)\Vert f_{n_0+1}(\y_{n_0})-f_{n_0+1}(\y_{n_0}')\Vert_{n_0+1}
\ge (1+\e)^2\Vert\y_{n_0}-\y_{n_0}'\Vert_{n_0} $$

\\Repeating this $k$ times we get for all $k\ge 0$

$$\Vert{\cal P}_{n_0}^{k}(\y_{n_0})-{\cal P}_{n_0}^{k}(\y_{n_0}')\Vert_{n_0+k}
\ge(1+\e)^k\Vert \y_{n_0}-\y_{n_0}'\Vert_{n_0} \Eq(6.33.4)$$

\\Now $\y_{n_0},\y_{n_0}'$ belong to $W_{n_0}^{s}$ and 
${\cal P}_{n_0}^{k}(\y_{n_0})$ is a member of the sequence 
${\bf s}_{n_0}\in {\bf E}(1/4)$. Therefore 

\\$\Vert{\cal P}_{n_0}^{k}(\y_{n_0})\Vert_{n_0+k}< 1/4 $. Therefore we have 
from \equ(6.33.4) the bound 
${1\over 2} >(1+\e)^k\Vert \y_{n_0}-\y_{n_0}'\Vert_{n_0}$.
By making $k$  arbitrarily large we get a  
contradiction because $\y_{n_0}\ne \y_{n_0}'$ under \equ(6.33.1).
Hence \equ(6.33) is true and the theorem 6.4 has been proved \bull

\vglue0.1cm
\textRed
\\The next theorem establishes the uniqueness of the critical mass at
the unit lattice scale.

\vglue0.1cm
\\{\it Theorem~6.6}: Let $n_0$ and $\m_{n_0}$ be as in Theorem~6.4.
Let $\tilde g_0, \m_0, R_0,{\bf w}_0$ belong to $\tilde{\cal D}_0$, 
defined in the beginning of subsection 6.2. with $R_0=0,{\bf w}_0=0$. Let
$U_1(1)=\{\tilde g_0 :\> 2^{(n_0 +5)}(\n\bar g)^{-1}|\tilde g_0| <1\}$.
\textRed Let $\e$ be sufficiently small depending on $L$ and $n_0$.
Then there exists an open ball $U_{0}\subset U_{1}(1)$ and a $C^{1}$ function
$h_0:\> U_{0}\rightarrow \math{R}$ such that for  
$\m_{0}=h_{0}(\tilde g_0)$ the RG map applied $n_0$ times gives the effective 
critical mass $\m_{n_0}$.

\textRed
\\{\it Remark}:   This is the first time in our estimates that $\e$ has 
been chosen to depend on $n_0$. Recall that in Lemma~6.5 $n_0$ was taken to be
sufficiently large depending on $\n$ and $L$. \textBlack 

\vglue0.2cm

\\{\it Proof}: Let $\y_n$ be as in \equ(6.2). Let
$r_n=2^{-(n_0-n+5)}$ and $\l_n=L^{-{(3+\e)\over 2}(n_0-n)}$. 
Let $\tilde E_n$, $1\le n\le n_0$, be the Banach space consisting of $\y_n$
with norm

$$\Vert \y_n\Vert_{n}=\max (r_n^{-1}(\n\bar g)^{-1}|\tilde g_n|,\> 
r_n^{-1}\l_n^{-1}{\bar g}^{-(2-\d)}|\m_n|,\> 
{\bar g}^{-({11\over 4}-\eta)}|||R_n|||, 
\tilde c_L^{-1}\Vert \tilde{\bf w}_n\Vert_n) \Eq(6.33.6)  $$ 

\\We have $\tilde E_n \subset E_n$. $\tilde E_n(1)$ is the open unit ball
in $\tilde E_n$. Note that $\tilde E_0(1)$ coincides with $\tilde{\cal D}_0$
as defined in the beginning of subsection 6.2, for $R_0=0$ and ${\bf w}_0=0$,
and $\tilde E_{n_0}(1)=E_{n_0}({1\over 32})$.
Then by Theorem~5.1 and Lemma~5.9 for $1\le n\le n_0$ we have each RG map
$f_n:\> \tilde E_n(1)\rightarrow \tilde E_{n+1}(1)$. Moreover each such map
is (norm) analytic. Define the composition of maps 

$${\cal P}_0^{n_0}=f_{n_0}\circ f_{n_0-1}\circ \cdot\cdot f_1 :\>
\tilde E_0(1)\rightarrow \tilde E_{n_0}(1)  \Eq(6.33.7)  $$

\\${\cal P}_0^{n_0}$ is the composition of a finite number of analytic maps
and therefore analytic. We consider the equation
$\y_{n_0}={\cal P}_0^{n_0}(\y_{0})$ in the direction $\m$:

$$\m_{n_0}=({\cal P}_0^{n_0})_{\m}(\y_{0}) \Eq(6.33.8)  $$

\\with $\y_{0}=(\tilde g_0, \m_0, 0, \tilde{\bf w}_0)$ with
${\bf w}_0=0$, ( recall that $\tilde{\bf w}_0={\bf w}_0-{\bf w}_{*}$).

\\We will solve \equ(6.33.6) for $\m_0$ for fixed $\m_{n_0}$ using 
the (Banach space) implicit function theorem. 
Let $x=(\tilde g_{0}, \m_{n_0})$ and $y=\m_0$. We have set ${\bf w}_0=R_0=0$. 
Let $V_1$ be the Banach space of  elements $x$ with norm

$$\Vert x\Vert=\max (r_{0}^{-1}(\n\bar g)^{-1}|\tilde g_0|,\> 
r_{n_0}^{-1}{\bar g}^{-(2-\d)}|\m_{n_0}|)$$ 

\\Let $V_2$ be the Banach space of elements $y$ with norm

$$\Vert y\Vert=r_0^{-1}\l_0^{-1}{\bar g}^{-(2-\d)}|\m_0|$$

\\Let $V_j(r)$ be the open ball in $V_j$ of radius $r$, centered at the
origin. Define

$$F(x,y)=({\cal P}_0^{n_0})_{\m}(\y_{0})-\m_{n_0} \Eq(6.33.9) $$

\\Solving \equ(6.33.8) is equivalent to solving $F(x,y)=0$ for $y$.

\\Recall that $\y_{0}=(\tilde g_0, \m_0, 0, \tilde{\bf w}_0)$. We have
$F(0,0)=0$ and $F(\cdot,\cdot):\> V_{1}(1)\times V_{2}(1)\rightarrow 
\math{R}$ is an analytic map and therefore $C^{2}$. Taking a $y$ derivative
of $F(x,y)$ gives $D_{y}F(x,y)=D_{\m_0}({\cal P}_0^{n_0})_{\m}(\y_{0})$. We
will prove that the linear map 

$$D_{y}F(0,0): V_2\rightarrow \math{R}$$

\\is injective. It is easy to see that

$$\eqalign{({\cal P}_0^{n_0})_{\m}(\y_{0})&=L^{{(3+\e)\over 2}n_0}\Bigl( \m_0 +
L^{-{(3+\e)\over 2}}\sum_{j=0}^{n_0-1} L^{-{(3+\e)\over 2}j}
\tilde \r_j(\y_j) \Bigr)\cr
\tilde\r_j(\y_j) &=\tilde\r_j\circ ({\cal P}_0^{j})_{\m}(\y_{0})\cr} 
\Eq(6.34.1)$$

\\$\tilde\r_j$ was defined earlier in \equ(6.10) and is analytic in $E_j(1)$. 
The map $\tilde\r_j\circ ({\cal P}_0^{j})_{\m}$ is analytic since it is a 
composition of analytic maps. Let $\m_0\in V_2({1\over 4})$. Let $\g$
be the closed contour $\g=\{\m:\> \m-\m_0=Re^{i\theta}\}$ with 
$R={1\over 4}r_0\l_0{\bar g}^{(2-\d)}$. We estimate the $\m$ derivative
of $\tilde\r_j\circ ({\cal P}_0^{j})_{\m}(\y_{0})$ by using the Cauchy integral
formula integrating along the contour $\g$ enclosing a pole at $\m_0$
together with the estimate for $\tilde\r_j(\y_j)$ given in \equ(6.15) which is 
valid in $E_{j}(1)$. The latter is guaranteed by our choice of contour.
We have

$$\Bigl| D_{\m_0}\tilde \r_j\circ ({\cal P}_0^{j})_{\m}(\y_{0})\Bigr|_{\tilde g_0=\m_0=0} \le {c_L {\bar g}^{2}\over R} \le c_{L,n_0}{\bar g}^{\d} $$

\\Using this estimate we get from \equ(6.34.1)

$$D_{\m_0}({\cal P}_0^{n_0})_{\m}(\y_{0})\Bigr|_{\tilde g_0=\m_0=0}
\ge  L^{{(3+\e)\over 2}n_0}(1-c'_{L,n_0}{\bar g}^{\d}) $$

\\Taking $\e$ sufficiently small depending on $L$ and $n_0$ makes  
$\bar g$ sufficiently small so as to ensure
  
\\$D_{\m_0}({\cal P}_0^{n_0})_{\m}(\y_{0})\Bigr|_{\tilde g_0=\m_0=0}
\ge {1\over 2}$. Therefore the map $D_{y}F(0,0): V_2\rightarrow \math{R}$
is injective. Hence by the implicit function theorem there exists a ball
$\tilde U_0$ containing $x=0$ with $\tilde U_0\subset V_1(1)$, and a 
$C^1$ function $\tilde h_0$
in $\tilde U_0$ with $\tilde h_0(x) \in V_2({1\over 2})$ such that 
$F(x,\tilde h_0(x))=0$. For $\e$ sufficiently small depending on $L$ and 
$n_0$ we have $\m_{n_0}\in \tilde U_0$.
This completes the proof of the theorem because for $\m_{n_0}\in \tilde U_0$, 
$\tilde U_0$ restricts to the ball $U_0$ 
and correspondingly $\tilde h_0$  restricts to the desired function $h_0$. 
\bull

\vglue0.3cm

\\Theorem~6.4 and Theorem~6.6 put together completes our construction of the
stable manifold starting from the unit lattice. \textBlack  
 
\vglue0.3cm

\\Finally we remark that as a consequence of theorem~6.4 we have    
$\y_{n}\in E_n(1/4)$, $\forall n\ge n_0$. This implies 
that $|\tilde g_{n}| < {1\over 4}\n {\bar g}$, $\forall n\ge n_0$. By 
construction the same statement is also true for $0\le n\le n_0$.
Whence for all $n\ge 0$

$$(1-{1\over 4}\n) {\bar g}< g_{n} <  (1+{1\over 4}\n) {\bar g} \Eq(6.33.5) $$

\\We have $0<\n< {1\over 2} $. Therefore the effective coupling constant 
generated by the discrete RG flow is uniformly bounded away from $0$ at all 
RG scales.

\vglue.3truecm  

\\{\bf Acknowledgements} : \textRed
We thank an anonymous referee for detecting a 
non-uniqueness in the definition of norms involving fermions which occured
in an earlier version of this paper. We also thank him for his numerous
comments, suggestions and questions  which have helped us to improve the 
paper. \textBlack
One of us (PKM) is especially grateful to
David Brydges for many fruitful conversations during the course of this work.
He thanks G\'erard Menessier for helpful conversations and Erhard Seiler
for his comments on an earlier version of the manuscript.

\vglue.5truecm

\\{\bf References}

\vglue.3truecm
\\[AR] A. Abdesselam: A Complete Renormalization Group Trajectory Between
Two Fixed Points, Commun. Math. Phys (to be published), 
http://arXiv:math-ph/0610018.

\vglue.3truecm
\\[Bal1] Tadeusz Balaban: $({\rm Higgs})_{2,3}$ quantum fields in a 
finite volume. I. A lower bound,
Commun. Math. Phys.
{\bf 85}, 
603--626 (1982).

\vglue.3truecm
\\[Bal2] Tadeusz Balaban: $({\rm Higgs})_{2,3}$ quantum fields in a 
finite volume. II. An upper bound,
Commun. Math. Phys.
{\bf 86}, 
555--594 (1982).

\vglue.3truecm
\\[Bal3] Tadeusz Balaban: 
Renormalization group approach to lattice gauge theories. I.
Generation of effective actions in a small field approximation and a
coupling constant renormalization in four dimensions, 
Commun. Math. Phys.
{\bf 109}, 
249--301 (1987).

\textRed
\vglue.3cm
\\[BKL] J. Bricmont, A. Kupiainen and R. Lefevere:
Renormalizing the renormalization group pathologies,
Physics Reports
{\bf 348} 
5--31 (2001) 
\textBlack

\vglue.3truecm
\\[BDH-est] D. Brydges, J. Dimock and T.R. Hurd: 
Estimates on Renormalization Group Transformation,
Canad. J. Math. 
{\bf 50},
756--793 (1998),  no. 4.

\vglue.3truecm
\\[BDH-eps] D. Brydges, J. Dimock and T.R. Hurd: 
A Non-Gaussian Fixed Point for $\phi^4$ in 4-$\epsilon$ Dimensions,
Commun. Math. Phys.
{\bf 198}, 
111--156 (1998).

\vglue.3truecm  
\\[BGM] D. Brydges, G. Guadagni and P. K. Mitter: Finite range
Decomposition of Gaussian Processes,
J.Statist.Phys.
{\bf 115},
415--449 (2004)

\vglue.3truecm
\\[BEI] David Brydges, Steven N. Evans, John Z. Imbrie: 
Self-Avoiding Walk on a Hierarchical Lattice in Four Dimensions, 
The Annals of Probability,
{\bf 20},
82--124 (1992). 

\vglue.3truecm
\\[BI1]David C. Brydges and John Z. Imbrie:
End-to-End Distance from the Green's 
Function for a Hierarchical 
Self-Avoiding Walk in Four Dimensions, Commun. Math. Phys.
{\bf 239},
523--547 (2003).

\vglue.3truecm
\\[BI2] David C. Brydges and John Z. Imbrie: 
Green's Function for a Hierarchical 
Self-Avoiding Walk in Four Dimensions, Commun. Math. Phys.
{\bf 239},
549--584 (2003). 

\vglue.3truecm
\\[BM] David C. Brydges and   P. K. Mitter:
On the convergence to the continuum of finite range lattice covariances,
(in preparation).

\vglue.3truecm
\\[BMS] D.C. Brydges, P. K. Mitter and B. Scoppola : 
Critical $(\Phi^{4})_{3,\epsilon}$,
Commun. Math. Phys.
{\bf 240},
281--327 (2003). 

\textRed

\vglue.3truecm
\\[BS] David C. Brydges and Gordon Slade : (in preparation).

\textBlack

\vglue.3truecm 
\\[BY] D. Brydges and H.T. Yau: 
Grad $\phi$ Perturbations of Massless Gaussian Fields,
Commun. Math. Phys.
{\bf 129},
351--392 (1990).

\vglue.3truecm
\\[F] W. Feller,
An introduction to probability theory and its applications, Vol.2, 
John Wiley and Sons, Hoboken, N.J.,
1968.

\vglue.3truecm
\\[GK1] K. Gawedzki and A. Kupiainen:
A rigorous block spin approach to massless field theories,
Commun. Math. Phys.
{\bf 77},
31-64 (1980)

\vglue.3truecm
\\[GK2] K. Gawedzki and A. Kupiainen:
A rigorous block spin approach to massless field theories,
Ann.Phys.
{\bf 147},
198 (1980)

\vglue.3truecm
\\[GK3] K. Gawedzki and A. Kupiainen:
Massless $(\phi)^{4}_4$ theory: Rigorous control of a renormalizable
asmptotically free model,
Commun. Math. Phys.
{\bf 99},
197-252 (1985).

\textRed
\vglue.3truecm
\\[G] R. B. Griffiths:
Nonanalytic behaviour above the critical point in a random Ising
ferromagnet,
Phys. Rev. Lett.
{\bf 23},
17-20 (1969).

\vglue.3truecm
\\[GP1] R. B. Griffiths and P. A. Pearce:
Position-space renormalization group transformations: some proofs
and some problems,
Phys. Rev. Lett.
{\bf 41},
917-920 (1978).

\vglue.3truecm
\\[GP2] R. B. Griffiths and P. A. Pearce:
Mathematical properties of position-space renormalization group 
transformations,
J. Stat. Phys.
{\bf 20},
499-545 (1979).

\textBlack

\vglue.3truecm
\\[I] M.C. Irwin:
On the stable manifold theorem,
Bull. London Math. Soc.
{\bf 2},
196-198 (1970).

\vglue.3truecm
\\[KG] A. N. Kolmogoroff and B. V. Gnedenko,
Limit Distributions of sums of independant random variables,
Addison Wesley, Cambridge,Mass
1954.

\vglue.3truecm
\\[Mc] A. J. McKane:
Reformulation of $n\rightarrow 0$ models using supersymmetry,
Phys.Lett.
A {\bf 41},
22--44 (1980)

\vglue.3truecm
\\[PS] G. Parisi and N. Sourlas:
Self-avoiding walk and supersymmetry,
J.Phys.Lett.
{\bf 41},
L403--L406 (1980)

\vglue.3cm
\\[S] M. Shub:
Global Stability of Dynamical Systems,
Springer-Verlag, New York,
1987.

\vglue.3cm
\\[WK] K. G. Wilson and J. Kogut: 
The Renormalization Group and the $\epsilon$ expansion.
Phys.Rep.(Sect C of Phys. Lett.)
{\bf 12},
75-200 (1974).

\ciao